\documentclass{article}

\usepackage{amsmath}
\usepackage{amssymb}
\usepackage{graphicx}
\usepackage{authblk}
\usepackage{hyperref}
\usepackage{setspace}
\usepackage{fullpage}

\title{Non-Vacuum Solutions, Gravitational Collapse and Discrete Singularity Theorems in Wolfram Model Systems}
\author[1, 2]{Jonathan Gorard}
\affil[1]{Cardiff University, Cardiff, UK\footnote{\href{mailto:gorardj@cardiff.ac.uk}{gorardj@cardiff.ac.uk}}}
\affil[2]{University of Cambridge, Cambridge, UK\footnote{\href{mailto:jg865@cantab.ac.uk}{jg865@cantab.ac.uk}}}

\begin{document}

\maketitle

\begin{abstract}
The celebrated geodesic congruence equation of Raychaudhuri, together with the resulting singularity theorems of Penrose and Hawking that it enabled, yield a highly general set of conditions under which a spacetime (or, more generically, a pseudo-Riemannian manifold) is expected to become geodesically incomplete. However, the proofs of these theorems traditionally depend upon a collection of assumptions about the continuum spacetime (and, in the physical case, the stress-energy distribution defined over it), including its global structure, its energy conditions, the existence of trapped null surfaces and the various volume/intersection properties of geodesic congruences, that are inherently difficult to translate to the case of discrete spacetime formalisms, such as causal set theory or the Wolfram model. Some, such as the discrete analog of Raychaudhuri's equation for the volumes of geodesic congruences, are subtle to formulate due to intrinsic differences in the behavior of discrete vs. continuous geodesics; for others, such as the definition of a trapped null surface, no appropriate translation is known for the general discrete case due to a lack of a priori coordinate information. It is therefore a non-trivial question to ask whether (and to what extent) there exist equivalently general conditions under which one expects discrete spacetimes to become geodesically incomplete, and how these conditions might differ from those in the continuum. This article builds upon previous work, in which the conformal and covariant Z4 (CCZ4) formulation of the Cauchy problem for the Einstein field equations, with constraint-violation damping, was defined in terms of Wolfram model evolution over discrete (spatial) hypergraphs for the case of \textit{vacuum} spacetimes, and proceeds to consider a minimal extension to the \textit{non-vacuum} case by introducing a massive scalar field distribution, defined in either spherical or axial symmetry. Under appropriate assumptions, this scalar field distribution admits a physical interpretation as a collapsing (and, in the axially-symmetric case, uniformly rotating) dust, and we are able to show, through a combination of rigorous mathematical analysis and explicit numerical simulation, that the resulting discrete spacetimes converge asymptotically to either non-rotating Schwarzschild black hole solutions or maximally-rotating (extremal) Kerr black hole solutions, respectively. Although the assumptions used in obtaining these preliminary results are very strong, they nevertheless offer hope that a more general, perhaps ultimately ``Penrose-like'', singularity theorem may be provable in the discrete spacetime case too.
\end{abstract}

\section{Introduction}

Singularities have been discussed as an essential feature of general relativity more-or-less since its inception: the first non-trivial solution to the Einstein field equations (discovered independently by Schwarzschild\cite{schwarzschild} and Droste\cite{droste} within a year of Einstein's original publication of the theory in 1915), namely the Schwarzschild metric for a spherically-symmetric matter distribution of mass $M$, was identified by Hilbert\cite{hilbert} as containing singularities (i.e. points at which certain components of the metric tensor would become divergent) at the coordinate values ${r = 0}$ and ${r = 2M}$. The singularity at ${r = 2M}$ can easily be shown to be a byproduct of the choice of coordinate system, and hence non-physical, as first pointed out by Lema\^itre\cite{lemaitre}. Specifically, although it is present in the case of Schwarzschild's spherical coordinates, there exist many alternative coordinatizations of the Schwarzschild geometry (such as Eddington-Finkelstein coordinates or Gullstrand-Painlev\'e coordinates) for which the solution is perfectly regular at ${r = 2M}$ or its equivalent. However, the singularity at ${r = 0}$ appears somehow to be more fundamental: at this point, the Kretschmann scalar ${K = R_{\mu \nu \rho \sigma} R^{\mu \nu \rho \sigma}}$ also diverges, and since $K$ is a quadratic scalar invariant of the Riemann curvature tensor (and thus invariant under arbitrary diffeomorphism transformations of spacetime), the singularity at ${r = 0}$ cannot be merely a coordinate artifact, and must instead be considered an intrinsic feature of the geometry that is independent of coordinates. In 1939, Oppenheimer and Snyder\cite{oppenheimer} considered the case of an idealized, spherically-symmetric star undergoing continual gravitational collapse, showing that, to an outside observer in Schwarzschild coordinates, the stellar radius would appear to approach the coordinate singularity at ${r = 2M}$ asymptotically, although for an infalling observer in, say, Eddington-Finkelstein coordinates, the stellar radius would appear to cross the coordinate horizon at ${r = 2M}$ within finite proper time and thus continue collapsing indefinitely. Hence, at least within the Oppenheimer-Snyder model of gravitational collapse, the singularity at ${r = 0}$ seems to be physically realized. Nor are such singularities purely a byproduct of spherically-symmetric spacetimes: the axially-symmetric Kerr geometry (as well as the more general Kerr-Newman class of electrovacuum solutions to the Einstein-Maxwell equations) exhibits a spatially-extended, ``ring-like'' singularity, as well as a pair of Schwarzschild-like coordinate singularities, at least in the oblate spheroidal coordinates of Boyer and Lindquist\cite{boyer}.

However, one might still have argued, perhaps perfectly reasonably, that the apparent existence of such ``physical'' singularities was nevertheless a consequence of the unphysically high degree of symmetry that these solutions possessed (either spherical in the case of Schwarzschild or axial in the case of Kerr), and that the singularity structure would not remain stable under perturbations away from axial symmetry. In 1964, Penrose\cite{penrose} showed that this was not, in fact, the case. To begin, Penrose proposed the first mathematically rigorous definition of a generic spacetime singularity: a spacetime ${\mathcal{M}}$ contains a singularity if it is not \textit{geodesically complete}, meaning that there exist timelike or null geodesics ${\gamma}$ in ${\mathcal{M}}$ that cannot be extended indefinitely either into the future, or into the past, or both (i.e. there exists a limit to how far one can smoothly continue the domain of definition of the time parameter ${\tau}$ of ${\gamma}$, for the case of timelike geodesics, or the affine parameter ${\lambda}$ of ${\gamma}$, for the case of null geodesics). Next, Penrose assumed a matter distribution exhibiting a sufficiently strong gravitational field that a \textit{trapped null/timelike surface} (i.e. a compact, spacelike surface at which all outward-pointing null/timelike geodesics become convergent) is induced, but with no a priori assumptions regarding its symmetry. Penrose proceeded to prove that, so long as the spacetime is globally hyperbolic (and thus admits a foliation into non-overlapping spacelike hypersurfaces) and the matter distribution obeys either the \textit{null energy condition} ${T_{\mu \nu} X^{\mu} X^{\nu} \geq 0}$ (for the case of null vector fields ${\mathbf{X}}$) or the \textit{strong energy condition} ${\left( T_{\mu \nu} - \frac{1}{2} T g_{\mu \nu} \right) X^{\mu} X^{\nu} \geq 0}$ (for the case of timelike vector fields ${\mathbf{X}}$), the resulting spacetime is necessarily geodesically incomplete. Such energy conditions would naturally be satisfied by any perfect fluid distribution with non-negative density and pressure, as in the case of an idealized fluid model for a collapsing star. Penrose's original singularity theorem principally concerned the case of \textit{future} geodesic incompleteness relevant for gravitational collapse models; Hawking\cite{hawking} subsequently considered the time-reversed case of \textit{past} geodesic incompleteness relevant for the cosmology of expanding universes (in which the \textit{dominant energy condition} must assumed - for a perfect fluid, this corresponds to the density being at least as large as the absolute value of the pressure). Many other singularity theorems, concerning different energy conditions or involving different assumptions on the global structure of the spacetime, are now known.

Penrose's proof is remarkably simple, in large part because it builds so heavily upon the earlier work of Raychaudhuri\cite{raychaudhuri} (much of it independently discovered by Landau\cite{hawking2}) concerning the behavior of \textit{geodesic congruences} (i.e. families of geodesics sharing the same affine parameter ${\lambda}$ or time parameter ${\tau}$) in spacetime. Raychaudhuri's equation relates the divergence of a geodesic congruence ${\mathbf{X}}$ to (amongst other things) the so-called \textit{Raychaudhuri scalar} ${R_{\mu \nu} X^{\mu} X^{\nu}}$ (otherwise known as the trace of the \textit{tidal tensor}, or the trace of the \textit{electrogravitic tensor} in the context of the \textit{Bel decomposition} of the Riemann curvature tensor\cite{matte}\cite{bel}). The essence of the proof is then as follows (stated here for the case of null geodesic congruences, but with a straightforward extension to timelike ones). The Raychaudhuri equation implies that if one starts with a congruence of (initially parallel) null geodesics emanating from some spacetime region, and that congruence starts to \textit{converge} (i.e. the volume of the congruence begins to decrease), then it will continue to converge for as long as the Raychaudhuri scalar is non-negative. By the Einstein field equations, the non-negativity of the Raychaudhuri scalar ${R_{\mu \nu} X^{\mu} X^{\nu} \geq 0}$ is equivalent to the satisfaction of the null energy condition ${T_{\mu \nu} X^{\mu} X^{\nu} \geq 0}$ on the stress-energy tensor, which holds by assumption. Moreover, the initial convergence of the null geodesics is guaranteed by the assumed presence of the trapped null surface. Taken together, these imply that the null geodesic congruence will ``collapse'' to have zero volume within some finite value of the affine parameter ${\lambda}$: a crucial lemma known as the \textit{focusing theorem}. This ``collapse'' of the null geodesic congruence implies that all neighboring null geodesics must intersect with one another in some way. However, if two null geodesics intersect, then they cannot lie on the boundary of the proper future of the initial spacetime region (indeed, one way to \textit{define} the boundary of the proper future is that it is the collection of all null geodesics in the congruence that do not intersect). Thus, if all null geodesics in the congruence intersect with at least one other null geodesic, then the initial spacetime region must have no proper future boundary. Hence, the spacetime is (null) geodesically incomplete, which completes the proof. We see immediately that the Penrose singularity theorem is thus a kind of ``physical analog'' of the Bonnet-Myers theorem from Riemannian geometry\cite{myers}, by which any Riemannian manifold whose Ricci curvature is bounded below by a positive constant must either be compact or geodesically incomplete.

In the case of discrete quantum gravity theories such as causal set theory\cite{bombelli}\cite{bombelli2}\cite{sorkin}\cite{sorkin2} or the Wolfram model\cite{wolfram}\cite{wolfram2}\cite{gorard}\cite{gorard2}, in which the underlying structure of spacetime is given in terms of some fundamentally combinatorial data structure (such as a partially ordered set, a directed acyclic graph or a time-ordered sequence of hypergraphs), the situation is somewhat more complicated. Although in many instances agreement with standard (continuum) general relativity is established in cases where an appropriate continuum limit exists (for instance, via the Benincasa-Dowker action from discrete d'Alembertians in causal set theory\cite{benincasa}, or the discrete Einstein-Hilbert action from Ollivier-Ricci curvature in Wolfram model systems\cite{gorard}\cite{gorard3}\cite{gorard4}), far less is known about the relativistic properties of such theories at the sub-continuum scale, including the validity of the singularity theorems. Indeed, it is far from obvious that singularities are even a well-defined concept in a (finite) discrete spacetime model, since if the number of gravitational degrees of freedom is necessarily finite, then all components and projections of discrete metric and curvature tensors will also be finite. Even the supposedly idealized cases that do not require full singularity theorems in the continuum, such as the spherically-symmetric Oppenheimer-Snyder stellar collapse model, do not admit straightforward translations to the discrete spacetime case, due to the difficulties of enforcing the requisite symmetries. It is possible to construct discrete spacetimes that are compatible with continuous symmetry groups in the continuum limit (for instance, causal sets produced via \textit{Poisson sprinkling} into a ${3 + 1}$-dimensional Minkowski spacetime are provably compatible with the action of the restricted Lorentz group ${SO^{+} \left( 1, 3 \right)}$, in the limit of infinite sprinkling density\cite{bombelli3}), but these symmetries do not hold generically when the process is truncated after a finite number of steps. Thus, even if one constructs the Cauchy initial data for the Einstein field equations by sprinkling into a Riemannian manifold exhibiting the requisite spherical symmetry, the inherent randomness of the sprinkling process will have the effect of perturbing the initial data away from perfect spherical symmetry, hence making the methods of Oppenheimer and Snyder fundamentally inapplicable. It is therefore an extremely important, yet highly non-trivial, question to ask whether a result akin to the Penrose singularity theorem can be proven in the discrete case, and what kinds of assumptions (analogous to energy conditions and global causality conditions in the continuum case) might be necessary for such a proof to be possible. Though we do not claim to be able to answer this question with any generality as of yet, this is nevertheless the question with which the present article is concerned.

The Wolfram model is a discrete spacetime formalism in which Cauchy initial data is specified in the form of a \textit{spatial hypergraph} (i.e. the discrete analog of a spacelike hypersurface), with dynamics determined by \textit{hypergraph rewriting rules}. The causal interactions between these hypergraph rewrites generate a partially-ordered set, which is typically represented as a directed acyclic graph known as a \textit{causal graph}; subject to certain assumptions (such as \textit{causal invariance} and \textit{asymptotic dimension preservation}), the combinatorial structure of this causal graph is known to converge to the conformal structure of a Lorentzian manifold obeying the Einstein field equations in the continuum limit\cite{gorard}\cite{gorard4}. Causal set theory is a deeply related discrete spacetime model in which spacetime structure is also represented as a partially-ordered set; indeed, Wolfram model evolution may be interpreted as endowing causal set theory with an explicit algorithmic dynamics\cite{bolognesi}\cite{bolognesi2}\cite{gorard3} (since the transitive reduction of a causal graph generated via Wolfram model evolution corresponds to the Hasse diagram of some corresponding causal set). The problem of defining singularities within such discrete spacetime models turns out to be surprisingly straightforward, since Penrose's notion of (null) geodesic incompleteness may be imported more-or-less ``wholesale'': spacetime geodesics become directed paths in the causal graph, with the question of geodesic completeness being the question of whether such paths can always be extended, or, more concretely, the question of future/past geodesic incompleteness is equivalent to the question of whether there exist vertices within some subgraph of the causal graph (corresponding to the geodesic congruence) whose out/in-components are empty. However, appropriate discrete translations for much of the remaining mathematical apparatus of the Penrose singularity theorem remain elusive, or at the very least obscure. For instance, geodesic ``collisions'' are far more common in discrete spacetimes than in continuous ones, even in cases where there is no net convergence of the geodesic congruence, for the simple reason that there are fewer degrees of freedom (i.e. there are fewer ``spaces'' for the geodesic to occupy within a discrete spacetime, so pairs of geodesics are inherently much more likely to occupy the same ``space'' by ``accident''). This makes the intersection and volume properties of geodesic congruences in discrete spacetimes noticeably and qualitatively different than in continuous ones, rendering the appropriate discrete analog of the Raychaudhuri equation somewhat non-obvious. Although considerable attention has been paid to the structure of discrete vacuum spacetimes (especially in the Wolfram model case), far less is known about non-vacuum solutions, and there has as of yet been no systematic investigation of discrete analogs of standard general relativistic energy conditions. Finally, the lack of any a priori inner product structure (although such structures can be defined, albeit not uniquely) on causal graphs makes Penrose's original definition of trapped null surfaces hard to utilize directly, though equivalent descriptions in terms of cross-sectional areas of null congruences may be more amenable to immediate discretization.

The main result presented within this article is that, for a massive scalar field ``bubble collapse'' problem obeying one of two approximate spatial symmetries, when formulated and discretized as a Wolfram model evolution problem, the resulting discrete spacetime converges to one of two standard black hole spacetimes (namely either Schwarzschild or extremal Kerr). Specifically, if the initial scalar field distribution is approximately spherically-symmetric, then the resulting discrete spacetime is asymptotically Schwarzschild, whereas if the distribution is instead approximately axially-symmetric, then the resulting discrete spacetime is asymptotically extremal Kerr (reducing to Schwarzschild in degenerate cases). The significance of the word ``approximately'' in the above is in reference to the symmetry discretization problem referenced previously; although the Cauchy initial data is exactly spherically/axially-symmetric analytically, the discretization procedure inevitably has the effect of perturbing the initial hypersurface away from exact spatial symmetry, and so one must show that the resulting convergence remains stable under perturbations of this general form. We show this stable convergence property using both rigorous mathematical analysis and explicit numerical simulation (by means of a newly-developed, high-resolution, hypergraph-based numerical relativity code called \href{https://github.com/JonathanGorard/Gravitas}{\textsc{Gravitas}}, featuring totally unstructured adaptive mesh refinement), and demonstrate the expected agreement between the analytical and numerical results. Our rationale for choosing a massive scalar field as the underlying stress-energy model (rather than the dust or perfect fluid models commonly used in relativistic astrophysics) is to enable more direct comparison with pure Wolfram model evolution: in recent work\cite{gorard5}, we showed how a massless scalar field theory (obeying the discrete Klein-Gordon equation) could be defined over an arbitrary Wolfram model system, building upon the previous work of Dowker and Glaser\cite{dowker}, Sorkin\cite{sorkin3} and Johnston\cite{johnston} in the context of causal set theory. Following an ansatz proposed by Dowker et al.\cite{dowker2}, as well as a more direct approach outlined by Johnston\cite{johnston2}\cite{johnston3}, the discrete massless Green's functions can naturally be extended to the massive case. To the best of our knowledge, no comparable proposal has yet been made for equipping arbitrary causal sets/Wolfram model evolutions with matter fields consistent with either the dust or perfect fluid forms of the continuum stress-energy tensor. However, as we shall show through the course of this article, the \textit{WKB approximation} of Wentzel\cite{wentzel}, Kramers\cite{kramers} and Brillouin\cite{brillouin} may nevertheless be applied to find a range of parameter values in which a collapsing massive scalar field bubble may be interpreted as a non-rotating, collapsing ball of dust (described by the Lema\^itre-Tolman-Bondi metric\cite{lemaitre}\cite{tolman}\cite{bondi}) in the spherically-symmetric case, or a spinning, collapsing disk of dust (described by the Weyl-Lewis-Papapetrou metric\cite{weyl}\cite{lewis}\cite{papapetrou}) in the axially-symmetric case.

In Section \ref{sec:Section1}, we begin by introducing the fully covariant and conformally-invariant Z4 formulation (CCZ4) of the Einstein field equations due to Alic et al. \cite{alic} and Bona et al.\cite{bona} used by \href{https://github.com/JonathanGorard/Gravitas}{\textsc{Gravitas}} in the formulation of the Cauchy problem for general relativity (along with the relevant gauge conditions, adapted for the spherically-symmetric and axially-symmetric spacetimes simulated in this article). We also outline the various numerical algorithms used in the discrete evolution of these equations, including the fourth-order Runge-Kutta scheme for the time evolution (with appropriate modification of the finite-difference stencils to make them suitable for general hypergraphs with totally unstructured topology), the generalized local adaptive mesh refinement (AMR) algorithm, based on the approach of Berger and Colella\cite{berger}, for coarsening and refining the hypergraph topology, and the higher-order weighted essentially non-oscillatory (WENO) scheme for extrapolating boundary values. In Section \ref{sec:Section2}, we follow the approach of Gon\c{c}alves and Moss\cite{goncalves} to show that the WKB approximation may be used to treat a minimally-coupled massive scalar field in spherical symmetry as a non-rotating inhomogeneous dust described by the Lema\^itre-Tolman-Bondi metric, at least in the limit as the mass of the field goes to infinity, and thus we rigorously derive a sufficient condition for the spherically-symmetric massive scalar field ``bubble collapse'' problem to yield the idealized stellar collapse solution of Oppenheimer and Snyder. This condition can be derived analytically for a ``top hat'' initial density distribution of the scalar field, and numerically for the (more physical) exponential initial density distribution. We also present the numerical solutions to the massive scalar field bubble collapse problem in spherical symmetry, showing agreement with the analytic predictions, within this section. However, the collapse of an axially-symmetric massive scalar field to an extremal Kerr black hole is inherently more complicated, since the exterior solution for a spinning disk of dust is \textit{not} extremal Kerr, but rather the Weyl-Lewis-Papapetrou metric (in contrast to the spherically-symmetric case, in which the exterior solutions for both the collapsing dust and the resulting non-rotating black hole are described by the Schwarzschild geometry, by virtue of Birkhoff's theorem), so one must attempt to construct some kind of smooth transition between the two distinct geometries. In Section \ref{sec:Section3}, we follow the methods of Neugebauer and Meinel\cite{neugebauer}\cite{neugebauer2}\cite{neugebauer3} (based on an earlier conjecture of Bardeen and Wagoner\cite{bardeen}\cite{bardeen2}) to prove that the collapse of a minimally-coupled massive scalar field bubble in axial symmetry (and thus, by the WKB approximation, the collapse of a uniformly-spinning disk of dust) may be formulated as a boundary-value problem for the Ernst equation, whose solution can be obtained as a special case of the Jacobi inversion problem for Abelian/hyperelliptic integrals. The full solution may be given in terms of so-called \textit{ultraelliptic functions} (i.e. hyperelliptic functions of two variables) and \textit{ultraelliptic integrals}, and ultimately depends upon both the the angular velocity of the disk and the relative redshift from the disk's center, with the \textit{Maclaurin disk} solution (i.e. a gaseous astrophysical disk under Newtonian gravity) and the extremal Kerr solution being obtained in the appropriate ``Newtonian'' and ``exterior'' limits, respectively. The numerical solutions to the massive scalar field bubble collapse problem in axial symmetry, illustrating agreement with the analytic solutions, are also presented within this section. In Section \ref{sec:Section4}, we outline how the approach of Johnston may be used to equip an arbitrary causal graph produced by Wolfram model evolution with the Green's function for a massive scalar field, thus allowing us to make a direct comparison between the numerical results obtained within the preceding sections and those produced via ``pure'' Wolfram model evolution (without any underlying PDE system), using a hypergraph rewriting rule that provably satisfies the Einstein field equations in the continuum limit. Finally, in Section \ref{sec:Section5}, some potential directions for future research are proposed, including the extension to more complex matter/stress-energy models involving more sophisticated equations of state, the extension to more general classes of perturbations away from spherical or axial symmetry, and the more speculative possibility of extending the kinds of global topological techniques of Penrose in order to prove a fully generalized singularity theorem for arbitrary discrete spacetimes.

Note that all of the code necessary to reproduce all of the results presented within this article is fully open source and freely available as part of the \href{https://github.com/JonathanGorard/Gravitas}{\textsc{Gravitas} package} on GitHub. In particular, \textsc{Gravitas} features in-built functionality for performing the canonical ${3 + 1}$ metric decomposition and computing the corresponding evolution equations, constraint equations, gauge satisfaction equations, etc. (e.g. through \href{https://github.com/JonathanGorard/Gravitas/blob/main/ADMDecomposition.wl}{\textit{ADMDecomposition}}), discretizing the resulting evolution using adaptive hypergraph methods (e.g. through \href{https://github.com/JonathanGorard/Gravitas/blob/main/DiscreteHypersurfaceDecomposition.wl}{\textit{DiscreteHypersurfaceDecomposition}}), coupling arbitrary spacetimes to arbitrary matter/stress-energy distributions (e.g. through \href{https://github.com/JonathanGorard/Gravitas/blob/main/StressEnergyTensor.wl}{\textit{StressEnergyTensor}}) and solving the Einstein field equations, both numerically and analytically (e.g. through \href{https://github.com/JonathanGorard/Gravitas/blob/main/SolveEinsteinEquations.wl}{\textit{SolveEinsteinEquations}} and \href{https://github.com/JonathanGorard/Gravitas/blob/main/SolveVacuumEinsteinEquations.wl}{\textit{SolveVacuumEinsteinEquations}}). \textsc{Gravitas} also contains an extensive library of in-built metrics, geometries, gauge choices and coordinate systems. Several functions are also fully-documented and exposed through the \textit{Wolfram Function Repository}, including \href{https://resources.wolframcloud.com/FunctionRepository/resources/MetricTensor/}{\textit{MetricTensor}} for representing arbitrary spacetime metrics, \href{https://resources.wolframcloud.com/FunctionRepository/resources/RicciTensor}{\textit{RicciTensor}} for computing Ricci curvature tensors (and their projections) for arbitrary spacetimes, \href{https://resources.wolframcloud.com/FunctionRepository/resources/MultiwaySystem}{\textit{MultiwaySystem}} for evolving arbitrary Wolfram model (multiway) systems and computing their corresponding causal graphs, etc. Note also that, throughout this article, we adopt the general convention that Greek indices (i.e. ${\mu}$, ${\nu}$, ${\rho}$, ${\sigma}$, etc.) correspond to spacetime coordinates, while Latin indices (i.e. $i$, $j$, $k$, $l$, etc.) correspond to spatial coordinates. Einstein summation convention is assumed throughout, unless otherwise specified.

\section{Governing Equations, Discretization Scheme and Numerics}
\label{sec:Section1}

The starting point for our numerical implementation is the canonical ${3 + 1}$ decomposition of the spacetime metric tensor ${g_{\mu \nu}}$ into the standard ADM line element of Arnowitt, Deser and Misner\cite{arnowitt}:

\begin{equation}
d s^2 = - \alpha^2 d t^2 + \gamma_{i j} \left( d x^{i} + \beta^{i} dt \right) \left( d x^{j} + \beta^{j} dt \right),
\end{equation}
where the lapse function ${\alpha}$ and the shift vector ${\boldsymbol\beta}$ are the ADM gauge variables (defining a foliation of a globally hyperbolic spacetime into non-overlapping spacelike hypersurfaces), and ${\gamma_{i j}}$ denotes the induced spatial metric tensor on the spacelike hypersurfaces of constant time. Within this foliation, the future-pointing (null) unit normal vector ${\mathbf{n}}$ to the spacelike hypersurfaces is given in terms of the contravariant derivative ${\nabla^{\mu}}$ of the time coordinate $t$:

\begin{equation}
n^{\mu} = - \alpha \nabla^{\mu} t,
\end{equation}
with the corresponding \textit{time vector} ${\mathbf{t}}$ given by:

\begin{equation}
t^{\mu} = \alpha n^{\mu} + \beta^{\mu}.
\end{equation}
This decomposition allows us to formulate the Einstein field equations as a Cauchy initial-value problem on spacelike hypersurfaces, defined by a system of 12 independent hyperbolic partial differential equations: 6 for the spatial metric tensor ${\gamma_{i j}}$:

\begin{equation}
\partial_t \gamma_{i j} = - 2 \alpha K_{i j} + \nabla_{i} \beta_{j} + \nabla_{j} \beta_{i},
\end{equation}
and 6 for the extrinsic curvature tensor on spacelike hypersurfaces ${K_{i j}}$:

\begin{equation}
\partial_t K_{i j} = \alpha \left( R_{i j}^{\left( 3 \right)} - 2 K_{i k} K_{j}^{k} + K_{i j} K \right) - \nabla_{i} \nabla_{j} \alpha - 8 \pi \left[ S_{i j} - \frac{1}{2} \gamma_{i j} \left( S - \tau \right) \right] + \beta^{k} \partial_{k} K_{i j} + K_{i k} \partial_{j} \beta^{k} + K_{k j} \partial_{i} \beta^{k}.
\end{equation}
In the above, ${K_{i j}}$, i.e. the extrinsic curvature tensor, may be defined abstractly in terms of the Lie derivative of the spatial metric tensor ${\gamma_{i j}}$ along the normal vector ${\mathbf{n}}$:

\begin{equation}
K_{i j} = - \frac{1}{2} \mathcal{L}_{\mathbf{n}} \gamma_{i j} = - \gamma_{i}^{k} \gamma_{j}^{l} \nabla_{k}^{\left( 4 \right)} n_{l} = - \nabla_{j}^{\left( 3 \right)} n_{i},
\end{equation}
$K$ denotes its trace (i.e. the mean curvature):

\begin{equation}
K = \gamma^{i j} K_{i j},
\end{equation}
and ${S_{i j}}$, $S$ and ${\tau}$ are simple functions of the stress-energy tensor ${T_{\mu \nu}}$:

\begin{equation}
\tau = T_{\mu \nu} n^{\mu} n^{\nu}, \qquad S_{i j} = T_{\mu \nu} \gamma_{i}^{\mu} \gamma_{j}^{\nu}, \qquad S = \gamma^{i j} S_{i j}.
\end{equation}
The resulting system is closed by imposing appropriate conditions on the ADM Hamiltonian/energy constraint $H$ and the three ADM momentum constraints ${M_i}$, given by:

\begin{equation}
H = R^{\left( 3 \right)} - K_{i j} K^{i j} + K^2 = 0,
\end{equation}
and:

\begin{equation}
M_i = \gamma^{j l} \left( \partial_l K_{i j} - \partial_i K_{j l} - \Gamma_{j l}^{m} K_{m i} + \Gamma_{j i}^{m} K_{m l} \right) = 0,
\end{equation}
respectively. Here, as elsewhere, the bracketed integers are used to distinguish between spatial vs. spacetime variants of expressions and operations: ${\nabla^{\left( 3 \right)}}$ designates a covariant derivative over a spacelike hypersurface, ${\nabla^{\left( 4 \right)}}$ designates a covariant derivative over the entire spacetime, ${R^{\left( 3 \right)}}$ designates the Ricci scalar curvature restricted to a spacelike hypersurface, etc. Unbracketed expressions and operations are assumed to refer to their spacetime variants unless otherwise specified.

As first proposed by Bona, Ledvinka, Palenzuela and \u{Z}\'a\u{c}ek\cite{bona2}, one can construct an explicitly covariant formulation of the Einstein field equations with Lagrangian density\cite{bona3}:

\begin{equation}
\mathcal{L} = \sqrt{g} g^{\mu \nu} \left[ R_{\mu \nu} + 2 \nabla_{\mu} Z_{\nu} \right],
\end{equation}
for some 4-vector ${\mathbf{Z}}$, such that ADM Hamiltonian and momentum constraints reduce to the trivial algebraic condition ${Z_{\mu} = 0}$, and the Einstein field equations themselves take the form:

\begin{equation}
R_{\mu \nu} + \nabla_{\mu} Z_{\nu} + \nabla_{\nu} Z_{\mu} = 8 \pi \left( T_{\mu \nu} - \frac{1}{2} T g_{\mu \nu} \right).
\end{equation}
This is commonly known as the \textit{(fully covariant) Z4 formulation}. The components of the 4-vector ${Z_{\mu}}$ may therefore be thought of as quantifying the degree to which the ADM constraints are violated; as shown by Gundlach, Mart\'in-Garc\'ia, Calabrese and Hinder\cite{gundlach}, one can introduce damping terms into the covariant Z4 equations so as to formulate the evolution equations as a so-called ``${\lambda}$-system'' based on the techniques of Brodbeck, Frittelli, H\"ubner and Reula\cite{brodbeck}. More specifically, if one replaces the Z4 evolution equations with the damped form:

\begin{equation}
R_{\mu \nu} + \nabla_{\mu} Z_{\nu} + \nabla_{\nu} Z_{\mu} - \kappa_1 \left[ t_{\mu} Z_{\nu} + t_{\nu} Z_{\mu} - \left( 1 + \kappa_2 \right) g_{\mu \nu} t^{\sigma} Z_{\sigma} \right] = 8 \pi \left( T_{\mu \nu} - \frac{1}{2} T g_{\mu \nu} \right),
\end{equation}
where ${\kappa_1 \geq 0}$ and ${\kappa_2}$ are real constants, and ${t^{\mu}}$ is any non-vanishing null vector field (this effectively translates to a weak constraint on the permitted choices of ADM gauge conditions), then the surface in phase space on which the constraint equations ${Z_{\mu} = 0}$ are satisfied will provably form a set of attractor solutions for the corresponding ${\lambda}$-system, thus guaranteeing symmetric hyperbolicity of the (constrained) evolution equations. The explicit time evolution equations for the (damped) Z4 system, in terms of the canonical ${3 + 1}$/ADM decomposition, can now be expressed reasonably succinctly via Lie derivatives along the shift vector ${\boldsymbol\beta}$. The 6 evolution equations for the spatial metric tensor ${\gamma_{i j}}$ remain unchanged:

\begin{equation}
\left( \partial_t - \mathcal{L}_{\boldsymbol\beta} \right) \gamma_{i j} = - 2 \alpha K_{i j},
\end{equation}
although the 6 evolution equations for the extrinsic curvature tensor ${K_{i j}}$ now inherit a non-trivial dependency on the ${\mathbf{Z}}$ 4-vector (as well as the damping terms ${\kappa_1}$ and ${\kappa_2}$):

\begin{multline}
\left( \partial_t \mathcal{L}_{\boldsymbol\beta} \right) K_{i j} = - \nabla_i \alpha_j + \alpha \left[ R_{i j} + \nabla_i Z_j + \nabla_j Z_i - 2 K_{i}^{l} K_{l j} + \left( K - 2 \Theta \right) K_{i j} - \kappa_1 \left( 1 + \kappa_2 \right) \Theta \gamma_{i j} \right]\\
- 8 \pi \alpha \left[ S_{i j} - \frac{1}{2} \left( S - \tau \right) \gamma_{i j} \right],
\end{multline}
where ${\Theta}$ is simply the projection of the ${\mathbf{Z}}$ 4-vector in the normal direction ${\mathbf{n}}$:

\begin{equation}
\Theta = n_{\mu} Z^{\mu} = \alpha Z^0.
\end{equation}
This projection itself is subject to an evolution equation:

\begin{equation}
\left( \partial_t - \mathcal{L}_{\boldsymbol\beta} \right) \Theta = \frac{\alpha}{2} \left[ R + 2 \nabla_j Z^j + \left( K - 2 \Theta \right) K - K^{i j} K_{i j} - 2 \frac{Z_j \alpha_j}{\alpha} - 2 \kappa_1 \left( 2 + \kappa_2 \right) \Theta - 16 \pi \tau \right],
\end{equation}
as indeed is the overall ${\mathbf{Z}}$ 4-vector:

\begin{equation}
\left( \partial_t - \mathcal{L}_{\boldsymbol\beta} \right) Z_i = \alpha \left[ \nabla_j \left( K_{i}^{j} - \delta_{i}^{j} K \right) + \partial_i \Theta - 2 K_{i}^{j} Z_j - \Theta \frac{\alpha_i}{\alpha} - \kappa_1 Z_i - 8 \pi S_i \right],
\end{equation}
which together close the system. Here, the ${\alpha_i}$ terms correspond to the constants ${\alpha_i \neq 0}$ that appear in the general definition of the ${\lambda}$-system for constraint quantities in the Einstein field equations:

\begin{equation}
\frac{\partial \lambda}{\partial t} = \alpha_0 \mathcal{C} - \beta_0 \lambda, \qquad \frac{\partial \lambda^i}{\partial t} = \alpha_1 \mathcal{C}^i - \beta_1 \lambda^i,
\end{equation}

\begin{equation}
\frac{\partial \lambda_{k}^{i j}}{\partial t} = \alpha_3 \mathcal{C}_{k}^{i j} - \beta_3 \lambda_{k}^{i j}, \qquad \frac{\partial \lambda_{k l}^{i j}}{\partial t} = \alpha_4 \mathcal{C}_{k l}^{i j} - \beta_4 \lambda_{k l}^{i j},
\end{equation}
where the ${\beta_i > 0}$ are also constants, ${\mathcal{C}}$, ${\mathcal{C}^i}$, ${\mathcal{C}_{k}^{i j}}$ and ${\mathcal{C}_{k l}^{i j}}$ are scalar, vector, rank-3 tensor and rank-4 tensor-valued constraint quantities, respectively, and ${\lambda}$, ${\lambda^i}$, ${\lambda_{k}^{i j}}$ and ${\lambda_{k l}^{i j}}$ are corresponding tensorial quantities with the same symmetries. Note also that the functions of the stress-energy tensor ${S_{i j}}$, ${S_i}$ and ${\tau}$ are subtly different for the covariant Z4 system from their counterparts in conventional ADM (indeed, one of them has been now been promoted from a scalar to a covector):

\begin{equation}
S_{i j} = T_{i j}, \qquad S_i = n_{\nu} T_{i}^{\nu}, \qquad \tau = n_{\mu} n_{\nu} T^{\mu \nu}.
\end{equation}

The fully conformal \text{and} covariant formulation of Z4 (otherwise known as CCZ4) is thus obtained by performing a rescaling of the spatial metric tensor ${\gamma_{i j}}$ by a conformal factor ${\phi}$\cite{bernuzzi}\cite{hilditch}:

\begin{equation}
\tilde{\gamma}_{i j} = \phi^2 \gamma_{i j},
\end{equation}
chosen such as to guarantee unit determinant for the conformal spatial metric ${\tilde{\gamma}_{i j}}$:

\begin{equation}
\phi = \left( \mathrm{det} \left( \gamma_{i j} \right) \right)^{- \frac{1}{6}}
\end{equation}
The trace-free components ${\tilde{A}_{i j}}$ of the total extrinsic curvature tensor ${K_{i j}}$ on spacelike hypersurfaces then become (after conformal rescaling):

\begin{equation}
\tilde{A}_{i j} = \phi^2 \left( K_{i j} \right)^{TF} = \phi^2 \left( K_{i j} - \frac{1}{3} K \gamma_{i j} \right) = \phi^2 \left( K_{i j} - \frac{1}{3} K_{i j} \gamma^{i j} \gamma_{i j} \right).
\end{equation}
By introducing the Christoffel symbols ${\tilde{\Gamma}_{j k}^{i}}$ associated to the conformal spatial metric tensor ${\gamma_{i j}}$, as well as the pseudovector ${\tilde{\Gamma}^{i}}$ derived from the contraction of this conformal connection, namely:

\begin{equation}
\tilde{\Gamma}_{j k}^{i} = \frac{1}{2} \tilde{\gamma}^{i l} \left( \partial_j \tilde{\gamma}_{k l} + \partial_k \tilde{\gamma}_{j l} - \partial_l \tilde{\gamma}_{j k} \right),
\end{equation}
and:

\begin{equation}
\tilde{\Gamma}^i = \tilde{\gamma}^{j k} \tilde{\Gamma}_{j k}^{i} = \tilde{\gamma}_{i j} \tilde{\gamma}^{k l} \partial_l \tilde{\gamma}_{j k},
\end{equation}
respectively, we can further decompose the Ricci curvature tensor ${R_{i j}^{\left( 3 \right)}}$ on spacelike hypersurfaces into a sum:

\begin{equation}
R_{i j}^{\left( 3 \right)} = \tilde{R}_{i j}^{\left( 3 \right)} + \tilde{R}_{i j}^{\phi \left( 3 \right)}
\end{equation}
of a rank-2 tensor ${R_{i j}^{\left( 3 \right)}}$ depending only upon spatial derivatives of the conformal spatial metric tensor ${\tilde{\gamma}_{i j}}$:

\begin{equation}
\tilde{R}_{i j}^{\left( 3 \right)} = - \frac{1}{2} \tilde{\gamma}^{l m} \partial_l \partial_m \tilde{\gamma}_{i j} + \tilde{\gamma}_{k \left( i \right.} \partial_{\left. j \right)} \tilde{\Gamma}^k + \tilde{\Gamma}^k \tilde{\Gamma}_{\left( i j \right) k} + \tilde{\gamma}^{l m} \left[ 2 \tilde{\Gamma}_{l \left( i \right.}^{k} \tilde{\Gamma}_{\left. j \right) k m} + \tilde{\Gamma}_{i m}^{k} \tilde{\Gamma}_{k j l} \right],
\end{equation}
and a second rank-2 tensor ${\tilde{R}_{i j}^{\phi \left( 3 \right)}}$ depending only upon derivatives of the conformal factor ${\phi}$:

\begin{equation}
\tilde{R}_{i j}^{\phi \left( 3 \right)} = \frac{1}{\phi^2} \left[ \phi \left( \tilde{\nabla}_i \tilde{\nabla}_j \phi + \tilde{\gamma}_{i j} \tilde{\nabla}^{l} \tilde{\nabla}_l \phi \right) - 2 \tilde{\gamma}_{i j} \tilde{\nabla}^{l} \phi \tilde{\nabla}_{l} \phi \right].
\end{equation}
The fully conformal and covariant Z4 formulation now decomposes into a system of 6 evolution equations for the conformal spatial metric tensor ${\tilde{\gamma}_{i j}}$ (written here explicitly, without Lie derivatives):

\begin{equation}
\partial_t \tilde{\gamma}_{i j} = - 2 \alpha \tilde{A}_{i j} + 2 \tilde{\gamma}_{k \left( i \right.} \partial_{\left. j \right)} \beta^k - \frac{2}{3} \tilde{\gamma}_{i j} \partial_k \beta^k + \beta^k \partial_k \tilde{\gamma}_{i j},
\end{equation}
6 evolution equations for the trace-free components of the (conformal) extrinsic curvature tensor ${\tilde{A}_{i j}}$:

\begin{multline}
\partial_t \tilde{A}_{i j} = \phi^2 \left[ - \nabla_i \nabla_j \alpha + \alpha \left( R_{i j} + \nabla_i Z_j + \nabla_j Z_i - 8 \pi S_{i j} \right) \right]^{TF} + \alpha \tilde{A}_{i j} \left( K - 2 \Theta \right) - 2 \alpha \tilde{A}_{i l} \tilde{A}_{j}^{l}\\
+ 2 \tilde{A}_{k \left( i \right.} \partial_{\left. j \right)} \beta^k - \frac{2}{3} \tilde{A}_{i j} \partial_k \beta^k + \beta^k \partial_k \tilde{A}_{i j},
\end{multline}
and single evolution equations for the trace of the extrinsic curvature tensor $K$:

\begin{equation}
\partial_t K = - \nabla^i \nabla_i \alpha + \alpha \left( R + 2 \nabla_i Z^i + K^2 - 2 \Theta K \right) + \beta^j \partial_j K - 3 \alpha \kappa_1 \left( 1 + \kappa_2 \right) \Theta + 4 \pi \alpha \left( S - 3 \tau \right),
\end{equation}
the conformal factor ${\phi}$:

\begin{equation}
\partial_t \phi = \frac{1}{3} \alpha \phi K - \frac{1}{3} \phi \partial_k \beta^k + \beta^k \partial_k \phi,
\end{equation}
and the projection ${\Theta}$ of the ${\mathbf{Z}}$ 4-vector in the normal direction ${\mathbf{n}}$:

\begin{equation}
\partial_t \Theta = \frac{1}{2} \alpha \left( R + 2 \nabla_i Z^i - \tilde{A}_{i j} \tilde{A}^{i j} + \frac{2}{3} K^2 - 2 \Theta K \right) - Z^i \partial_i \alpha + \beta^k \partial_k \Theta - \alpha \kappa_1 \left( 2 + \kappa_2 \right) \Theta - 8 \pi \alpha \tau.
\end{equation}
Finally, in order to close the system, we must introduce a set of 4 evolution equations for the components of the pseudovector ${\tilde{\Gamma}^{i}}$ derived from the conformal connection coefficients ${\tilde{\Gamma}_{j k}^{i}}$; for simplicity, we represent these as equations for the components ${\hat{\Gamma}^i}$, defined as:

\begin{equation}
\hat{\Gamma}^i = \tilde{\Gamma}^i + 2 \tilde{\gamma}^{i j} Z_j,
\end{equation}
namely:

\begin{multline}
\partial_t \hat{\Gamma}^i = 2 \alpha \left( \tilde{\Gamma}_{j k}^{i} \tilde{A}^{j k} - 3 \tilde{A}^{j k} \frac{\partial_j \phi}{\phi} - \frac{2}{3} \tilde{\gamma}^{i j} \partial_j K \right) + 2 \tilde{\gamma}^{k i} \left( \alpha \partial_k \Theta - \Theta \partial_k \alpha - \frac{2}{3} \alpha K Z_k \right) - 2 \tilde{A}^{i j} \partial_j \alpha\\
+ \tilde{\gamma}^{k l} \partial_k \partial_l \beta^i + \frac{1}{3} \tilde{\gamma}^{i k} \partial_k \partial_l \beta^l + \frac{2}{3} \tilde{\Gamma}^i \partial_k \beta^k - \tilde{\Gamma}^k \partial_k \beta^i + 2 \kappa_3 \left( \frac{2}{3} \tilde{\gamma}^{i j} Z_j \partial_k \beta^k - \tilde{\gamma}^{j k} Z_j \partial_k \beta^i \right) + \beta^k \partial_k \hat{\Gamma}^i\\
- 2 \alpha \kappa_1 \tilde{\gamma}^{i j} Z_j - 16 \pi \alpha \tilde{\gamma}^{i j} S_j.
\end{multline}

Note that the new ${\kappa_3}$ parameter in the evolution equations for the conformal connection coefficients ${\hat{\Gamma}^i}$ above is currently unconstrained; although ${\kappa_3 = 1}$ yields a fully covariant formulation (at the cost of numerical stability in the case of black hole spacetimes) and ${\kappa_3 = \frac{1}{2}}$ yields a non-covariant formulation which is nevertheless numerically stable, we choose instead to follow the prescription of Alic, Kastaun and Rezzolla\cite{alic2} and eliminate the numerical instabilities by choosing ${\kappa_1 \to \frac{\kappa_1}{\alpha}}$ as the choice of damping parameter, which preserves spatial covariance in all terms depending on ${\kappa_1}$, and full spacetime covariance for all other terms. Moreover, for simulations involving spherical symmetry, we employ the \textit{maximal slicing} gauge condition\cite{alcubierre}\cite{alcubierre2} on the lapse function ${\alpha}$:

\begin{equation}
\nabla^2 \alpha = \alpha K_{i j} K^{i j},
\end{equation}
and the more general \textit{${1 + log}$ slicing} gauge condition\cite{alcubierre3} for the axially-symmetric case:

\begin{equation}
\partial_t \alpha = -2 \alpha \left( K - 2 \Theta \right) + \beta^k \partial_k \alpha,
\end{equation}
due to their known stability (and singularity-avoidance) properties for black hole spacetimes. By the same token, we use the \textit{gamma-driver} gauge conditions\cite{anninos}\cite{smarr} on the components of the shift vector ${\beta^i}$:

\begin{equation}
\partial_t \beta^i = \eta_1 B^i + \beta^k \partial_k \beta^i, \qquad \text{ where } \qquad \partial_t B^i = \mu_{\beta_1} \alpha^{\mu_{\beta_2}} \partial_t \hat{\Gamma}^i - \beta^k \partial_k \hat{\Gamma}^i + \beta^k \partial_k B^i - \eta_2 B^i,
\end{equation}
or, in the simplified case (without any advection terms):

\begin{equation}
\partial_t \beta^i = \eta_1 B^i, \qquad \text{ where } \qquad \partial_t B^i = \mu_{\beta_1} \alpha^{\mu_{\beta_2}} \partial_t \hat{\Gamma}^i - \eta_2 B^i
\end{equation}
where ${B^i}$ is an auxiliary vector field, and ${\eta_1}$, ${\eta_2}$, ${\mu_{\beta_1}}$ and ${\mu_{\beta_2}}$ are scalar parameters, chosen here to be ${\eta_1 = \frac{3}{4}}$, ${\mu_{\beta_1} = 1}$, ${\mu_{\beta_2} = 0}$ and ${\eta_2 = 1}$ (i.e. the standard \textit{hyperbolic driver} conditions). The standard conformal \textit{BSSN formalism} of Baumgarte, Shapiro, Shibata and Nakamura\cite{nakamura} is recovered in the special case where ${\Theta = 0}$ and ${Z^i = 0}$, the trace of the extrinsic curvature tensor $K$ is replaced with ${K^{BSSN}}$, defined by:

\begin{equation}
K^{BSSN} = K - 2 \Theta,
\end{equation}
and the Hamiltonian constraint is used to eliminate any explicit dependence on the (spatial) Ricci scalar $R$ from the evolution equation for the trace of the extrinsic curvature tensor $K$. The consistency conditions on the two additional scalar fields introduced by the conformal rescaling of the covariant Z4 system (namely ${\mathrm{det} \left( \tilde{\gamma}_{i j} \right)}$ and ${\mathrm{tr} \left( \tilde{A}_{i j} \right)}$):

\begin{equation}
\mathrm{det} \left( \tilde{\gamma}_{i j} \right) = 1, \qquad \text{ and } \qquad \mathrm{tr} \left( \tilde{A}_{i j} \right) = 0,
\end{equation}
are enforced explicitly as part of the numerical integration algorithm at each time step (by subtracting out the trace from ${\tilde{A}_{i j}}$ and rescaling the conformal spatial metric tensor ${\tilde{\gamma}_{i j}}$ accordingly). The singularity-avoidance properties of our chosen gauge conditions remove any need for us to employ techniques such as \textit{excision} or \textit{punctures} for the case of the symmetrical black hole spacetimes simulated within this article.

With regards to the governing equations for our numerical simulations, all that remains is to specify how our spacetime may be equipped with a minimally-coupled (massive) scalar field ${\Phi}$ obeying the (massive) Klein-Gordon equation:

\begin{equation}
\left( \Box + m^2 \right) \Phi = 0,
\end{equation}
or for a more general potential ${V \left( \Phi \right)}$ (with the massive case hence corresponding to ${V \left( \Phi \right) = m^2 \bar{\Phi} \Phi}$):

\begin{equation}
\Box \Phi + \frac{d V}{d \Phi} = 0,
\end{equation}
which, in curved spacetime, becomes:

\begin{equation}
g^{\mu \nu} \nabla_{\mu} \nabla_{\nu} \Phi = \frac{d V}{d \Phi}.
\end{equation}
The solution to the Klein-Gordon equation yields a stress-energy tensor ${T^{\mu \nu}}$ with components:

\begin{equation}
T^{\mu \nu} = 2 \nabla^{\mu} \bar{\Phi} \nabla^{\nu} \Phi - g^{\mu \nu} \left( \nabla^{\rho} \bar{\Phi} \nabla_{\rho} \Phi - m^2 \bar{\Phi} \Phi \right),
\end{equation}
in the massive case, or, in lowered-index form ${T_{\mu \nu}}$:

\begin{equation}
T_{\mu \nu} = \nabla_{\mu} \Phi \nabla_{\nu} \Phi - \frac{1}{2} g_{\mu \nu} \left( \nabla_{\rho} \Phi \nabla^{\rho} \Phi + 2 V \right),
\end{equation}
in the case of a general potential. Within the initial-value formulation of the Einstein field equations afforded by the ADM decomposition, the (hyperbolic) evolution equation for this scalar field ${\Phi}$ is then given by:

\begin{equation}
\partial_t \Phi = \alpha \Pi_{m} + \beta^i \partial_i \Phi,
\end{equation}
where ${\Pi_m}$ is the (negative of the) conjugate momentum of the scalar field given by:

\begin{equation}
\Pi_m = \frac{1}{\alpha} \left( \partial_t \Phi - \beta^i \partial_i \Phi \right).
\end{equation}
However, we can immediately see that this ``definition'' of the (negative) conjugate momentum ${\Pi_m}$ is really just a trivial restatement of the ``evolution equation'' for the scalar field ${\Phi}$; therefore, the ``true'' evolution equation for the scalar field must be given in terms of an evolution equation for its (negative) conjugate momentum ${\Pi_m}$ directly, instead, i.e:

\begin{equation}
\partial_t \Pi_m = \beta^i \partial_i \Pi_m + \gamma^{i j} \left( \alpha \partial_j \partial_i \Phi + \partial_j \Phi \partial_i \alpha \right) + \alpha \left( K \Pi_m - \gamma^{i j} \Gamma_{i j}^{k} \partial_k \Phi + \frac{d V}{d \Phi} \right).
\end{equation}

\begin{figure}[ht]
\centering
\includegraphics[width=0.495\textwidth]{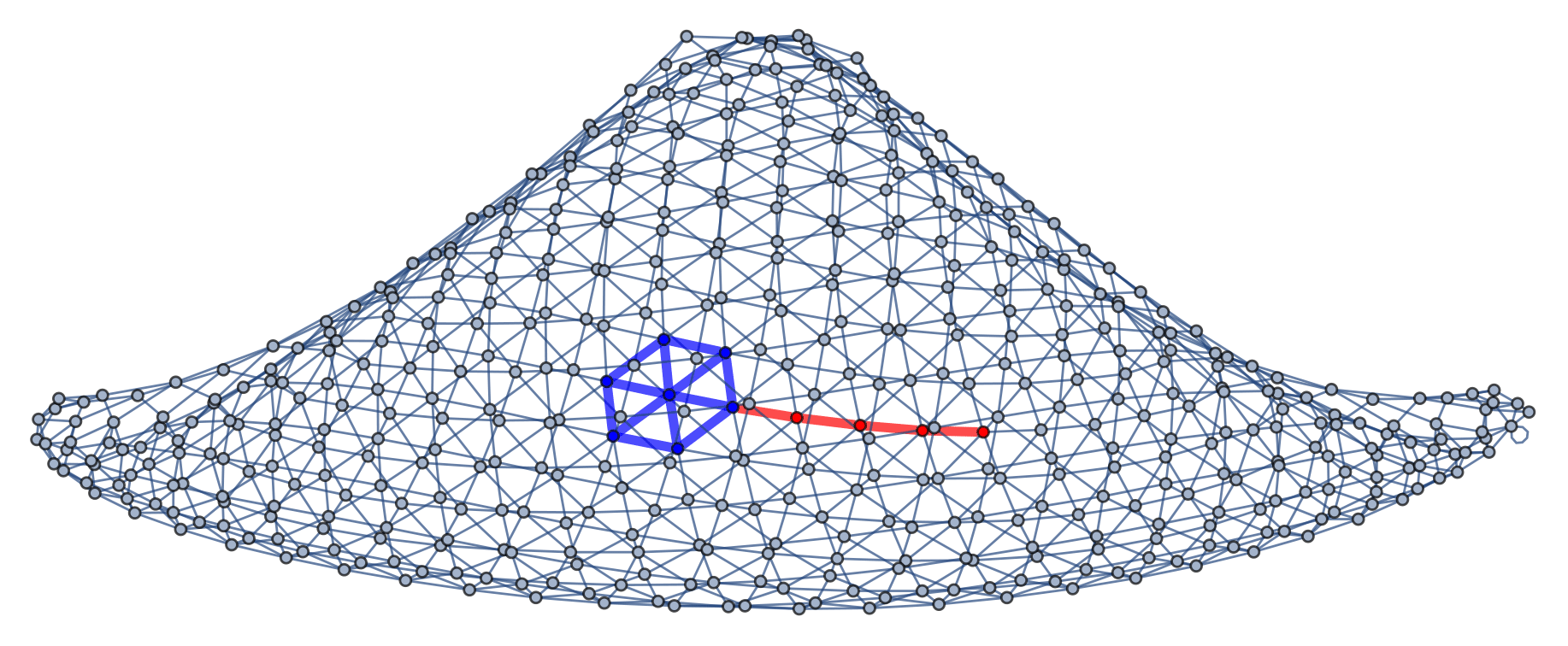}
\caption{Interpolating a value for the endpoint of the red geodesic in a spatial hypergraph with a two-dimensional Riemannian manifold-like limiting structure, as generated by the hypergraph rewriting rule ${\left\lbrace \left\lbrace x, x, y \right\rbrace, \left\lbrace x, z, w \right\rbrace \right\rbrace \to \left\lbrace \left\lbrace w, w, v \right\rbrace, \left\lbrace v, w, y \right\rbrace, \left\lbrace z, y, v \right\rbrace \right\rbrace}$, using a generalized bicubic interpolation algorithm applied to the 7 vertices contained within the blue subhypergraph.}
\label{fig:Figure35}
\end{figure}

\begin{figure}[ht]
\centering
\includegraphics[width=0.495\textwidth]{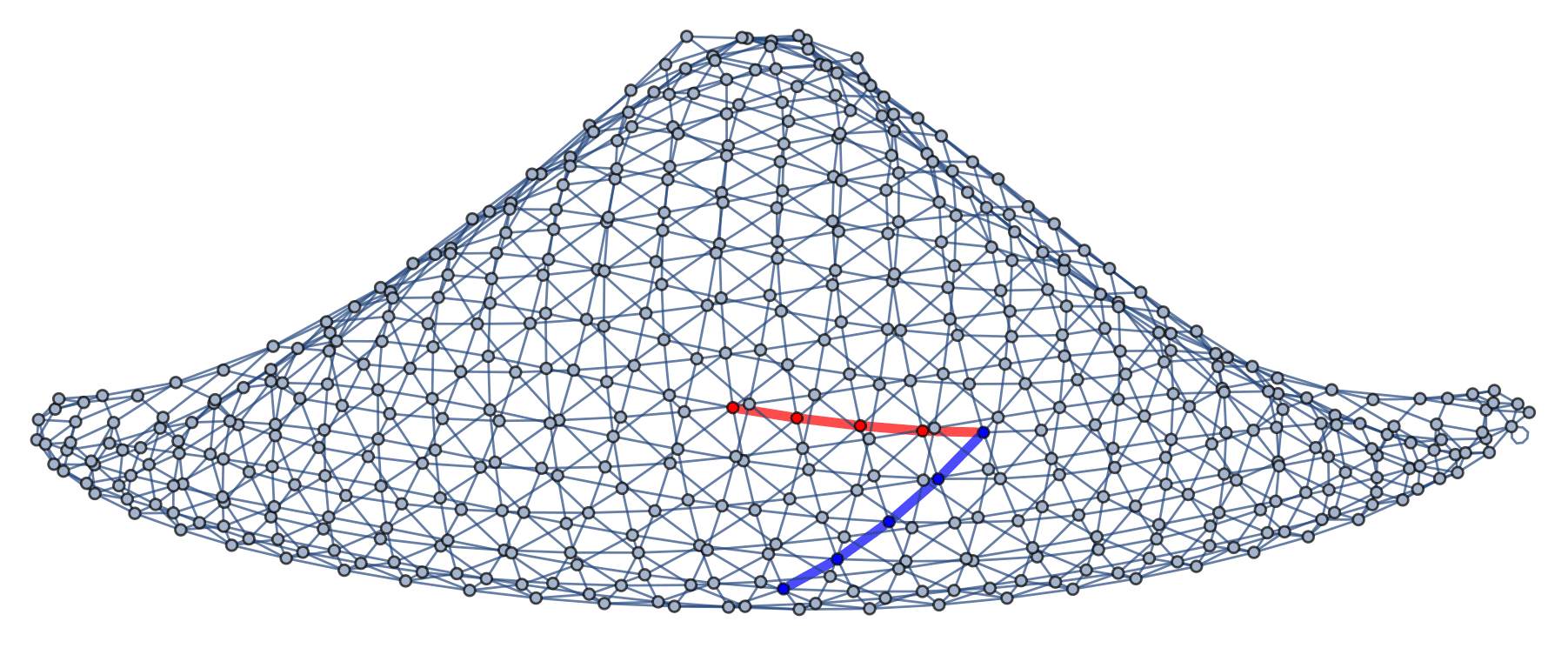}
\includegraphics[width=0.495\textwidth]{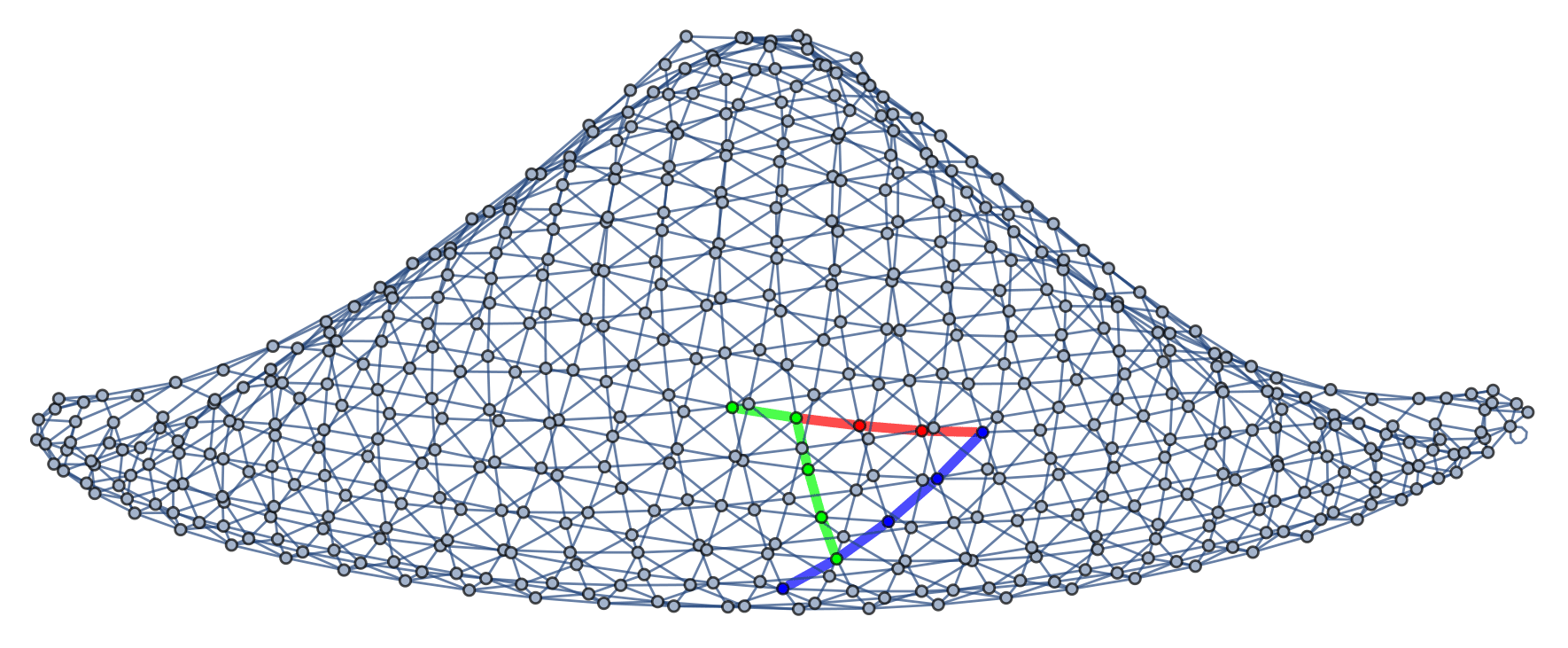}
\caption{Computing an inner product of two discrete geodesics (shown in red and blue) in a spatial hypergraph with a two-dimensional Riemannian manifold-like limiting structure, as generated by the hypergraph rewriting rule ${\left\lbrace \left\lbrace x, x, y \right\rbrace, \left\lbrace x, z, w \right\rbrace \right\rbrace \to \left\lbrace \left\lbrace w, w, v \right\rbrace, \left\lbrace v, w, y \right\rbrace, \left\lbrace z, y, v \right\rbrace \right\rbrace}$, by discrete projection/``dropping a perpendicular'' (shown in green), yielding a normalized inner product value of ${\frac{3}{4}}$.}
\label{fig:Figure36}
\end{figure}

In order to evolve the resulting system of equations using an explicit finite-difference time-stepping method, it is first necessary to introduce a means of defining finite-difference numerical stencils over hypergraphs of arbitrary topology (i.e. hypergraphs with no a priori spatial coordinate structure). As outlined in previous work\cite{gorard4}, this can be achieved by applying a generalized bicubic:

\begin{equation}
f \left( x, y \right) = \sum_{i = 0}^{3} \sum_{j = 0}^{3} a_{i j} x^i y^j,
\end{equation}
or tricubic\cite{lekien}:

\begin{equation}
f \left( x, y, z \right) = \sum_{i = 0}^{3} \sum_{j = 0}^{3} \sum_{k = 0}^{3} a_{i j k} x^i y^j z^k,
\end{equation}
interpolation scheme to the neighborhoods at the endpoints of hypergraph geodesics, depending upon whether the hypergraph has a limiting 2-dimensional or 3-dimensional structure, respectively (and where the number of ``$a$'' coefficients is therefore either ${2^4}$ or ${2^6}$), as shown in Figure \ref{fig:Figure35}. By equipping the hypergraph with a local inner product structure using discrete projections, as shown in Figure \ref{fig:Figure36} (with the axioms of linearity and conjugate symmetry being satisfied whenever the hypergraph converges to a Riemannian manifold in the continuum limit), one can therefore define a local set of orthonormal coordinate axes on the hypergraph, over which a typical finite-difference scheme can be formulated. For the simulation results presented within this article, we choose to use an explicit fourth-order Runge-Kutta numerical scheme, which, for an initial value problem of the form:

\begin{equation}
\frac{d \Phi \left( \mathbf{x}, t \right)}{d t} = F \left( t, \Phi \left( \mathbf{x}, t \right) \right), \qquad \text{ where } \qquad \Phi \left( \mathbf{x}, t_0 \right) = \Phi_0,
\end{equation}
where ${\Phi \left( \mathbf{x}, t \right)}$ is a generic \textit{state vector}, i.e:

\begin{equation}
\Phi \left( \mathbf{x}, t \right) = \left( \phi_1 \left( \mathbf{x}, t \right), \phi_2 \left( \mathbf{x}, t \right), \phi_3 \left( \mathbf{x}, t \right), \dots \right),
\end{equation}
we can evolve the system forwards in units of a discrete time step ${\Delta t}$ (chosen so as to be compatible with the Courant condition, and therefore numerically stable) as\cite{kreiss}\cite{levy}:

\begin{equation}
\Phi \left( \mathbf{x}, t_{n + 1} \right) = \Phi \left( \mathbf{x}, t_n \right) + \frac{1}{6} \Delta t \left( k_1 + 2 k_2 + 2 k_3 + k_4 \right), \qquad \text{ where } \qquad t_{n + 1} = t_n + \Delta t.
\end{equation}
Such an initial-value ODE problem can be obtained from a system of non-linear (hyperbolic) PDEs, such as the Einstein field equations, of the general form:

\begin{equation}
\frac{\partial \Phi \left( \mathbf{x}, t \right)}{\partial t} = \mathcal{F} \left( \Phi \left( \mathbf{x}, t \right) \right),
\end{equation}
where ${\mathcal{F}}$ is a non-linear operator acting on ${\Phi}$, by computing a new set of fluxes between vertices in the hypergraph (thus linearizing the problem) at each time step. Note that this assumes a set of local spatial coordinates ${\mathbf{x}}$ and a global time coordinate $t$ for the hypergraphs, which can be obtained using the geometrical construction described above.

In the explicit time evolution formula above, we choose:

\begin{equation}
k_1 = F \left( t_n, \phi \left( \mathbf{x}, t_n \right) \right), \qquad k_4 = F \left( t_n + \Delta t, \Phi \left( \mathbf{x}, t_n \right) + \Delta t k_3 \right),
\end{equation}
and:

\begin{equation}
k_2 = F \left( t_n + \frac{\Delta t}{2}, \Phi \left( \mathbf{x}, t_n \right) + \Delta t \frac{k_1}{2} \right), \qquad k_3 = F \left( t_n + \frac{\Delta t}{2}, \Phi \left( \mathbf{x}, t_n \right) + \Delta t \frac{k_2}{2} \right),
\end{equation}
which completes the temporal discretization of the scheme. For spatial discretization, we choose a coordinate structure such that ${\Delta x = x_{i + 1} - x_i}$, ${\Delta y = y_{j + 1} - y_j}$ and ${\Delta z = z_{k + 1} - z_k}$ (i.e. the three discrete coordinates are indexed by $i$, $j$ and $k$, respectively), leading to a discrete form of the state vector ${\Phi_{i, j, k}^{n}}$ defined at each vertex and for each discrete time step. The spatial derivatives may then be computed by means of the fourth-order centered finite-difference stencils of Zlochower, Baker, Campanelli and Lousto\cite{zlochower}, with first derivatives (in $x$) defined by:

\begin{equation}
\partial_x F \left( t_n, \Phi_{i, j, k}^{n} \right) = \frac{1}{12 \Delta x} \left( F \left( t_n, \Phi_{i - 2, j, k}^{n} \right) - 8 F \left( t_n, \Phi_{i - 1, j , k}^{n} \right) + 8 F \left( t_n, \Phi_{i + 1, j, k}^{n} \right) - F \left( t_n, \Phi_{i + 2, j, k}^{n} \right) \right),
\end{equation}
and likewise for all other first derivatives. Note that, whenever advection terms (i.e. terms of the form ${\beta^i \partial_i F \left( t_n, \Phi_{i, j, k}^{n} \right)}$) are present, one must instead use a fourth-order upwind finite-difference scheme:

\begin{multline}
\partial_x F \left( t_n, \Phi_{i, j, k}^{n} \right) = \frac{1}{12 \Delta x} \left( - F \left( t_n, \Phi_{i - 3, j, k}^{n} \right) + 6 F \left( t_n, \Phi_{i - 2, j, k}^{n} \right) - 18 F \left( t_n, \Phi_{i - 1, j, k}^{n} \right) \right.\\
\left. + 10 F \left( t_n, \Phi_{i, j, k}^{n} \right) + 3 F \left( t_n, \Phi_{i + 1, j, k}^{n} \right) \right),
\end{multline}
whenever ${\beta^x < 0}$, and:

\begin{multline}
\partial_x F \left( t_n, \Phi_{i, j, k}^{n} \right) = \frac{1}{12 \Delta x} \left( F \left( t_n, \Phi_{i + 3, j, k}^{n} \right) - 6 F \left( t_n, \Phi_{i + 2, j, k}^{n} \right) + 18 F \left( t_n, \Phi_{i + 1, j, k}^{n} \right) \right.\\
\left. - 10 F \left( t_n, \Phi_{i, j, k}^{n} \right) - 3 F \left( t_n, \Phi_{i - 1, j, k}^{n} \right) \right),
\end{multline}
whenever ${\beta^x > 0}$, and likewise for all other first derivatives. The finite-difference stencils for second derivatives (both for derivatives purely in $x$, and for mixed derivatives in $x$ and $y$) can be deduced by applying the first derivative stencils sequentially, in any order, for instance yielding:

\begin{multline}
\partial_{x x} F \left( t_n, \Phi_{i, j, k}^{n} \right) = \frac{1}{12 \left( \Delta x \right)^2} \left( -F \left( t_n, \Phi_{i + 2, j, k}^{n} \right) + 16 F \left( t_n, \Phi_{i + 1, j, k}^{n} \right) - 30 F \left( t_n, \Phi_{i, j, k}^{n} \right) \right.\\
\left. + 16 F \left( t_n, \Phi_{i - 1, j, k}^{n} \right) - F \left( t_n, \Phi_{i - 2, j, k}^{n} \right) \right),
\end{multline}
and:

\begin{multline}
\partial_{x y} F \left( t_n, \Phi_{i, j, k}^{n} \right) = \frac{1}{144 \Delta x \Delta y} \left[ F \left( t_n, \Phi_{i - 2, j - 2, k}^{n} \right) - 8 F \left( t_n, \Phi_{i - 1, j - 2, k}^{n} \right) + 8 F \left( t_n, \Phi_{i + 1, j - 2, k}^{n} \right) \right. \\
\left. - F \left( t_n, \Phi_{i + 2, j - 2, k}^{n} \right) - 8 \left( F \left( t_n, \Phi_{i - 2, j - 1, k}^{n} \right) - 8 F \left( t_n, \Phi_{i - 1, j - 1, k}^{n} \right) + 8 F \left( t_n, \Phi_{i + 1, j - 1, k}^{n} \right) \right. \right.\\
\left. \left. - F \left( t_n, \Phi_{i + 2, j - 1, k}^{n} \right) \right) + 8 \left( F \left( t_n, \Phi_{i - 2, j + 1, k}^{n} \right) - 8 F \left( t_n, \Phi_{i - 1, j + 1, k}^{n} \right) + 8 F \left( t_n, \Phi_{i + 1, j + 1, k}^{n} \right) \right. \right.\\
\left. \left. - F \left( t_n, \Phi_{i + 2, j + 1, k}^{n} \right) \right) - \left( F \left( t_n, \Phi_{i - 2, j + 2, k}^{n} \right) - 8 F \left( t_n, \Phi_{i - 1, j + 2, k}^{n} \right) + 8 F \left( t_n, \Phi_{i + 1, j + 2, k}^{n} \right) \right. \right.\\
\left. \left. - F \left( t_n, \Phi_{i + 2, j + 2, k}^{n} \right) \right) \right],
\end{multline}
respectively, for the fourth-order centered case, and likewise for the fourth-order upwind case (and for all other second derivatives). We also incorporate a dissipation term of Kreiss-Oliger type\cite{kreiss2} into the definition of the (discrete) time derivative for the fourth-order finite-difference scheme, in order to reduce the probability of spurious high-frequency modes destabilizing the numerical solution:

\begin{multline}
\partial_t \Phi_{i, j, k} \to \partial_t \Phi_{i, j, k} + \frac{\sigma}{64 \Delta x} \left( \Phi_{i + 3, j, k} - 6 \Phi_{i + 2, j, k} \right.\\
\left. + 15 \Phi_{i + 1, j, k} - 20 \Phi_{i, j, k} + 15 \Phi_{i - 1, j, k} - 6 \Phi_{i - 2, j, k} + \Phi_{i - 3, j, k} \right),
\end{multline}
for derivatives in the $x$-direction, and likewise for all other directional derivatives.

\begin{figure}[ht]
\centering
\includegraphics[width=0.495\textwidth]{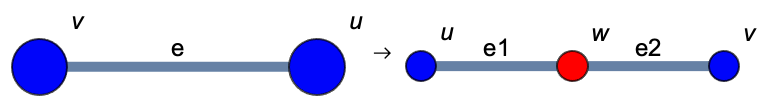}
\includegraphics[width=0.495\textwidth]{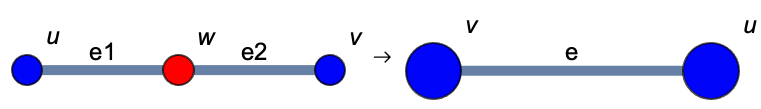}
\caption{An illustration of the refinement/coarsening procedure for hypergraphs, wherein a hyperedge ${e = \left\lbrace u, v \right\rbrace}$ is \textit{subdivided} into a pair of hyperedges ${e_1 = \left\lbrace u, w \right\rbrace}$, ${e_2 = \left\lbrace w, v \right\rbrace}$ for some newly-generated vertex $w$ (i.e. refinement, on the left), or a pair of hyperedges ${e_1 = \left\lbrace u, w \right\rbrace}$, ${e_2 = \left\lbrace w, v \right\rbrace}$ is \textit{smoothed} into a single hyperedge ${e = \left\lbrace u, v \right\rbrace}$ (i.e. coarsening, on the right).}
\label{fig:Figure37}
\end{figure}

As our adaptive refinement algorithm, we use the hypergraph generalization of the local adaptive mesh refinement (AMR) algorithm proposed by Berger and Colella\cite{berger}, based on previous methods developed by Berger and Oliger\cite{berger2} and Gropp\cite{gropp}. Broadly speaking, we introduce a \textit{tagging function} ${f \left( x, y, z \right)}$ based on whether the ${L_2}$ norm of the change in some scalar field ${\phi}$ (which, for the simulations presented within this article, will always be chosen to be a Riemannian scalar curvature invariant) exceeds a pre-defined threshold ${\sigma \left( \phi \right)}$ across a given vertex or subhypergraph:

\begin{align}
f \left( x, y, z \right) = \begin{cases}
1, \qquad &\text{ if } \sqrt{\sum\limits_{i = 1}^{3} \left( \phi \left( \mathbf{x} + \Delta x \hat{\mathbf{x}}_i \right) - \phi \left( \mathbf{x} - \Delta x \hat{\mathbf{x}}_i \right) \right)^2} > \sigma \left( \phi \right),\\
0, \qquad &\text{ otherwise},
\end{cases}
\end{align}
such that ${f \left( x, y, z \right) = 1}$ signifies that a given vertex or subhypergraph is tagged for refinement, and ${f \left( x, y, z \right) = 0}$ otherwise. The entire hypergraph is then partitioned into subhypergraphs, the \textit{signatures} ${X \left( x \right)}$, ${Y \left( y \right)}$ and ${Z \left( z \right)}$ are computed for each subhypergraph:

\begin{equation}
X \left( x \right) = \int f \left( x, y, z \right) dy dz, \qquad Y \left( y \right) = \int f \left( x, y, z \right) dx dx, \qquad Z \left( z \right) = \int f \left( x, y, z \right) dx dy,
\end{equation}
and the Laplacians ${\partial_{x}^{2} X \left( x \right)}$, ${\partial_{y}^{2} Y \left( y \right)}$ and ${\partial_{z}^{2} Z \left( z \right)}$ of these signatures determine the axis that will separate the tagged and untagged vertices in the direction orthogonal to the signature function (i.e. the axis of partition), since this will be the local coordinate direction which maximizes ${\Delta \left( \partial_{i}^{2} X_i \right)}$ (and hence which induces an appropriate inflection point in the Laplacian). If the signature is identically zero in any direction (i.e. if ${X_i \left( x_i \right) = 0}$), then that direction is chosen as the axis of partition instead. So long as any given subhypergraph satisfies the two principal axioms of hierarchical grid/hypergraph nesting (i.e. that fine subhypergraphs must always be contained within the neighborhood of a vertex in the next coarsest subhypergraph, and that a vertex that is not at the boundary of the domain at refinement level $l$ must be separated from a vertex at refinement level ${l - 2}$ by at least one vertex at refinement level ${l - 1}$, in any direction in the neighborhood), and the ratio of tagged vertices to total vertices is at least ${\epsilon}$ (for some predefined ${0 < \epsilon < 1}$), the partitioning procedure terminates; otherwise, it continues on recursively subdividing subhypergraphs into further subhypergraphs and the procedure starts again. The refinement/coarsening procedure for arbitrary vertices/hypergraphs, which can be enacted via the operations of \textit{hyperedge subdivision} and \textit{hyperedge smoothing}, respectively, is illustrated in Figure \ref{fig:Figure37}.

Following refinement, the time evolution scheme for the finite-difference method, which may be represented in a very explicit (and manifestly conservative) form as:

\begin{multline}
\Phi_{i, j, k}^{n + 1} = \Phi_{i, j, k}^{n} - \frac{\Delta t}{\Delta x} \left( F \left( \Phi_{i + \frac{1}{2}, j, k}^{n} \right) - F \left( \Phi_{i - \frac{1}{2}, j, k}^{n} \right) \right) - \frac{\Delta t}{\Delta y} \left( G \left( \Phi_{i, j + \frac{1}{2}, k}^{n} \right) -G \left( \Phi_{i, j - \frac{1}{2}, k}^{n} \right) \right)\\
- \frac{\Delta t}{\Delta x} \left( H \left( \Phi_{i, j, k + \frac{1}{2}}^{n} \right) - H \left( \Phi_{i, j, k - \frac{1}{2}}^{n} \right) \right),
\end{multline}
with $F$, $G$ and $H$ denoting the inter-vertex flux functions projected in the $x$, $y$ and $z$ directions, respectively, and with ${\Phi_{i \pm \frac{1}{2}, j, k}}$, ${\Phi_{i, j \pm \frac{1}{2}, k}}$ and ${\Phi_{i, j, k \pm \frac{1}{2}}}$ denoting the extrapolated values of the state vector ${\Phi}$ at the $x$-boundaries, $y$-boundaries and $z$-boundaries, respectively, must now be modified so as to account for the presence of new boundaries between coarse and fine subhypergraphs. Whenever a coarse vertex (i.e. a vertex at refinement level ${l - 1}$) is overlaid by a refined subhypergraph (i.e. a subhypergraph at refinement level $l$), the values of the coarse state vector ${\Phi^{coarse}}$ (i.e. the state vector at refinement level ${l - 1}$) are defined in terms of (conservative) averages of the values ${\Phi^{fine}}$ in the fine subhypergraph (i.e. the state vectors at refinement level $l$):

\begin{equation}
\left( \frac{1}{r^3} \sum_{p = 0}^{r - 1} \sum_{q = 0}^{r - 1} \sum_{s = 0}^{r - 1} \Phi_{m + p, n + q, o + s}^{fine} \right) \to \Phi_{i, j, k}^{coarse},
\end{equation}
where $i$, $j$ and $k$ are the discrete coordinates of the coarse vertex, where the same vertex spans the discrete intervals ${\left[ m, m + r - 1 \right]}$, ${\left[ n, n + r - 1 \right]}$ and ${\left[ o, o + r - 1 \right]}$ within the fine subhypergraph. On the other hand, whenever a coarse vertex (at refinement level ${l - 1}$) is \textit{adjacent} to an interface with a fine subhypergraph (at refinement level $l$), it is necessary to modify the time evolution scheme to be of the following general form (assuming that the coarse-to-fine interface is projected in the $x$ direction, with the obvious modifications for the other coordinate directions):

\begin{multline}
\Phi_{i, j, k} \left( t + \Delta t_{coarse} \right) = \Phi_{i, j, k} - \frac{\Delta t_{coarse}}{\Delta x} \left[ F \left( \Phi_{i + \frac{1}{2}, j, k} \left( t \right) \right) \right.\\
\left. - \frac{1}{r^3} \sum_{q = 0}^{r - 1} \sum_{p = 0}^{r - 1} \sum_{s = 0}^{r - 1} F \left( \Phi_{m + \frac{1}{2}, n + p, o + s} \left( t + q \Delta t_{fine} \right) \right) \right] - \frac{\Delta t_{coarse}}{\Delta y} \left[ G \left( \Phi_{i, j + \frac{1}{2}, k} \left( t \right) \right) - G \left( \Phi_{i, j - \frac{1}{2}, k} \left( t \right) \right) \right]\\
- \frac{\Delta t_{coarse}}{\Delta z} \left[ H \left( \Phi_{i, j, k + \frac{1}{2}} \left( t \right) \right) - H \left( \Phi_{i, j, k - \frac{1}{2}} \left( t \right) \right) \right],
\end{multline}
with ${\Delta t_{coarse}}$ and ${\Delta t_{fine}}$ denoting the stable time steps at refinement levels ${l - 1}$ and $l$, respectively, scaled using the same refinement ratios as the vertex sizes themselves, i.e:

\begin{equation}
\frac{\Delta t_l}{\Delta x_l} = \frac{\Delta t_{l - 1}}{\Delta x_{l - 1}} = \cdots = \frac{\Delta t_1}{\Delta x_1},
\end{equation}
and with ${\Delta x}$, ${\Delta y}$ and ${\Delta z}$ designating the spatial size of the vertices in the coarse subhypergraph.

\begin{figure}[ht]
\centering
\includegraphics[width=0.395\textwidth]{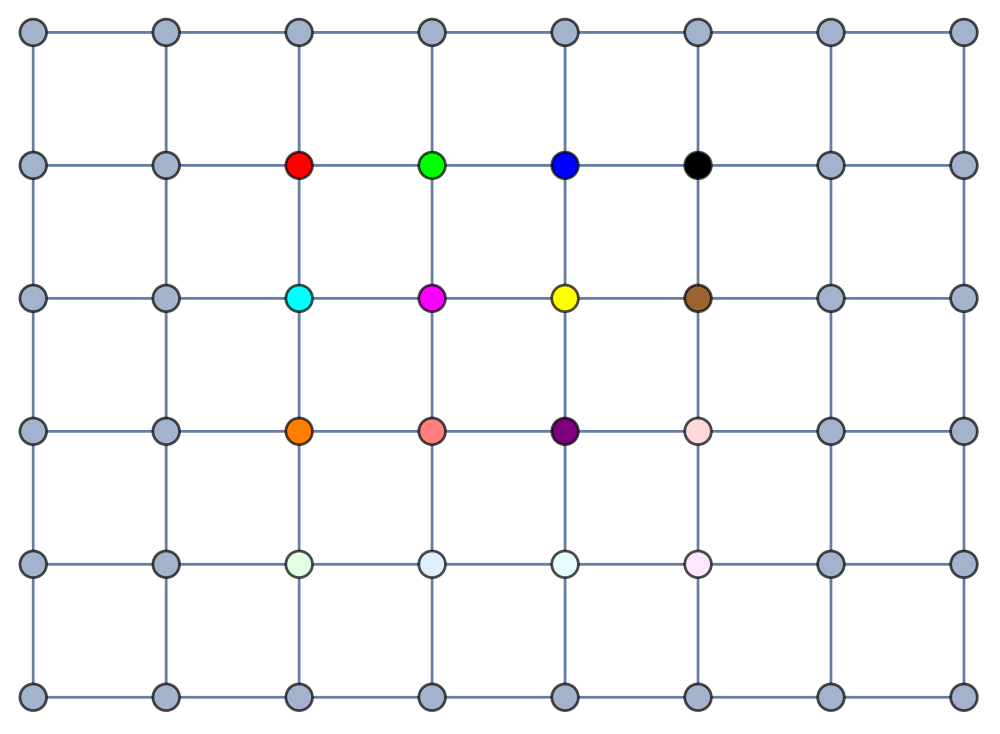}\hspace{0.1\textwidth}
\includegraphics[width=0.445\textwidth]{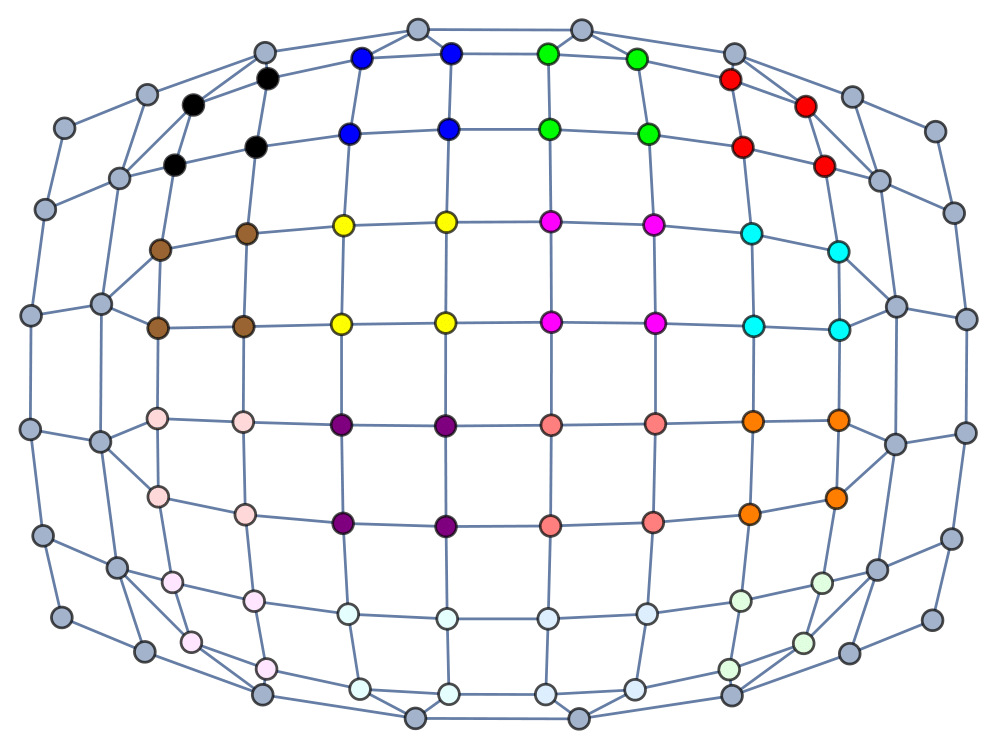}
\caption{An illustration of the refinement procedure for (structured) hypergraphs, wherein a collection of twelve colored vertices in a two-dimensional quadrilateral mesh are marked for refinement (left), and are consequently replaced with finer (2-by-2) two-dimensional grids (right).}
\label{fig:Figure38}
\end{figure}

\begin{figure}[ht]
\centering
\includegraphics[width=0.495\textwidth]{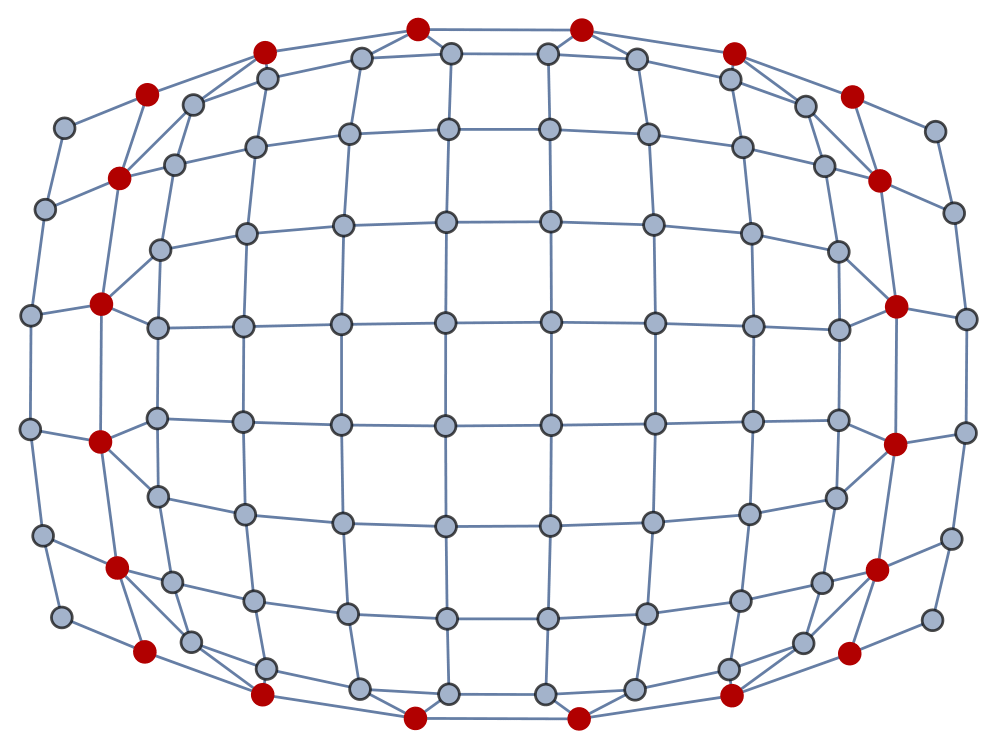}
\caption{An illustration of which coarse vertices (highlighted in dark red) must have their finite-difference evolution scheme modified by virtue of being adjacent to interfaces with a fine grid/subhypergraph, following refinement of the two-dimensional quadrilateral mesh shown above.}
\label{fig:Figure39}
\end{figure}

This modification can be implemented as a corrector step to the na\"ive flux computation in the coarse subhypergraph, whereby the coarse fluxes are subtracted from ${\Phi_{i, j, k}^{coarse} \left( t + \Delta t_{coarse} \right)}$ and replaced with the corresponding fine ones. One starts by constructing a tensor ${\delta F}$ of fluxes through all coarse subhypergraph boundaries that are also outer boundaries of fine subhypergraphs, i.e. in the $x$ direction:

\begin{equation}
\delta F^{old} \left( \Phi_{i + \frac{1}{2}, j, k} \right) = - F^{coarse} \left( \Phi_{i + \frac{1}{2}, j, k} \right),
\end{equation}
and likewise for the other coordinate directions. After each stable time step ${\Delta t_{fine}}$ in the fine subhypergraph, we then add a sum of all the fine subhypergraph fluxes through the boundary at discrete coordinates ${\left( i + \frac{1}{2}, j, k \right)}$:

\begin{equation}
\delta F^{new} \left( \Phi_{i + \frac{1}{2}, j, k} \right) = \delta F^{old} \left( \Phi_{i + \frac{1}{2}, j, k} \right) + \frac{1}{r^2} \sum_{p = 0}^{r - 1} \sum_{q = 0}^{r - 1} F \left( \Phi_{m + \frac{1}{2}, n + p, o + s} \right),
\end{equation}
and likewise for the other discrete coordinate projections ${\left( i, j + \frac{1}{2}, k \right)}$ and ${\left( i, j, k + \frac{1}{2} \right)}$. After $r$ such time steps ${\Delta t_{fine}}$ have elapsed, the flux tensor ${\delta F^{new} \left( \Phi_{i + \frac{1}{2}, j, k} \right)}$ can be used to correct the solutions over the coarse subhypergraph so as to match those computed using the modified time evolution formula presented above; for instance, for a vertex with discrete coordinates ${\left( i + 1, j, k \right)}$ only one correction is needed:

\begin{equation}
\Phi_{i + 1, j, k}^{coarse \left( new \right)} = \Phi_{i + 1, j, k}^{coarse \left( old \right)} + \frac{\Delta t_{coarse}}{\Delta x_{coarse}} \delta F^{new} \left( \Phi_{i + \frac{1}{2}, j, k} \right),
\end{equation}
whereas two corrections would be needed for the vertex at discrete coordinates ${\left( i + 2, j, k \right)}$:

\begin{equation}
\Phi_{i + 2, j, k}^{coarse \left( new \right)} = \Phi_{i + 2, j, k}^{coarse \left( old \right)} + \frac{\Delta t_{coarse}}{\Delta x_{coarse}} \delta F^{new} \left( \Phi_{i + \frac{1}{2}, j, k} \right) - \frac{\Delta t_{coarse}}{\Delta x_{coarse}} \delta F^{new} \left( \Phi_{i + \frac{3}{2}, j, k} \right),
\end{equation}
etc. An example of a refinement of a two-dimensional spatial hypergraph with a regular quadrilateral grid structure into a collection of 2-by-2 two-dimensional grids is shown in Figure \ref{fig:Figure38}, along with the corresponding configuration of coarse vertices for which the finite-difference evolution scheme must be modified to account for the adjacencies between coarse and fine subhypergraphs in Figure \ref{fig:Figure39}.

In order to extrapolate the boundary values of the state vector ${\Phi}$, i.e. ${\Phi_{i \pm \frac{1}{2}, j, k}}$, ${\Phi_{i, j \pm \frac{1}{2}, k}}$ and ${\Phi_{i, j, k \pm \frac{1}{2}}}$ (and hence also to compute the inter-vertex flux functions ${F \left( \Phi_{i \pm \frac{1}{2}, j, k} \right)}$, ${G \left( \Phi_{i, j \pm \frac{1}{2}, k} \right)}$ and ${H \left( \Phi_{i, j, k \pm \frac{1}{2}} \right)}$), we choose to use a \textit{weighted, essentially non-oscillatory} (WENO) spatial reconstruction scheme\cite{jiang}\cite{jiang2} with the required fourth-order accuracy. For this purpose, we first choose a set of ${M + 1}$ linearly-independent polynomials in the \textit{nodal basis}, denoted ${\left\lbrace \Psi_{l} \right\rbrace_{l = 1}{M + 1}}$; the Laguerre polynomials ${\ell_{j} \left( x \right)}$ constitute a natural such choice:

\begin{equation}
\ell_j \left( x \right) = \prod_{1 \leq m \leq M + 1, m \neq j} \left( \frac{x - x_m}{x_j - x_m} \right) = \left( \frac{\left( x - x_1 \right)}{\left( x_j - x_1 \right)} \right) \cdots \left( \frac{\left( x - x_{j - 1} \right)}{\left( x_j - x_{j - 1} \right)} \right) \left( \frac{\left( x - x_{j  + 1} \right)}{\left( x_j - x_{j + 1} \right)} \right) \cdots \left( \frac{\left( x - x_{M + 1} \right)}{\left( x_j - x_{M + 1} \right)} \right),
\end{equation}
where ${1 \leq j \leq M + 1}$, interpolating between the set of ${M + 1}$ \textit{nodal points} ${\left( x_j, y_j, z_j \right)}$ from ${\left( x_1, y_1, z_1 \right)}$ to ${\left( x_{M + 1}, y_{M + 1}, z_{M + 1} \right)}$ in such a way that no two ${x_j}$ are ever the same. In this context, the term \textit{nodal basis} indicates that the polynomials must all be of degree $M$ (as the Laguerre polynomials are), and that the basis functions have been rescaled so as to fit within the unit reference interval ${\left[ 0, 1 \right]}$ using the coordinate transformation ${\left( x, y, z \right) \to \left( \xi, \eta, \zeta \right)}$:

\begin{equation}
\xi \left( x, i \right) = \frac{1}{\Delta x_i} \left( x - x_{i - \frac{1}{2}} \right), \qquad \eta \left( y, j \right) = \frac{1}{\Delta y_j} \left( y - y_{j - \frac{1}{2}} \right), \qquad \zeta \left( z, k \right) = \frac{1}{\Delta z_k} \left( z - z_{k - \frac{1}{2}} \right).
\end{equation}
We denote the ${M + 1}$ nodal points ${\left( x_j, y_j, z_j \right)}$ using the shorthand ${\left\lbrace x_k \right\rbrace_{k = 1}^{M + 1}}$, such that:

\begin{equation}
\psi_{l} \left( x_k \right) = \delta_{l k}, \qquad \text{ where } \qquad l, k = 1, 2, \dots, M + 1.
\end{equation}
Using the numerical stencils ${\mathcal{S}_{i, j, k}}$, defined for each of the Cartesian coordinate directions:

\begin{equation}
\mathcal{S}_{i, j, k}^{s, x} = \bigcup_{e = i - L}^{i + R} I_{e, j, k}, \qquad \mathcal{S}_{i, j, k}^{s, y} = \bigcup_{e = j - L}^{j + R} I_{i, e, k}, \qquad \mathcal{S}_{i, j, k}^{s, z} = \bigcup_{e = k - L}^{k + R} I_{i, j, e},
\end{equation}
where ${I_{i, j, k}}$ denotes the vertex with discrete coordinates ${\left( i, j, k \right)}$, and where $L$ and $R$ designate the spatial extent of the numerical stencil to the left and right, respectively, it is now possible to perform the required spatial reconstruction. Since we wish in this particular case for the spatial reconstruction to be fourth-order accurate (i.e. ${M + 1 = 4}$), this implies that the basis polynomials $M$ will all be of odd degree (i.e. ${M = 3}$), and so four numerical stencils are used: two centered (with ${s = 0}$, ${L = \left\lfloor \frac{M}{2} \right\rfloor + 1}$, ${R = \left\lfloor \frac{M}{2} \right\rfloor}$ and ${s = 1}$, ${L = \left\lfloor \frac{M}{2} \right\rfloor}$, ${R = \left\lfloor \frac{M}{2} \right\rfloor + 1}$, respectively), one left-sided (with ${s = 2}$, ${L = M}$, ${R = 0}$) and one right-sided (with ${s = 3}$, ${L = 0}$, ${R = M}$). The spatial reconstruction procedure itself works by taking the second-order boundary extrapolation step from the SLIC/MUSCL-Hancock finite-volume approaches\cite{sweby}\cite{toro}, namely (assuming extrapolation in the $x$ coordinate direction):

\begin{equation}
\Phi_{i + \frac{1}{2}, j, k}^{L} = \Phi_{i, j, k}^{n} + \frac{1}{2} \phi_{i - \frac{1}{2}, j, k}^{+} \left( \Phi_{i, j, k}^{n} - \Phi_{i - 1, j, k}^{n} \right),
\end{equation}
for left boundary extrapolation and:

\begin{equation}
\Phi_{i + \frac{1}{2}, j, k}^{R} = \Phi_{i + 1, j, k}^{n} - \frac{1}{2} \phi_{i + \frac{3}{2}, j, k}^{-} \left( \Phi_{i + 2, j, k}^{n} - \Phi_{i + 1, j, k} \right),
\end{equation}
for right boundary extrapolation, with (diagonal) limiter functions ${\phi^{\pm}}$, and replacing the linearized approximation scheme on the right-hand side with a higher-degree spatial reconstruction polynomial ${\mathbf{w}_{h}^{s, x} \left( x, t^n \right)}$. This polynomial (again assuming reconstruction in the $x$ coordinate direction) may be expanded in terms of the nodal basis polynomials ${\Psi_{l} \left( \xi \right)}$ as:

\begin{equation}
\mathbf{w}_{h}^{s, x} \left( x, t^n \right) = \sum_{p = 0}^{M} \Psi_{p} \left( \xi \right) \hat{\mathbf{w}}_{i, j, k, p}^{n, s} = \Psi_{p} \left( \xi \right) \hat{\mathbf{w}}_{i, j, k, p}^{n, s},
\end{equation}
where we have made use of the Einstein summation convention within the second equality. By imposing the weak (i.e. integral) form of the conservation equations across every vertex within the numerical stencil ${\mathcal{S}_{i, j, k}^{s, x}}$, we obtain the following (linear) system of integral equations:

\begin{equation}
\forall I_{e, j, k} \in \mathcal{S}_{i, j, k}^{s, x}, \qquad \frac{1}{\Delta x_e} \int_{x_{e - \frac{1}{2}}}^{x_{e + \frac{1}{2}}} \Psi_{p} \left( \xi \left( x \right) \right) \hat{\mathbf{w}}_{i, j, k, p}^{n, s} dx = \bar{\Phi}_{e, j, k}^{n},
\end{equation}
where ${\bar{\Phi}_{i, j, k}^{n}}$ denotes the (spatial) average of the solution vector ${\Phi}$, when integrated over the vertex ${I_{i, j, k}}$ at time ${t^n}$. The coefficients ${\hat{\mathbf{w}}_{i, j, k, p}^{n, s}}$ of the reconstruction polynomials for each stencil may thus be obtained by solving the resulting linear system, and so the reconstruction for the entire vertex can be performed by taking the following non-linear combination of the polynomials for each stencil:

\begin{equation}
\mathbf{w}_{h}^{x} = \Psi_{p} \left( \xi \right) \hat{\mathbf{w}}_{i, j, k, p}^{n}, \qquad \text{ where } \qquad \hat{\mathbf{w}}_{i, j, k, p}^{n} = \sum_{s = 1}^{N_s} \omega_s \hat{\mathbf{w}}_{i, j, k, p}^{n, s},
\end{equation}
where we set ${N_s = 4}$ since the polynomial degree $M$ is odd, and the data-dependent non-linear weights ${\omega_s}$ are defined by:

\begin{equation}
\omega_s = \frac{\tilde{\omega}_s}{\sum\limits_q \tilde{\omega}_q}, \qquad \text{ where } \qquad \tilde{\omega}_s = \frac{\lambda_s}{\left( \sigma_s + \epsilon \right)^r}.
\end{equation}
In the above, we choose ${\sigma_s}$ to be an \textit{oscillation indicator function}:

\begin{equation}
\sigma_s = \Sigma_{l m} \tilde{\mathbf{w}}_{l}^{n, s} \tilde{\mathbf{w}}_{m}^{n, s},
\end{equation}
where ${\Sigma_{l m}}$ is an \textit{oscillation indicator matrix}\cite{dumbser}:

\begin{equation}
\Sigma_{l m} = \sum_{\alpha = 1}^{M} \int_{0}^{1} \left( \frac{\partial^{\alpha} \Phi_l \left( \xi \right)}{\partial \xi^{\alpha}} \right) \left( \frac{\partial^{\alpha} \Psi_m \left( \xi \right)}{\partial \xi^{\alpha}} \right) d \xi,
\end{equation}
so as to damp the effects of spurious numerical oscillations in the solution (hence making the resulting scheme \textit{essentially non-oscillatory}, as required). Based on the outcomes of numerical experimentation, for the remainder of this article we choose ${\lambda_s = 1}$ and ${\lambda_s = 10^5}$ for one-sided and centered stencils, respectively, ${\epsilon = 10^{-4}}$ (to prevent division by zero in the case of smooth solutions) and ${r = 8}$. For a more complete description of the hypergraph-based numerical algorithms used within this article, we invite the reader to consult \cite{gorard4}.

\section{Massive Scalar Field Collapse to a Non-Rotating Schwarzschild Black Hole}
\label{sec:Section2}

Our approach here will be to follow the analysis of Gon\c{c}alves and Moss\cite{goncalves}, in which the WKB approximation of Wentzel\cite{wentzel}, Kramers\cite{kramers} and Brillouin\cite{brillouin} is applied to the case of a minimally-coupled massive scalar field in spherical symmetry, in order to show that, in the limit of a scalar field of infinite mass, the resulting spacetime geometry is described by the Lema\^itre-Tolman-Bondi metric\cite{lemaitre}\cite{tolman}\cite{bondi} for a non-rotating inhomogeneous (collapsing or expanding) dust. From this analysis, it becomes possible to deduce sufficient conditions on the field parameters for a spherically-symmetric massive scalar field ``bubble collapse'' problem to yield the same idealized stellar collapse solution as that analyzed by Oppenheimer and Snyder\cite{oppenheimer}. These conditions may be derived analytically for the toy case of an idealized ``top hat'' initial density distribution of the scalar field, but for the more physically plausible case of an exponential initial density distribution, a numerical approach must be adopted instead. To a great extent, the analysis is simplified by the fact that Birkhoff's theorem guarantees that the exterior spacetime geometry for a spherically-symmetric ball of collapsing dust and the exterior spacetime geometry for the resulting uncharged, non-rotating black hole are identical: they are both described by the Schwarzschild metric. This allows us to neglect any considerations regarding the existence of a smooth transition between the relevant geometries (in contrast to the axially-symmetric case).

We begin by considering the line element for the most general possible spherically-symmetric metric, i.e. the line element for a spacetime whose isometry group contains a subgroup of the special orthogonal group ${SO \left( 3 \right)}$, such that the orbits of this group are all 2-spheres, namely:

\begin{equation}
ds^2 = -e^{2 \Psi \left( \tau, r \right)} d \tau^2 + e^{-2 \Lambda \left( \tau, r \right)} dr^2 + R^2 \left( \tau, r \right) d \Omega^2,
\end{equation}
where $r$ is the ordinary radial coordinate, ${R \left( \tau, r \right)}$ is the circumferential radial coordinate (such that the proper circumference at radius ${R \left( \tau, r \right)}$ is always ${2 \pi R \left( \tau, r \right)}$), ${\tau}$ is the proper time, ${\Psi \left( \tau, r \right)}$ and ${\Lambda \left( \tau, r \right)}$ are arbitrary functions, and ${d \Omega^2}$ designates the induced metric on the 2-sphere, i.e:

\begin{equation}
d \Omega^2 = d \theta^2 + \sin \left( \theta \right) d \phi^2,
\end{equation}
for colatitude coordinate ${\theta}$ and longitude coordinate ${\phi}$. If we rescale the proper time coordinate ${\tau}$ to correspond instead to the proper time $t$ for an observer that is comoving with the radial coordinate $r$:

\begin{equation}
t = \int_{0}^{\tau} e^{\Psi \left( \tau^{\prime}, r \right)} d \tau^{\prime},
\end{equation}
then we can rewrite the metric line element in Gaussian polar coordinates as:

\begin{equation}
ds^2 = -dt^2 + e^{- 2 \Lambda \left( t, r \right)} dr^2 + R^2 \left( t, r \right) d \Omega^2.
\end{equation}
The non-zero components of the Einstein curvature tensor ${G_{\mu \nu}}$ for such a spacetime are then given, up to redundancies due to symmetry, by the time-time component ${G_{t t}}$:

\begin{multline}
G_{t t} = \frac{1}{R^{2} \left( t, r \right)} \left[ - R \left( t, r \right) e^{2 \Lambda \left( t, r \right)} \left( 2 \partial_r R \left( t, r \right) \partial_r \Lambda \left( t, r \right) + 2 \partial_{r r} R \left( t, r \right) + \frac{1}{R \left( t, r \right)} \left( \partial_r R \left( t, r \right) \right)^2 \right) \right.\\
\left. - 2 \partial_t R \left( t, r \right) \partial_t \Lambda \left( t, r \right) R \left( t, r \right) + 1 + \left( \partial_t R \left( t, r \right) \right)^2 \right],
\end{multline}
the radial-time component ${G_{r t}}$:

\begin{equation}
G_{r t} = - \frac{2}{R \left( t, r \right)} \left[ \partial_{r t} R \left( t, r \right) + \partial_r R \left( t, r \right) \partial_t \Lambda \left( t, r \right) \right],
\end{equation}
the radial-radial component ${G_{r r}}$:

\begin{equation}
G_{r r} = - \frac{1}{R^{2} \left( t, r \right)} \left[ e^{-2 \Lambda \left( t, r \right)} \left( 2 \partial_{t t} R \left( t, r \right) R \left( t, r \right) + \left( \partial_t R \left( t, r \right) \right)^2 + 1 \right) - \left( \partial_r R \left( t, r \right) \right)^2 \right],
\end{equation}
and the colatitude-colatitude component ${G_{\theta \theta}}$ (or, equivalently, the longitude-longitude component ${G_{\phi \phi}}$):

\begin{multline}
G_{\theta \theta} = \frac{G_{\phi \phi}}{\sin^{2} \left( \theta \right)} = R \left( t, r \right) \left[ \partial_t R \left( t, r \right) \partial_t \Lambda \left( t, r \right) + \partial_r R \left( t, r \right) \partial_r \Lambda \left( t, r \right) e^{2 \Lambda \left( t, r \right)} + \partial_{r r} R \left( t, r \right) e^{2 \Lambda \left( t, r \right)} \right.\\
\left. - \partial_{t t} R \left( t, r \right) + \partial_{t t} \Lambda \left( t, r \right) R \left( t, r \right) - \left( \partial_t \Lambda \left( t, r \right) \right)^2 R \left( t, r \right) \right].
\end{multline}

As stated previously, we take as our matter content a minimally-coupled, real-valued massive scalar field ${\Phi \left( t, r \right)}$ with mass ${\mu}$, defined by the equation of motion:

\begin{equation}
\left( - \nabla_{\mu} \nabla^{\mu} + \mu^2 \right) \Phi \left( t, r \right) = \left( - \Box + \mu^2 \right) \Phi \left( t, r \right) = 0,
\end{equation}
where ${\Box}$ designates the covariant wave operator:

\begin{equation}
\Box = \nabla_{\mu} \nabla^{\mu}.
\end{equation}
When written out explicitly in terms of our spherically-symmetric metric in Gaussian polar coordinates, this equation of motion becomes:

\begin{multline}
\partial_{t t} \Phi \left( t, r \right) - e^{2 \Lambda \left( t, r \right)} \partial_{r r} \Phi \left( t, r \right) + \mu^2 \Phi \left( t, r \right) + \left[ \partial_t \Lambda \left( t, r \right) + \frac{2}{R \left( t, r \right)} \partial_t R \left( t, r \right) \right] \partial_t \Phi \left( t, r \right)\\
- e^{2 \Lambda \left( t, r \right)} \left[ \partial_r \Lambda \left( t, r \right) - \frac{2}{R \left( t, r \right)} \partial_r R \left( t, r \right) \right] \partial_r \Phi \left( t, r \right) = 0,
\end{multline}
such that the non-zero components of the stress energy tensor ${T_{\mu \nu}}$ for the scalar field ${\Phi \left( t, r \right)}$ become, up to redundancies due to symmetry, the time-time component ${T_{t t}}$:

\begin{equation}
T_{t t} = \frac{1}{2} \left( \partial_t \Phi \left( t, r \right) \right)^2 + \frac{1}{2} e^{2 \Lambda \left( t, r \right)} \left( \partial_r \Phi \left( t, r \right) \right)^2 + \frac{1}{2} \mu^2 \Phi \left( t, r \right),
\end{equation}
the radial-time component ${T_{r t}}$:

\begin{equation}
T_{r t} = \partial_t \Phi \left( t, r \right) \partial_r \Phi \left( t, r \right),
\end{equation}
the mixed-index radial-radial component ${T_{r}^{r}}$:

\begin{equation}
T_{r}^{r} = \frac{1}{2} \left( \partial_t \Phi \left( t, r \right) \right)^2 + \frac{1}{2} e^{2 \Lambda \left( t, r \right)} \left( \partial_r \Phi \left( t, r \right) \right)^2 - \frac{1}{2} \mu^2 \Phi \left( t, r \right),
\end{equation}
and the colatitude-colatitude component ${T_{\theta \theta}}$ (or, equivalently, the longitude-longitude component ${T_{\phi \phi}}$):

\begin{equation}
T_{\theta \theta} = \frac{T_{\phi \phi}}{\sin^{2} \left( \theta \right)} = \frac{1}{2} R^2 \left( t, r \right) \left[ \left( \partial_t \Phi \left( t, r \right) \right)^2 - e^{2 \Lambda \left( t, r \right)} \left( \partial_r \Phi \left( t, r \right) \right)^2 - \mu^2 \Phi \left( t, r \right) \right].
\end{equation}
Following Gon\c{c}alves and Moss\cite{goncalves}, we now recast two of the Einstein field equations (corresponding to the radial-time component ${G_{r t}}$ and the radial-radial component ${G_{r r}}$ of the Einstein curvature tensor, respectively) in terms of purely first derivatives of two scalar functions, namely ${k \left( t, r \right)}$ and ${m \left( t, r \right)}$, of the form:

\begin{equation}
k \left( t, r \right) = 1 - e^{2 \Lambda \left( t, r \right)} \left( \partial_r R \left( t, r \right) \right)^2, \qquad \text{ and } \qquad m \left( t, r \right) = \frac{1}{2} R \left( t, r \right) \left[ \left( \partial_t R \left( t, r \right) \right)^2 + k \left( t, r \right) \right],
\end{equation}
thus yielding:

\begin{equation}
\partial_t k \left( t, r \right) = 8 \pi R \left( t, r \right) \partial_r R \left( t, r \right) T_{t}^{r},
\end{equation}
and:

\begin{equation}
\partial_t m \left( t, r \right) = 4 \pi R^2 \left( t, r \right) \partial_r R \left( t, r \right) T_{t}^{r} - 4 \pi R^2 \left( t, r \right) \partial_t R \left( t, r \right) T_{r}^{r},
\end{equation}
respectively. The overall system of equations can now be closed by also incorporating the aforementioned second-order equation of motion for the scalar field ${\Phi \left( t, r \right)}$ in Gaussian polar coordinates:

\begin{multline}
\partial_{t t} \Phi \left( t, r \right) - e^{2 \Lambda \left( t, r \right)} \partial_{r r} \Phi \left( t, r \right) + \mu^2 \Phi \left( t, r \right) + \left[ \partial_t \Lambda \left( t, r \right) + \frac{2}{R \left( t, r \right)} \partial_t R \left( t, r \right) \right] \partial_t \Phi \left( t, r \right)\\
- e^{2 \Lambda \left( t, r \right)} \left[ \partial_r \Lambda \left( t, r \right) - \frac{2}{R \left( t, r \right)} \partial_r R \left( t, r \right) \right] \partial_r \Phi \left( t, r \right) = 0.
\end{multline}
A third Einstein field equation (corresponding to the time-time component ${G_{t t}}$ of the Einstein curvature tensor) can further be used to derive the following constraint on the spatial derivative of ${m \left( t, r \right)}$:

\begin{equation}
\partial_r m \left( t, r \right) = 4 \pi R^2 \left( t, r \right) \partial_r R \left( t, r \right) T_{t t} - 4 \pi R^2 \left( t, r \right) \partial_t R \left( t, r \right) T_{r t},
\end{equation}
from which we are therefore able to reconstruct the initial value of ${m \left( 0, r \right)}$, given the Cauchy initial data for the spacetime defined on a spacelike hypersurface with proper time ${t = 0}$.

From here, we seek to apply the standard WKB approximation of Wentzel, Kramers and Brillouin\cite{wentzel}\cite{kramers}\cite{brillouin} (as conventionally used in semiclassical approximations to the Schr\"odinger equation in non-relativistic quantum mechanics) in order to derive a wavelike solution to the equation of motion for the scalar field ${\Phi \left( t, r \right)}$. Specifically, the WKB approximation applies whenever one has an $n$-th order differential equation with coefficients ${c_i \left( x \right)}$ depending on a spatial parameter $x$, and in which the $n$-th order derivative is multiplied by a scalar parameter ${\epsilon}$

\begin{equation}
\epsilon \frac{d^n f \left( x \right)}{d x^n} + c_{n - 1} \left( x \right) \frac{d^{n - 1} f \left( x \right)}{d x^{n - 1}} + \cdots + c_{1} \left( x \right) \frac{d f \left( x \right)}{dx} + c_{0} \left( x \right) f \left( x \right) = 0,
\end{equation}
in which case, in the limit as ${\epsilon \to 0}$, one has the following ansatz for ${f \left( x \right)}$ in the form of an asymptotic series expansion:

\begin{equation}
f \left( x \right) = \lim_{\delta \to 0} \left[ \exp \left( \frac{1}{\delta} \sum_{i = 0}^{\infty} \delta^i S_i \left( x \right) \right) \right],
\end{equation}
with expansion terms ${S_i \left( x \right)}$ yet to be determined, and in which the relative asymptotic scaling of the parameters ${\delta}$ and ${\epsilon}$ is determined by the equation in question. Therefore, if we let the parameter ${\delta}$ correspond now to the Compton wavelength of the scalar field ${\Phi \left( t, r \right)}$, i.e. ${\frac{1}{\mu}}$ (for scalar field mass ${\mu}$, assuming natural units with ${\hbar = c = 1}$), and if we assume that the function ${\Phi \left( t, r \right)}$ is of compact support, with a finite radius ${\lambda}$ in which the value of the field ${\Phi \left( t, r \right)}$ is non-vanishing, then the WKB approximation applies whenever the scalar field is sufficiently massive, such that the Compton wavelength is much less than the radius of support:

\begin{equation}
\frac{1}{\mu} \ll \lambda.
\end{equation}
This allows us to perform an asymptotic series expansion for a scalar amplitude function ${\Psi \left( t, r \right)}$, such that the following wavelike solution ansatz for ${\Phi \left( t, r \right)}$:

\begin{equation}
\Phi \left( t, r \right) = \frac{1}{\mu} \Psi \left( t, r \right) \cos \left( \mu t \right),
\end{equation}
holds, with the non-zero components of the stress-energy tensor ${T_{\mu \nu}}$ for the scalar field ${\Phi \left( t, r \right)}$ therefore given (via simple differentiation) in terms of the amplitude function ${\Psi \left( t, r \right)}$; up to redundancies due to symmetry, these are the time-time component ${T_{t t}}$:

\begin{multline}
T_{t t} = \frac{1}{2} \Psi^2 \left( t, r \right) - \frac{1}{2 \mu} \Psi \left( t, r \right) \partial_t \Psi \left( t, r \right) \sin \left( 2 \mu t \right)\\
+ \frac{1}{4 \mu^2} \left[ \left( \partial_t \Psi \left( t, r \right) \right)^2 e^{4 \Lambda \left( t, r \right)} \left( \partial_r \Psi \left( t, r \right) \right)^2 \right] \left[ 1 + \cos \left( 2 \mu t \right) \right],
\end{multline}
the radial-time component ${T_{r t}}$:

\begin{equation}
T_{r t} = - \frac{1}{2 \mu} \Psi \left( t, r \right) \partial_r \Psi \left( t, r \right) \sin \left( 2 \mu t \right) + \frac{1}{2 \mu^2} \partial_t \Psi \left( t, r \right) \partial_r \Psi \left( t, r \right) \left[ 1 + \cos \left( 2 \mu t \right) \right],
\end{equation}
the mixed-index radial-radial component ${T_{r}^{r}}$:

\begin{multline}
T_{r}^{r} = - \frac{1}{2} \Psi^2 \left( t, r \right) \cos \left( 2 \mu t \right) - \frac{1}{2 \mu} \Psi \left( t, r \right) \partial_t \Psi \left( t, r \right) \sin \left( 2 \mu t \right)\\
+ \frac{1}{4 \mu^2} \left[ \left( \partial_t \Psi \left( t, r \right) \right)^2 + e^{4 \Lambda \left( t, r \right)} \left( \partial_r \Psi \left( t, r \right) \right)^2 \right] \left[ 1 + \cos \left( 2 \mu t \right) \right],
\end{multline}
and the colatitude-colatitude component ${T_{\theta \theta}}$ (or, equivalently, the longitude-longitude component ${T_{\phi \phi}}$):

\begin{multline}
T_{\theta \theta} = \frac{T_{\phi \phi}}{\sin^2 \left( \theta \right)} = \frac{1}{2} R^2 \left( t, r \right) \left[ - \Psi^2 \left( t, r \right) \cos \left( 2 \mu t \right) - \frac{1}{\mu} \Psi \left( t, r \right) \partial_t \Psi \left( t, r \right) \sin \left( 2 \mu t \right) \right.\\
\left. + \frac{1}{2 \mu^2} \left[ \left( \partial_t \Psi \left( t, r \right) \right)^2 - e^{4 \Lambda \left( t, r \right)} \left( \partial_r \Psi \left( t, r \right) \right)^2 \right] \left[ 1 + \cos \left( 2 \mu t \right) \right] \right].
\end{multline}

If we now write out the asymptotic expansion for the scalar amplitude function ${\Psi \left( t, r \right)}$ as an explicit trigonometric series of the general form:

\begin{multline}
\Psi \left( t, r \right) = \Psi_0 \left( t, r \right) + \Psi_{0 1}^{\left( cos \right)} \left( t, r \right) \cos \left( \mu t \right) + \Psi_{0 1}^{\left( sin \right)} \left( t, r \right) \sin \left( \mu t \right)\\
+ \sum_{m = 1}^{\infty} \sum_{n = 2}^{2m} \frac{1}{\mu^m} \left[ \Psi_{m n}^{\left( cos \right)} \left( t, r \right) \cos \left( n \mu t \right) + \Psi_{m n}^{\left( sin \right)} \left( t, r \right) \sin \left( n \mu t \right) \right],
\end{multline}
for undetermined coefficient functions ${\Psi_{m n}^{\left( cos \right)} \left( t, r \right)}$ and ${\Psi_{m n}^{\left( sin \right)} \left( t, r \right)}$, then, up to second-order in the Compton wavelength expansion parameter ${\frac{1}{\mu}}$, we obtain:

\begin{equation}
\Psi \left( t, r \right) = \Psi_{0 1} \left( t, r \right) \cos \left( \mu t \right) + \frac{1}{\mu^2} \Psi_{2 2}^{\left( cos \right)} \left( t, r \right) \cos \left( 2 \mu t \right) + O \left( \frac{1}{\mu^3} \right).
\end{equation}
We can also write out the explicit trigonometric series expansions of the scalar functions ${k \left( t, r \right)}$ and ${m \left( t, r \right)}$ (as derived from the radial-time component ${G_{r t}}$ and the radial-radial component ${G_{r r}}$ component of the Einstein curvature tensor, respectively) in much the same way, yielding:

\begin{multline}
k \left( t, r \right) = k_0 \left( t, r \right) + k_{0 1}^{\left( cos \right)} \left( t, r \right) \cos \left( \mu t \right) + k_{0 1}^{\left( sin \right)} \left( t, r \right) \sin \left( \mu t \right)\\
+ \sum_{m = 1}^{\infty} \sum_{n = 2}^{2 m} \frac{1}{\mu^m} \left[ k_{m n}^{\left( cos \right)} \left( t, r \right) \cos \left( n \mu t \right) + k_{m n}^{\left( sin \right)} \left( t, r \right) \sin \left( n \mu t \right) \right],
\end{multline}
for coefficient functions ${k_{m n}^{\left( cos \right)} \left( t, r \right)}$ and ${k_{m n}^{\left( sin \right)} \left( t, r \right)}$, and:

\begin{multline}
m \left( t, r \right) = m_0 \left( t, r \right) + m_{0 1}^{\left( cos \right)} \left( t, r \right) \cos \left( \mu t \right) + m_{0 1}^{\left( sin \right)} \left( t, r \right) \sin \left( \mu t \right)\\
+ \sum_{m = 1}^{\infty} \sum_{n = 2}^{2m} \frac{1}{\mu^m} \left[ m_{m n}^{\left( cos \right)} \left( t, r \right) \cos \left( n \mu t \right) + m_{m n}^{\left( sin \right)} \left( t, r \right) \sin \left( n \mu t \right) \right],
\end{multline}
for coefficient functions ${m_{m n}^{\left( cos \right)} \left( t, r \right)}$ and ${m_{m n}^{\left( sin \right)} \left( t, r \right)}$, respectively, from which we obtain (up to second order in the expansion parameter ${\frac{1}{\mu}}$):

\begin{equation}
k \left( t, r \right) = k_0 \left( t, r \right) + \frac{1}{\mu^2} k_{2 2}^{\left( cos \right)} \left( t, r \right) \cos \left( 2 \mu t \right) + O \left( \frac{1}{\mu^3} \right),
\end{equation}
and:

\begin{equation}
m \left( t, r \right) = m_0 \left( t, r \right) + \frac{1}{\mu} m_{1 2}^{\left( sin \right)} \left( t, r \right) \sin \left( 2 \mu t \right) + \frac{1}{\mu^2} m_{2 2}^{\left( cos \right)} \left( t, r \right) \cos \left( 2 \mu t \right) + O \left( \frac{1}{\mu^3} \right),
\end{equation}
respectively. Due to the known definitional relationship between the functions ${k \left( t, r \right)}$ and the ${m \left( t, r \right)}$, and the circumferential radial coordinate ${R \left( t, r \right)}$, we can therefore deduce an analogous expansion for the ${R \left( t, r \right)}$ coordinate up to second order in the expansion parameter ${\frac{1}{\mu}}$ also:

\begin{equation}
R \left( t, r \right) = R_0 \left( t, r \right) + \frac{1}{\mu^2} R_{2 2}^{\left( cos \right)} \left( t, r \right) \cos \left( 2 \mu t \right) + O \left( \frac{1}{\mu^3} \right).
\end{equation}

At this point, we are able to substitute these explicit trigonometric expansions for the scalar functions ${k \left( t, r \right)}$ and ${m \left( t, r \right)}$ into the first-order differential equations derived above, namely:

\begin{equation}
\partial_t k \left( t, r \right) = 8 \pi R \left( t, r \right) \partial_r R \left( t, r \right) T_{t}^{r},
\end{equation}
and:

\begin{equation}
\partial_t m \left( t, r \right) = 4 \pi R^2 \left( t, r \right) \partial_r R \left( t, r \right) T_{t}^{r} - 4 \pi R^2 \left( t, r \right) \partial_t R \left( t, r \right) T_{r}^{r},
\end{equation}
respectively, yielding simply, up to first-order in the expansion parameter ${\frac{1}{\mu}}$:

\begin{equation}
\partial_t k_0 \left( t, r \right) = 0, \qquad \text{ and } \qquad \partial_t m_0 \left( t, r \right) = 0,
\end{equation}
respectively. Moreover, we can rearrange the definition of the scalar function ${m \left( t, r \right)}$ in order to obtain a first-order differential equation for the circumferential radial coordinate ${R \left( t, r \right)}$ as well:

\begin{equation}
m \left( t, r \right) = \frac{1}{2} R \left( t, r \right) \left[ \left( \partial_t R \left( t, r \right) \right)^2 + k \left( t, r \right) \right], \qquad \implies \qquad \partial_t R \left( t, r \right) = \pm \frac{\sqrt{2 m \left( t, r \right) - k \left( t, r \right) R \left( t, r \right)}}{\sqrt{R \left( t, r \right)}},
\end{equation}
into which we can substitute its own trigonometric expansion in much the same way, obtaining (up to first-order in the expansion parameter ${\frac{1}{\mu}}$):

\begin{equation}
\partial_t R_0 \left( t, r \right) = \pm \sqrt{\frac{2 m_0 \left( t, r \right)}{R \left( t, r \right)} - k_0 \left( t, r \right)}.
\end{equation}
On the other hand, if we substitute the trigonometric expansion for the coordinate ${R \left( t, r \right)}$ into the constraint equation for the spatial derivative of the function ${m \left( t, r \right)}$ that is used in the reconstruction of the initial value of ${m \left( 0, r \right)}$ from the Cauchy data, namely:

\begin{equation}
\partial_r m \left( t, r \right) = 4 \pi R^2 \left( t, r \right) \partial_r R \left( t, r \right) T_{t t} - 4 \pi R^2 \left( t, r \right) \partial_t R \left( t, r \right) T_{r t},
\end{equation}
then we also find (up to first-order in the expansion parameter ${\frac{1}{\mu}}$):

\begin{equation}
\partial_r m_0 \left( t, r \right) = 2 \pi R_{0}^{2} \left( t, r \right) \partial_r R_0 \left( t, r \right) \Psi_{0}^{2} \left( t, r \right).
\end{equation}
Therefore, we infer from the definitions of the scalar functions ${k \left( t, r \right)}$ and ${m \left( t, r \right)}$ that, up to first-order in ${\frac{1}{\mu}}$, the overall metric line element is of the following form (in Gaussian polar coordinates):

\begin{equation}
d s^2 = - dt^2 + \frac{\left( \partial_r R_0 \left( t, r \right) \right)^2}{1 - k_0 \left( t, r \right)} dr^2 + R_{0}^{2} \left( t, r \right) d \Omega^2, \qquad \text{ with } \qquad d \Omega^2 = d \theta^2 + \sin \left( \theta \right) d \phi^2,
\end{equation}
with the scalar field amplitude parameter ${\Psi_0 \left( t, r \right)}$ given by:

\begin{equation}
\Psi_0 \left( t, r \right) = \pm \sqrt{ \frac{\partial_r m_0 \left( t, r \right)}{2 \pi R_{0}^{2} \left( t, r \right) \partial_r R_0 \left( t, r \right)}}.
\end{equation}

As pointed out by Gon\c{c}alves and Moss\cite{goncalves}, this metric has exactly the form of the Lema\^itre-Tolman-Bondi metric\cite{lemaitre}\cite{tolman}\cite{bondi} for a spherically-symmetric distribution of inhomogeneous dust that is either expanding or contracting uniformly due to gravity, namely:

\begin{equation}
d s^2 = - dt^2 + \frac{\left( \partial_r R_0 \left( t, r \right) \right)^2}{1 + 2 E \left( t, r \right)} dr^2 + R_{0}^{2} \left( t, r \right) d \Omega^2,
\end{equation}
where we have introduced the parameter ${E \left( t, r \right)}$ to represent the specific energy (i.e. energy per unit mass) of the dust particles located at the comoving coordinate radius $r$ at time $t$:

\begin{equation}
E \left( t, r \right) = - \frac{k_0 \left( t, r \right)}{2} > - \frac{1}{2},
\end{equation}
obeying the same equation of motion derived above for the scalar field amplitude ${\Psi_0 \left( t, r \right)}$, namely:

\begin{equation}
\partial_t R_{0} \left( t, r \right) = \pm \sqrt{ \frac{2 m_0 \left( t, r \right)}{R_0 \left( t, r \right)} + 2 E \left( t, r \right)}.
\end{equation}
We can verify explicitly that this indeed corresponds to a dust solution, since the pressure contribution to the stress-energy tensor vanishes identically (i.e. ${p \left( t, r \right) = 0}$), leaving only an energy density contribution ${\rho \left( t, r \right)}$ of the form:

\begin{equation}
\rho \left( t, r \right) = \frac{\partial_r m_0 \left( t, r \right)}{4 \pi R_{0}^{2} \left( t, r \right) \partial_r R_0 \left( t, r \right)},
\end{equation}
with the scalar function ${m_0 \left( t, r \right)}$ now playing the role of the total gravitational mass contained within the sphere of comoving coordinate radius $r$ at time $t$. The equation of motion for ${R_0 \left( t, r \right)}$ can now be integrated analytically in terms of the parameter ${\eta}$, revealing that it has exactly three solutions, corresponding to the cases of \textit{hyperbolic} evolution (i.e. ${E \left( t, r \right) > 0}$):

\begin{equation}
R_0 \left( t, r \right) = \frac{m_0 \left( t, r \right)}{2 E \left( t, r \right)} \left( \cosh \left( \eta \right) - 1 \right), \qquad \text{ where } \qquad \left( \sinh \left( \eta \right) - \eta \right) = \frac{\left( 2 E \left( t, r \right) \right)^{\frac{3}{2}} \left( t - t_0 \left( r \right) \right)}{m_0 \left( t, r \right)},
\end{equation}
\textit{parabolic} evolution (i.e. ${E \left( t, r \right) = 0}$):

\begin{equation}
R_0 \left( t, r \right) = \left( \frac{9 m_0 \left( t, r \right) \left( t - t_0 \left( r \right) \right)^2}{2} \right)^{\frac{1}{3}},
\end{equation}
and \textit{elliptic} evolution (i.e. ${E \left( t, r \right) < 0}$):

\begin{equation}
R_0 \left( t, r \right) = \frac{m_0 \left( t, r \right)}{2 \left\lvert E \left( t, r \right) \right\rvert} \left( 1 - \cos \left( \eta \right) \right), \qquad \text{ where } \qquad \left( \eta - \sin \left( \eta \right) \right) \frac{\left( 2 \left\lvert E \left( t, r \right) \right\rvert \right)^{\frac{3}{2}} \left( t - t_0 \left( r \right) \right)}{m_0 \left( t, r \right)},
\end{equation}
respectively, where ${t_0 \left( r \right)}$ is an arbitrary scalar function designating the initial time parameter for the world lines of dust particles located at the comoving coordinate radius $r$.

The Schwarzschild metric within a geodesic coordinate system can then be obtained as a limiting case of the Lema\^itre-Tolman-Bondi metric by setting the mass parameter ${m_0 \left( t, r \right)}$ to be constant. For instance, if we take the specific energy of the dust to vanish (i.e. ${E \left( t, r \right) = 0}$), then we effectively transform the Schwarschild metric in the Schwarzschild coordinate system ${\left\lbrace t, r \right\rbrace}$ (with Schwarzschild radius ${r_s = 2 m_0}$):

\begin{equation}
ds^2 = - \left( 1 - \frac{r_s}{r} \right) dt^2 + \frac{d r^2}{1 - \frac{r_s}{r}} + r^2 \left( d \theta^2 + \sin^2 \left( \theta \right) d \phi^2 \right),
\end{equation}
to the geodesic coordinate system ${\left\lbrace \tau, \rho \right\rbrace}$:

\begin{equation}
d \tau = dt + \sqrt{\frac{r_s}{r}} \left( 1 - \frac{r_s}{r} \right)^{-1} dr, \qquad \text{ and } \qquad d \rho = dt + \sqrt{\frac{r}{r_s}} \left( 1 - \frac{r_s}{r} \right)^{-1} dr,
\end{equation}
in which the geodesics with constant ${\rho}$ coordinate are all timelike, parametrized by proper time ${\tau}$, corresponding to the trajectories of particles in free fall that begin at infinity with zero velocity, yielding the following form of the Schwarzschild metric in the Lema\^itre coordinate system\cite{lemaitre}:

\begin{equation}
ds^2 = - d \tau^2 + \frac{r_s}{r} d \rho^2 + r^2 \left( d \theta^2 + \sin^2 \left( \theta \right) d \phi^2 \right), \qquad \text{ where } \qquad r = \left[ \frac{3}{2} \left( \rho - \tau \right) \right]^{\frac{2}{3}} r_{s}^{\frac{1}{3}}.
\end{equation}
On the other hand, if we set the specific energy of the dust to be of the form:

\begin{equation}
E \left( t, r \right) = - \frac{1}{2 \left( 1 + r^2 \right)},
\end{equation}
then we obtain instead a coordinate system in which the Schwarzschild radial coordinate $r$ is transformed to the Novikov radial coordinate ${r_{nov}}$, dependent upon the initial radial height ${r_{max}}$ of the particle on its trajectory:

\begin{equation}
r_{nov} = \sqrt{\frac{r_{max}}{r_s} - 1},
\end{equation}
with the time coordinate ${\tau}$ still corresponding to the proper time of the falling particle, yielding the Novikov form of the Schwarzschild metric, representing the trajectories of particles in free fall that begin at an arbitrary radial distance:

\begin{equation}
ds^2 = - d \tau^2 + \frac{r_{nov} + 1}{r^2} \left( \frac{\partial r}{\partial r_{nov}} \right)^2 d r_{nov}^2 + r^2 \left( d \theta^2 + \sin^2 \left( \theta \right) d \phi^2 \right),
\end{equation}
from which the coordinates $r$ and ${\tau}$ can be recovered by solving the following (implicitly time-dependent) pair of equations for ${\eta}$:

\begin{equation}
\tau = m_0 \left( r_{nov}^{2} + 1 \right)^{\frac{3}{2}} \left( \eta + \sin \left( \eta \right) \right), \qquad \text{ and } \qquad r = m_0 \left( r_{nov}^{2} + 1 \right) \left( \eta + \cos \left( \eta \right) \right).
\end{equation}
Similarly, the Friedmann-Lema\^itre-Robertson-Walker metric\cite{friedmann}\cite{lemaitre}\cite{robertson}\cite{walker} for a homogeneous and isotropic universe:

\begin{equation}
d \tau^2 = - dt^2 + a^2 \left( t \right) \left[ \frac{dr^2}{1 - kr^2} + r^2 \left( d \theta^2 + \sin^2 \left( \theta \right) d \phi^2 \right) \right],
\end{equation}
with dimensionless scale parameter ${a \left( t \right)}$ and discrete curvature parameter $k$, is recovered in the limiting case of the Lema\^itre-Tolman-Bondi metric in which the initial time parameter ${t_0 \left( r \right)}$ is constant, with the hyperbolic, parabolic and elliptic evolution cases (corresponding to ${E \left( t, r \right) > 0}$, ${E \left( t, r \right) = 0}$ and ${E \left( t, r \right) < 0}$, respectively) yielding discrete curvature values of ${k = -1}$, ${k = 0}$ and ${k = 1}$, respectively.

Above, we showed that the trigonometric expansions for the scalar functions ${k \left( t, r \right)}$, ${m \left( t, r \right)}$ and ${R \left( t, r \right)}$, namely:

\begin{equation}
k \left( t, r \right) = k_0 \left( t, r \right) + \frac{1}{\mu^2} k_{2 2}^{\left( cos \right)} \left( t, r \right) \cos \left( 2 \mu t \right) + O \left( \frac{1}{\mu^3} \right),
\end{equation}
\begin{equation}
m \left( t, r \right) = m_0 \left( t, r \right) + \frac{1}{\mu} m_{1 2}^{\left( sin \right)} \left( t, r \right) \sin \left( 2 \mu t \right) + \frac{1}{\mu^2} m_{2 2}^{\left( cos \right)} \left( t, r \right) \cos \left( 2 \mu t \right) + O \left( \frac{1}{\mu^3} \right),
\end{equation}
and:

\begin{equation}
R \left( t, r \right) = R_0 \left( t, r \right) + \frac{1}{\mu^2} R_{2 2}^{\left( cos \right)} \left( t, r \right) \cos \left( 2 \mu t \right) + O \left( \frac{1}{\mu^3} \right),
\end{equation}
respectively, could be applied to yield, up to first-order in the expansion parameter ${\frac{1}{\mu}}$, the following conditions on the partial derivatives ${\partial_t k_0 \left( t, r \right)}$, ${\partial_t m_0 \left( t, r \right)}$, ${\partial_t R_0 \left( t, r \right)}$ and ${\partial_r m_0 \left( t, r \right)}$:

\begin{equation}
\partial_t k_0 \left( t, r \right) = 0, \qquad \partial_t m_0 \left( t, r \right) = 0, \qquad \partial_t R_0 \left( t, r \right) = \pm \sqrt{\frac{2 m_0 \left( t, r \right)}{R \left( t, r \right)} - k_0 \left( t, r \right)},
\end{equation}
and:

\begin{equation}
\partial_r m_0 \left( t, r \right) = 2 \pi R_{0}^{2} \left( t, r \right) \partial_r R_0 \left( t, r \right) \Psi_{0}^{2} \left( t, r \right),
\end{equation}
respectively. Moving now up to second-order in the expansion parameter ${\frac{1}{\mu}}$, we obtain the following sub-leading corrections to the functions ${k \left( t, r \right)}$, ${m \left( t, r \right)}$ and ${R \left( t, r \right)}$:

\begin{equation}
k_{2 2} \left( t, r \right) = 2 \pi \frac{R_0 \left( t, r \right)}{\partial_r R_0 \left( t, r \right)} \left( 1 - k_0 \left( t, r \right) \right) \Psi_0 \left( t, r \right) \partial_r \Psi \left( t, r \right),
\end{equation}

\begin{equation}
m_{1 2} \left( t, r \right) = \pi R_{0}^{2} \left( t, r \right) \partial_t R_0 \left( t, r \right) \Psi_{0}^{2} \left( t, r \right),
\end{equation}
and:

\begin{equation}
R_{2 2} \left( t, r \right) = - \frac{1}{2} R_0 \left( t, r \right) \Psi_{0}^{2} \left( t, r \right),
\end{equation}
respectively. Without loss of generality, we shall assume the validity of the WKB approximation whenever the magnitudes of the sub-leading corrections ${k_{2 2} \left( t, r \right)}$, ${m_{1 2} \left( t, r \right)}$ and ${R_{2 2} \left( t, r \right)}$ are no greater than half of the magnitudes of the leading-order terms ${k_0 \left( t, r \right)}$, ${m_0 \left( t, r \right)}$ and ${R_0 \left( t, r \right)}$ (the multiplicative factor of ${\frac{1}{2}}$ can be modified by simply changing the value of the scalar field mass parameter ${\mu}$). Since the single largest correction comes from the ${m_{1 2} \left( t, r \right)}$ contribution to ${m \left( t, r \right)}$, we obtain the following region in the ${\left( t, r \right)}$-plane:

\begin{equation}
\frac{m_{1 2} \left( t, r \right)}{\mu m_0 \left( t, r \right)} \leq \frac{1}{2},
\end{equation}
bounded by the curve:

\begin{equation}
\partial_t R_0 \left( t, r \right) \partial_r m_0 \left( t, r \right) = \mu m_0 \left( t, r \right) \partial_r R_0 \left( t, r \right),
\end{equation}
within which the WKB approximation holds. Thus, we henceforth restrict ourselves to points in our spacetime for which the entirety of the (interior) past light cone lies within this region.

Returning now to the equation of motion for the circumferential radial coordinate ${R_0 \left( t, r \right)}$ (up to first order in the expansion parameter ${\frac{1}{\mu}}$), expressed in terms of the scalar function ${k_0 \left( t, r \right)}$:

\begin{equation}
\partial_t R_0 \left( t, r \right) = \pm \sqrt{\frac{2 m_0 \left( t, r \right)}{R_0 \left( t, r \right)} - k_0 \left( t, r \right)},
\end{equation}
we use the parametric analytic integration of ${R_0 \left( t, r \right)}$ presented previously to deduce that:

\begin{equation}
t \left( \eta, r \right) = t_0 \left( r \right) + \frac{m_0 \left( t, r \right)}{k_{0}^{\frac{3}{2}} \left( t, r \right)} \left( \eta + \sin \left( \eta \right) \right), \qquad \text{ and } \qquad R_0 \left( \eta, r \right) = \frac{2 m_0 \left( t, r \right)}{k_0 \left( t, r \right)} \cos^2 \left( \frac{\eta}{2} \right),
\end{equation}
subject to the hypothesis that ${\partial_r R_0 \left( t, r \right) \neq 0}$, thus ensuring that the constant-time ``shells'' of the dust solution do not intersect, and therefore that the evolution remains globally hyperbolic. If the Cauchy initial data are time-symmetric and of the general form:

\begin{equation}
R_0 \left( 0, r \right) = r, \qquad \text{ and } \qquad \partial_t R_0 \left( t, r \right) = 0,
\end{equation}
then this fixes the choice of radial coordinate $r$, as well as the choice of initial time parameter for world lines ${t_0 \left( r \right)}$ and the scalar function ${k_0 \left( t, r \right)}$:

\begin{equation}
t_0 \left( r \right) = 0, \qquad \text{ and } \qquad k_0 \left( t, r \right) = \frac{2 m_0 \left( t, r \right)}{r},
\end{equation}
such that one has:

\begin{equation}
t \left( \eta, r \right) = \sqrt{\frac{r^3}{8 m_0 \left( t, r \right)}} \left( \eta + \sin \left( \eta \right) \right), \qquad \text{ and } \qquad R_0 \left( \eta, r \right) = r \cos^2 \left( \frac{\eta}{2} \right),
\end{equation}
where ${\eta \in \left[ 0, \pi \right]}$. Therefore, assuming an initial density profile ${\rho \left( 0, r \right)}$:

\begin{equation}
\rho \left( 0, r \right) = T_{t t} \left( 0, r \right),
\end{equation}
then the solution ${m_0 \left( t, r \right)}$ can be reconstructed (for any time $t$) by means of the following integral:

\begin{equation}
m_0 \left( t, r \right) = 4\pi \int_{0}^{r} \rho \left( 0, r^{\prime} \right) \left( r^{\prime} \right)^2 d r^{\prime} = 4 \pi \int_{0}^{r} T_{t t} \left( 0, r^{\prime} \right) \left( r^{\prime} \right)^2 d r^{\prime}.
\end{equation}
Recalling that ${m_0 \left( t, r \right)}$ designates the gravitational mass contained inside the comoving sphere of coordinate radius $r$ at time $t$, if this gravitational mass parameter converges to a finite value in the limit as ${r \to \infty}$, then this value consequently corresponds to the ADM mass $M$ of the spacetime:

\begin{equation}
\lim_{r \to \infty} \left[ m_0 \left( t, r \right) \right] = M.
\end{equation}
For the purposes of the numerical tests presented within this article, we will generally employ an exponential initial density profile of the form:

\begin{equation}
\rho \left( 0, r \right) = T_{t t} \left( 0, r \right) = \rho_0 \exp \left( - \left( \frac{r}{\lambda} \right)^3 \right),
\end{equation}
whose solution is therefore given by (after evaluating the integral for ${m_0 \left( t, r \right)}$):

\begin{equation}
m_0 \left( t, r \right) = M \left( 1 - \exp \left( - \left( \frac{r}{\lambda} \right)^3 \right) \right),
\end{equation}
where ${\rho_0}$ denotes an initial density constant, and ${\lambda}$ denotes (as previously) the radius of support, i.e. the radius within which the scalar field ${\Phi \left( t, r \right)}$ does not vanish.

We can determine the radial position of the resulting event horizon within this Lema\^itre-Tolman-Bondi solution by treating the parameter ${\eta}$ as an explicit function of the radial coordinate, i.e. ${\eta \left( r \right)}$, yielding the following equation for the radial lightlike geodesics:

\begin{equation}
\frac{d \eta \left( r \right)}{\partial r} = \frac{1}{R_0 \left( t, r \right)} \left[ \pm \left( \frac{r}{2m} - 1 \right)^{- \frac{1}{2}} \partial_r R_0 \left( t, r \right) - \frac{1}{2} \gamma \left( r \right) \left( \eta \left( r \right) + \sin \left( \eta \left( r \right) \right) \right) \right],
\end{equation}
where we have introduced a function ${\gamma \left( r \right)}$ whose form depends upon the initial mass distribution ${m_0 \left( t, r \right)}$:

\begin{equation}
\gamma \left( r \right) = \frac{3}{2} - \frac{r \partial_r m_0 \left( t, r \right)}{2 m_0 \left( t, r \right)}.
\end{equation}
Since the resulting black hole should have a Schwarzschild radius equal to twice the ADM mass ${r_s = 2M}$, we impose the following boundary condition at spatial infinity:

\begin{equation}
\lim_{r \to \infty} \left[ R_0 \left( t, r \right) \right] = 2M;
\end{equation}
moreover, we see that the mass distribution collapses to form a spacelike singularity at proper time ${t = \frac{\pi}{\sqrt{8 M}}}$. This solution allows us to compute the time and space derivatives of the circumferential radial coordinate ${R_0 \left( t, r \right)}$ analytically as:

\begin{equation}
\partial_t R_0 \left( t, r \right) = - \sqrt{\frac{2 m_0 \left( t, r \right)}{r}} \tan \left( \frac{\eta \left( r \right)}{2} \right),
\end{equation}
and:

\begin{equation}
\partial_r R_0 \left( t, r \right) = \cos^2 \left( \frac{\eta \left( r \right)}{2} \right) + \frac{1}{2} \gamma \left( r \right) \left( \eta \left( r \right) + \sin \left( \eta \left( r \right) \right) \right) \tan \left( \frac{\eta \left( r \right)}{2} \right),
\end{equation}
respectively. We require that the initial mass distribution should satisfy ${\gamma \left( r \right) \geq 0}$, since otherwise there exists the possibility that ${\partial_r R_0 \left( t, r \right) = 0}$, causing the constant-time ``shells'' of the dust solution to intersect, and thereby also causing the equation for the scalar field amplitude parameter ${\Psi_0 \left( t, r \right)}$:

\begin{equation}
\Psi_0 \left( t, r \right) = \pm \sqrt{\frac{\partial_r m_0 \left( t, r \right)}{2 \pi R_{0}^{2} \left( t, r \right) \partial_r R_0 \left( t, r \right)}},
\end{equation}
to diverge, due to a failure of global hyperbolicity.

It is instructive at this point to compare this solution against the case of a massive scalar field ${\Phi \left( t, r \right)}$ with a constant initial density profile ${\rho \left( 0, r \right) = T_{t t} \left( 0, r \right)}$, as considered by Oppenheimer and Snyder\cite{oppenheimer}, whose solution is therefore given by (after evaluating the integral for ${m_0 \left( t, r \right)}$):

\begin{equation}
m_0 \left( t, r \right) = \begin{cases}
M \left( \frac{r}{\lambda} \right)^3, \qquad & \text{ if } 0 < r < \lambda,\\
M, \qquad &\text{ if } r \geq \lambda.
\end{cases}
\end{equation}
This ``top-hat'' form of the potential causes the WKB approximation for the amplitude of the scalar field ${\Psi \left( t, r \right)}$ to break down along the boundary ${r = \lambda}$ due to the presence of the discontinuity in the solution, although this can be rectified by making the amplitude piecewise-linear near the boundary:

\begin{equation}
\Psi \left( t, r \right) = \begin{cases}
\Psi_0 \left( t, r \right), \qquad &\text{ if } r \leq \lambda - \frac{1}{\mu},\\
\Psi_1 \left( t, r \right), \qquad &\text{ if } \lambda - \frac{1}{\mu} < r < \lambda,\\
0, \qquad &\text{ if } r \geq \lambda,
\end{cases}
\end{equation}
such that ${\Psi_0 \left( t, r \right)}$ is constant and ${\Psi_1 \left( t, r \right)}$ is no steeper than linear (in $r$):

\begin{equation}
\Psi_0 \left( t, r \right) = O \left( 1 \right), \qquad \text{ and } \qquad \partial_r \Psi_1 \left( t, r \right) = O \left( 1 \right).
\end{equation}
Recall that we previously ascertained the form of the curve bounding the region within which the WKB approximation remained valid, namely:

\begin{equation}
\partial_t R_0 \left( t, r \right) \partial_r m_0 \left( t, r \right) = \mu m_0 \left( t, r \right) \partial_r R_0 \left( t, r \right).
\end{equation}
Hence, if we now parametrize this boundary curve using the scalar function ${\eta_{*} \left( r \right)}$, then we can use the explicitly-computed time and space derivatives of the circumferential radial coordinate derived above, namely:

\begin{equation}
\partial_t R_0 \left( t, r \right) = - \sqrt{\frac{2 m_0 \left( t, r \right)}{r}} \tan \left( \frac{\eta_{*} \left( r \right)}{2} \right),
\end{equation}
and:

\begin{equation}
\partial_r R_0 \left( t, r \right) = \cos^2 \left( \frac{\eta_{*} \left( r \right)}{2} \right) + \frac{1}{2} \gamma \left( r \right) \left( \eta_{*} \left( r \right) + \sin \left( \eta_{*} \left( r \right) \right) \right) \tan \left( \frac{\eta_{*} \left( r \right)}{2} \right),
\end{equation}
to obtain the following simple constraint on the parameter ${\eta_{*} \left( r \right)}$:

\begin{equation}
\tan \left( \frac{\eta_{*} \left( r \right)}{2} \right) = \frac{\mu}{3} \sqrt{\frac{\lambda^3}{2 M}} \cos^2 \left( \frac{\eta_{*} \left( r \right)}{2} \right).
\end{equation}
From this, we can conclude that the parameter ${\eta_{*} \left( r \right)}$ must be constant, and hence the proper time coordinate ${t \left( \eta_{*}, r \right)}$ must also be correspondingly constant, inside the region of support for the scalar field ${\Phi \left( t, r \right)}$ (i.e. for all ${r < \lambda}$). We also know that the Lema\^itre-Tolman-Bondi solution holds outside the region of support (i.e. for all ${r > \lambda}$), and consequently the WKB approximation must break down somewhere along a curve with a fixed proper time coordinate ${t_{*}}$ and ${r \in \left[ 0, \lambda \right]}$; thus, if we wish (as indeed we do) to guarantee that the WKB approximation holds everywhere outside the event horizon of the resulting black hole, it is necessary that this curve should lie entirely \textit{within} the Lema\^itre-Tolman-Bondi event horizon.

Outside the region of support for the scalar field ${\Phi \left( t, r \right)}$ (i.e. for ${r \geq \lambda}$), we know that the event horizon of the Lema\^itre-Tolman-Bondi solution lies along the circumferential radial line ${R = 2 M}$, and therefore, if the event horizon is described by the curve ${\eta_{EH} \left( r \right)}$, then we have:

\begin{equation}
\forall r \geq \lambda, \qquad \cos^2 \left( \frac{\eta_{EH} \left( r \right)}{2} \right) = \frac{2M}{r}.
\end{equation}
The endpoint of the curve along which the WKB approximation fails to hold will lie on the interior of the region bounded by ${\eta_{EH} \left( r \right)}$ whenever:

\begin{equation}
\eta_{*} \left( r \right) = \eta_{EH} \left( \lambda \right) \qquad \iff \qquad \tan \left( \frac{\eta_{*} \left( r \right)}{2} \right) > \sqrt{\frac{\lambda}{2M}  -1}.
\end{equation}
From the aforementioned constraint on the parameter ${\eta_{*} \left( r \right)}$, namely:

\begin{equation}
\tan \left( \frac{\eta_{*} \left( r \right)}{2} \right) = \frac{\mu}{3} \sqrt{\frac{\lambda^3}{2M}} \cos^2 \left( \frac{\eta_{*} \left( r \right)}{2} \right),
\end{equation}
we can moreover deduce that:

\begin{equation}
\tan \left( \frac{\eta_{*} \left( r \right)}{2} \right) \left( 1 + \tan^2 \left( \frac{\eta_{*} \left( r \right)}{2} \right) \right) = \frac{\mu}{3} \sqrt{\frac{\lambda^3}{2M}},
\end{equation}
yielding the desired breakdown of the WKB approximation on the interior of the Lema\^itre-Tolman-Bondi event horizon ${\eta_{EH} \left( r \right)}$ whenever the following inequality is satisfied:

\begin{equation}
\mu M > \frac{3}{2} \sqrt{1 - \frac{2M}{\lambda}}, \qquad \text{ such that } \qquad \lim_{\lambda \to \infty} \left[ \mu M \right] > \frac{3}{2}.
\end{equation}
Thus, any constant configuration of a massive scalar field ${\Phi \left( t, r \right)}$ defined within a region ${r < \lambda}$ (with ${\lambda}$ denoting the radius of support of the field), with total ADM mass $M$, must collapse to form a black hole whenever this inequality is satisfied.

\begin{figure}[ht]
\centering
\includegraphics[width=0.325\textwidth]{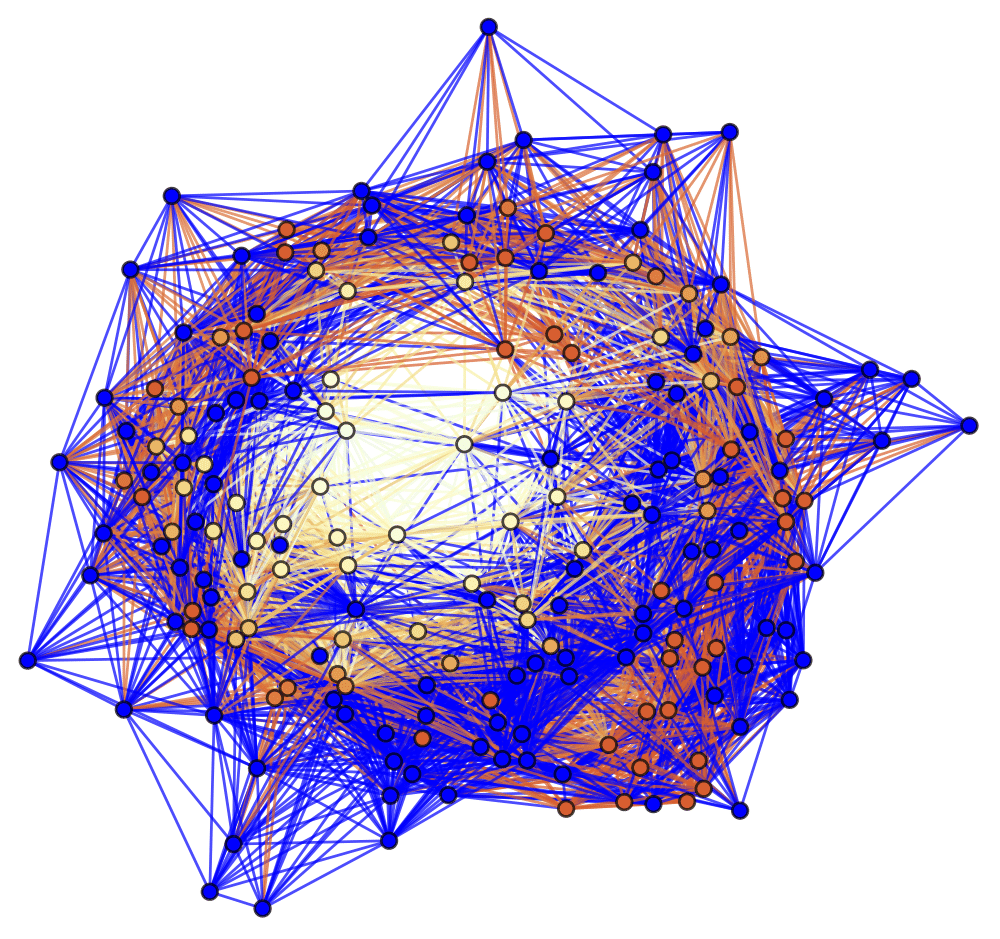}
\includegraphics[width=0.325\textwidth]{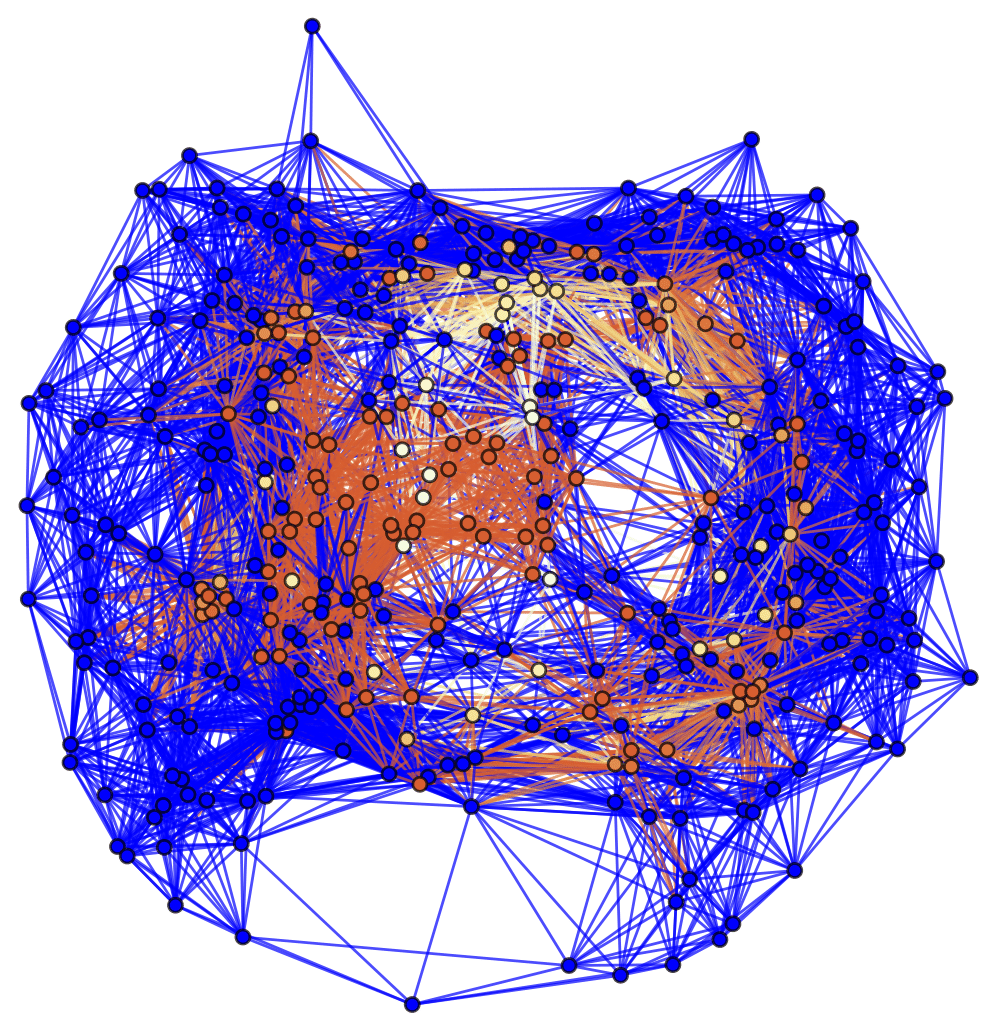}
\includegraphics[width=0.325\textwidth]{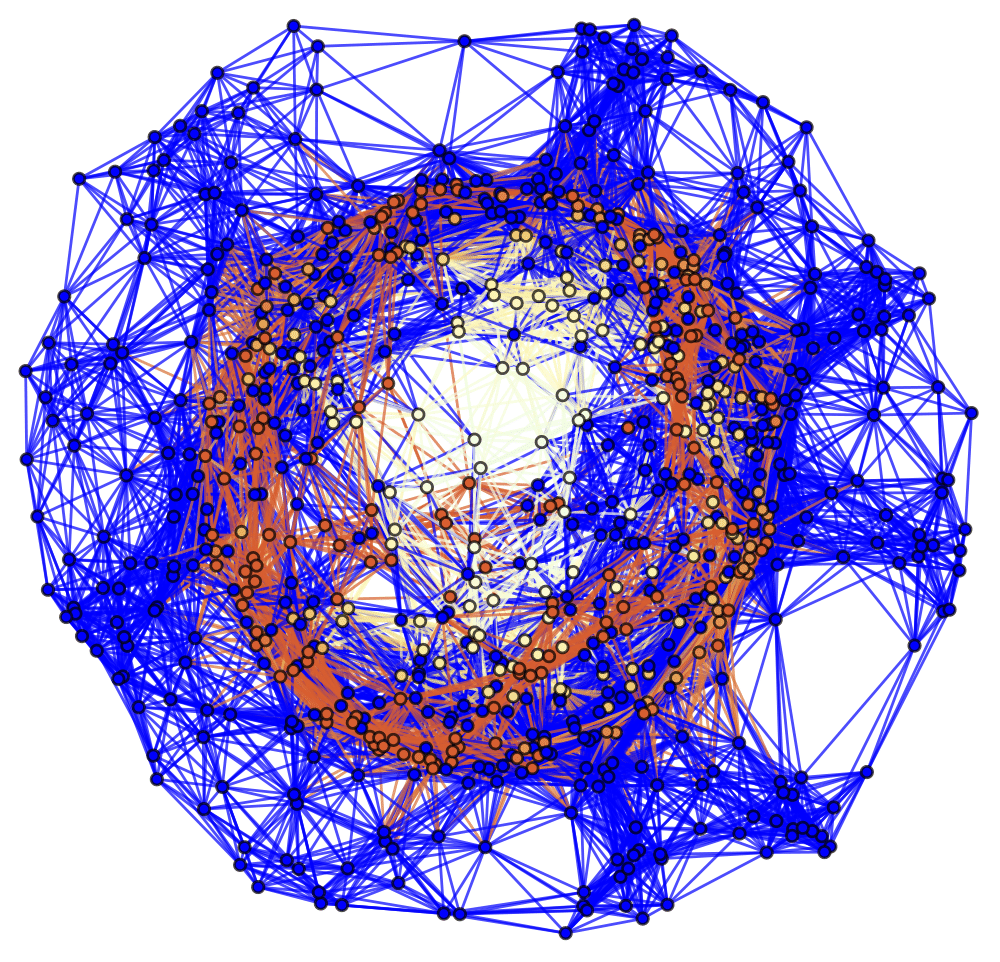}
\caption{Spatial hypergraphs corresponding to the initial hypersurface configuration of the massive scalar field ``bubble collapse'' to a non-rotating Schwarzschild black hole test, with an exponential initial density distribution, at time ${t = 0 M}$, with resolutions of 200, 400 and 800 vertices, respectively. The hypergraphs have been adapted using the local curvature in the Schwarzschild conformal factor ${\psi}$, and colored according to the value of the scalar field ${\Phi \left( t, r \right)}$.}
\label{fig:Figure1}
\end{figure}

\begin{figure}[ht]
\centering
\includegraphics[width=0.325\textwidth]{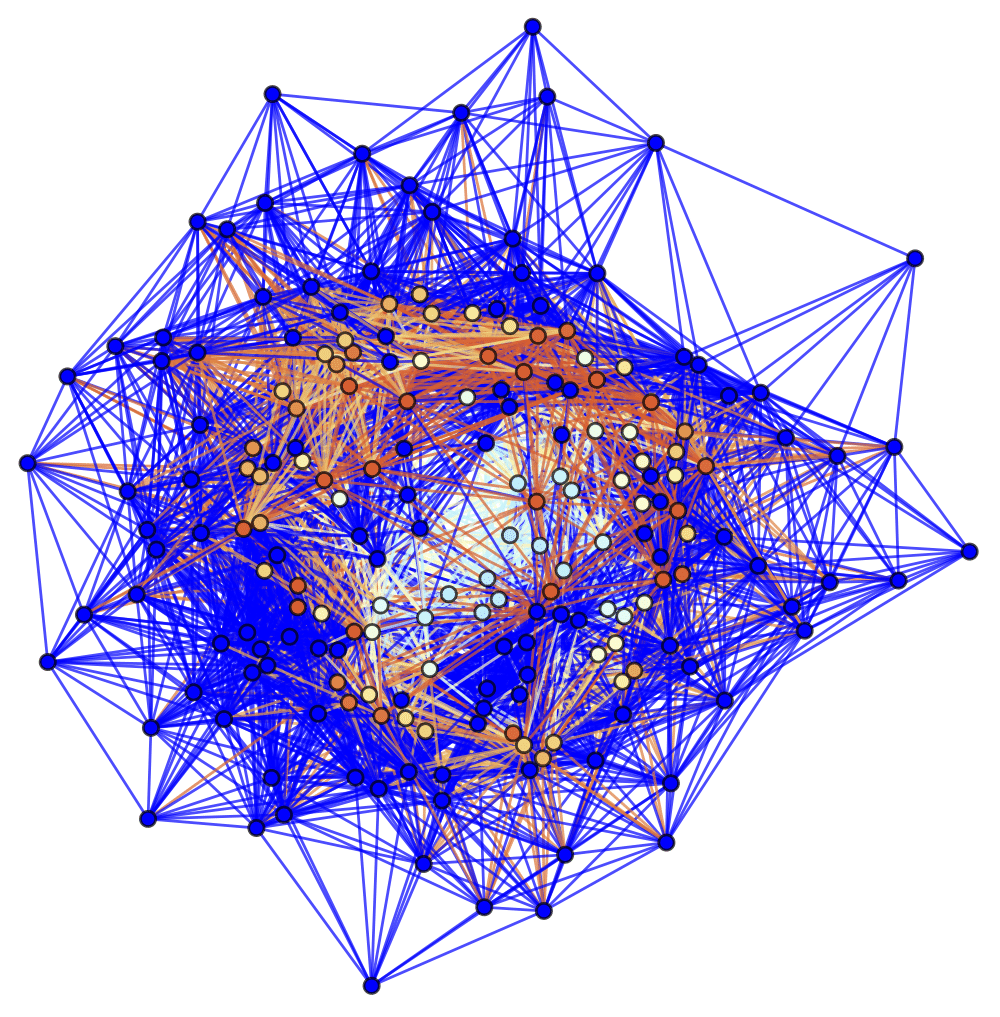}
\includegraphics[width=0.325\textwidth]{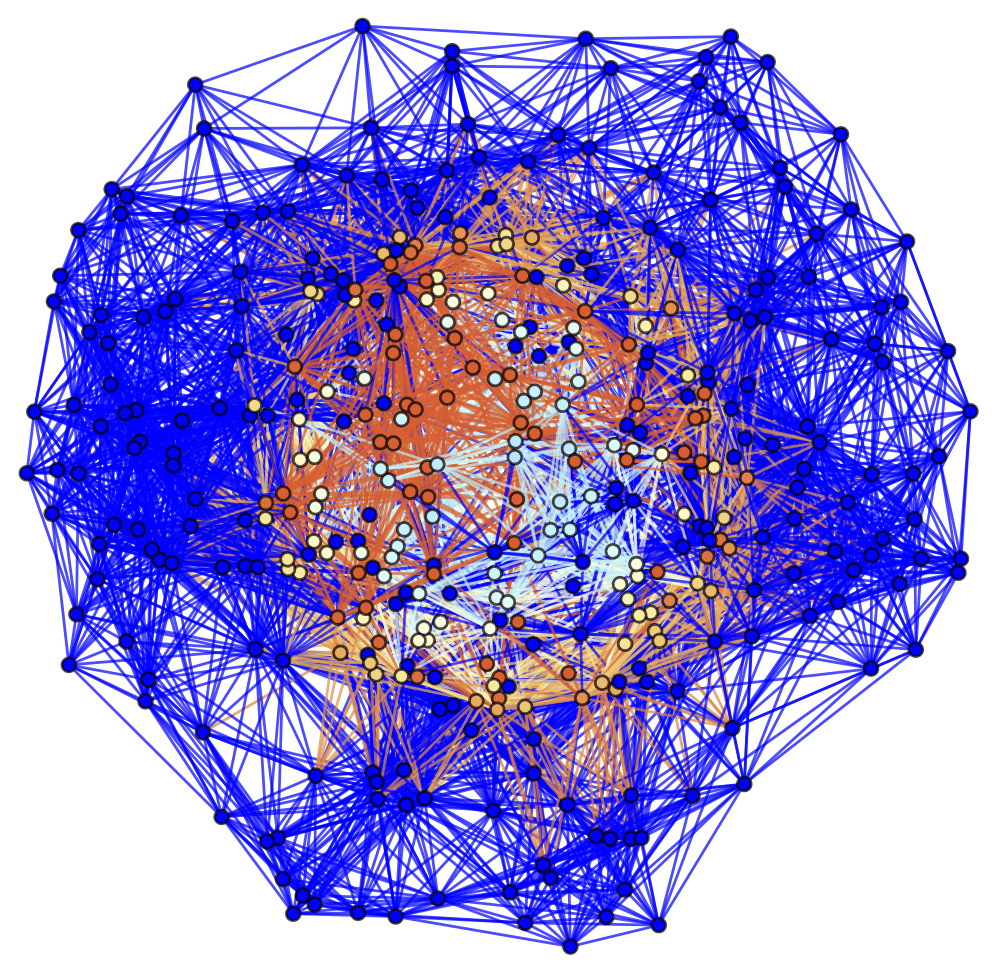}
\includegraphics[width=0.325\textwidth]{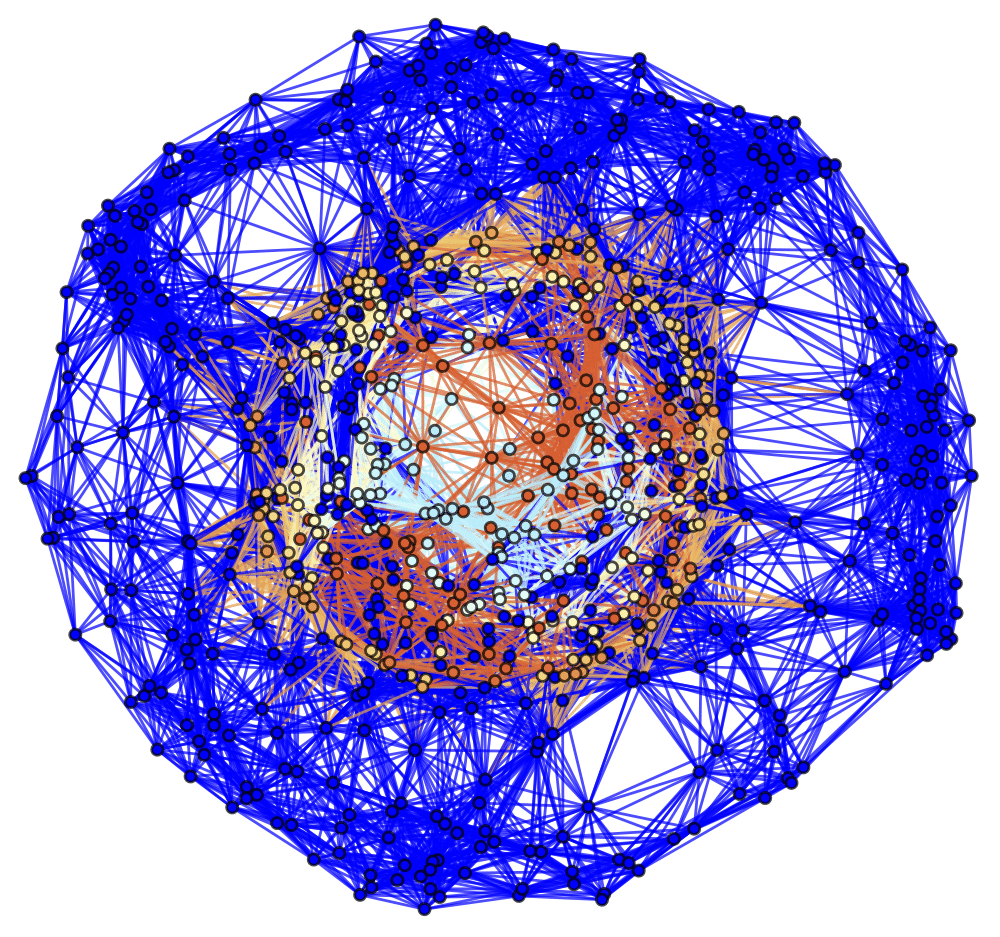}
\caption{Spatial hypergraphs corresponding to the first intermediate hypersurface configuration of the massive scalar field ``bubble collapse'' to a non-rotating Schwarzschild black hole test, with an exponential initial density distribution, at time ${t = 1.5 M}$, with resolutions of 200, 400 and 800 vertices, respectively. The hypergraphs have been adapted using the local curvature in the Schwarzschild conformal factor ${\psi}$, and colored according to the value of the scalar field ${\Phi \left( t, r \right)}$.}
\label{fig:Figure2}
\end{figure}

\begin{figure}[ht]
\centering
\includegraphics[width=0.325\textwidth]{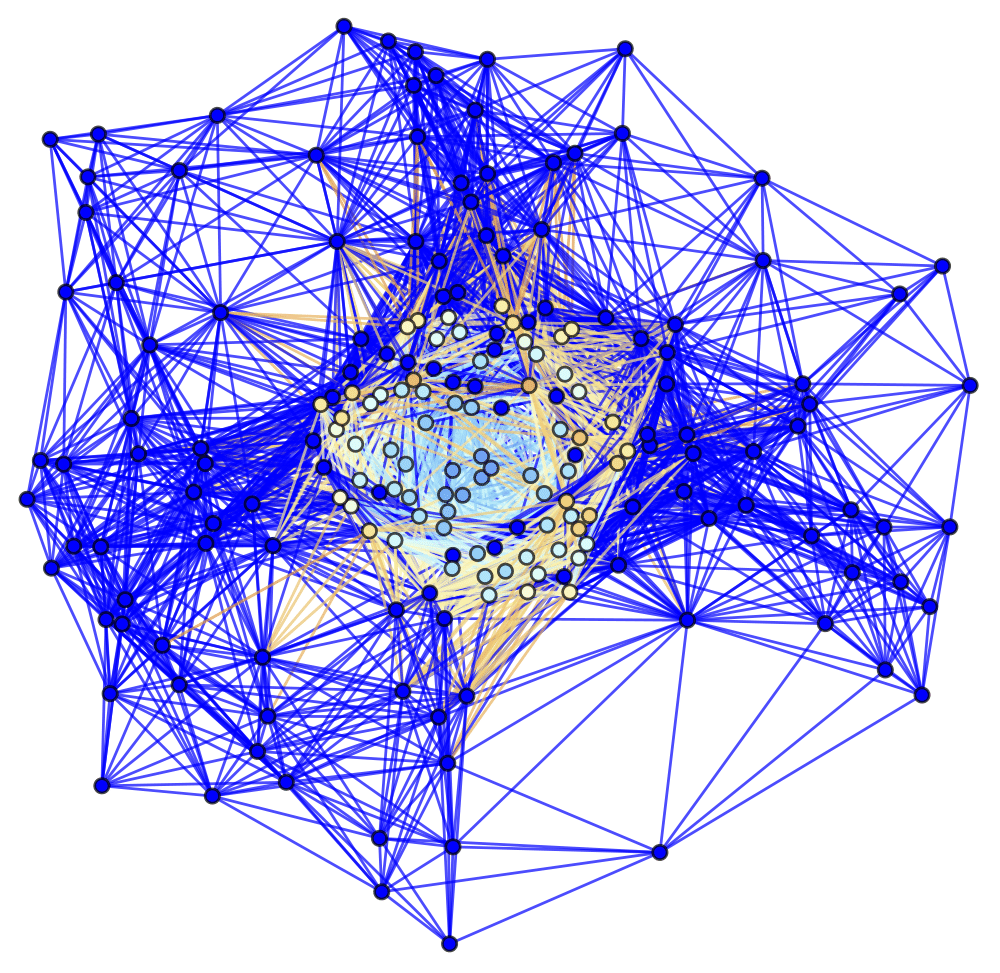}
\includegraphics[width=0.325\textwidth]{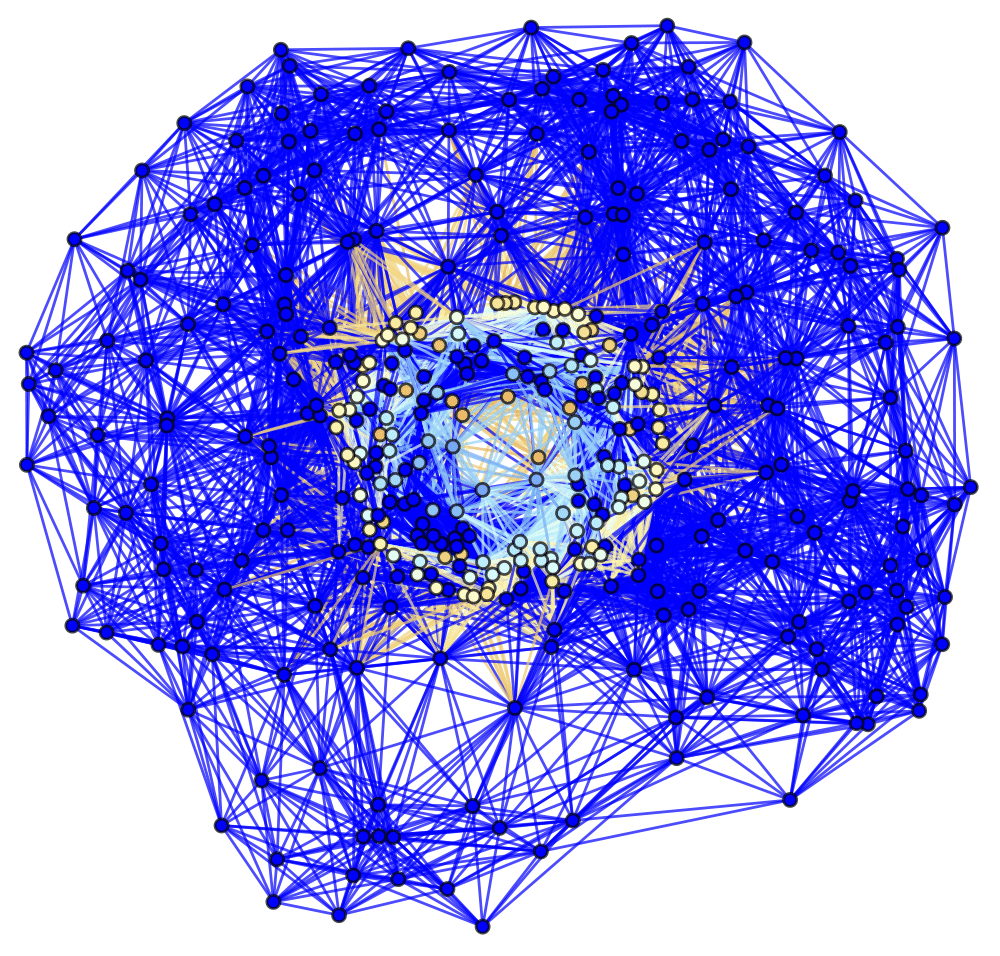}
\includegraphics[width=0.325\textwidth]{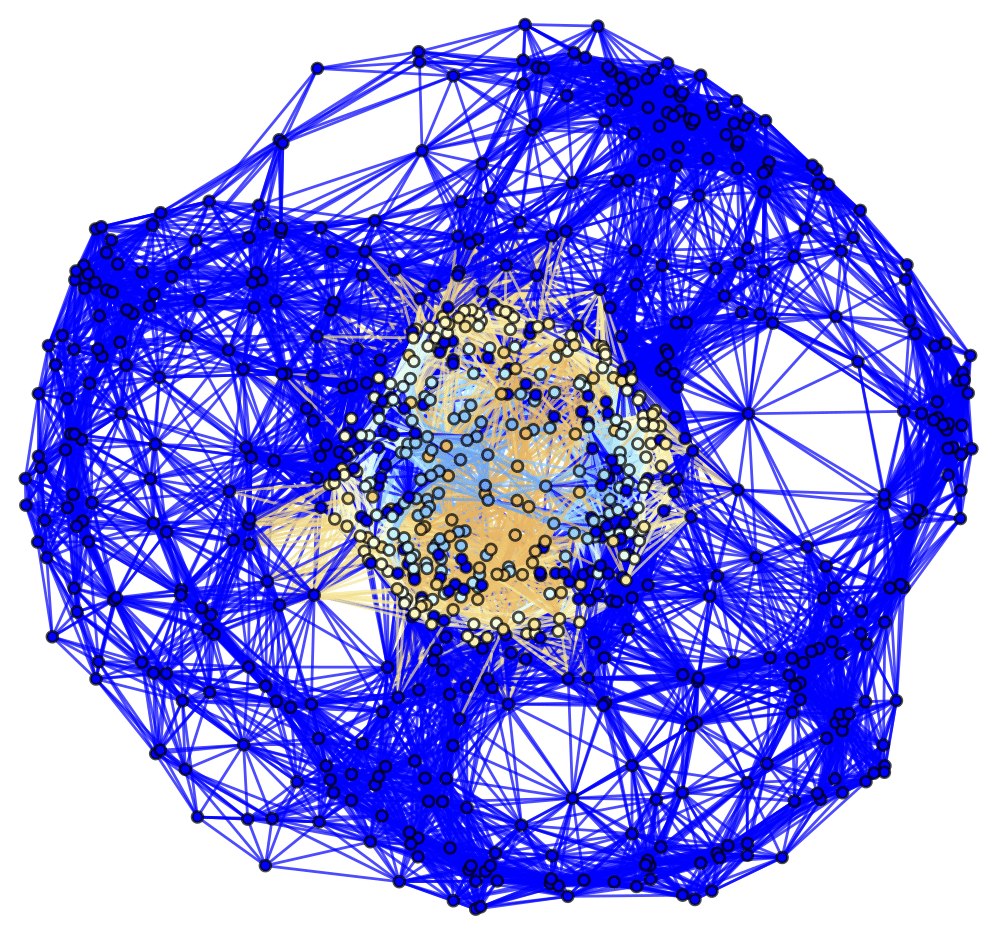}
\caption{Spatial hypergraphs corresponding to the second intermediate hypersurface configuration of the massive scalar field ``bubble collapse'' to a non-rotating Schwarzschild black hole test, with an exponential initial density distribution, at time ${t = 3 M}$, with resolutions of 200, 400 and 800 vertices, respectively. The hypergraphs have been adapted using the local curvature in the Schwarzschild conformal factor ${\psi}$, and colored according to the value of the scalar field ${\Phi \left( t, r \right)}$.}
\label{fig:Figure3}
\end{figure}

\begin{figure}[ht]
\centering
\includegraphics[width=0.325\textwidth]{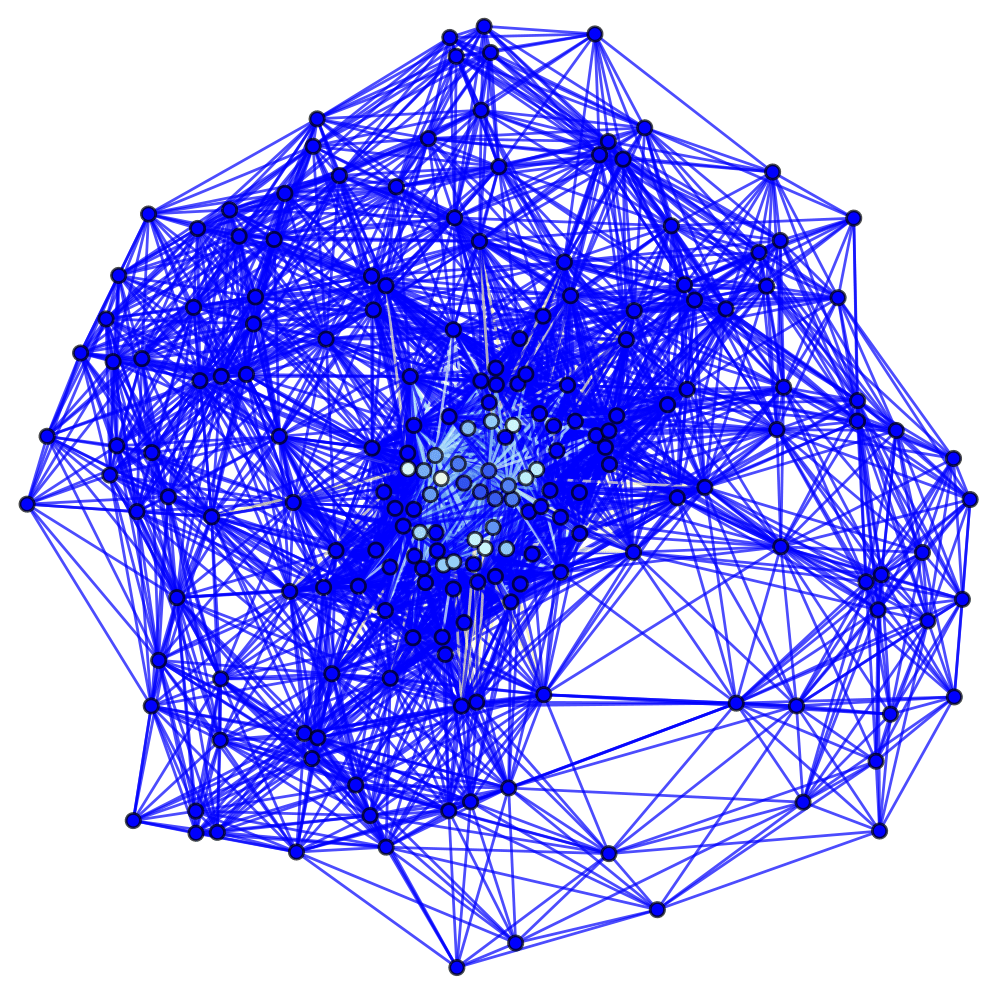}
\includegraphics[width=0.325\textwidth]{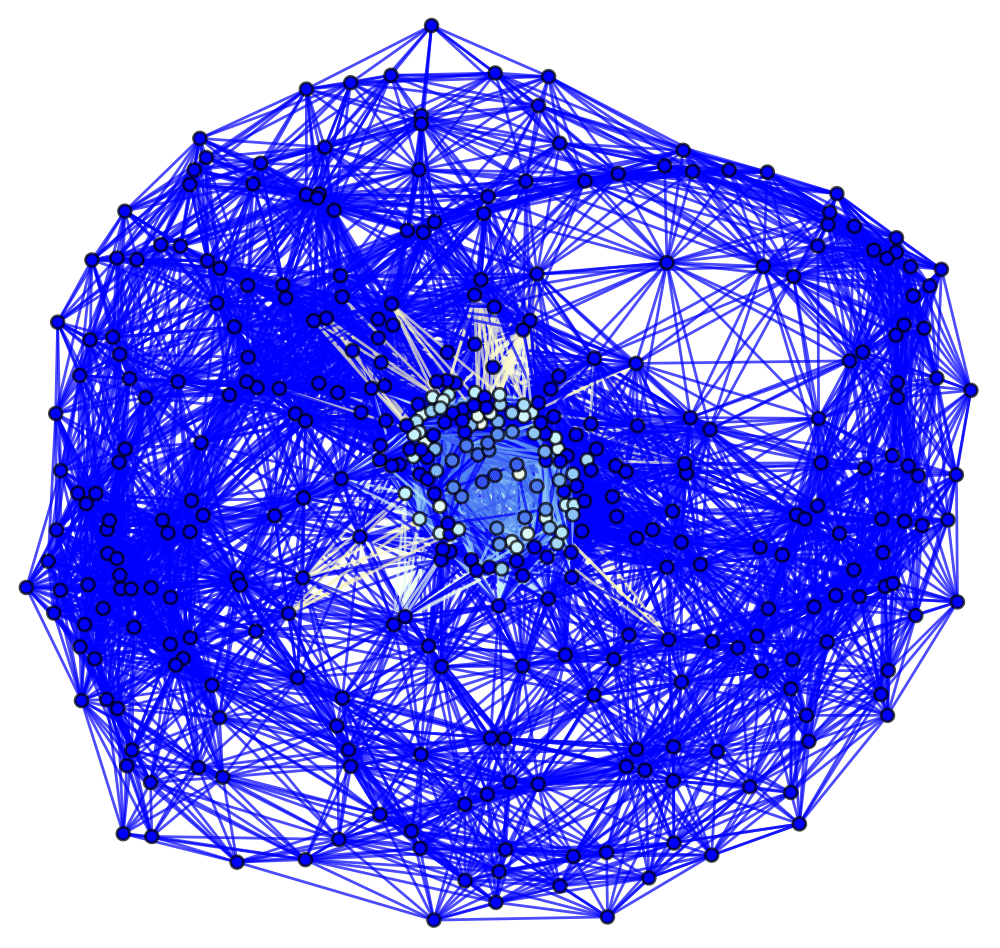}
\includegraphics[width=0.325\textwidth]{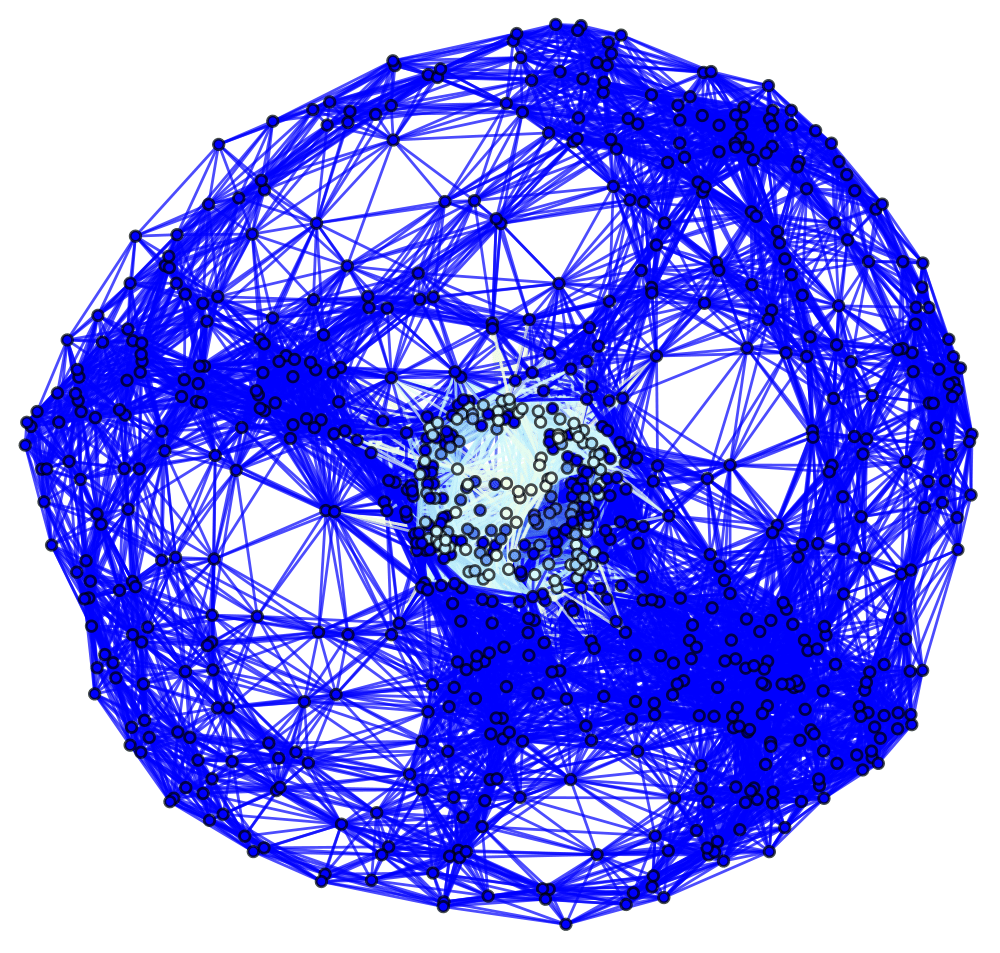}
\caption{Spatial hypergraphs corresponding to the final hypersurface configuration of the massive scalar field ``bubble collapse'' to a non-rotating Schwarzschild black hole test, with an exponential initial density distribution, at time ${t = 4.5 M}$, with resolutions of 200, 400 and 800 vertices, respectively. The hypergraphs have been adapted using the local curvature in the Schwarzschild conformal factor ${\psi}$, and colored according to the value of the scalar field ${\Phi \left( t, r \right)}$.}
\label{fig:Figure4}
\end{figure}

\begin{figure}[ht]
\centering
\includegraphics[width=0.325\textwidth]{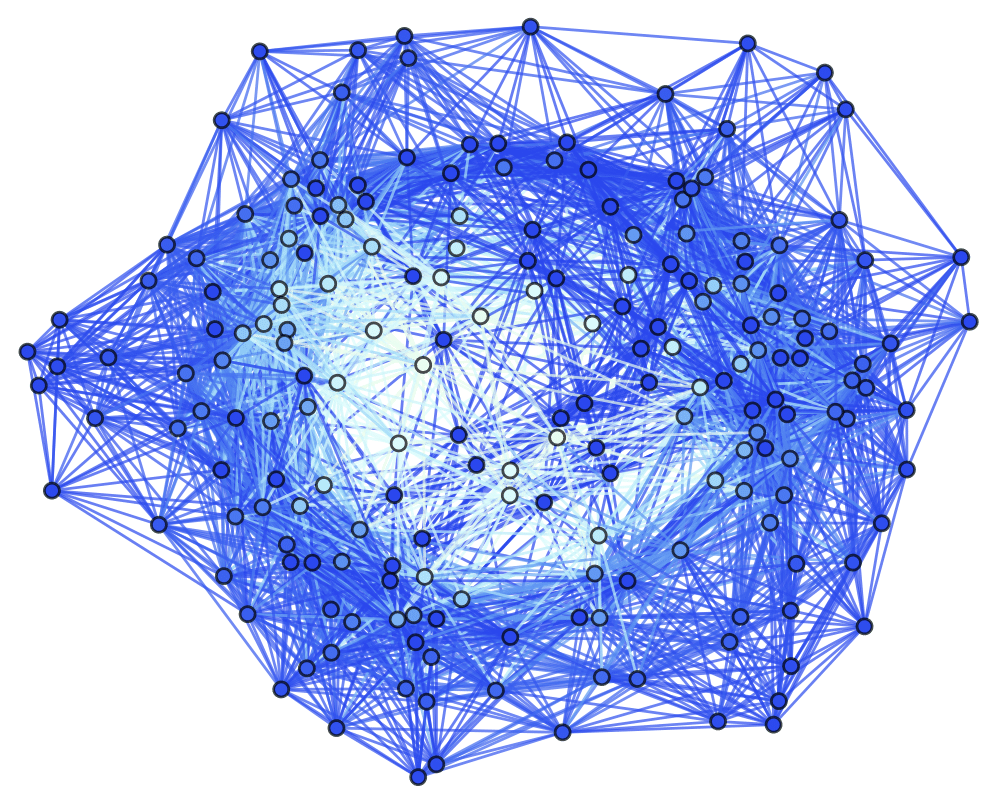}
\includegraphics[width=0.325\textwidth]{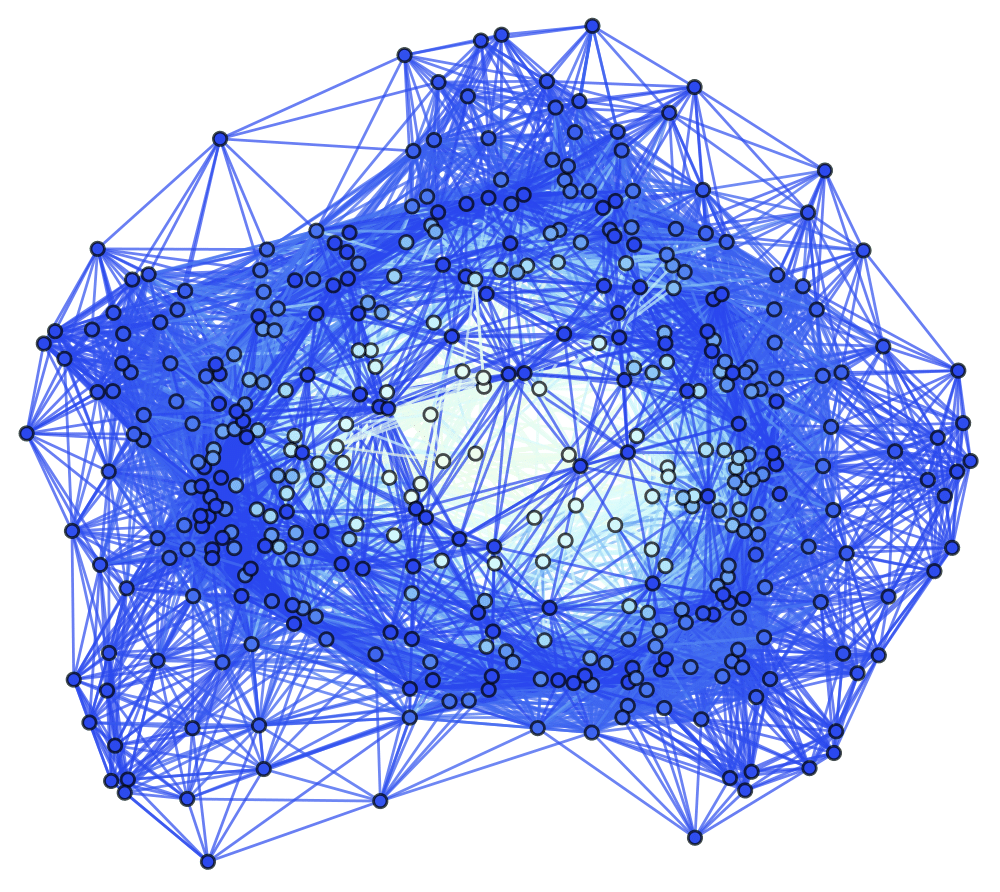}
\includegraphics[width=0.325\textwidth]{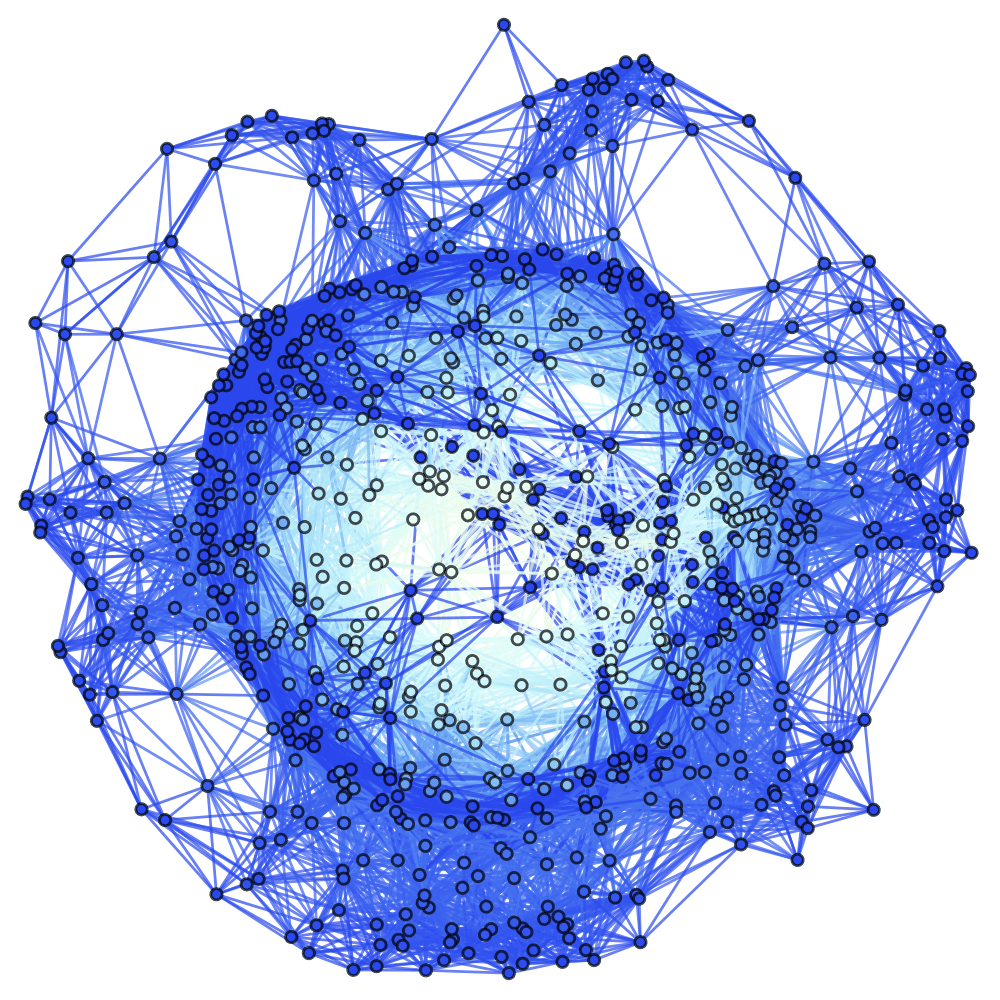}
\caption{Spatial hypergraphs corresponding to the initial hypersurface configuration of the massive scalar field ``bubble collapse'' to a non-rotating Schwarzschild black hole test, with an exponential initial density distribution, at time ${t = 0 M}$, with resolutions of 200, 400 and 800 vertices, respectively. The hypergraphs have been adapted and colored using the local curvature in the Schwarzschild conformal factor ${\psi}$.}
\label{fig:Figure5}
\end{figure}

\begin{figure}[ht]
\centering
\includegraphics[width=0.325\textwidth]{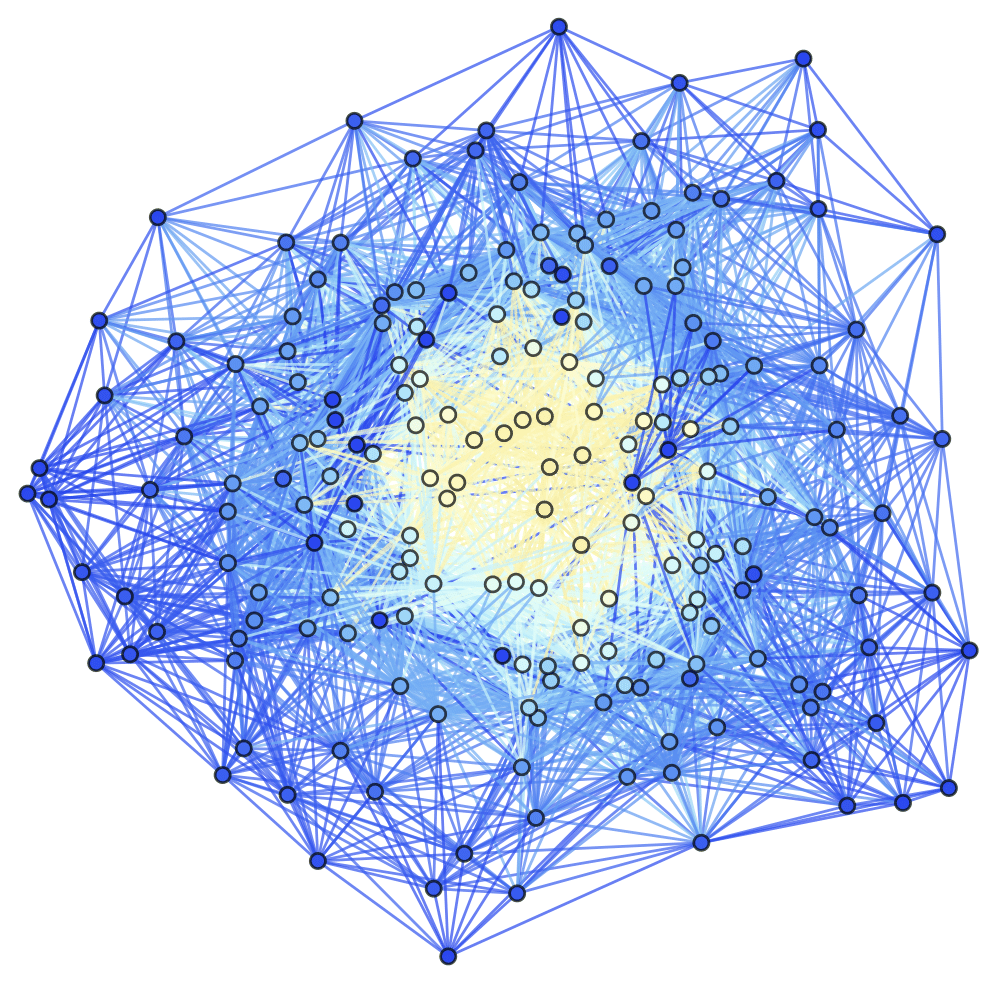}
\includegraphics[width=0.325\textwidth]{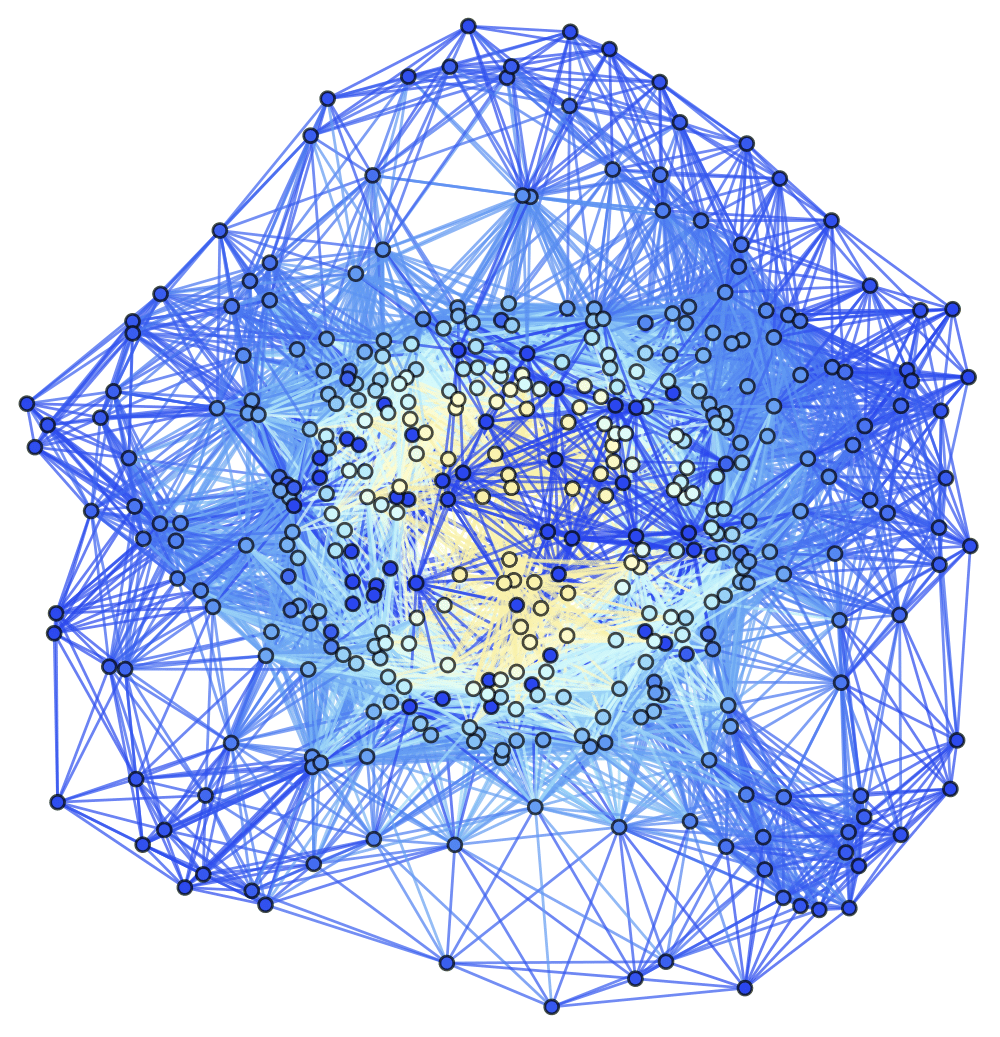}
\includegraphics[width=0.325\textwidth]{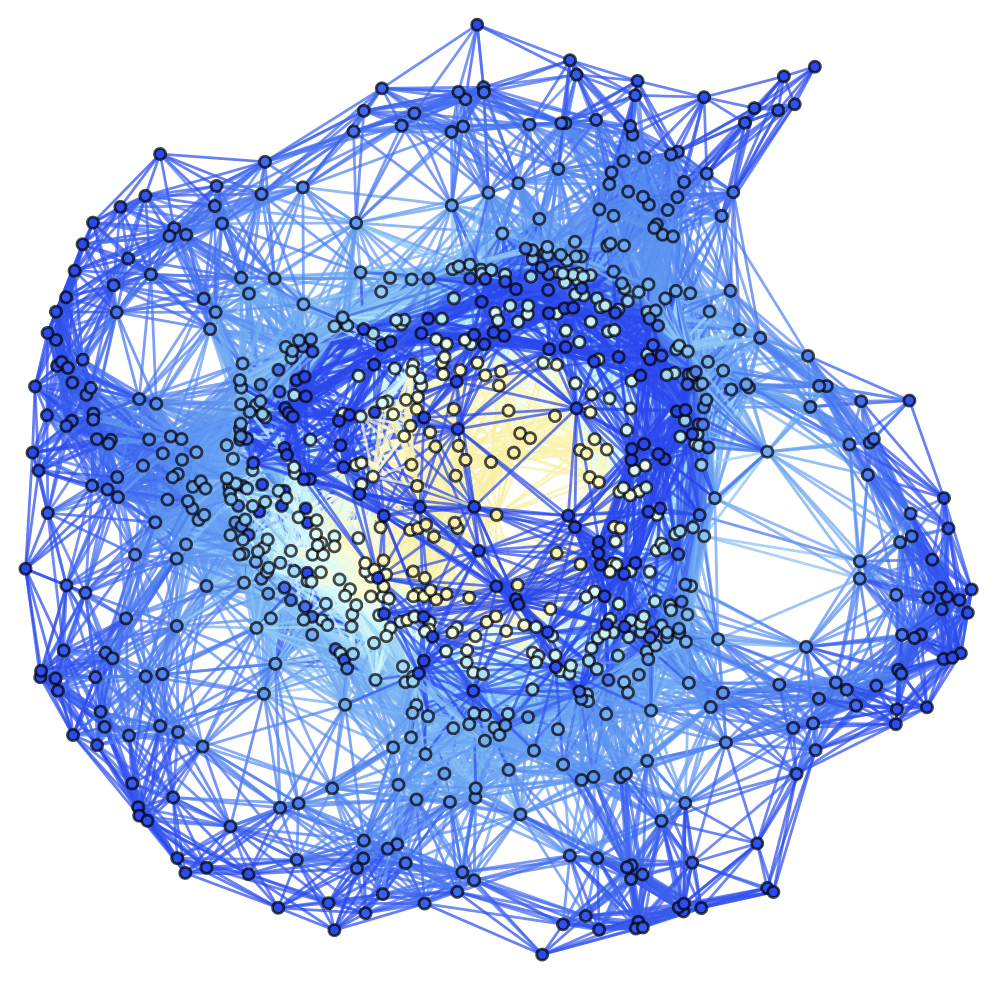}
\caption{Spatial hypergraphs corresponding to the first intermediate hypersurface configuration of the massive scalar field ``bubble collapse'' to a non-rotating Schwarzschild black hole test, with an exponential initial density distribution, at time ${t = 1.5 M}$, with resolutions of 200, 400 and 800 vertices, respectively. The hypergraphs have been adapted and colored using the local curvature in the Schwarzschild conformal factor ${\psi}$.}
\label{fig:Figure6}
\end{figure}

\begin{figure}[ht]
\centering
\includegraphics[width=0.325\textwidth]{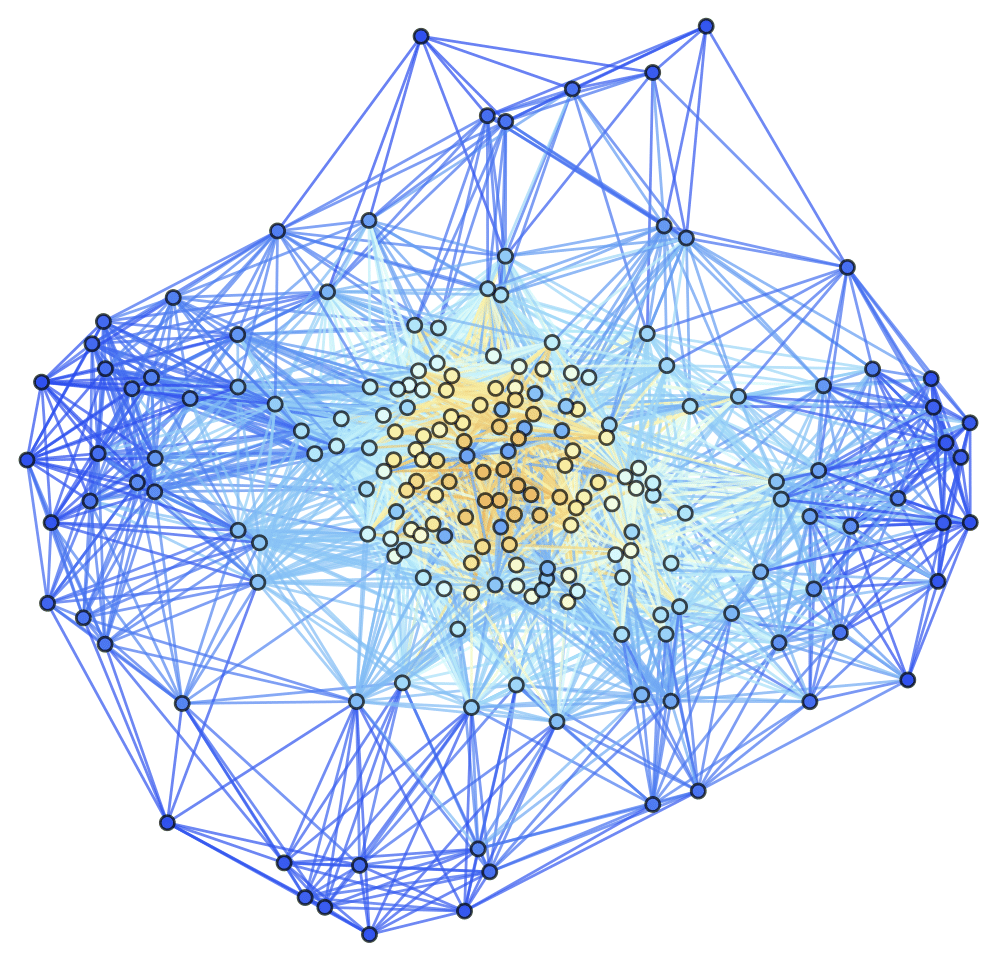}
\includegraphics[width=0.325\textwidth]{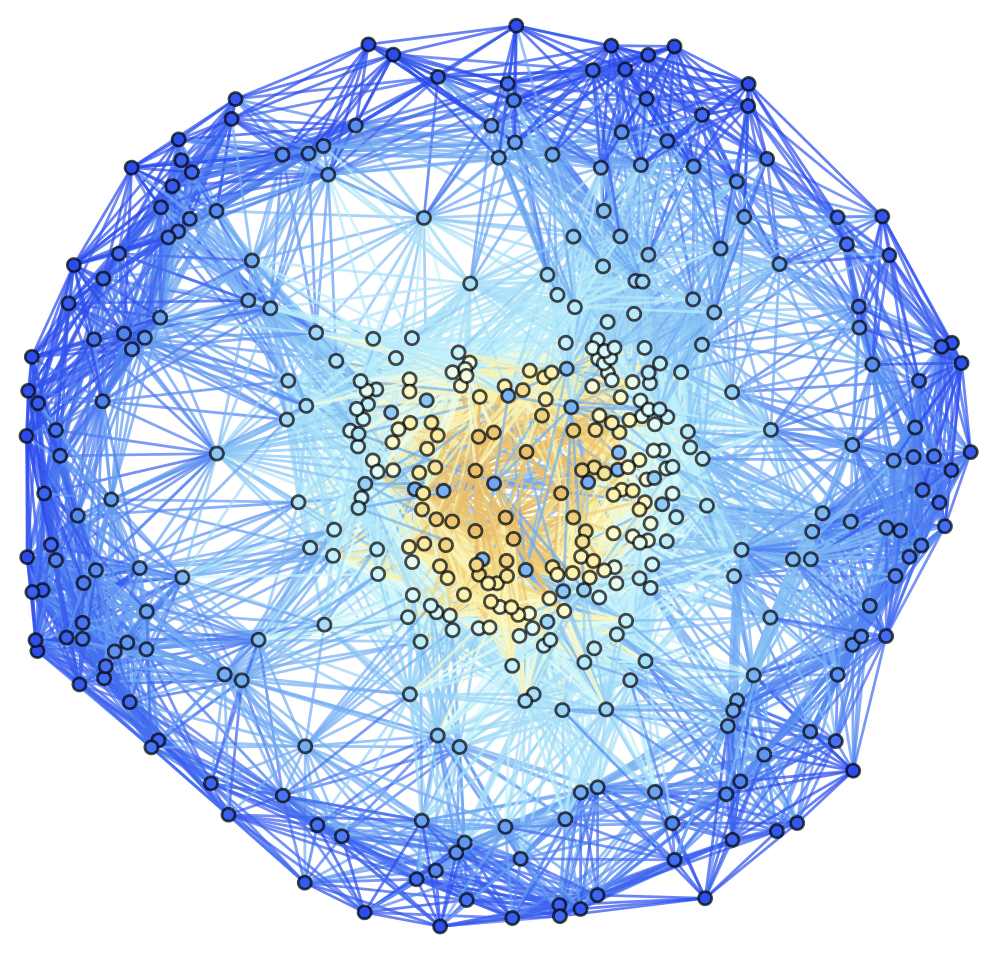}
\includegraphics[width=0.325\textwidth]{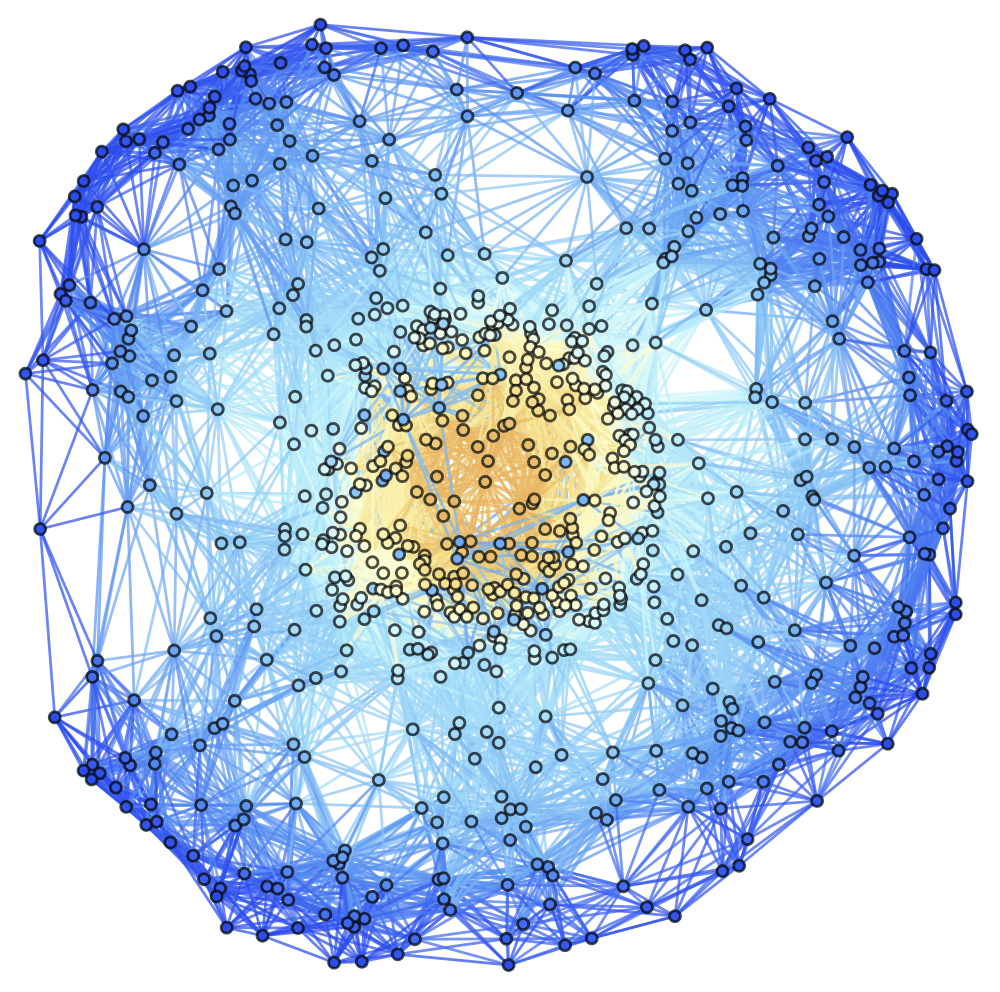}
\caption{Spatial hypergraphs corresponding to the second intermediate hypersurface configuration of the massive scalar field ``bubble collapse'' to a non-rotating Schwarzschild black hole test, with an exponential initial density distribution, at time ${t = 3 M}$, with resolutions of 200, 400 and 800 vertices, respectively. The hypergraphs have been adapted and colored using the local curvature in the Schwarzschild conformal factor ${\psi}$.}
\label{fig:Figure7}
\end{figure}

\begin{figure}[ht]
\centering
\includegraphics[width=0.325\textwidth]{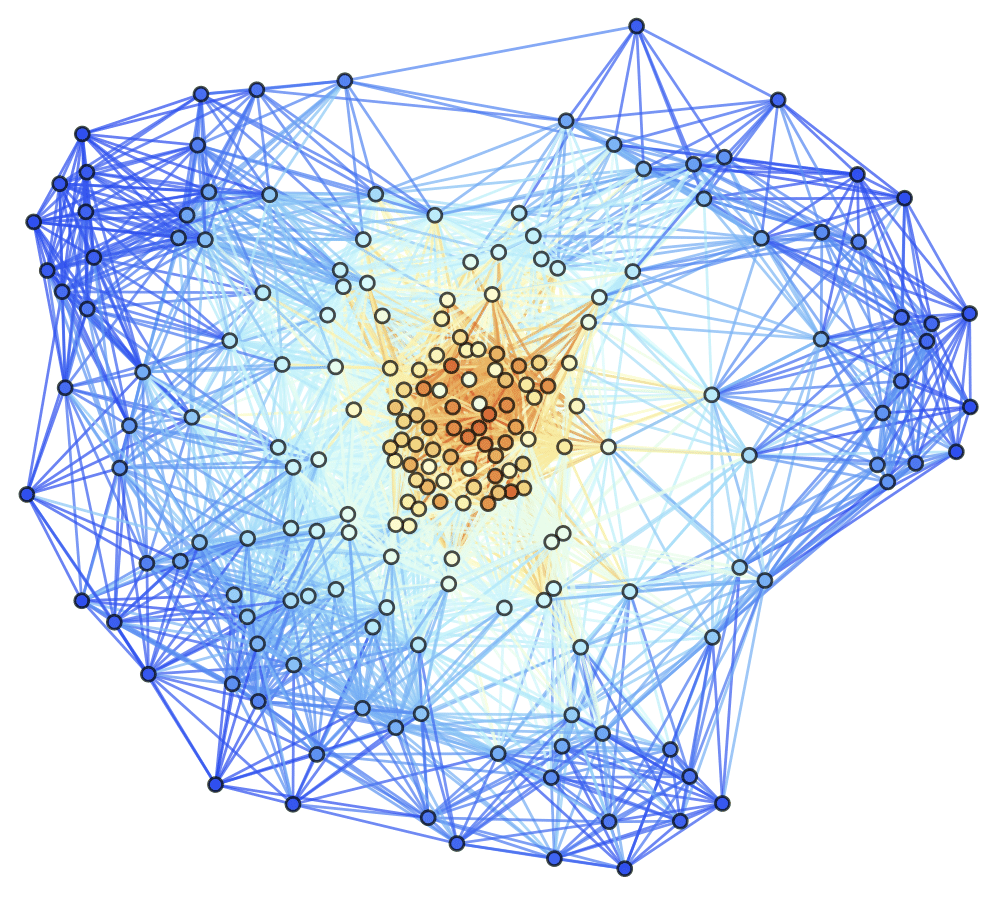}
\includegraphics[width=0.325\textwidth]{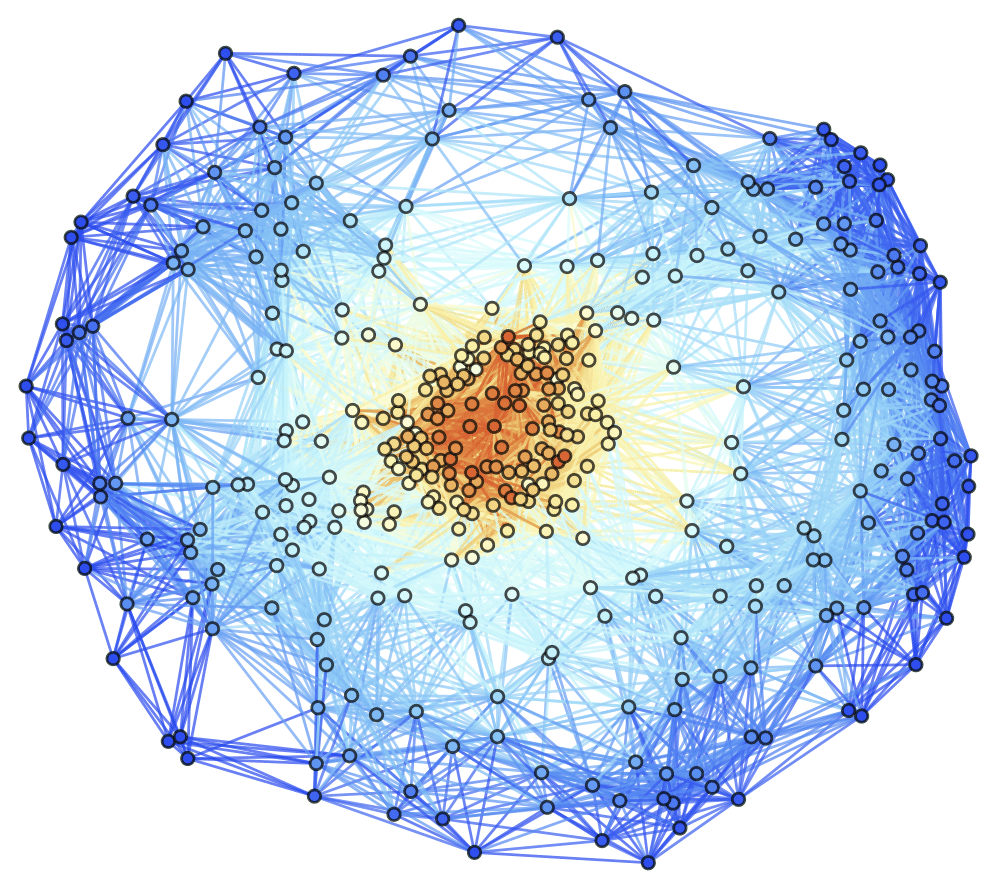}
\includegraphics[width=0.325\textwidth]{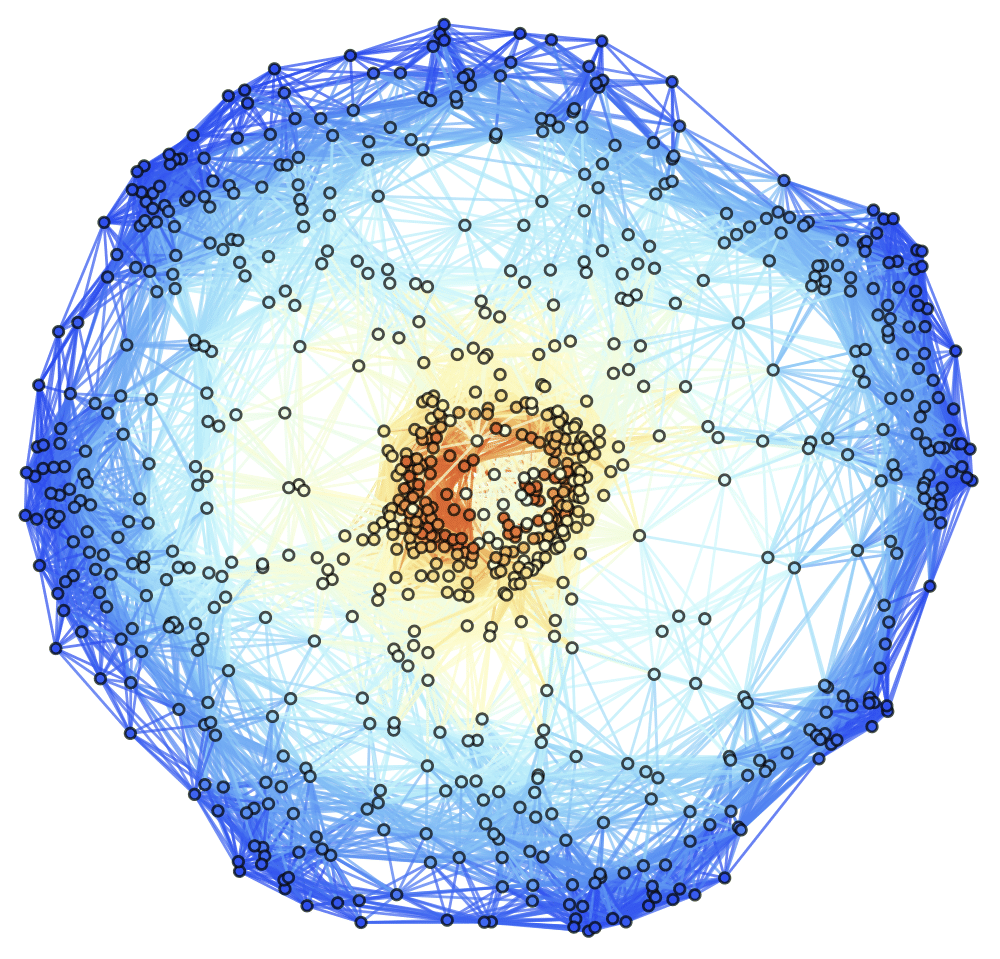}
\caption{Spatial hypergraphs corresponding to the final hypersurface configuration of the massive scalar field ``bubble collapse'' to a non-rotating Schwarzschild black hole test, with an exponential initial density distribution, at time ${t = 4.5 M}$, with resolutions of 200, 400 and 800 vertices, respectively. The hypergraphs have been adapted and colored using the local curvature in the Schwarzschild conformal factor ${\psi}$.}
\label{fig:Figure8}
\end{figure}

\begin{figure}[ht]
\centering
\includegraphics[width=0.325\textwidth]{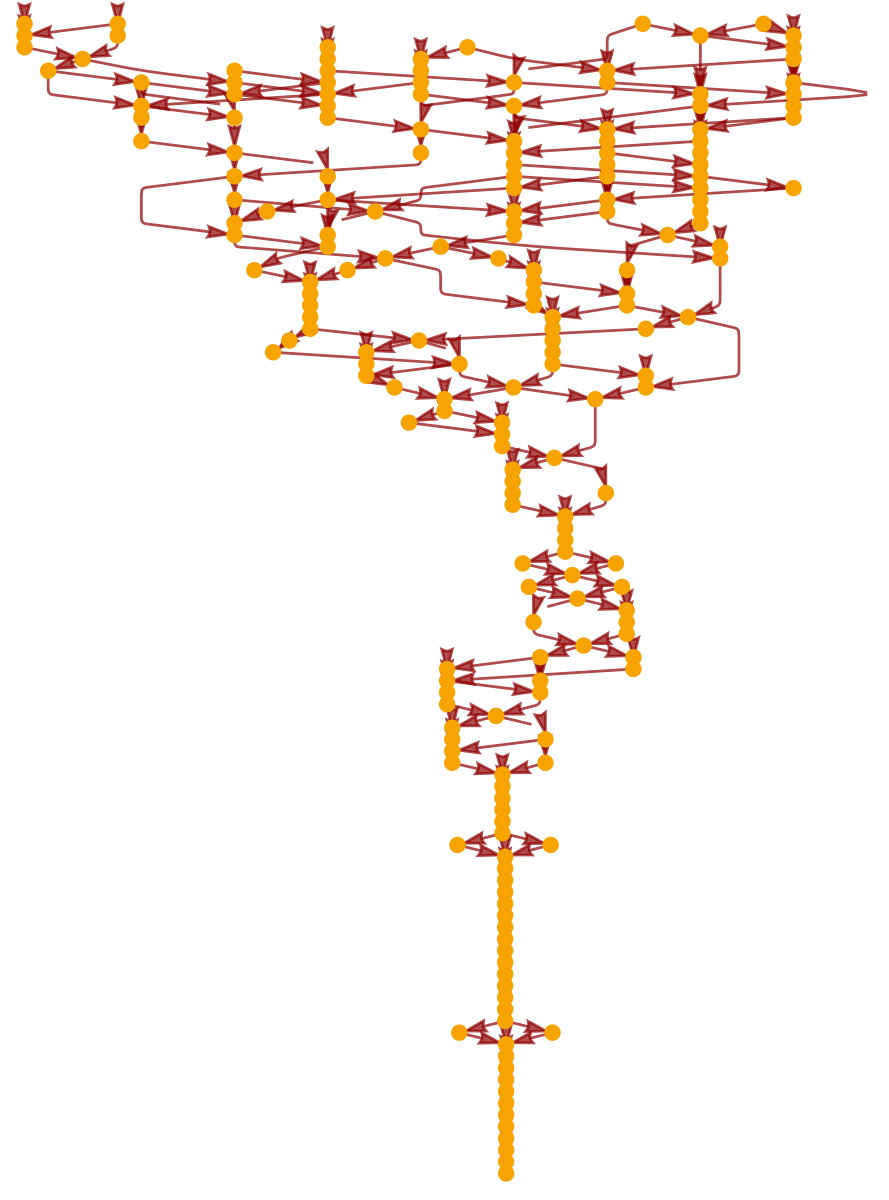}
\includegraphics[width=0.325\textwidth]{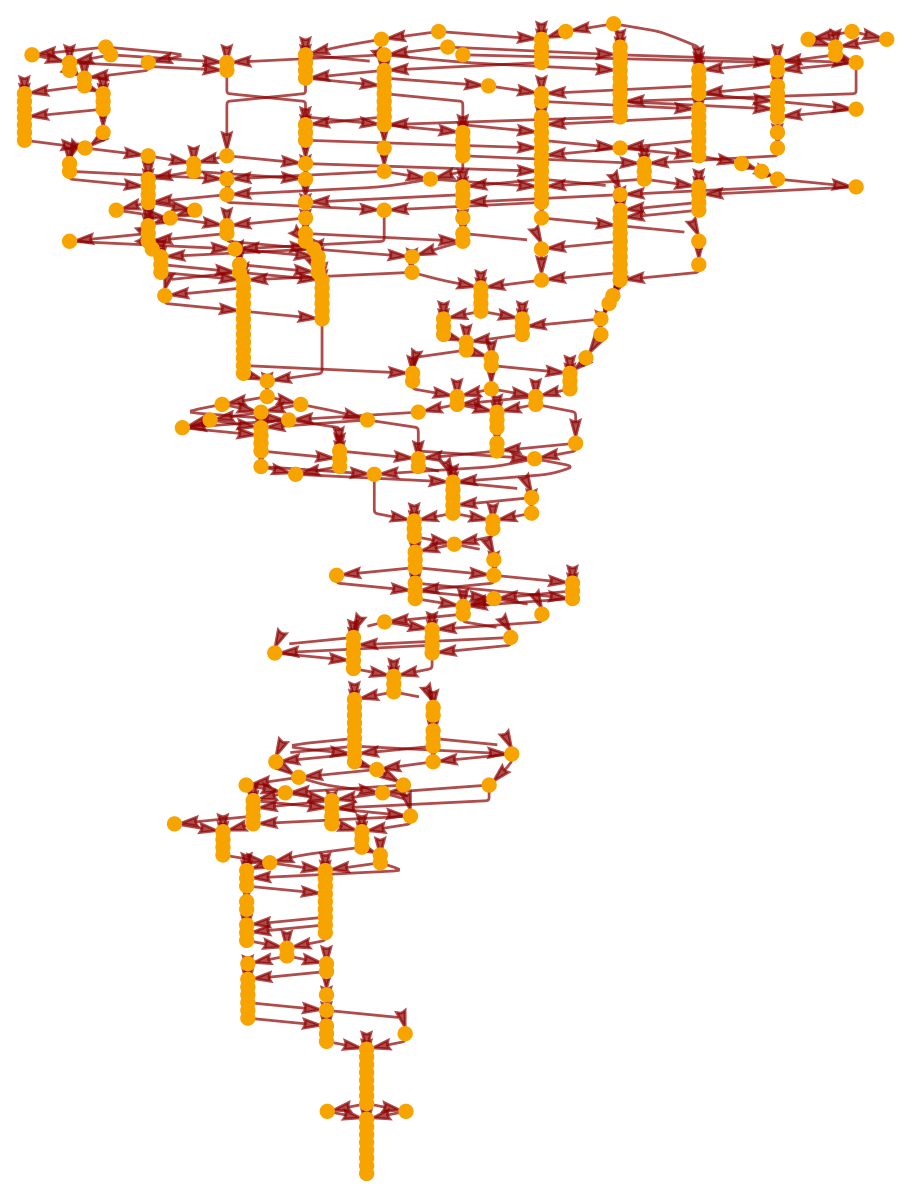}
\includegraphics[width=0.325\textwidth]{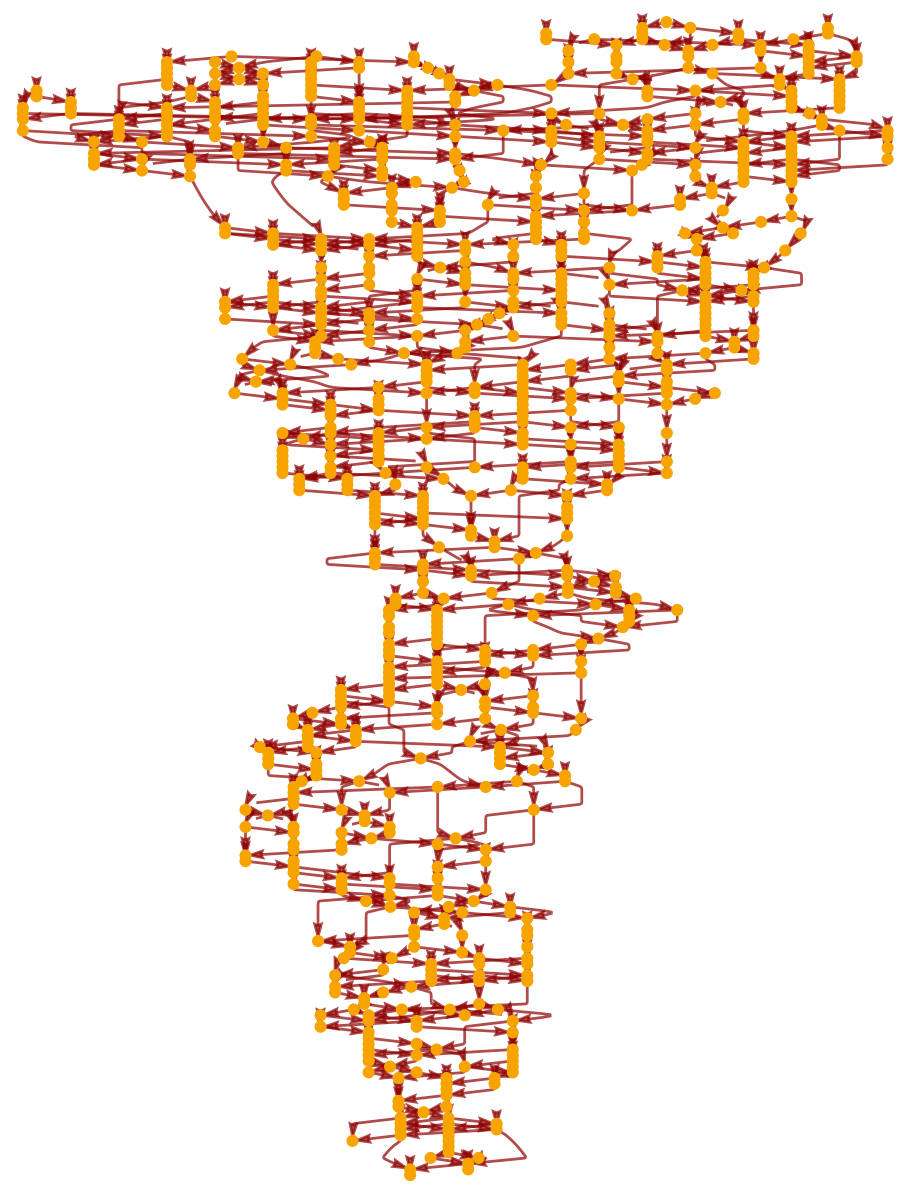}
\caption{Causal graphs corresponding to the discrete characteristic structure of the massive scalar field ``bubble collapse'' to a non-rotating Schwarzschild black hole test, with an exponential initial density distribution, at time ${t = 4.5 M}$, with resolutions of 200, 400 and 800 hypergraph vertices, respectively.}
\label{fig:Figure9}
\end{figure}

\begin{figure}[ht]
\centering
\includegraphics[width=0.325\textwidth]{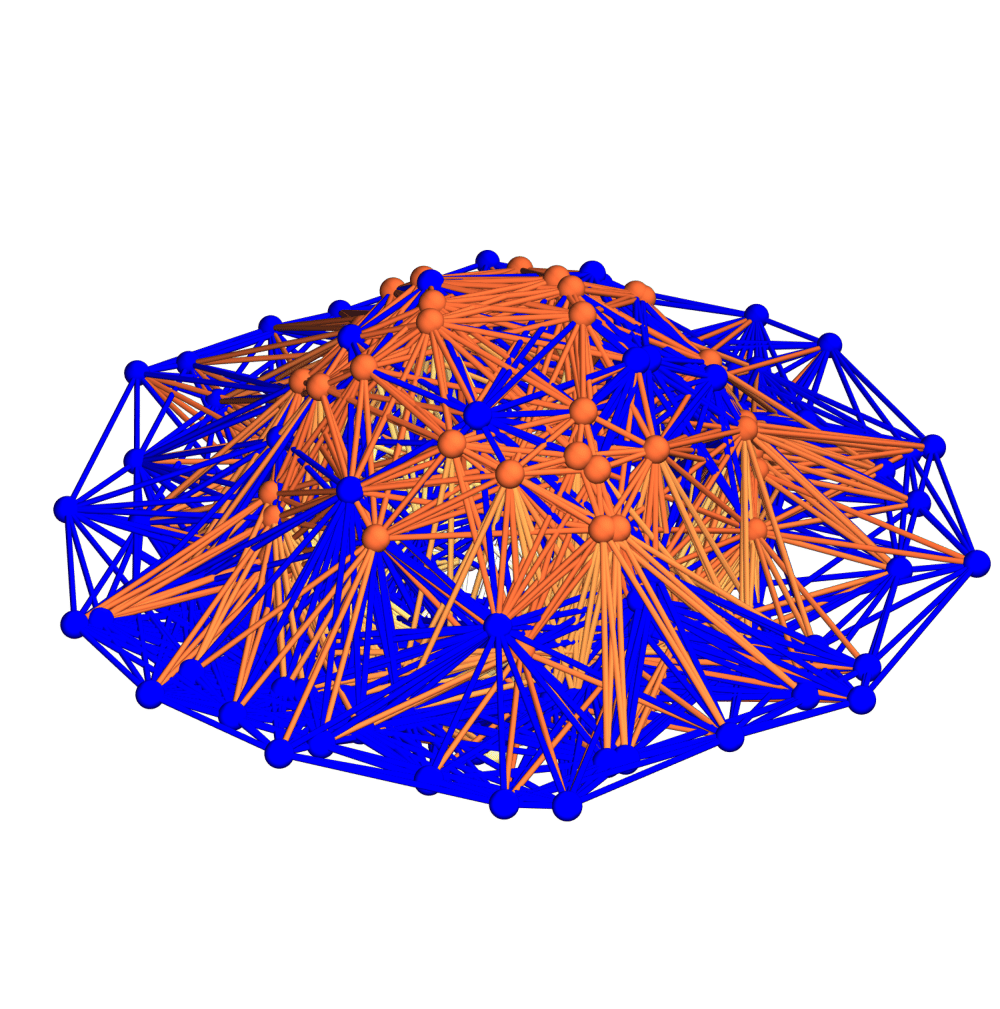}
\includegraphics[width=0.325\textwidth]{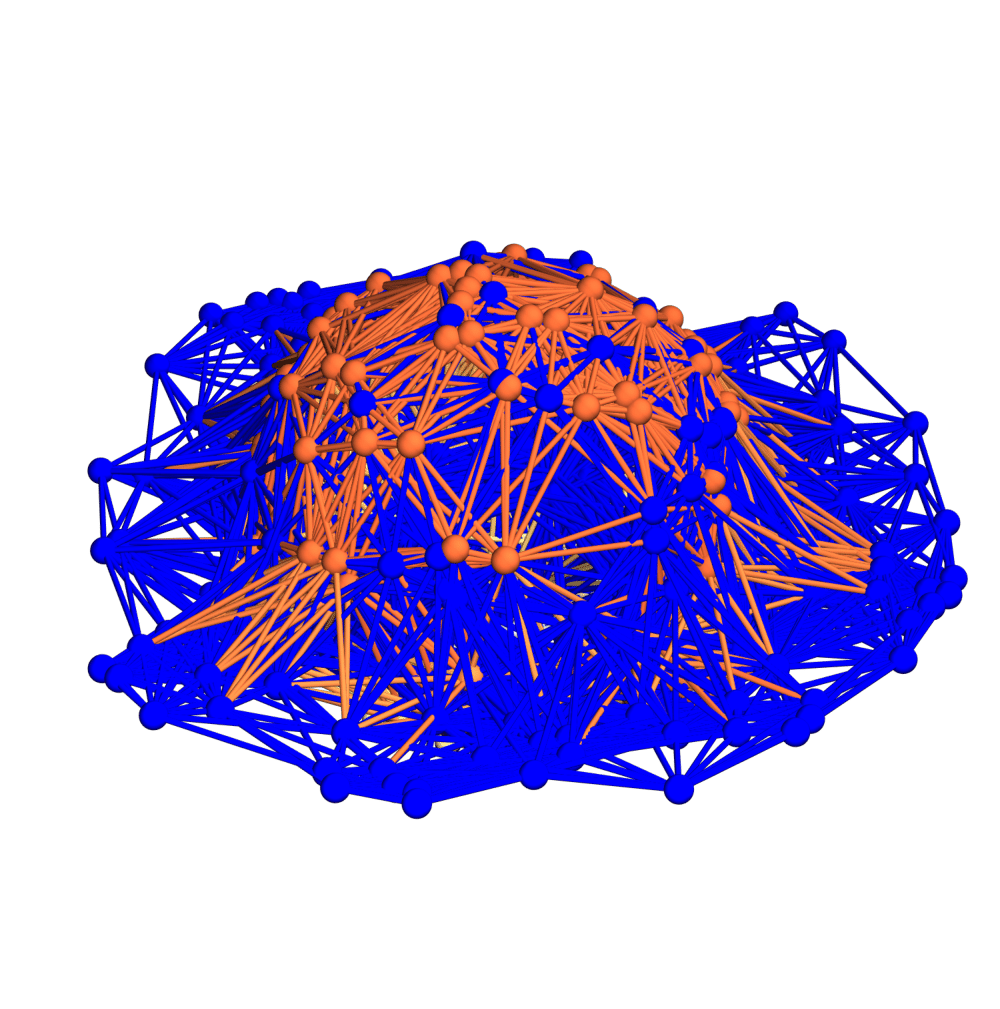}
\includegraphics[width=0.325\textwidth]{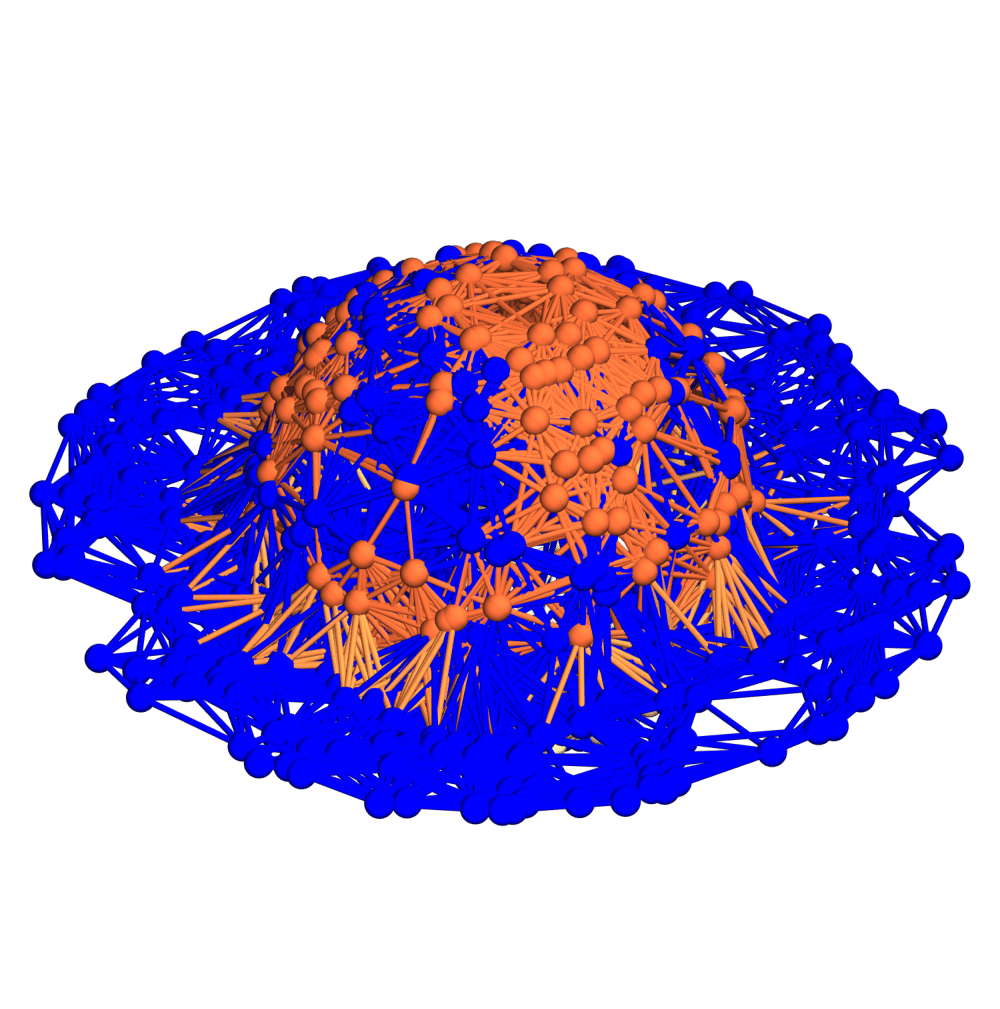}
\caption{Spatial hypergraphs corresponding to projections along the $z$-axis of the initial hypersurface configuration of the massive scalar field ``bubble collapse'' to a non-rotating Schwarzschild black hole test, with an exponential initial density distribution, at time ${t = 0 M}$, with resolutions of 200, 400 and 800 vertices, respectively. The vertices have been assigned spatial coordinates according to the profile of the Schwarzschild conformal factor ${\psi}$ through a spatial slice perpendicular to the $z$-axis, and the hypergraphs have been adapted using the local curvature in ${\psi}$, and colored according to the value of the scalar field ${\Phi \left( t, r \right)}$.}
\label{fig:Figure10}
\end{figure}

\begin{figure}[ht]
\centering
\includegraphics[width=0.325\textwidth]{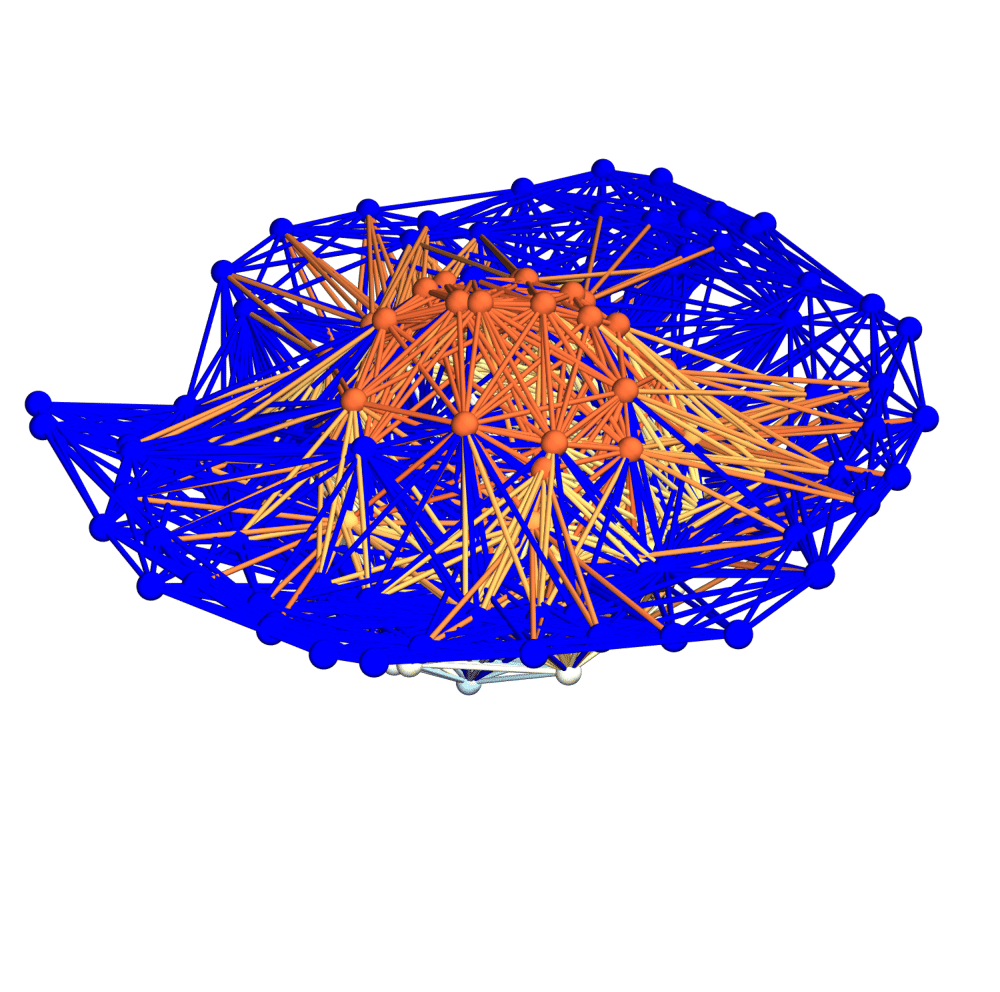}
\includegraphics[width=0.325\textwidth]{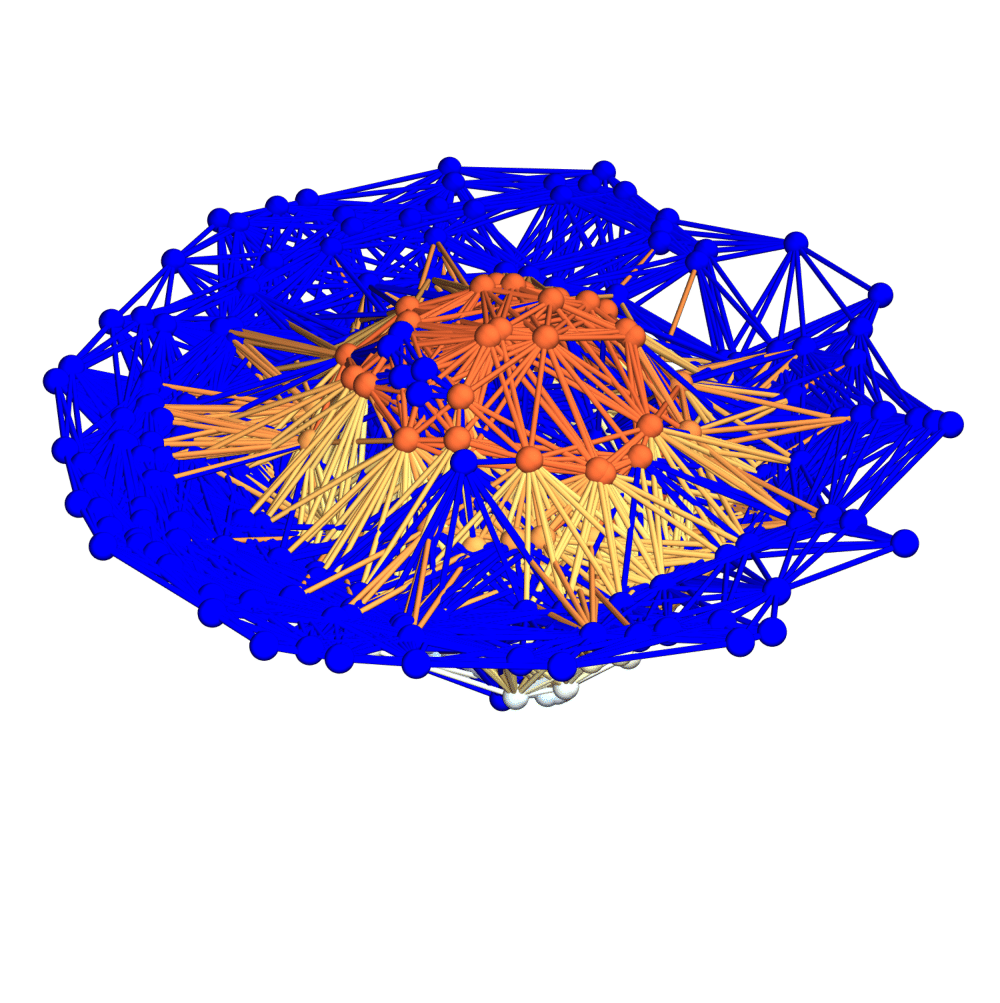}
\includegraphics[width=0.325\textwidth]{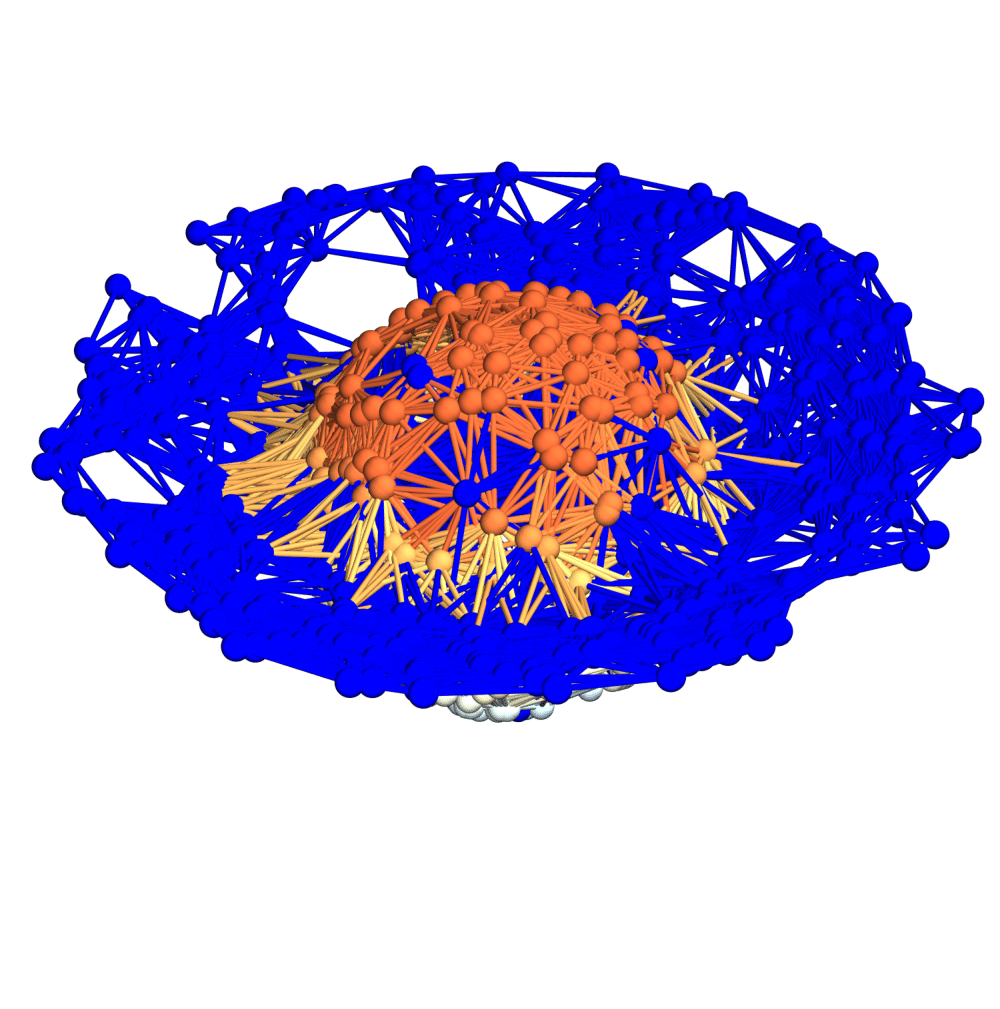}
\caption{Spatial hypergraphs corresponding to projections along the $z$-axis of the first intermediate hypersurface configuration of the massive scalar field ``bubble collapse'' to a non-rotating Schwarzschild black hole test, with an exponential initial density distribution, at time ${t = 1.5 M}$, with resolutions of 200, 400 and 800 vertices, respectively. The vertices have been assigned spatial coordinates according to the profile of the Schwarzschild conformal factor ${\psi}$ through a spatial slice perpendicular to the $z$-axis, and the hypergraphs have been adapted using the local curvature in ${\psi}$, and colored according to the value of the scalar field ${\Phi \left( t, r \right)}$.}
\label{fig:Figure11}
\end{figure}

\begin{figure}[ht]
\centering
\includegraphics[width=0.325\textwidth]{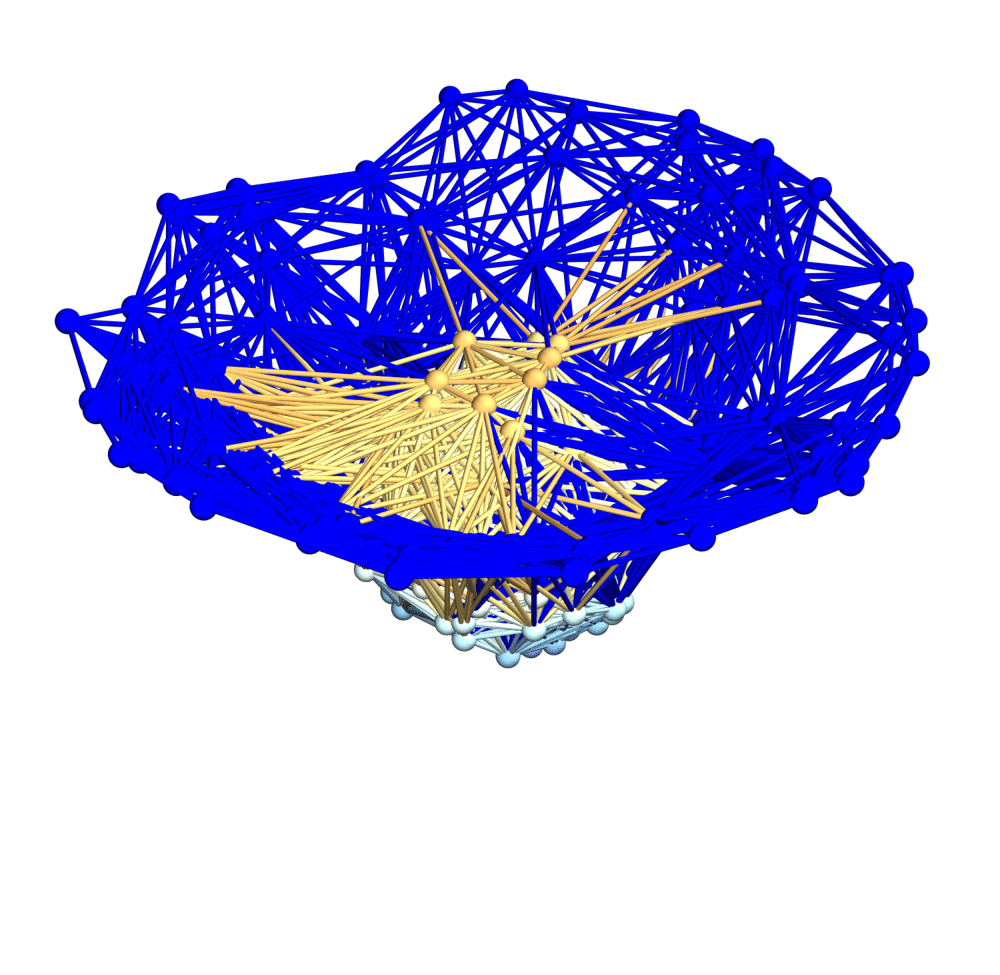}
\includegraphics[width=0.325\textwidth]{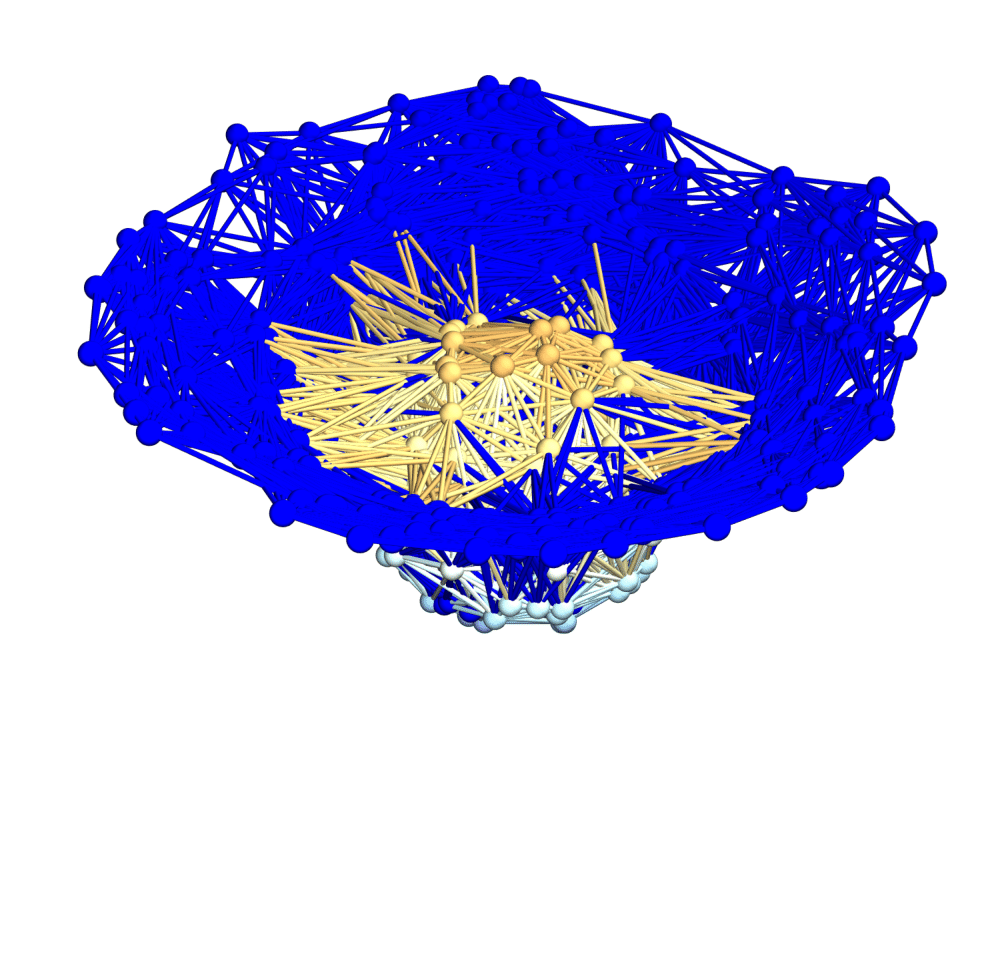}
\includegraphics[width=0.325\textwidth]{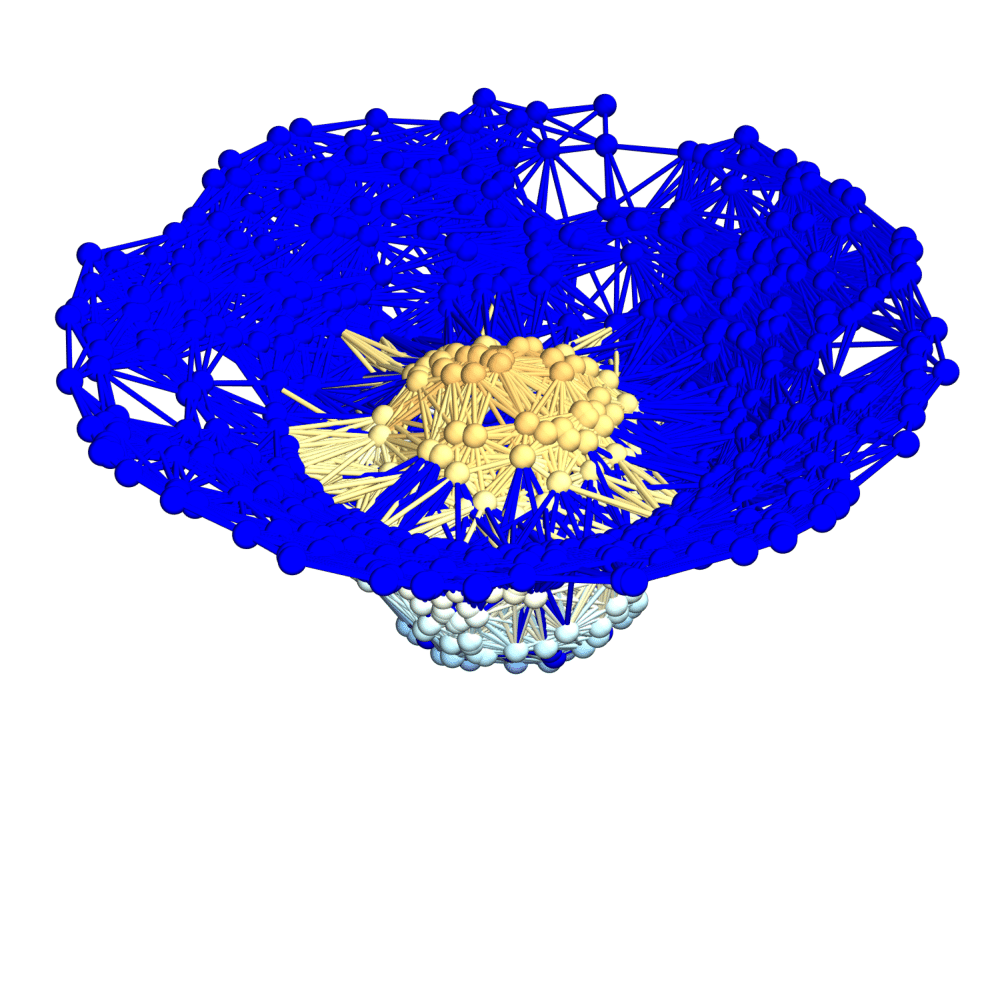}
\caption{Spatial hypergraphs corresponding to projections along the $z$-axis of the second intermediate hypersurface configuration of the massive scalar field ``bubble collapse'' to a non-rotating Schwarzschild black hole test, with an exponential initial density distribution, at time ${t = 3 M}$, with resolutions of 200, 400 and 800 vertices, respectively. The vertices have been assigned spatial coordinates according to the profile of the Schwarzschild conformal factor ${\psi}$ through a spatial slice perpendicular to the $z$-axis, and the hypergraphs have been adapted using the local curvature in ${\psi}$, and colored according to the value of the scalar field ${\Phi \left( t, r \right)}$.}
\label{fig:Figure12}
\end{figure}

\begin{figure}[ht]
\centering
\includegraphics[width=0.325\textwidth]{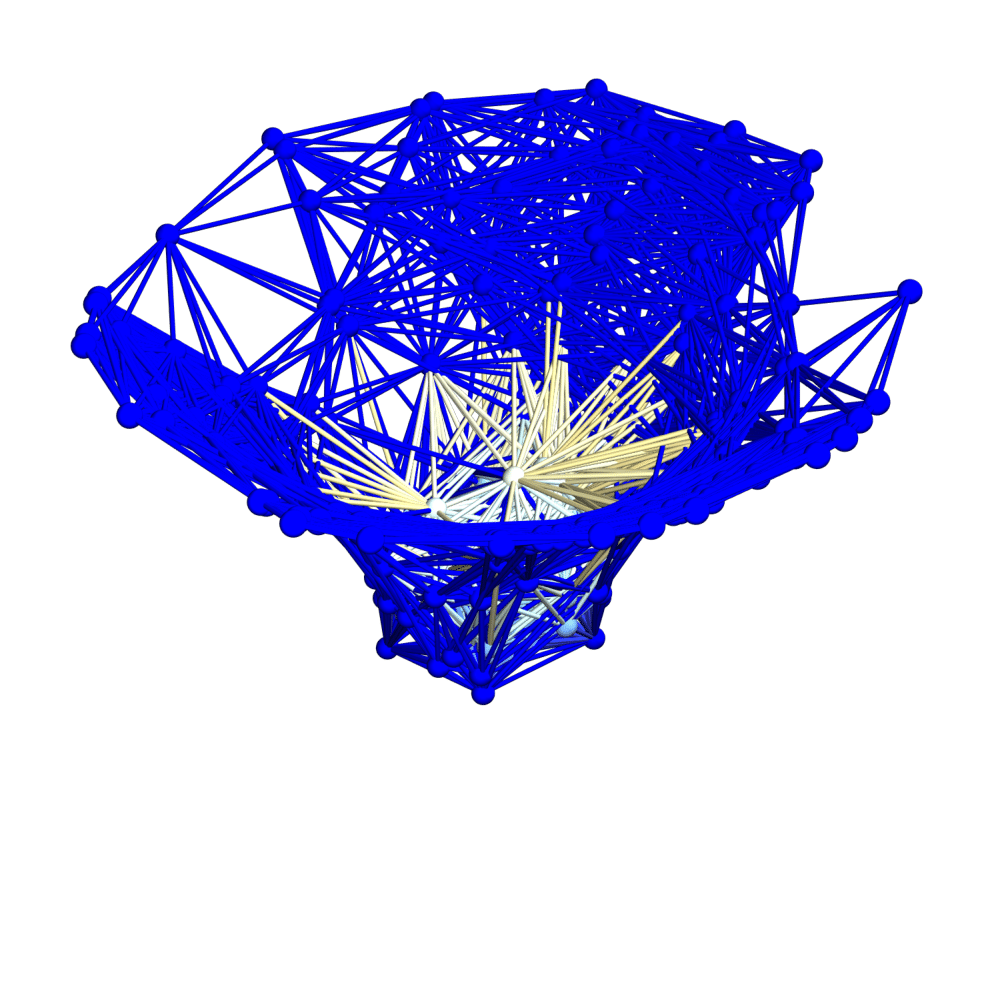}
\includegraphics[width=0.325\textwidth]{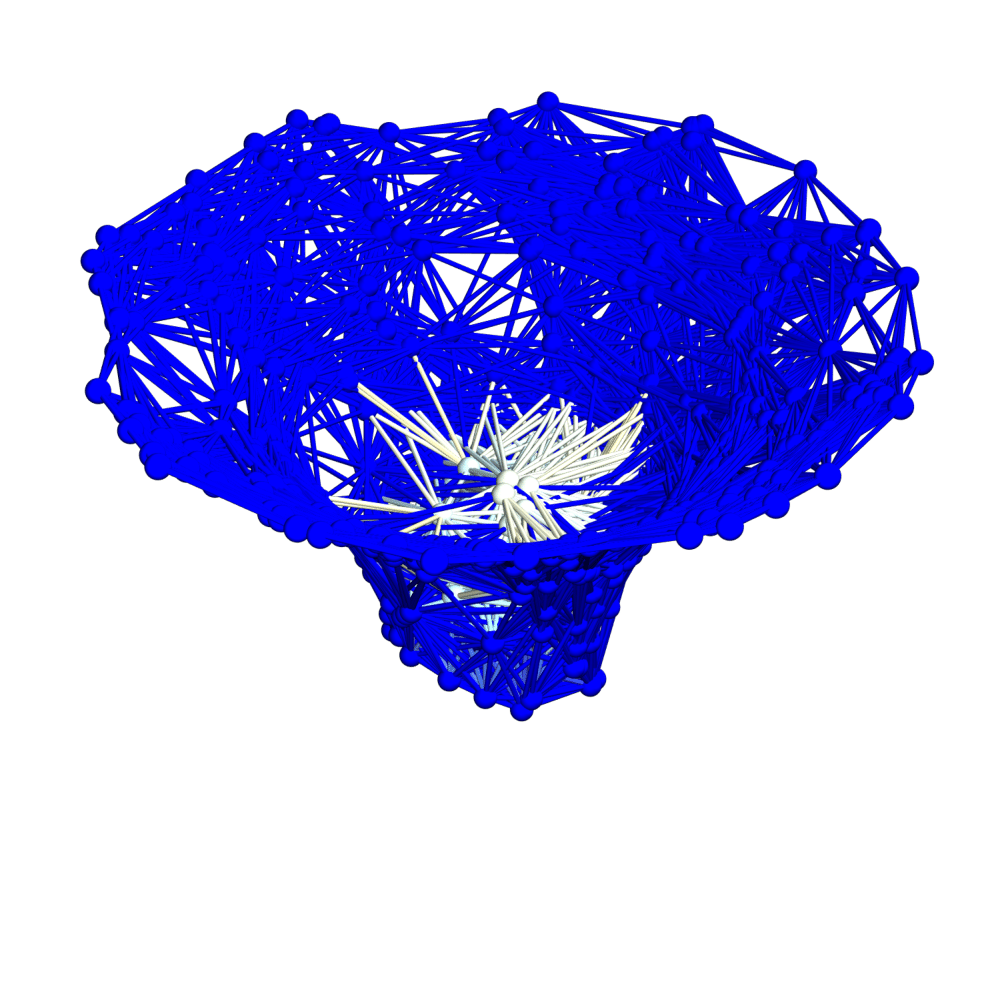}
\includegraphics[width=0.325\textwidth]{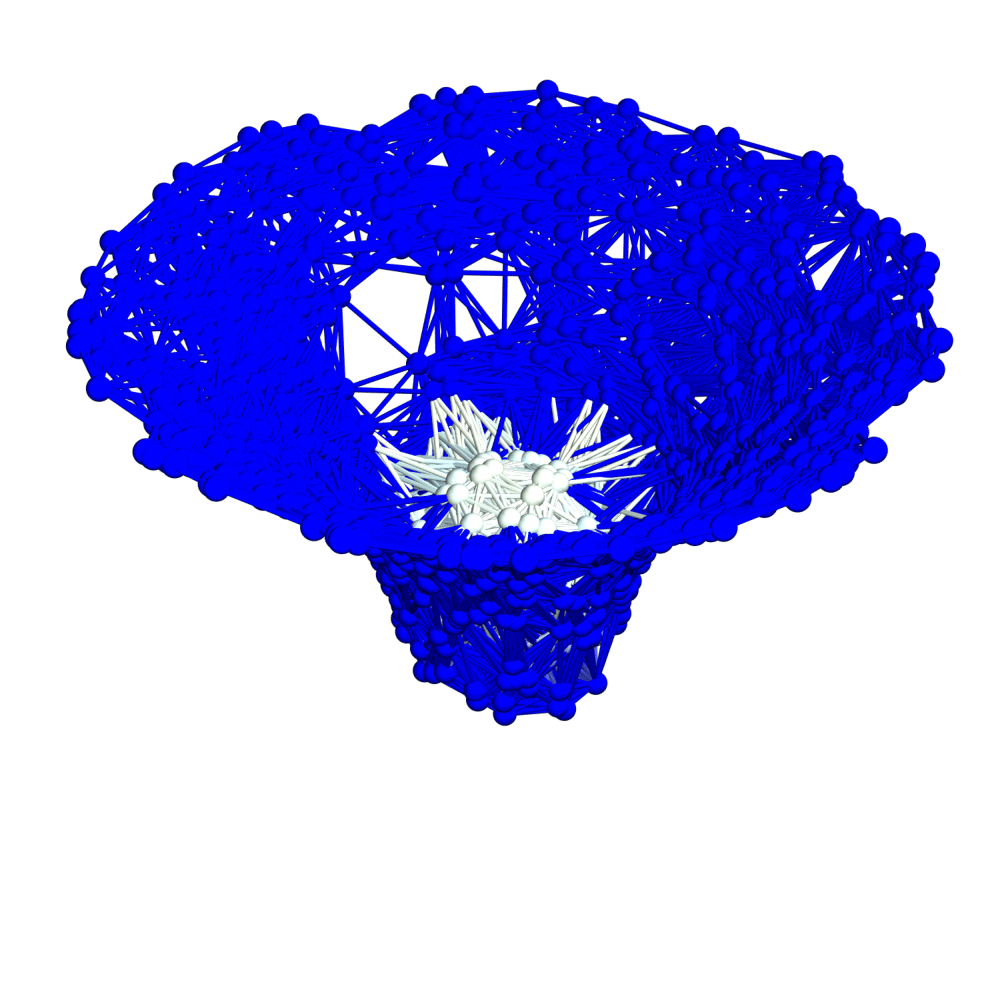}
\caption{Spatial hypergraphs corresponding to projections along the $z$-axis of the final hypersurface configuration of the massive scalar field ``bubble collapse'' to a non-rotating Schwarzschild black hole test, with an exponential initial density distribution, at time ${t = 4.5 M}$, with resolutions of 200, 400 and 800 vertices, respectively. The vertices have been assigned spatial coordinates according to the profile of the Schwarzschild conformal factor ${\psi}$ through a spatial slice perpendicular to the $z$-axis, and the hypergraphs have been adapted using the local curvature in ${\psi}$, and colored according to the value of the scalar field ${\Phi \left( t, r \right)}$.}
\label{fig:Figure13}
\end{figure}

\begin{figure}[ht]
\centering
\includegraphics[width=0.325\textwidth]{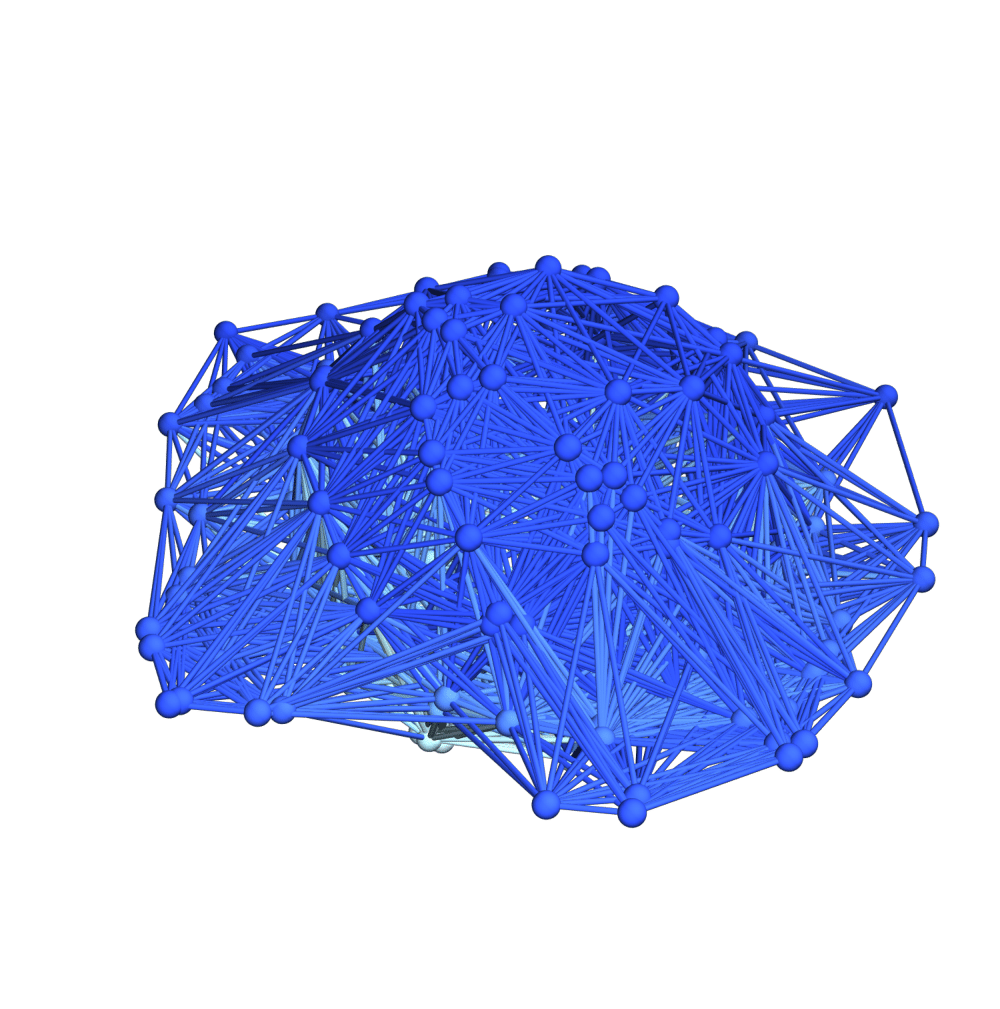}
\includegraphics[width=0.325\textwidth]{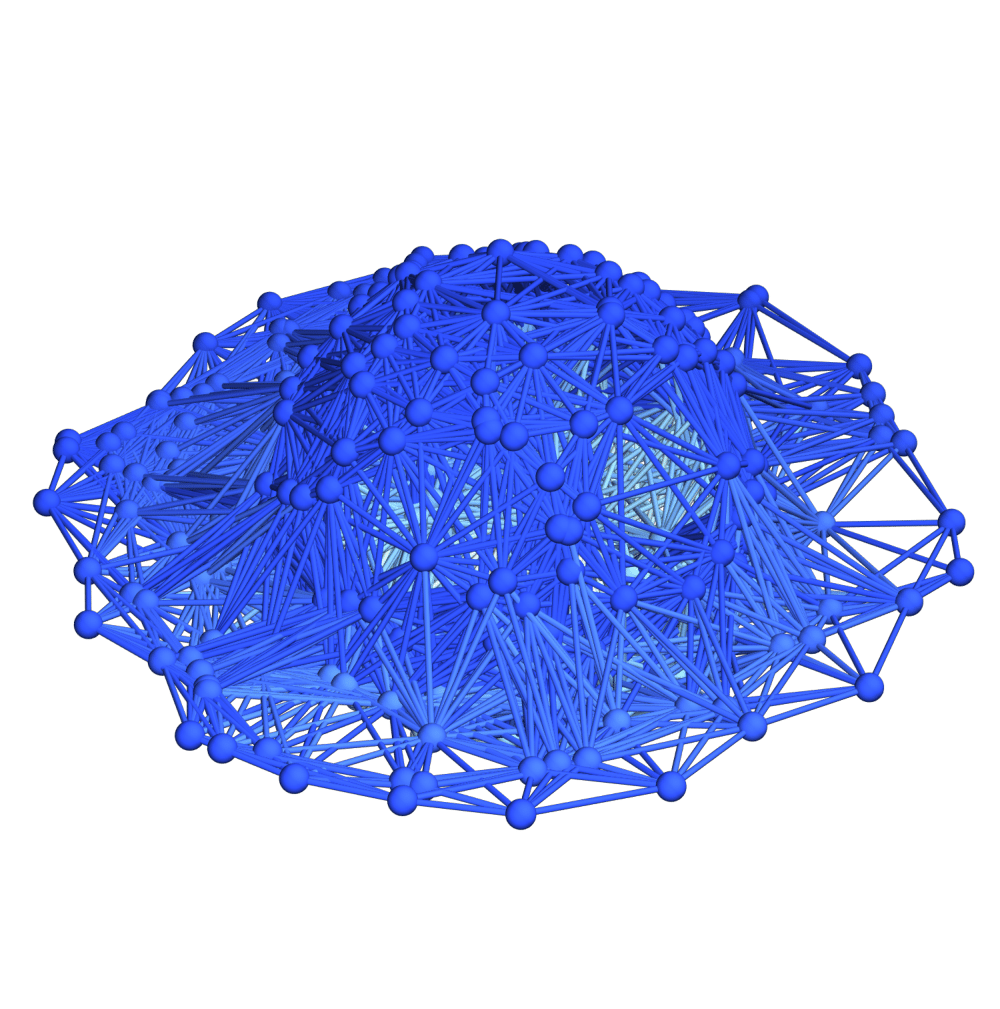}
\includegraphics[width=0.325\textwidth]{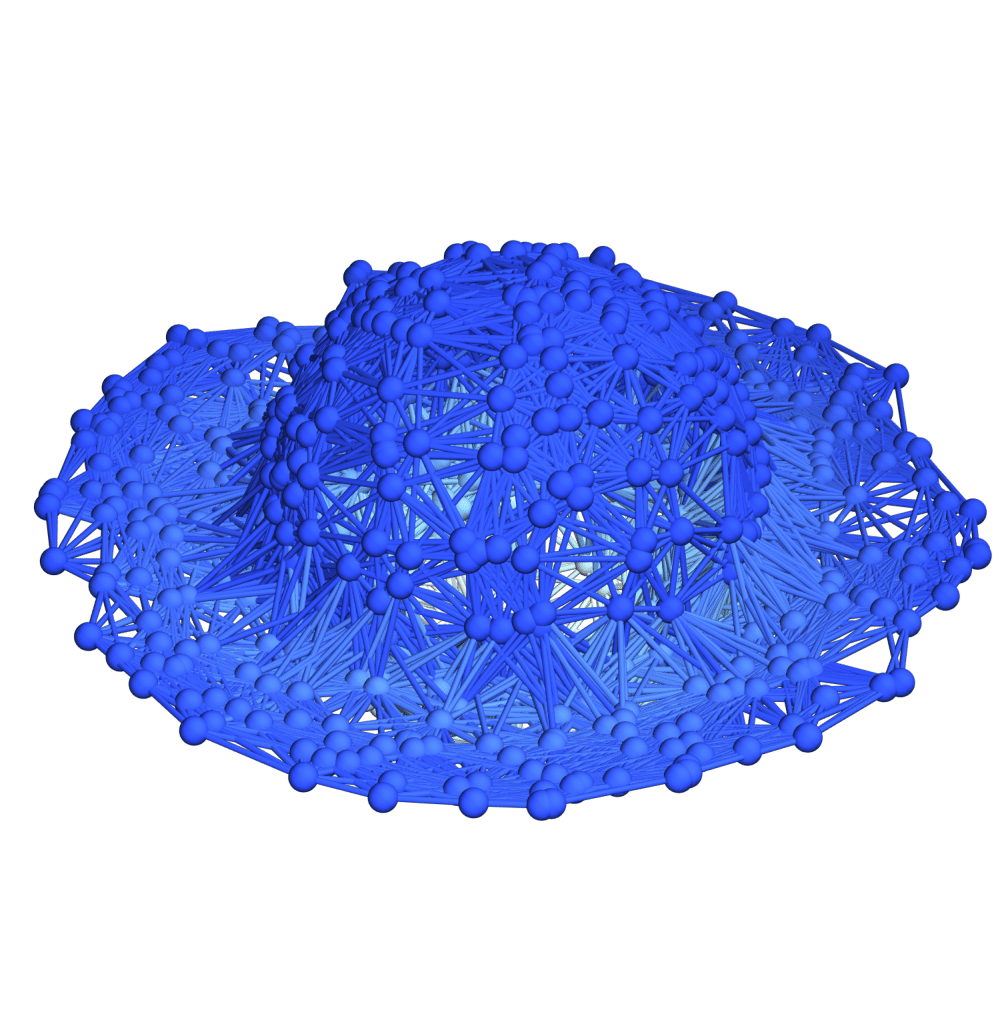}
\caption{Spatial hypergraphs corresponding to projections along the $z$-axis of the initial hypersurface configuration of the massive scalar field ``bubble collapse'' to a non-rotating Schwarzschild black hole test, with an exponential initial density distribution, at time ${t = 0 M}$, with resolutions of 200, 400 and 800 vertices, respectively. The vertices have been assigned spatial coordinates according to the profile of the Schwarzschild conformal factor ${\psi}$ through a spatial slice perpendicular to the $z$-axis, and the hypergraphs have been adapted and colored using the local curvature in ${\psi}$.}
\label{fig:Figure14}
\end{figure}

\begin{figure}[ht]
\centering
\includegraphics[width=0.325\textwidth]{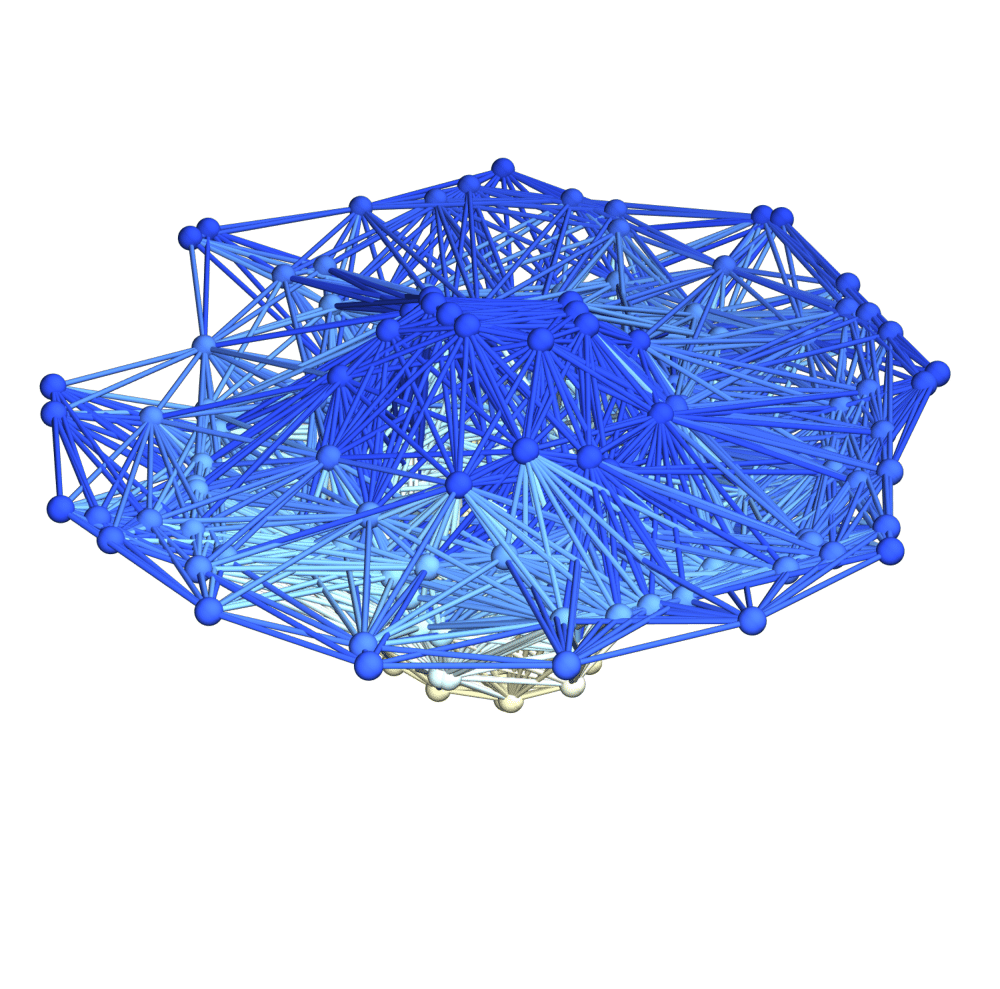}
\includegraphics[width=0.325\textwidth]{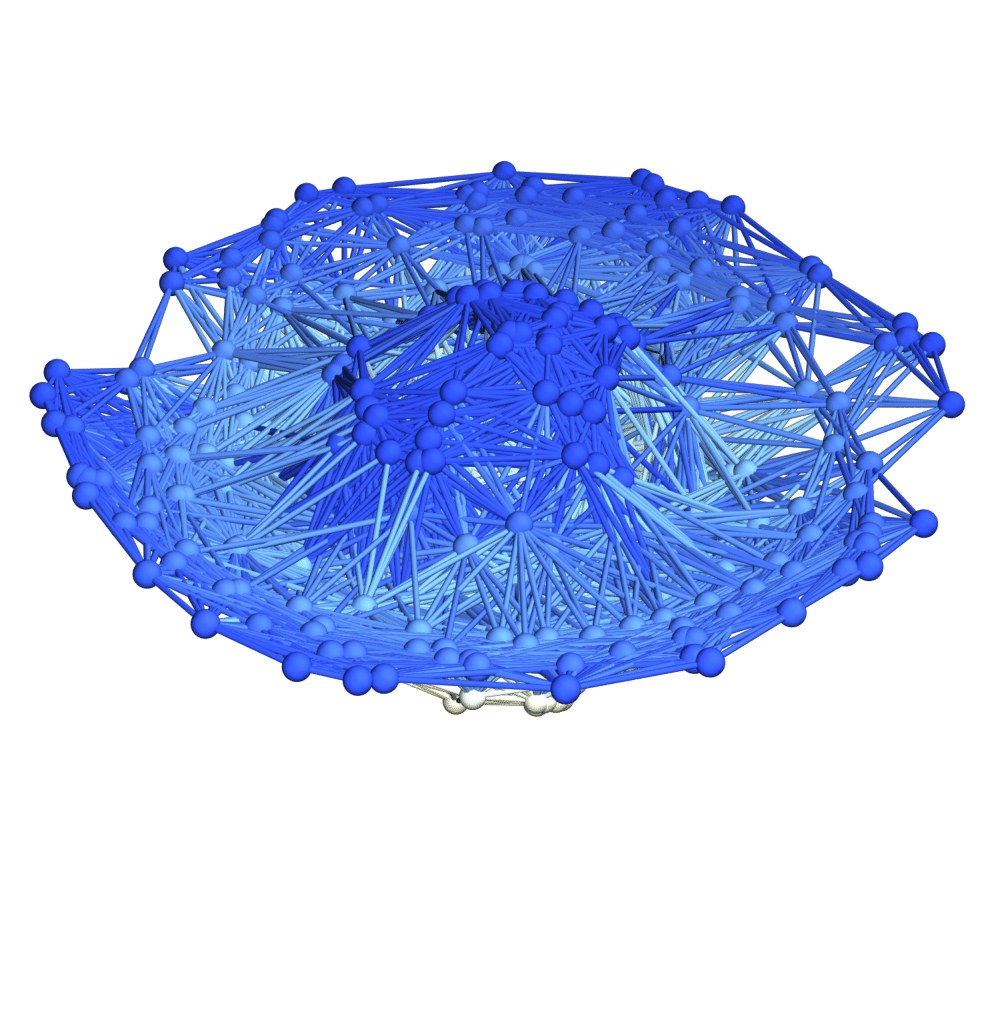}
\includegraphics[width=0.325\textwidth]{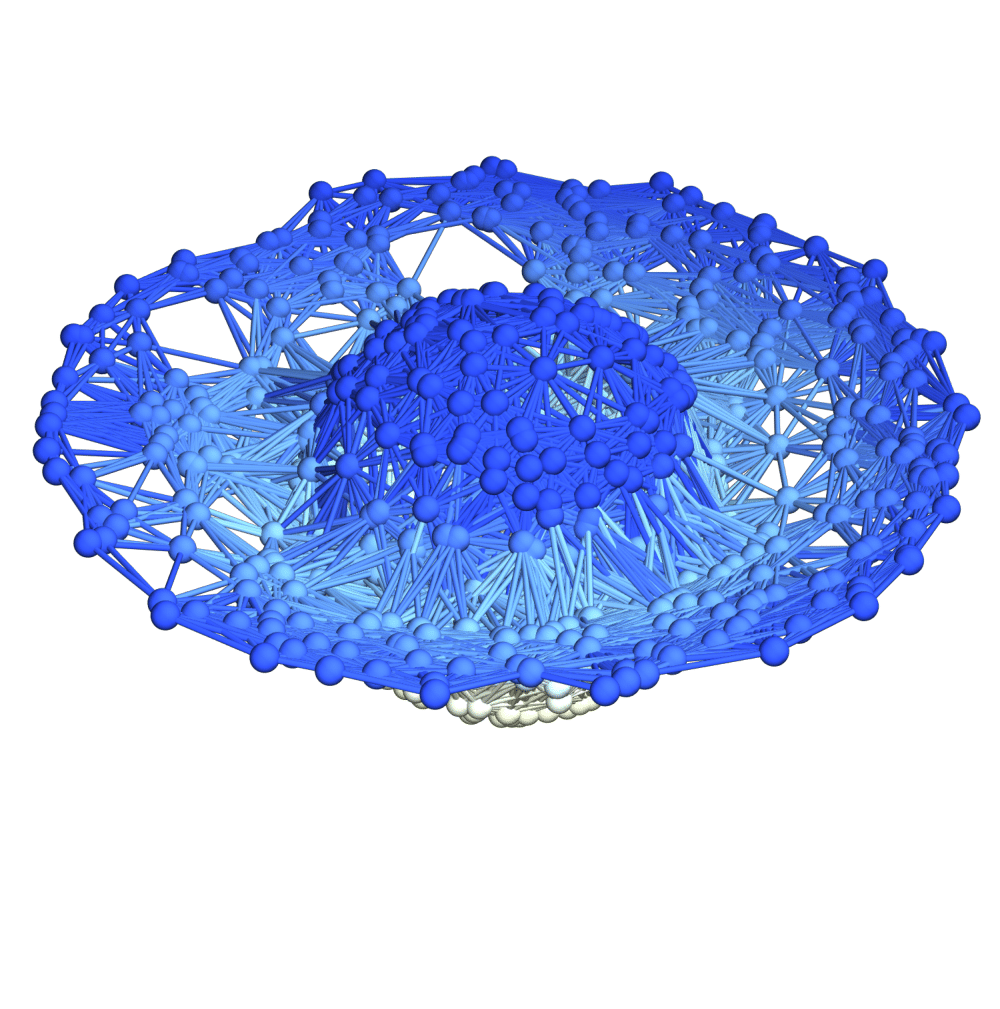}
\caption{Spatial hypergraphs corresponding to projections along the $z$-axis of the first intermediate hypersurface configuration of the massive scalar field ``bubble collapse'' to a non-rotating Schwarzschild black hole test, with an exponential initial density distribution, at time ${t = 1.5 M}$, with resolutions of 200, 400 and 800 vertices, respectively. The vertices have been assigned spatial coordinates according to the profile of the Schwarzschild conformal factor ${\psi}$ through a spatial slice perpendicular to the $z$-axis, and the hypergraphs have been adapted and colored using the local curvature in ${\psi}$.}
\label{fig:Figure15}
\end{figure}

\begin{figure}[ht]
\centering
\includegraphics[width=0.325\textwidth]{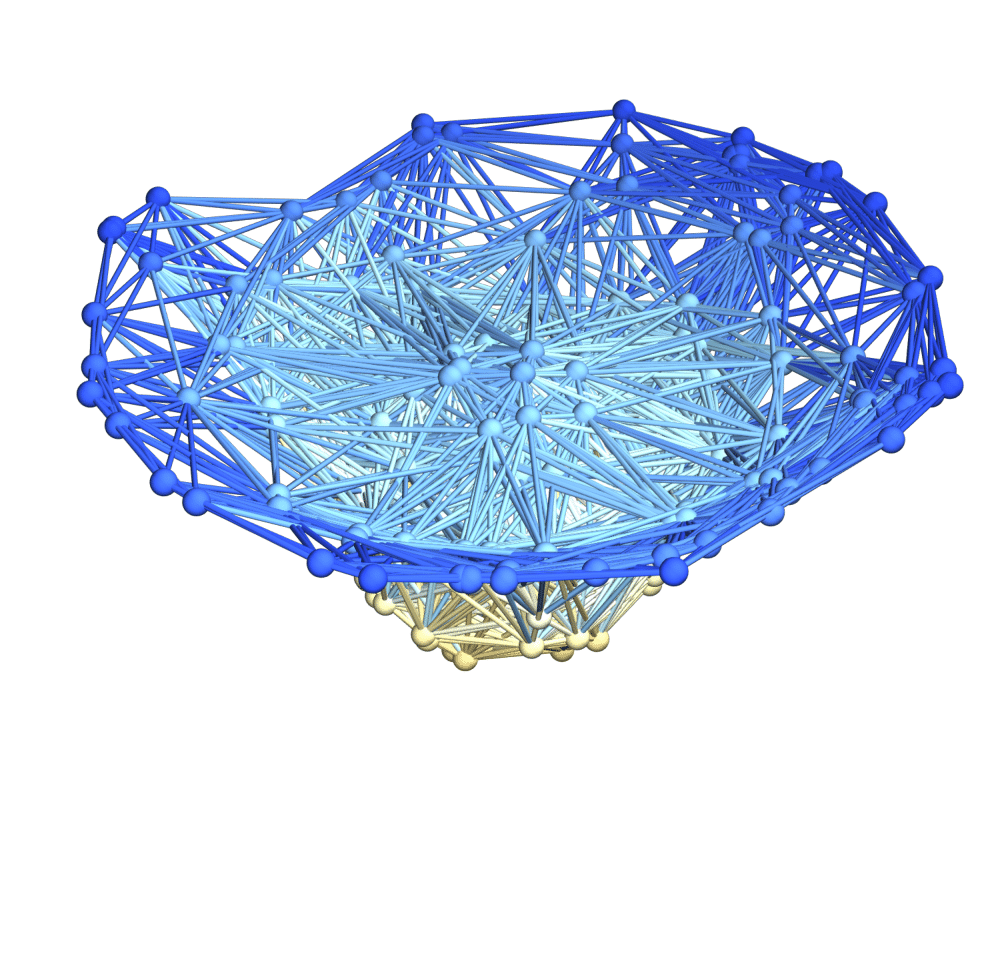}
\includegraphics[width=0.325\textwidth]{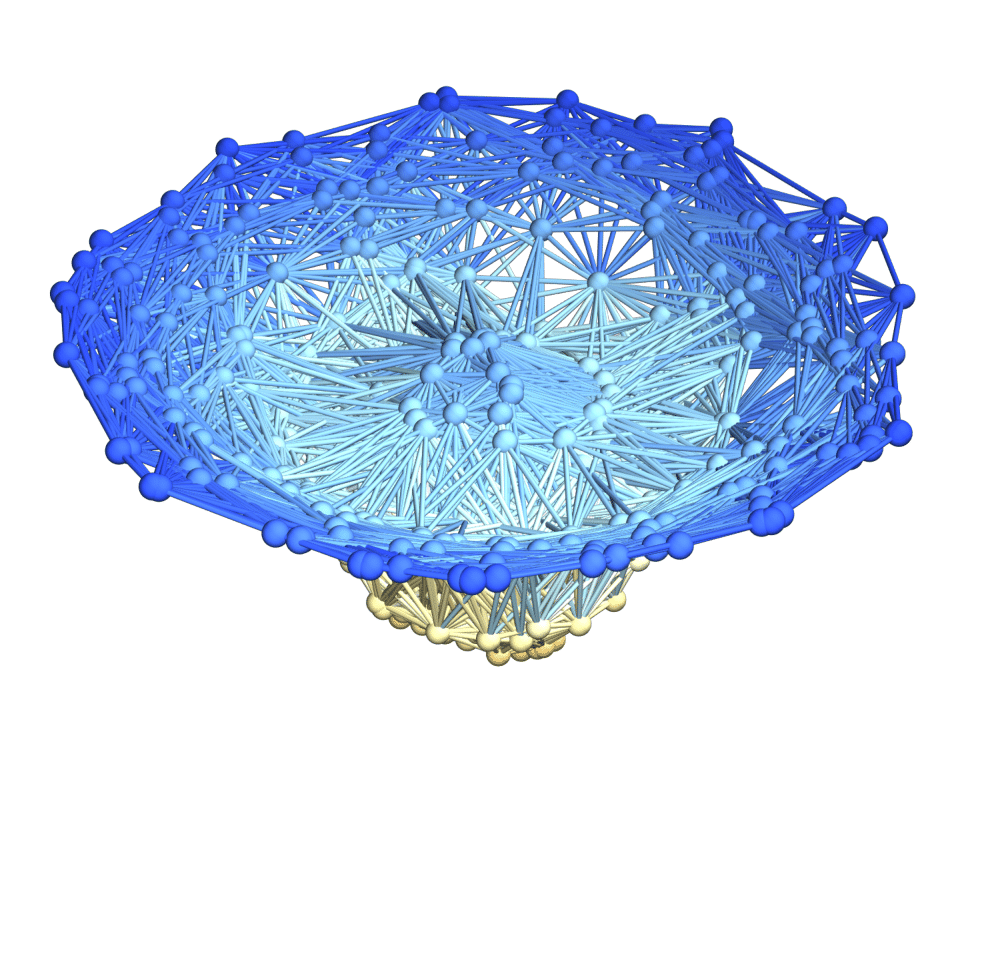}
\includegraphics[width=0.325\textwidth]{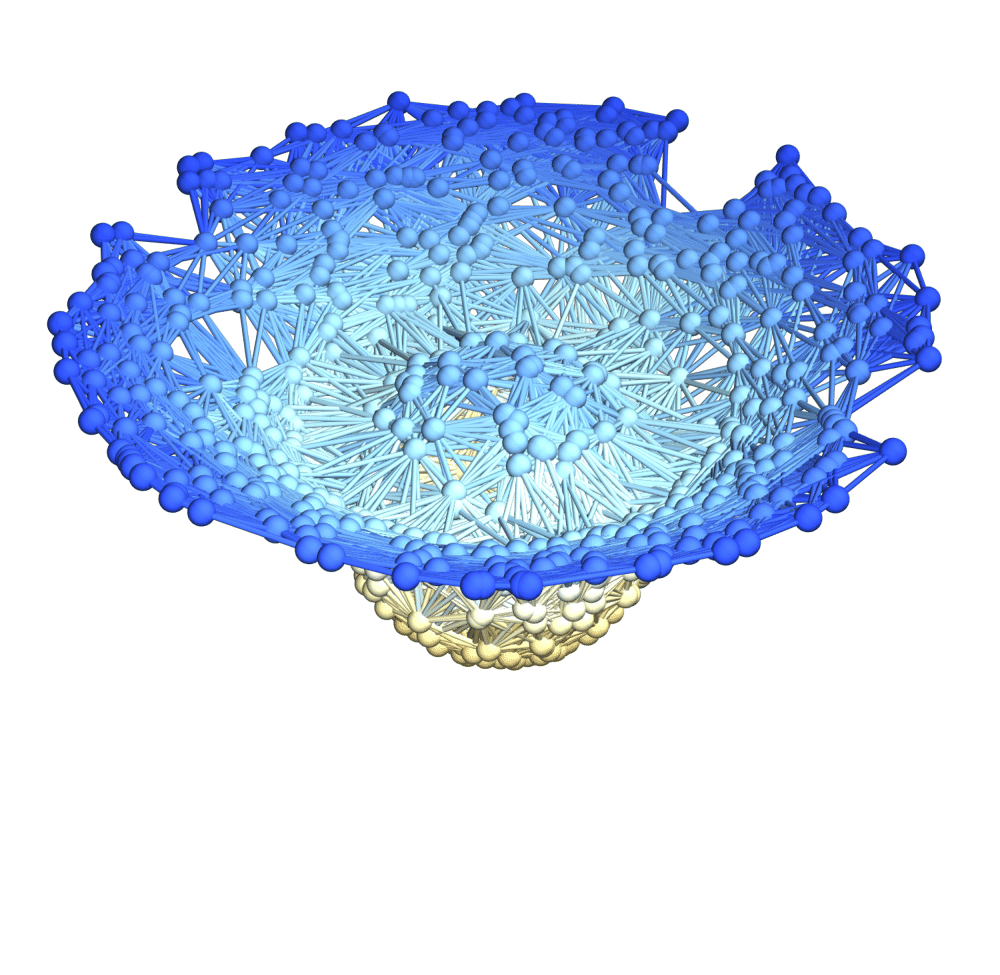}
\caption{Spatial hypergraphs corresponding to projections along the $z$-axis of the second intermediate hypersurface configuration of the massive scalar field ``bubble collapse'' to a non-rotating Schwarzschild black hole test, with an exponential initial density distribution, at time ${t = 3 M}$, with resolutions of 200, 400 and 800 vertices, respectively. The vertices have been assigned spatial coordinates according to the profile of the Schwarzschild conformal factor ${\psi}$ through a spatial slice perpendicular to the $z$-axis, and the hypergraphs have been adapted and colored using the local curvature in ${\psi}$.}
\label{fig:Figure16}
\end{figure}

\begin{figure}[ht]
\centering
\includegraphics[width=0.325\textwidth]{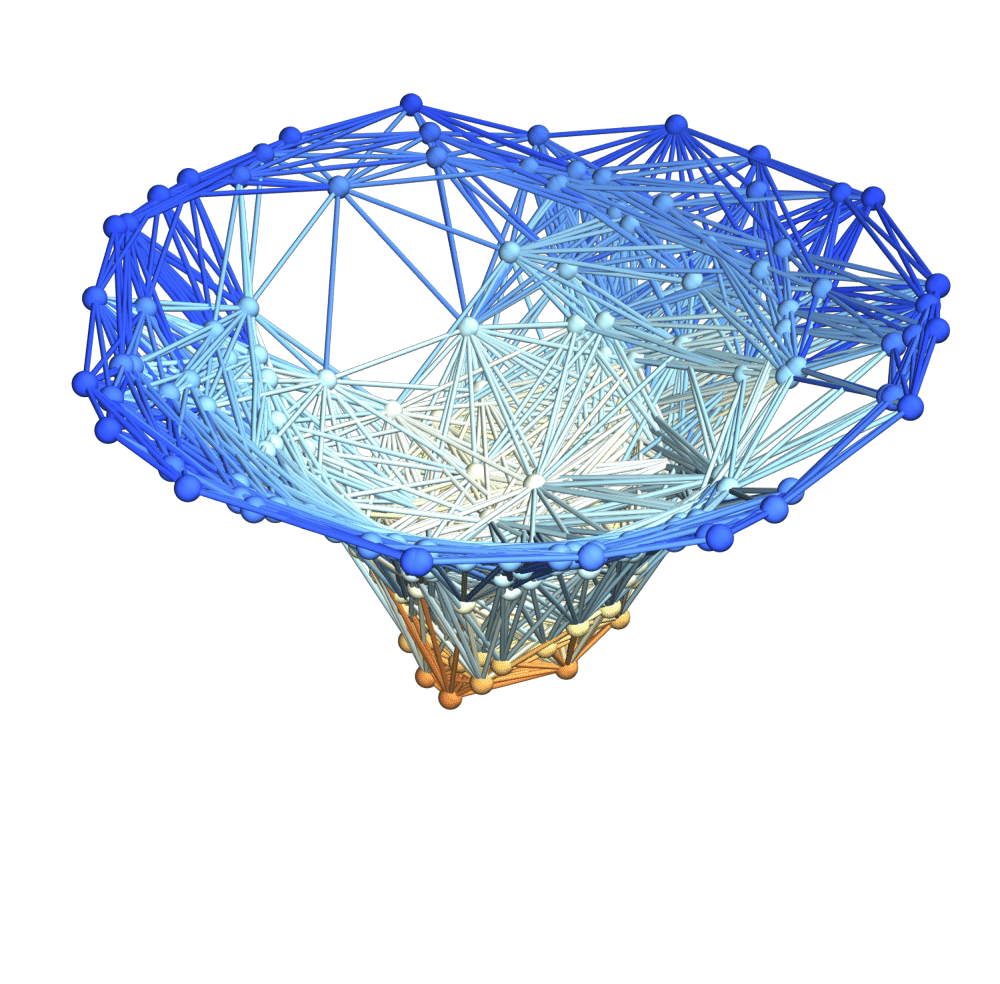}
\includegraphics[width=0.325\textwidth]{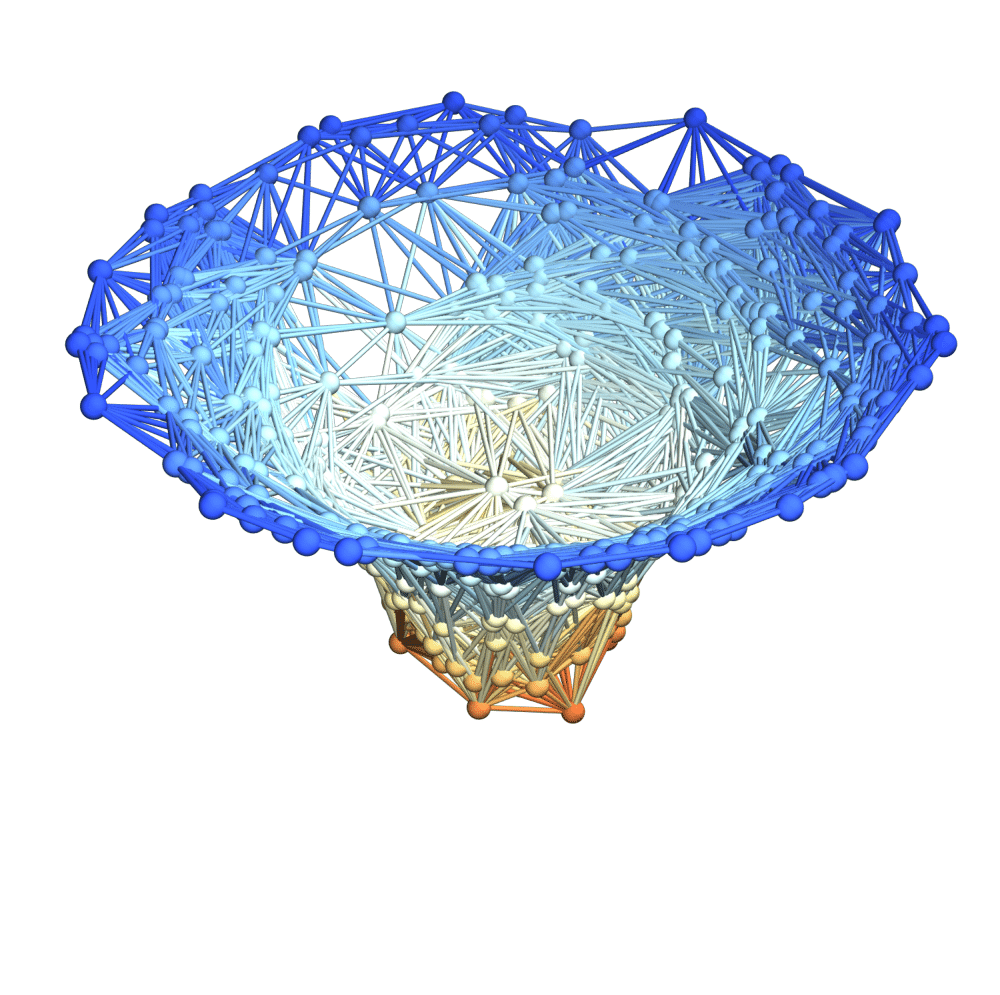}
\includegraphics[width=0.325\textwidth]{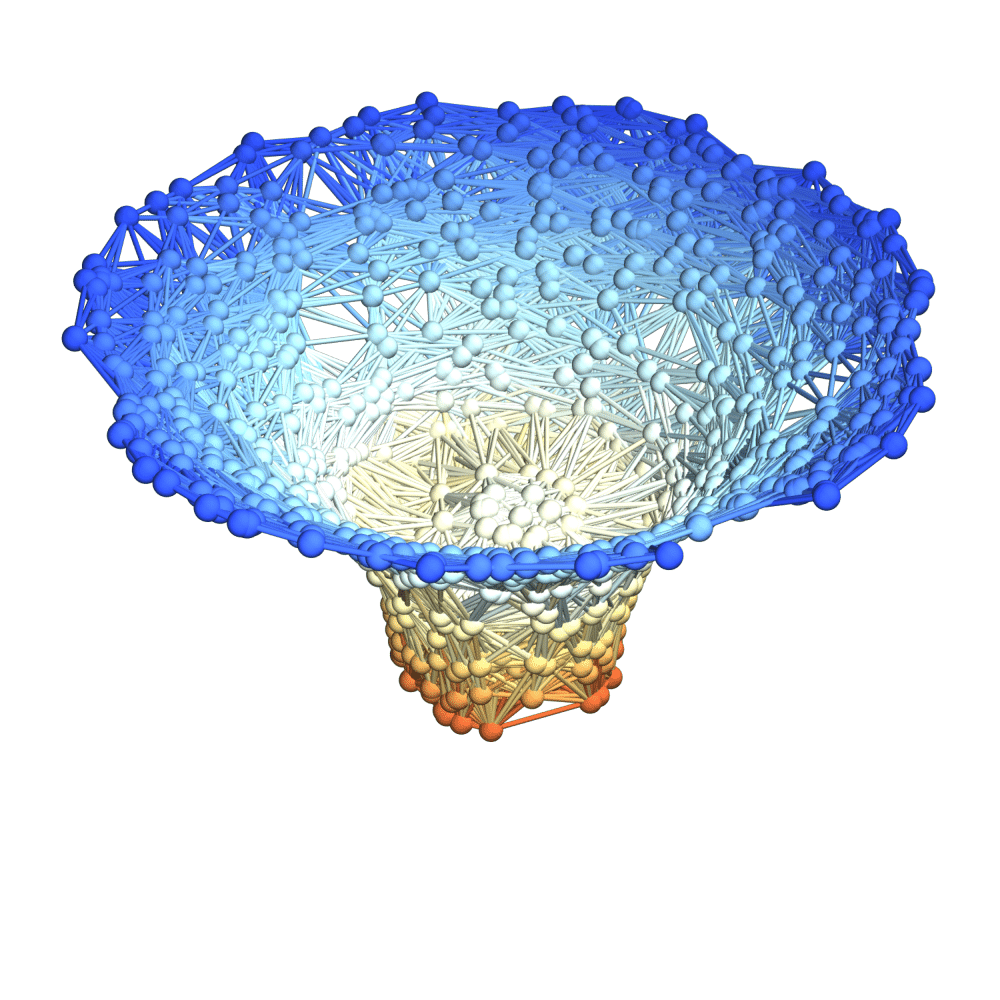}
\caption{Spatial hypergraphs corresponding to projections along the $z$-axis of the final hypersurface configuration of the massive scalar field ``bubble collapse'' to a non-rotating Schwarzschild black hole test, with an exponential initial density distribution, at time ${t = 4.5 M}$, with resolutions of 200, 400 and 800 vertices, respectively. The vertices have been assigned spatial coordinates according to the profile of the Schwarzschild conformal factor ${\psi}$ through a spatial slice perpendicular to the $z$-axis, and the hypergraphs have been adapted and colored using the local curvature in ${\psi}$.}
\label{fig:Figure17}
\end{figure}

In order to test this hypothesis numerically for the case of the exponential initial density profile described above, we place the outermost boundary of the computational domain at radius ${60 M}$, and enforce the Sommerfeld (radiative) boundary condition:

\begin{equation}
\frac{\partial f \left( x_i, t \right)}{\partial t} = - \frac{v x_i}{r} \frac{\partial f \left( x_i, t \right)}{\partial x_i} - v \frac{f \left( x_i, t \right) - f_0 \left( x_i, t \right)}{r},
\end{equation}
for scalar fields $f$, with ${r = \sqrt{x_{1}^{2} + x_{2}^{2} + x_{3}^{2}}}$ denoting the radial distance parameter, ${f_0}$ denoting the values of the specified scalar field at the boundary of the spacetime (which, for the purposes of the tests presented in this article, are taken to be the Minkowski space reference values), and $v$ denoting the radiative velocity (henceforth, we take ${v = 1}$). We evolve the solution until a final time of ${t = 4.5 M}$, with intermediate checks at times ${t = 1.5 M}$ and ${t = 3 M}$; the initial, first intermediate, second intermediate and final hypersurface configurations, with the hypergraphs adapted using the Schwarzschild conformal factor ${\psi}$ and colored using the scalar field ${\Phi \left( t, r \right)}$, are shown in Figures \ref{fig:Figure1}, \ref{fig:Figure2}, \ref{fig:Figure3} and \ref{fig:Figure4}, respectively, with resolutions of 200, 400 and 800 vertices; similarly, Figures \ref{fig:Figure5}, \ref{fig:Figure6}, \ref{fig:Figure7} and \ref{fig:Figure8} show the initial, first intermediate, second intermediate and final hypersurface configurations, but with the hypergraphs \textit{both} adapted \textit{and} colored using the Schwarzschild conformal factor ${\psi}$, respectively. Figure \ref{fig:Figure9} shows the discrete characteristic structure of the solutions after time ${t = 4.5 M}$ (using directed acyclic causal graphs to show discrete characteristic lines). Projections along the $z$-axis of the initial, first intermediate, second intermediate and final hypersurface configurations, with vertices assigned spatial coordinates according to the profile of the Schwarzschild conformal factor ${\psi}$, are shown in Figures \ref{fig:Figure10}, \ref{fig:Figure11}, \ref{fig:Figure12} and \ref{fig:Figure13} (with hypergraphs colored using the scalar field ${\Phi \left( t, r \right)}$) and Figures \ref{fig:Figure14}, \ref{fig:Figure15}, \ref{fig:Figure16} and \ref{fig:Figure17} (with hypergraphs colored using the local curvature in ${\psi}$), respectively.

We confirm that the ADM mass of the overall spacetime (computed by integrating over a surface surrounding the boundary of asymptotic flatness) remains approximately constant, and that the linear and angular momenta of the resulting black hole converge to be approximately zero, as expected. The ${L_1}$, ${L_2}$ and ${L_{\infty}}$-norms of the (non-vacuum) Hamiltonian constraint $H$:

\begin{equation}
H = R - K_{i j} K^{i j} + K^2 - 16 \pi \rho,
\end{equation}
are computed using:

\begin{equation}
\left\lVert H \right\rVert_1 = \sum_{i} m_i \left\lvert H_i \right\rvert, \qquad \left\lVert H \right\rVert_2 = \sqrt{\sum_{i} m_i \left\lvert H_{i}^{2} \right\rvert}, \qquad \left\lVert H \right\rVert_{\infty} = \max_{i} \left[ m_i \left\lvert H \right\rvert \right],
\end{equation}
respectively, with ${m_i}$ designating the fraction of the overall volume of the computational domain being occupied by the $i$-th partitioned subhypergraph:

\begin{equation}
m_i = \frac{V_i}{V_{tot}},
\end{equation}
and likewise for the (non-vacuum) momentum constraints ${M_i}$

\begin{equation}
M_i = \gamma^{j l} \left( \partial_l K_{i j} \partial_i K_{j l} - \Gamma_{j l}^{m} K_{m i} + \Gamma_{j i}^{m} K_{m l} \right) - 8 \pi S_i.
\end{equation}
We also approximate the location of the event horizon of the resulting black hole in the moving puncture gauge with lapse ${\alpha = 0.3}$, confirming that this enables us to excise the spacetime region with lapse ${\alpha < 0.3}$ by setting $H$ to be identically zero (although we stress that, due to the aforementioned singularity-avoidance properties of the maximal-slicing gauge condition\cite{alcubierre}\cite{alcubierre2}, such an excision is strictly unnecessary). The convergence rates for the Hamiltonian constraint after time ${t = 4.5 M}$, with respect to the ${L_1}$, ${L_2}$ and ${L_{\infty}}$-norms, illustrating approximately fourth-order convergence of the finite-difference scheme, are shown in Table \ref{tab:Table1}.

\begin{table}[ht]
\centering
\begin{tabular}{|c|c|c|c|c|c|c|}
\hline
Vertices & ${\epsilon \left( L_1 \right)}$ & ${\epsilon \left( L_2 \right)}$ & ${\epsilon \left( L_{\infty} \right)}$ & ${\mathcal{O} \left( L_1 \right)}$ & ${\mathcal{O} \left( L_2 \right)}$ & ${\mathcal{O} \left( L_{\infty} \right)}$\\
\hline\hline
100 & ${7.36 \times 10^{-3}}$ & ${9.05 \times 10^{-3}}$ & ${2.94 \times 10^{-3}}$ & - & - & -\\
\hline
200 & ${4.83 \times 10^{-4}}$ & ${5.74 \times 10^{-4}}$ & ${3.59 \times 10^{-4}}$ & 3.93 & 3.98 & 3.03\\
\hline
400 & ${2.75 \times 10^{-5}}$ & ${5.30 \times 10^{-5}}$ & ${1.41 \times 10^{-5}}$ & 4.14 & 3.44 & 4.67\\
\hline
800 & ${2.56 \times 10^{-6}}$ & ${2.56 \times 10^{-6}}$ & ${6.28 \times 10^{-7}}$ & 3.42 & 4.37 & 4.49\\
\hline
1600 & ${1.77 \times 10^{-7}}$ & ${2.60 \times 10^{-7}}$ & ${3.53 \times 10^{-8}}$ & 3.85 & 3.30 & 4.15\\
\hline
\end{tabular}
\caption{Convergence rates for the massive scalar field ``bubble collapse'' to a non-rotating Schwarzschild black hole test, with respect to the ${L_1}$, ${L_2}$ and ${L_{\infty}}$-norms for the Hamiltonian constraint $H$ after time ${t = 4.5 M}$, showing approximately fourth-order convergence.}
\label{tab:Table1}
\end{table}

\clearpage

\section{Massive Scalar Field Collapse to a Maximally-Rotating (Extremal) Kerr Black Hole}
\label{sec:Section3}

With the case of a spherically-symmetric (non-rotating) scalar field distribution thus considered, we proceed now to consider the case of an axially-symmetric (rotating) scalar field distribution of the same general form. As mentioned previously, the primary complicating factor with the more general axially-symmetric case is that, unlike in the spherically-symmetric case (in which, by Birkhoff's theorem, all of the relevant geometries are asymptotically Schwarzschild), the exterior spacetime geometry for an axially-symmetric disk of uniformly-rotating, collapsing dust and the exterior spacetime geometry for the resulting maximally-rotating (extremal) Kerr black hole are distinct, being given by the Weyl-Lewis-Papepetrou metric\cite{weyl}\cite{lewis}\cite{papapetrou} and the Kerr metric, respectively. One is therefore forced to construct some kind of smooth transition between the two limiting geometries, at least if one wishes to obtain an analytical description of the collapse. All other aspects of the analysis of Gon\c{c}alves and Moss\cite{goncalves}, including, in particular, the application of the WKB approximation:

\begin{equation}
\Phi \left( t, r \right) = \frac{1}{\mu} \Psi \left( t, r \right) \cos \left( \mu t \right),
\end{equation}
 in the limit of vanishing Compton wavelength ${\frac{1}{\mu} \ll \lambda}$, in order to derive appropriate coordinate conditions for which the stress-energy tensor for an initially axially-symmetric massive scalar field ${\Phi \left( t, r \right)}$ may be approximated by the stress-energy tensor for a uniformly-rotating (and collapsing) disk of dust, may be employed without significant modification from their spherically-symmetric counterparts. Our approach to guaranteeing the necessary smooth transition between geometries will be based on the general methods developed by Neugebauer and Meinel\cite{neugebauer}\cite{neugebauer2}\cite{neugebauer3}, building upon the previous work of Bardeen and Wagoner\cite{bardeen}\cite{bardeen2}, which we briefly recapitulate here.

The line element for the spacetime region surrounding such an axially-symmetric distribution of dust can be expressed in the cylindrical Weyl-Lewis-Papapetrou coordinates ${\left( t, \rho, \varphi, \zeta \right)}$ as:

\begin{equation}
d s^2 = -e^{2 \nu \left( \rho, \zeta \right)} d t^2 + W^2 \left( \rho, \zeta \right) e^{-2 \nu \left( \rho, \zeta \right)} \left( d \varphi - \omega \left( \rho, \zeta \right) \right)^2 + e^{2 \alpha \left( \rho, \zeta \right)} \left( d \rho^2 + d \zeta^2 \right),
\end{equation}
where the metric potential functions ${\alpha \left( \rho, \zeta \right)}$, ${W \left( \rho, \zeta \right)}$, ${\nu \left( \rho, \zeta \right)}$ and ${\omega \left( \rho, \zeta \right)}$ depend upon the non-angular spatial coordinates ${\rho}$ and ${\zeta}$ only (i.e. significantly they are independent of the angular coordinate ${\varphi}$); these are sometimes known as \textit{Weyl's canonical coordinates}\cite{weyl}. In the absence of any rotation, i.e. in the limit as the angular velocity parameter ${\omega \left( \rho, \zeta \right) \to 0}$, this reduces to the standard line element for the static and axially-symmetric Weyl metric:

\begin{equation}
\lim_{\omega \left( \rho, \zeta \right) \to 0} \left[ d s^2 \right] = -e^{2 \nu \left( \rho, \zeta \right)} d t^2 + e^{-2 \nu \left( \rho, \zeta \right)} \rho^2 d \varphi^2 + e^{2 \alpha \left( \rho, \zeta \right) - 2 \nu \left( \rho, \zeta \right)} \left( d \rho^2 + d \zeta^2 \right).
\end{equation}
Along the axis of symmetry, i.e. in the limit as the radial coordinate ${\rho \to 0}$, the required existence of a local isometry to Minkowski space (the \textit{elementary flatness} condition) furthermore implies that:

\begin{equation}
\lim_{\rho \to 0} \left[ \frac{W \left( \rho, \zeta \right)}{\rho} e^{- \left( \alpha \left( \rho, \zeta \right) + \nu \left( \rho, \zeta \right) \right)} \right] = 1,
\end{equation}
which furnishes one with a natural set of boundary conditions on the metric potential functions ${\alpha \left( \rho, \zeta \right)}$, ${W \left( \rho, \zeta \right)}$ and ${\nu \left( \rho, \zeta \right)}$. The other natural set of boundary conditions arises from enforcing the \textit{asymptotic flatness} condition, namely the requirement that, at spatial infinity, i.e. in the limit as ${\rho^2 + \zeta^2 \to \infty}$, the resulting line element must converge asymptotically to the line element of the Minkowski metric in the cylindrical coordinate system:

\begin{equation}
\lim_{\rho^2 + \zeta^2 \to \infty} \left[ d s^2 \right] = -d t^2 + d \rho^2 + d \zeta^2 + \rho^2 d \varphi^2,
\end{equation}
implying the explicit boundary conditions:

\begin{equation}
\lim_{\rho^2 + \zeta^2 \to \infty} \left[ \alpha \left( \rho, \zeta \right) \right] = 0, \qquad \lim_{\rho^2 + \zeta^2 \to \infty} \left[ W \left( \rho, \zeta \right) \right] = \rho, \qquad \lim_{\rho^2 + \zeta^2 \to \infty} \left[ \nu \left( \rho, \zeta \right) \right] = 0,
\end{equation}
and:

\begin{equation}
\lim_{\rho^2 + \zeta^2 \to \infty} \left[ \omega \left( \rho, \zeta \right) \right] = 0.
\end{equation}

By introducing the alternative metric potential functions ${U \left( \rho, \zeta \right)}$, ${k \left( \rho, \zeta \right)}$ and ${a \left( \rho, \zeta \right)}$, relating them to the old functions ${\alpha \left( \rho, \zeta \right)}$, ${W \left( \rho, \zeta \right)}$, ${\nu \left( \rho, \zeta \right)}$ and ${\omega \left( \rho, \zeta \right)}$ as follows:

\begin{equation}
\alpha \left( \rho, \zeta \right) = k \left( \rho, \zeta \right) - U \left( \rho, \zeta \right), \qquad \text{ and } \qquad \frac{e^{2 \nu \left(\rho, \zeta \right)}}{W \left( \rho, \zeta \right)} \pm \omega \left( \rho, \zeta \right) = \frac{1}{W \left( \rho, \zeta \right) e^{-2 U \left( \rho, \zeta \right)} \mp a \left( \rho, \zeta \right)},
\end{equation}
it becomes possible to rewrite the axially-symmetric spacetime line element in the equivalent form:

\begin{equation}
d s^2 = - e^{2 U \left( \rho, \zeta \right)} \left( d t + a \left( \rho, \zeta \right) d \varphi \right)^2 + e^{-2 U \left( \rho, \zeta \right)} \left[ e^{2 k \left( \rho, \zeta \right)} \left( d \rho^2 + d \zeta^2 \right) + W^2 \left( \rho, \zeta \right) d \varphi^2 \right].
\end{equation}
In order to obtain the metric for a uniformly-rotating disk of dust, we set:

\begin{equation}
W \left( \rho, \zeta \right) = \rho,
\end{equation}
such that the metric can be recovered from the (complex) Ernst potential ${f \left( \rho, \zeta \right)}$:

\begin{equation}
f \left( \rho, \zeta \right) = e^{2 U \left( \rho, \zeta \right)} + i \psi \left( \rho, \zeta \right).
\end{equation}
In the above, we are assuming an oriented spacetime with a timelike Killing vector field $X$, such that the \textit{twist 1-form}, denoted ${\tau}$, is given by:

\begin{equation}
* \tau = \xi \wedge d \xi, \qquad \text{ where } \qquad \xi = X_i d x^i,
\end{equation}
in Cartesian coordinates. The function ${\psi \left( \rho, \zeta \right)}$ in the imaginary part of the Ernst potential ${f \left( \rho, \zeta \right)}$ arises from the fact that, within such a spacetime, one can (locally) write the twist 1-form as:

\begin{equation}
\tau = d \psi \left( \rho, \zeta \right),
\end{equation}
for some constant ${\psi}$ along $X$. The real part ${e^{2 U \left( \rho, \zeta \right)}}$ of the Ernst potential ${f \left( \rho, \zeta \right)}$ is then simply given by:

\begin{equation}
e^{2 U \left( \rho, \zeta \right)} = X_i X^i.
\end{equation}
The collapse of an axially-symmetric, uniformly-rotating disk of dust can therefore be solved as a boundary value problem for the non-linear Ernst equation:

\begin{equation}
\left[ \mathrm{Re} \left( f \left( \rho, \zeta \right) \right) \right] \left[ \partial_{\rho \rho} f \left( \rho, \zeta \right) + \frac{\partial_{\rho} f \left( \rho, \zeta \right)}{\rho} + \partial_{\zeta \zeta} f \left( \rho, \zeta \right) \right] = \left( \partial_{\rho} f \left( \rho, \zeta \right) \right)^2 + \left( \partial_{\zeta} f \left( \rho, \zeta \right) \right)^2,
\end{equation}
with the new metric potential functions ${k \left( \rho, \zeta \right)}$ and ${a \left( \rho, \zeta \right)}$ being calculable by means of the following integrals:

\begin{equation}
k \left( \rho, \zeta \right) = \int_{0}^{\rho} \tilde{\rho} \left[ \left( \partial_{\tilde{\rho}} U \left( \tilde{\rho}, \zeta \right) \right)^2 - \left( \partial_{\zeta} U \left( \tilde{\rho}, \zeta \right) \right)^2 + \frac{e^{-4 U \left( \tilde{\rho}, \zeta \right)}}{4} \left( \left( \partial_{\tilde{\rho}} \psi \left( \tilde{\rho}, \zeta \right) \right)^2 - \left( \partial_{\zeta} \psi \left( \tilde{\rho}, \zeta \right) \right)^2 \right) \right] d \tilde{\rho},
\end{equation}
and:

\begin{equation}
a \left( \rho, \zeta \right) = \int_{0}^{\rho} \tilde{\rho} e^{-4 U \left( \tilde{\rho}, \zeta \right)} \partial_{\zeta} \psi \left( \tilde{\rho}, \zeta \right) d \tilde{\rho},
\end{equation}
respectively.

Although the Kerr solution is conventionally parametrized in terms of a total mass $M$ and a formal ``angular momentum'' parameter $J$, we choose instead to parametrize the uniformly-rotating disk of dust in terms of a coordinate radius ${\rho_0}$ and a function ${\mu}$ of the angular velocity parameter ${\Omega}$ (not to be confused with the mass parameter for the scalar field, which will no longer be required):

\begin{equation}
\mu = 2 \Omega^2 \rho_{0}^{2} e^{-2 V_0}, \qquad \text{ where } \qquad V_0 = U \left( 0, 0 \right),
\end{equation}
with ${V_0}$ playing the role of a ``surface potential''. These parameters are chosen in part to avoid the otherwise counterintuitive relationship ${J \geq M^2}$ between the angular momentum and mass of the disk (which, in the case of the Kerr solution, would indicate supercriticality). If we now introduce the following, purely imaginary, function ${H \left( K \right)}$:

\begin{equation}
H \left( K \right) = \frac{\mu \log \left[ \sqrt{1 + \mu^2 \left( 1 + \frac{K^2}{\rho_{0}^{2}} \right)^2} + \mu \left( 1 + \frac{K^2}{\rho_{0}^{2}} \right) \right]}{\pi i \rho_{0}^{2} \sqrt{1 + \mu^2 \left( 1 + \frac{K^2}{\rho_{0}^{2}} \right)^2}}, \qquad \text{ with } \qquad \mathrm{Re} \left( H \left( K \right) \right) = 0,
\end{equation}
as well as the complex functions ${Z \left( K, \rho, \zeta \right)}$ and ${Z_1 \left( K, \rho, \zeta \right)}$:

\begin{equation}
Z \left( K, \rho, \zeta \right) = \sqrt{\left( K + i \left( \rho + i \zeta \right) \right) \left( K - i \left( \rho - i \zeta \right) \right) \left( K^2 - K_{1}^{2} \right) \left( K^2 - K_{2}^{2} \right)},
\end{equation}
and:

\begin{equation}
Z_1 \left( K, \rho, \zeta \right) = \sqrt{\left( K + i \left( \rho + i \zeta \right) \right) \left( K - i \left( \rho - i \zeta \right) \right)},
\end{equation}
such that ${\mathrm{Re} \left( Z_1 \left( K, \rho, \zeta \right) \right) < 0}$ whenever ${\rho}$ and ${\zeta}$ lie strictly outside the disk, then we can write the Ernst potential ${f \left( \rho, \zeta \right)}$ purely in terms of hyperelliptic integrals as:

\begin{equation}
f \left( \rho, \zeta \right) = \exp \left\lbrace \int_{K_1}^{K_a \left( \rho, \zeta \right)} \frac{K^2}{Z \left( K, \rho, \zeta \right)} dK + \int_{K_2}^{K_b \left( \rho, \zeta \right)} \frac{K^2}{Z \left( K, \rho, \zeta \right)} dK - \int_{-i \rho_0}^{i \rho_0} \frac{H \left( K \right)}{Z_1 \left( K, \rho, \zeta \right)} K^2 dK \right\rbrace,
\end{equation}
where the lower limits of integration ${K_1}$ and ${K_2}$ are related by complex conjugation:

\begin{equation}
K_1 = - \bar{K_2} = \rho_0 \sqrt{\frac{i - \mu}{\mu}}, \qquad \text{ with } \qquad \mathrm{Re} \left( K_1 \right) = - \mathrm{Re} \left( K_2 \right) < 0,
\end{equation}
and where the upper limits of integration ${K_a \left( \rho, \zeta \right)}$ and ${K_b \left( \rho, \zeta \right)}$ are specified in terms of \textit{ultraelliptic} functions (i.e. hyperelliptic functions of two variables, ${\rho}$ and ${\zeta}$), defined via:

\begin{equation}
\int_{K_1}^{K_a \left( \rho, \zeta \right)} \frac{dK}{Z \left( K, \rho, \zeta \right)} + \int_{K_2}^{K_b \left( \rho, \zeta \right)} \frac{dK}{Z \left( K, \rho, \zeta \right)} = \int_{-i \rho_0}^{i \rho_0} \frac{H \left( K \right)}{Z_1 \left( K, \rho, \zeta \right)} dK,
\end{equation}
and:

\begin{equation}
\int_{K_1}^{K_a \left( \rho, \zeta \right)} \frac{K}{Z \left( K, \rho, \zeta \right)} dK + \int_{K_2}^{K_b \left( \rho, \zeta \right)} \frac{K}{Z \left( K, \rho, \zeta \right)} dK = \int_{-i \rho_0}^{i \rho_0} \frac{H \left( K \right)}{Z_1 \left( K, \rho, \zeta \right)} K dK,
\end{equation}
respectively.

The problem of recovering the upper integration limits ${K_a \left( \rho, \zeta \right)}$ and ${K_b \left( \rho, \zeta \right)}$ from these integral relations is now an instance of the Jacobi inversion problem for Abelian integrals, which can be solved explicitly in terms of (hyperelliptic) theta functions\cite{belokolos}\cite{neugebauer4}. Specifically, if we define a theta function ${\theta \left( x, y ; p, q, \alpha \right)}$ using:

\begin{equation}
\theta \left( x, y ; p, q, \alpha \right) = \sum_{m = - \infty}^{\infty} \sum_{n = - \infty}^{\infty} \left( -1 \right)^{m + n} p ^{m^2} q^{n^2} e^{2 m x + 2 n y + 4 m n \alpha},
\end{equation}
then the expression for the Ernst potential ${f \left( \rho, \zeta \right)}$ in terms of hyperelliptic integrals may be reformulated as:

\begin{multline}
f \left( \rho, \zeta \right) = \frac{\theta \left( \alpha_0 u \left( \rho, \zeta \right) + \alpha_1 v \left( \rho, \zeta \right) - C_1 \left( \rho, \zeta \right), \beta_0 u \left( \rho, \zeta \right) + \beta_1 v \left( \rho, \zeta \right) - C_2 \left( \rho, \zeta \right); p, q, \alpha \right)}{\theta \left( \alpha_0 u \left( \rho, \zeta \right) + \alpha_1 v \left( \rho, \zeta \right) + C_1 \left( \rho, \zeta \right), \beta_0 u \left( \rho, \zeta \right) + \beta_1 v \left( \rho, \zeta \right) + C_2 \left( \rho, \zeta \right); p, a, \alpha \right)}\\
\times e^{- \left( \gamma_0 u \left( \rho, \zeta \right) + \gamma_1 v \left( \rho, \zeta \right) + \mu w \left( \rho, \zeta \right) \right)}, 
\end{multline}
with functions ${u \left( \rho, \zeta \right)}$, ${v \left( \rho, \zeta \right)}$ and ${w \left( \rho, \zeta \right)}$ given by the integrals:

\begin{equation}
u \left( \rho, \zeta \right) = \int_{-i \rho_0}^{i \rho_0} \frac{H \left( K \right)}{Z_1 \left( K, \rho, \zeta \right)} dK, \qquad v \left( \rho, \zeta \right) = \int_{-i \rho_0}^{i \rho_0} \frac{H \left( K \right)}{Z_1 \left( K, \rho, \zeta \right)} K dK,
\end{equation}
and:

\begin{equation}
w \left( \rho, \zeta \right) = \int_{-i \rho_0}^{i \rho_0} \frac{H \left( K \right)}{Z_1 \left( K, \rho, \zeta \right)} K^2 dK,
\end{equation}
respectively, and where the normalization parameters ${\alpha_0}$, ${\alpha_1}$, ${\beta_0}$, ${\beta_1}$, ${\gamma_0}$ and ${\gamma_1}$, as well as the moduli of the theta function $p$, $q$ and ${\alpha}$, and the pair of functions ${C_1 \left( \rho, \zeta \right)}$ and ${C_2 \left( \rho, \zeta \right)}$, are all defined over the two-sheeted hyperelliptic Riemann surface associated to the complex function ${Z \left( K, \rho, \zeta \right)}$. More precisely, this is the hyperelliptic Riemann surface with cuts between the branch points ${K_1}$ and ${\bar{K_1}}$, ${K_2}$ and ${\bar{K_2}}$, and ${-i \left( \rho - i \zeta \right)}$ and ${i \left( \rho + i \zeta \right)}$, and with upper/lower sheets defined by ${Z \left( K, \rho, \zeta \right) \to \mu K^3}$ in the limit as ${K \to \infty}$. Thus, for consistency, the integrals from ${K_1}$ to ${K_a \left( \rho, \zeta \right)}$ and from ${K_2}$ to ${K_b \left( \rho, \zeta \right)}$ must be evaluated along the same curve on this Riemann surface. The first four normalization parameters ${\alpha_0}$, ${\alpha_1}$, ${\beta_0}$ and ${\beta_1}$ can be obtained from ${d \omega_1}$ and ${d \omega_2}$, namely two normalized Abelian differentials of the first kind, i.e:

\begin{equation}
d \omega_1 = \alpha_0 \frac{d K}{Z \left( K, \rho, \zeta \right)} + \alpha_1 \frac{K}{Z \left( K, \rho, \zeta \right)} d K, \qquad \text{ and } \qquad d \omega_2 = \beta_0 \frac{d K}{Z \left( K, \rho, \zeta \right)} + \beta_1 \frac{K}{Z \left( K, \rho, \zeta \right)} d K,
\end{equation}
given by:

\begin{equation}
\oint_{a_m} d \omega_n = \pi i \delta_{m n}, \qquad \text{ with } \qquad m, n \in \left\lbrace 1, 2 \right\rbrace,
\end{equation}
which can then be solved as a system of linear algebraic equations in order to recover ${\alpha_0}$, ${\alpha_1}$, ${\beta_0}$ and ${\beta_1}$, represented in terms of closed curves (periods) ${a_1}$ and ${a_2}$. Likewise, the remaining two normalization parameters ${\gamma_0}$ and ${\gamma_1}$ can be obtained from ${d \omega_3}$, a normalized Abelian differential of the third kind, i.e:

\begin{equation}
d \omega_3 = \gamma_0 \frac{d K}{Z \left( K, \rho, \zeta \right)} + \gamma_1 \frac{K}{Z \left( K, \rho, \zeta \right)} d K + \mu \frac{K^2}{Z \left( K, \rho, \zeta \right)} d K,
\end{equation}
whose periods ${a_1}$ and ${a_2}$ vanish:

\begin{equation}
\oint_{a_j} d \omega_3 = 0, \qquad \text{ with } \qquad j \in \left\lbrace 1, 2 \right\rbrace,
\end{equation}
to yield another system of linear algebraic equations for ${\gamma_0}$ and ${\gamma_1}$. Moreover, the moduli of the theta function $p$, $q$ and ${\alpha}$ may be derived from the Riemann matrix:

\begin{equation}
\left( B_{i j} \right) = \begin{bmatrix}
\log \left( p \right) & 2 \alpha\\
2 \alpha & \log \left( q \right)
\end{bmatrix},
\end{equation}
given by:

\begin{equation}
B_{i j} = \oint_{b_i} d \omega_j, \qquad \text{ with } \qquad i, j \in \left\lbrace 1, 2 \right\rbrace,
\end{equation}
such that the real part of ${\left( B_{i j} \right)}$ is negative definite, with another pair of closed curves (periods) ${b_1}$ and ${b_2}$. Finally, the pair of functions ${C_1 \left( \rho, \zeta \right)}$ and ${C_2 \left( \rho, \zeta \right)}$ can be computed by means of the integral:

\begin{equation}
C_i \left( \rho, \zeta \right) = - \int_{- i \left( \rho - i \zeta \right)}^{\infty^{+}} d \omega_i, \qquad \text{ with } \qquad i \in \left\lbrace 1, 2 \right\rbrace,
\end{equation}
with ${+}$ denoting the upper sheet of the hyperelliptic Riemann surface.

We are now able to determine that the surface mass density of the uniformly rotating disk of dust, denoted ${\sigma \left( \mu, \rho \right)}$, and given by:

\begin{equation}
\sigma \left( \mu, \rho \right) = - \frac{\Omega}{2 \pi e^{V_0 \left( \mu \right)}} \frac{\psi_{0}^{\prime} \left( \mu \left[ 1 - \frac{\rho^2}{\rho_{0}^{2}} \right] \right)}{e^{V_0 \left( \mu \left[ 1 - \frac{\rho^2}{\rho_{0}^{2}} \right] \right)}},
\end{equation}
is strictly positive, where, in the above, the surface potential parameter ${V_0 \left( \mu \right)}$ now depends only upon the angular velocity-dependent parameter ${\mu}$:

\begin{equation}
V_0 \left( \mu \right) = - \frac{1}{2} \sinh^{-1} \left\lbrace \mu + \frac{1 + \mu^2}{\wp \left[ I \left( \mu \right) ; \frac{4}{3} \mu^2 - 4, \frac{8}{3} \mu \left( 1 + \frac{\mu^2}{9} \right) \right] - \frac{2}{3} \mu} \right\rbrace,
\end{equation}
with ${\wp \left( x ; g_2, g_3 \right)}$ being the Weierstrass elliptic function:

\begin{equation}
\int_{\wp \left( x; g_2, g_3 \right)}^{\infty} \frac{dt}{\sqrt{4 t^3 - g_2 t - g_3}} = x,
\end{equation}
and with the single-parameter function ${I \left( \mu \right)}$ given by:

\begin{equation}
I \left( \mu \right) = \frac{1}{\pi} \int_{0}^{\mu} \frac{\log \left( x + \sqrt{1 + x^2} \right)}{\sqrt{\left( 1 + x^2 \right) \left( \mu - x \right)}} dx.
\end{equation}
Moreover, the imaginary part of the Ernst potential ${\psi \left( \rho, \zeta \right)}$ now also depends only upon ${\mu}$, hence the use of the notation ${\psi_0 \left( \mu \right)}$:

\begin{equation}
\psi \left( \rho, \zeta \right) = \psi_0 \left( \mu \left[ 1 - \frac{\rho^2}{\rho_{0}^{2}} \right] \right), \qquad \text{ with } \qquad \zeta = 0^{+}.
\end{equation}
If ${\mu_0}$ denotes the first zero of the Weierstrass elliptic function expression in the denominator of the equation for the surface potential ${V_0 \left( \mu \right)}$, i.e:

\begin{equation}
\wp \left[ I \left( \mu \right) ; \frac{4}{3} \mu^2 - 4, \frac{8}{3} \mu \left( 1 + \frac{\mu^2}{9} \right) \right] - \frac{2}{3} \mu = 0,
\end{equation}
then we have that the solution is non-singular (i.e. regular) whenever:

\begin{equation}
0 < \mu < \mu_0 \approx 4.63, \qquad \iff \qquad 0 > V_0 \left( \mu \right) > - \infty.
\end{equation}
Note that both along the axis of symmetry (${\rho = 0}$) and within the plane of rotation of the disk (${\zeta = 0}$), the hyperelliptic integrals for the Ernst potential:

\begin{equation}
f \left( \rho, \zeta \right) = \exp \left\lbrace \int_{K_1}^{K_a \left( \rho, \zeta \right)} \frac{K^2}{Z \left( K, \rho, \zeta \right)} dK + \int_{K_2}^{K_b \left( \rho, \zeta \right)} \frac{K^2}{Z \left( K, \rho, \zeta \right)} dK - \int_{- i \rho_0}^{i \rho_0} \frac{H \left( K \right)}{Z_1 \left( K, \rho, \zeta \right)} K^2 dK \right\rbrace,
\end{equation}
as well as the corresponding hyperelliptic integrals for the integration limits, all reduce to ordinary elliptic ones. In the former case, with ${\rho = 0}$, the potential may be written out as the following ratio of elliptic integrals:

\begin{equation}
f \left( 0, \zeta \right) = \frac{2 \pi + \int_{-1}^{1} \frac{\beta \left( x \right)}{i x - \frac{\zeta}{\rho_0}} dx}{2 \pi + \int_{-1}^{1} \frac{\alpha \left( x \right)}{i x - \frac{\zeta}{\rho_0}} dx}, \qquad \text{ with } \qquad \zeta > 0,
\end{equation}
with the function ${\beta \left( x \right)}$ defined as satisfying the following integral equation:

\begin{multline}
\beta \left( x \right) = \left( 2 \mu \right)^{\frac{3}{2}} e^{- V_0} x \left( 1 - x^2 \right) - \mu^2 \left[ \left( 1 - x^2 \right)^2 \beta \left( x \right) \right.\\
\left. + \frac{1 - x^2}{\pi^2} \left( \int_{-1}^{1} \frac{1 - \left( x^{\prime} \right)^2}{x^{\prime} - x} d x^{\prime} \right) \left( \int_{-1}^{1} \frac{\beta \left( x^{\prime \prime} \right)}{x^{\prime \prime} - x^{\prime}} d x^{\prime \prime} \right) \right],
\end{multline}
with the two improper integrals designating the Cauchy principal values, and with the function ${\alpha \left( x \right)}$ depending algebraically upon both $x$ and ${\beta \left( x \right)}$ as follows:

\begin{equation}
\alpha \left( x \right) = \frac{\left( 1 - w \left( x \right) \right) \beta \left( x \right) + i \sqrt{4 w^2 \left( x \right) e^{-4 V_0} \left( \psi_{0}^{2} + 4 \Omega^2 \rho_{0}^{2} x^2 \right) - \left( e^{4 V_0} + w^2 \left( x \right) \right) \beta^2 \left( x \right)}}{i \psi_0 - 2 \Omega \rho_0 x},
\end{equation}
where, for the sake of notational convenience, we have introduced the scalar function ${w \left( x \right)}$ and the scalar constant ${\psi_0}$ as:

\begin{equation}
w \left( x \right) = 2 \Omega^2 \rho_{0}^{2} \left( 1 - x^2 \right), \qquad \text{ and } \qquad \psi_0 = - \sqrt{1 - e^{4 V_0} - 4 \Omega^2 \rho_{0}^{2}},
\end{equation}
respectively. Likewise in the latter case, with ${\zeta = 0}$ (within the disk), the metric potential functions ${U \left( \rho, \zeta \right)}$, ${a \left( \rho, \zeta \right)}$ and ${k \left( \rho, \zeta \right)}$ may be computed purely in terms of ordinary elliptic functions as:

\begin{equation}
e^{2 U \left( \rho, \zeta \right)} = \exp \left\lbrace 2 V_0 \left( \mu \left[ 1 - \frac{\rho^2}{\rho_{0}^{2}} \right] \right) \right\rbrace - \frac{\mu \rho^2}{2 \rho_{0}^{2}},
\end{equation}
\begin{equation}
\left( 1 + \Omega a \left( \rho, \zeta \right) \right) e^{2 U \left( \rho, \zeta \right)} = e^{V_0 \left( \mu \right)} \exp \left\lbrace V_0 \left( \mu \left[ 1 - \frac{\rho^2}{\rho_{0}^{2}} \right] \right) \right\rbrace,
\end{equation}
and:

\begin{equation}
e^{2 k \left( \rho, \zeta \right) - 2 U \left( \rho, \zeta \right)} = e^{-2 V_0 \left( \mu \right)} \exp \left\lbrace - \int_{\mu \left( 1 - \frac{\rho^2}{\rho_{0}^{2}} \right)}^{\mu} \frac{f_{0}^{\prime} \left( \tilde{\mu} \right) \bar{f}_{0}^{\prime} \left( \tilde{\mu} \right)}{f_0 \left( \tilde{\mu} \right) \bar{f}_0 \left( \tilde{\mu} \right)} d \tilde{\mu} \right\rbrace,
\end{equation}
respectively, at least in the all cases for which ${\rho \leq \rho_0}$, where the function ${f_0 \left( \mu \right)}$ represents the value of the Ernst potential ${f \left( \rho, \zeta \right)}$ at ${\rho = 0}$ and ${\zeta = 0^{+}}$:

\begin{equation}
f_0 \left( \mu \right) = e^{2 V_0 \left( \mu \right)} + i \psi_0 \left( \mu \right).
\end{equation}

The angular velocity parameter (considered as a function of both ${\mu}$ and the coordinate radius ${\rho_0}$) can hence be computed by combining the original definition of the ${\mu}$ parameter:

\begin{equation}
\mu = 2 \Omega^2 \left( \mu, \rho_0 \right) \rho_{0}^{2} e^{-2 V_0 \left( \mu \right)},
\end{equation}
with the expression for the surface potential parameter ${V_0 \left( \mu \right)}$ in terms of the Weierstrass elliptic function ${\wp \left( x ; g_2, g_3 \right)}$:

\begin{equation}
V_0 \left( \mu \right) = - \frac{1}{2} \sinh^{-1} \left\lbrace \mu + \frac{1 + \mu^2}{\wp \left[ I \left( \mu \right) ; \frac{4}{3} \mu^2 - 4, \frac{8}{3} \mu \left( 1 + \frac{\mu^2}{9} \right) \right] - \frac{2}{3} \mu} \right\rbrace.
\end{equation}
The Newtonian limit of the uniformly-rotating disk of dust configuration now corresponds to the case in which the parameter ${\mu \ll 1}$, in which case the solution reduces to the so-called ``Maclaurin disk'' solution (i.e. a gaseous self-gravitating astrophysical disk under Newtonian gravity). We can see this directly by expanding the solution for the Ernst potential (written with an explicit dependence on ${\mu}$) ${f \left( \rho, \zeta \right)}$ as a formal power series, with expansion parameter ${\sqrt{\mu}}$, around the point ${\mu = 0}$:

\begin{equation}
f \left( \mu, \rho, \zeta \right) = 1 + \sum_{n = 1}^{\infty} f_n \left( \rho, \zeta \right) \mu^{\frac{\left( n + 1 \right)}{2}}.
\end{equation}
From the expressions for the integration limits ${K_a \left( \rho, \zeta \right)}$ and ${K_b \left( \rho, \zeta \right)}$, as well as for the Ernst potential ${f \left( \rho, \zeta \right)}$ itself, in terms of hyperelliptic integrals, we therefore obtain the following leading-order terms in the respective expansions:

\begin{equation}
K_a \left( \rho, \zeta \right) - K_1 = O \left( \mu^{\frac{3}{2}} \right), \qquad K_b \left( \rho, \zeta \right) - K_2 = O \left( \mu^{\frac{3}{2}} \right),
\end{equation}
and:

\begin{equation}
f \left( \mu, \rho, \zeta \right) = \exp \left\lbrace - \mu \int_{-i}^{i} \frac{H \left( K - K_1 \right) \left( K - K_2 \right)}{Z \left( K, \rho, \zeta \right)} dK + O \left( \mu^3 \right) \right\rbrace.
\end{equation}
By treating ${K = O \left( 1 \right)}$ within this integral and expanding the complex function ${Z \left( K, \rho, \zeta \right)}$, we can straightforwardly determine the expansion coefficients ${f_n \left( \frac{\rho}{\rho_0}, \frac{\zeta}{\rho_0} \right)}$ in terms of elementary functions. For instance, if we introduce the elliptic coordinate system ${\left( \xi, \eta \right)}$ using:

\begin{equation}
\rho = \rho_0 \sqrt{1 + \xi^2} \sqrt{1 - \eta^2}, \qquad \text{ with } \qquad \zeta = \rho_0 \xi \eta,
\end{equation}
such that ${0 \leq \xi < \infty}$ and ${-1 \leq \eta \leq 1}$, then the integral for the first coefficient ${f_1 \left( \xi, \eta \right)}$ may be evaluated as:

\begin{equation}
f_1 \left( \xi, \eta \right) = \frac{1}{\pi i} \int_{-i}^{i} \frac{\left( 1 + K^2 \right)}{Z \left( K, \rho, \zeta \right)} dK = - \frac{1}{\pi} \left\lbrace \frac{4}{3} \cot^{-1} \left( \xi \right) + \left[ \xi - \left( \xi^2 + \frac{1}{3} \right) \cot^{-1} \left( \xi \right) \right] \left( 1 - 3 \eta^2 \right) \right\rbrace.
\end{equation}
The Newtonian (Maclaurin disk) limit ${U_{Mac} \left( \mu, \rho, \zeta \right)}$ as ${\mu \to 0}$, given by (with an explicit dependence of the speed of light $c$ reintroduced temporarily for the sake of dimensional clarity):

\begin{equation}
U_{Mac} \left( \mu, \rho, \zeta \right) = \frac{c^2 \mu f_1 \left( \rho, \zeta \right)}{2},
\end{equation}
can thus be recovered by considering the time-time component ${g_{t t}}$ of the metric tensor ${g_{\mu \nu}}$:

\begin{equation}
g_{t t} \left( \mu, \rho, \zeta \right) = - \mathrm{Re} \left( f \left( \mu, \rho, \zeta \right) \right) = - \left( 1 + \mu f_1 \left( \rho, \zeta \right) \right) = - \left( 1 + \frac{2 U_{Mac} \left( \mu, \rho, \zeta \right)}{c^2} \right),
\end{equation}
such that:

\begin{equation}
\mu = - \frac{2 U_{Mac} \left( \mu, 0, 0 \right)}{c^2} = \frac{2 \Omega^2 \left( \mu, \rho \right) \rho_{0}^{2}}{c^2},
\end{equation}
as required. On the other hand, the extremal Kerr black hole limit is recovered as ${\mu \to \mu_0 \approx 4.63}$, as we shall now see.

Recall first that the spacetime line element for a Kerr black hole in the Boyer-Lindquist coordinate system ${\left( r, \theta, \varphi \right)}$ (i.e. the three-dimensional orthogonal coordinate system on an oblate spheroid) is traditionally given as:

\begin{multline}
d s^2 = - \left( 1 - \frac{2 M r}{\Sigma} \right) d t^2 + \frac{\Sigma}{\Delta} d r^2 + \Sigma d \theta^2\\
+ \left( r^2 + a^2 + \frac{2 M r a^2}{\Sigma} \sin^2 \left( \theta \right) \right) \sin^2 \left( \theta \right) d \varphi^2 - \frac{4 M r a \sin^2 \left( \theta \right)}{\Sigma} d t d \varphi,
\end{multline}
with $a$ being the ratio of the black hole's angular momentum $J$ to its ADM mass $M$:

\begin{equation}
a = \frac{J}{M},
\end{equation}
and with ${\Sigma}$ and ${\Delta}$ being two characteristic length scales, namely:

\begin{equation}
\Sigma = r^2 + a^2 \cos^2 \left( \theta \right), \qquad \text{ and } \qquad \Delta = r^2 - 2M r + a^2,
\end{equation}
respectively. However, by introducing the additional parameters $W$, ${\nu}$ and ${\omega}$, defined by:

\begin{equation}
W^2 = \Delta \sin^2 \left( \theta \right), \qquad e^{2 \nu} = \frac{\Delta \Sigma}{\left( r^2 + a^2 \right)^2 - \Delta a^2 \sin^2 \left( \theta \right)}, \qquad \omega = \frac{2 M r a}{\Delta \Sigma} e^{2 \nu},
\end{equation}
this line element can be rewritten in the following compressed (but equivalent) form that is closer to the Weyl-Lewis-Papapetrou cylindrical metric:

\begin{equation}
d s^2 = - e^{2 \nu} d t^2 + \Sigma \left( \frac{d r^2}{\Delta} + d \theta^2 \right) + W^2 e^{-2 \nu} \left( d \varphi - \omega dt \right)^2.
\end{equation}
The stationary and axially-symmetric nature of the metric can be inferred from the lack of explicit dependence of the line element on time $t$ or the angular coordinate ${\varphi}$, respectively. The Schwarzschild event horizon (otherwise known as the \textit{interior} horizon) appears at the point where the radial-radial component of the metric tensor ${g_{r r}}$ becomes singular, namely:

\begin{equation}
\frac{1}{g_{r r}} = 0, \qquad \iff \qquad r_I = M + \sqrt{M^2 - a^2},
\end{equation}
i.e. the largest root of the quadratic equation ${\Delta = 0}$. Likewise, the boundary of the \textit{ergosphere} (otherwise known as the \textit{exterior} horizon) appears at the point where the time-time component of the metric tensor ${g_{t t}}$ changes sign, namely:

\begin{equation}
g_{t t} = 0, \qquad \iff \qquad r_E = M + \sqrt{M^2 - a^2 \cos^2 \left( \theta \right)}.
\end{equation}
Note that, in contrast to the metric for the uniformly-rotating disk of dust, this solution is only physical if:

\begin{equation}
a \leq M, \qquad \iff \qquad J \leq M^2.
\end{equation}
Within the ergosphere, i.e. for ${r_I < r < r_E}$, the discrepancy between the non-zero value of the Boyer-Lindquist shift vector ${\beta_{i}^{Kerr}}$ and the (approximately zero) value of the numerical shift vector ${\beta^i}$ gives rise to an off-diagonal distortion in the metric tensor, causing coordinates to be dragged in the angular direction ${\theta}$, implying that all observers must rotate in the same direction as the black hole itself:

\begin{equation}
\frac{\partial \varphi}{\partial t} > 0,
\end{equation}
which we interpret physically as \textit{frame-dragging}.

The trajectories of test particles in the Kerr geometry, with instantaneous 4-momentum ${\left( p_t, p_r, p_{\theta}, p_{\varphi} \right)}$, can be described in terms of three conserved quantities, namely the total energy ${E = - p_t}$, the angular momentum component parallel to the axis of symmetry ${L = p_{\varphi}}$, and the more abstract quantity $Q$, defined by:

\begin{equation}
Q = p_{\theta}^{2} + \cos^2 \left( \theta \right) \left[ a^2 \left( \mu^2 - p_{t}^{2} \right) + 
\frac{p_{\varphi}^{2}}{\sin^2 \left( \theta \right)} \right],
\end{equation}
with ${\mu}$ being the trivial fourth conserved quantity, i.e. the particle's rest mass (without loss of generality, we henceforth assume ${\mu = 0}$ for photons and ${\mu = 1}$ for all other particles). ${Q = 0}$ is both a necessary and sufficient condition for a particle moving within the ``equatorial plane'' ${\theta = \frac{\pi}{2}}$ to remain within the equatorial plane indefinitely, while ${Q > 0}$ is a necessary condition for a particle to cross the equatorial plane. By solving the equation for $Q$ first for the lower-index momentum quantities ${p_{\mu}}$, followed by the upper-index momentum quantities ${p^{\mu}}$, we obtain the following equations of motion for the orbital trajectories of the test particles:

\begin{equation}
\Sigma \frac{d t}{d \lambda} = -a \left( a E \sin^2 \left( \theta \right) - L \right) + \left( r^2 + a^2 \right) \frac{T}{\Delta}, \qquad \Sigma \frac{d r}{d \lambda} = \pm \sqrt{V_r}, \qquad,
\end{equation}
\begin{equation}
\Sigma \frac{d \theta}{d \lambda} = \pm \sqrt{V_{\theta}}, \qquad \Sigma \frac{d \varphi}{d \lambda} = - \left( a E - \frac{L}{\sin^2 \left( \theta \right)} \right) + a \frac{T}{\Delta},
\end{equation}
with ${\lambda}$ being a function of the proper time ${\tau}$ on the particle's world line ${\tau}$:

\begin{equation}
\lambda = \frac{\tau}{\mu},
\end{equation}
which thus becomes an affine parameter in the limit ${\mu \to 0}$, with the quantity $T$ being given by:

\begin{equation}
T = E \left( r^2 + a^2 \right) - L a,
\end{equation}
and with quantities ${V_r}$ and ${V_{\theta}}$ being ``effective potentials'' that govern the motions of particles in the $r$ and ${\theta}$ coordinate directions, respectively:

\begin{equation}
V_r = T^2 - \Delta \left[ \mu^2 r^2 + \left( L - a E \right)^2 + Q \right], \qquad V_{\theta} = Q - \cos^2 \left( \theta \right) \left[ a^2 \left( \mu^2 - E^2 \right) + \frac{L^2}{\sin^2 \left( \theta \right)} \right].
\end{equation}
For a test particle in a circular orbit within the equatorial plane ${\theta = \frac{\pi}{2}}$ at a fixed radius $r$, the derivative ${\frac{dr}{d \lambda}}$ must vanish identically, yielding:

\begin{equation}
\Sigma \frac{dr}{d \lambda} = \pm \sqrt{V_r}, \qquad \implies V_r \left( r \right) = V_{r}^{\prime} \left( r \right) = 0,
\end{equation}
where these equations can be solved simultaneously to give the total energy $E$ and angular momentum component $L$ as:

\begin{equation}
\frac{E}{\mu} = \frac{r^{\frac{3}{2}} - 2 M \sqrt{r} \pm a \sqrt{M}}{r^{\frac{3}{4}} \sqrt{r^{\frac{3}{2}} - r M \sqrt{r} \pm 2 a \sqrt{M}}}, \qquad \text{ and } \qquad \frac{L}{\mu} = \frac{\pm \sqrt{M} \left( r^2 \mp 2 a \sqrt{M r} + a^2 \right)}{r^{\frac{3}{4}} \sqrt{r^{\frac{3}{2}} - 3 M \sqrt{r} \pm 2 a \sqrt{M}}},
\end{equation}
respectively, implying that the coordinate angular velocity of the orbit ${\Omega}$ is simply:

\begin{equation}
\Omega = \frac{d \varphi}{dt} = \frac{\pm \sqrt{M}}{r^{\frac{3}{2}} \pm a \sqrt{M}},
\end{equation}
with the upper signs designating the direct orbits (i.e. corotating orbits with the black hole, such that ${L > 0}$), and with lower signs designating the retrograde orbits (i.e. counterrotating with the black hole, such that ${L < 0}$).

Evidently, such circular orbits can only exist if the denominator in the equations for ${\frac{E}{\mu}}$ and ${\frac{L}{\mu}}$ is real, i.e. if and only if the following inequality is satisfied.

\begin{equation}
r^\frac{3}{2} - 3 M \sqrt{r} \pm 2 a \sqrt{M} \geq 0.
\end{equation}
The limiting case of equality occurs only in the case of photons (since this is associated with infinite energy $E$ per unit rest mass ${\mu}$), whose orbits therefore form the innermost boundary, denoted ${r_{ph}}$, of all the circular orbits:

\begin{equation}
r_{ph} = 2M \left\lbrace 1 + \cos \left[ \frac{2}{3} \cos^{-1} \left( \mp \frac{a}{M} \right) \right] \right\rbrace.
\end{equation}
For all other orbits (with ${r > r_{ph}}$), the trajectories of certain particles may diverge to infinity in  an asymptotically hyperbolic fashion if they are given an infinitesimal outward perturbation; such unstable ``unbound'' orbits occur for ${\frac{E}{\mu} > 1}$ (with the instability arising from the fact that, although their trajectories are given by circular functions, their energies are now given by hyperbolic ones), with the limiting case of equality ${\frac{E}{\mu} = 1}$ occurring in the case of ``marginally bound'' orbits, denoted ${r_{mb}}$, forming the innermost boundary of all orbits with parabolic energy functions:

\begin{equation}
r_{mb} = 2M \mp a + 2 \sqrt{M} \sqrt{M \mp a}.
\end{equation}
Parabolic trajectories (which are a good approximation for all non-relativistic matter ${v \ll c}$) that intersect the region ${r < r_{mb}}$ will therefore inevitably fall into the black hole. On the other hand, bound orbits with ${r > r_{mb}}$ will be stable if and only if ${V_{r}^{\prime \prime} \left( r \right) \leq 0}$, yielding the equivalent conditions:

\begin{equation}
1 - \left( \frac{E}{\mu} \right)^2 \geq \frac{2}{3} \left( \frac{M}{r} \right), \qquad \iff \qquad r^2 - 6 M r \pm 8 a \sqrt{M} \sqrt{r} - 3 a^2 \geq 0.
\end{equation}
Thus, the ``marginally stable'' orbits, denoted ${r_{ms}}$, forming the innermost boundary of all stable orbits ${r > r_{ms}}$, will be given by:

\begin{equation}
r_{ms} = m \left\lbrace 3 + Z_2 \mp \sqrt{\left( 3 - Z_1 \right) \left( 3 + Z_1 + 2 Z_2 \right)} \right\rbrace,
\end{equation}
where, for compactness, we have introduced the constants ${Z_1}$ and ${Z_2}$:

\begin{equation}
Z_1 = 1 + \left( 1 - \frac{a^2}{M^2} \right)^{\frac{1}{3}} \left[ \left( 1 + \frac{a}{M} \right)^{\frac{1}{3}} + \left( 1 - \frac{a}{M} \right)^{\frac{1}{3}} \right], \qquad Z_2 = \sqrt{3 \frac{a^2}{M^2} + Z_{1}^{2}}.
\end{equation}

The two limiting cases for the Kerr black hole consequently correspond to the non-rotating ${a = 0}$ (${J = 0}$) case in which the solution simply reduces to the Schwarzschild metric, since the interior and exterior horizon radii now both coincide with the Schwarzschild radius (i.e. ${r_I = r_E = 2M}$) and so the ergosphere disappears, and the maximally-rotating ${a = M}$ (${J = M^2}$) case, which we refer to here as the ``extremal Kerr'' solution. For the extremal ${a = M}$ case, the interior and exterior horizons are maximally separated, so the ergosphere becomes maximally extended, since ${r_I = M}$ and, in the equatorial plane with ${\theta = \frac{\pi}{2}}$, ${r_E = 2M}$. The equations for the specific total energy ${\frac{E}{\mu}}$ and specific angular momentum ${\frac{L}{\mu}}$ given above reduce in this case to:

\begin{equation}
\frac{E}{\mu} = \frac{r \pm \sqrt{M} \sqrt{r} - M}{r^{\frac{3}{4}} \sqrt{\sqrt{r} \pm 2 \sqrt{M}}}, \qquad \text{ and } \qquad \frac{L}{\mu} = \frac{\pm M \left( r^{\frac{3}{2}} \pm \sqrt{M} r + M \sqrt{r} \mp M^{\frac{3}{2}} \right)}{r^{\frac{3}{4}} \sqrt{\sqrt{r} \pm 2 \sqrt{M}}},
\end{equation}
respectively. For the Schwarzschild (${a = 0}$) case, the photon orbit radius is given by ${r_{ph} = 3M}$, the marginally bound orbit radius is given by ${r_{mb} = 4M}$, and the marginally stable orbit radius is given by ${r_{ms} = 6M}$. On the other hand, for the extremal Kerr (${a = M}$) case, we can infer from the reduced equations above that the photon orbit radius is given by either ${r_{ph} = M}$ (for direct orbits) or ${r_{ph} = 4M}$ (for retrograde orbits), the marginally bound orbit radius is given by ${r_{mb} = M}$ (for direct orbits) or ${r_{mb} = \left( 3 + 2 \sqrt{2} \right) M \approx 5.83 M}$ (for retrograde orbits), and the marginally stable orbit radius is given by ${r_{ms} = M}$ (for direct orbits) or ${r_{ms} = 9M}$ (for retrograde orbits). Switching momentarily from the cylindrical coordinate system ${\left( \rho, \zeta \right)}$ to the spherical coordinate system ${\left( R, \theta \right)}$:

\begin{equation}
\rho = R \sin \left( \theta \right), \qquad \text{ and } \qquad \zeta = R \cos \left( \theta \right),
\end{equation}
this allows us to write the complex Ernst potential ${f \left( \rho, \zeta \right)}$ for the extremal Kerr solution in the following, rather elegant, form:

\begin{equation}
f \left( \rho, \zeta \right) = f \left( R, \theta \right) = \frac{2 \Omega \left( \mu, \rho_0 \right) R - 1 - i \cos \left( \theta \right)}{2 \Omega \left( \mu, \rho_0 \right) R + 1 - i \cos \left( \theta \right)},
\end{equation}
assuming a strictly positive radial coordinate value ${R > 0}$.

In order to see how this extremal Kerr form of the complex Ernst potential is recovered in the limit as ${\mu \to \mu_0 \approx 4.63}$ from the more general Ernst potential ${f \left( \rho, \zeta \right)}$ for a uniformly-rotating disk of dust:

\begin{equation}
f \left( \rho, \zeta \right) = \exp \left\lbrace \int_{K_1}^{K_a \left( \rho, \zeta \right)} \frac{K^2}{Z \left( K, \rho, \zeta \right)} d K + \int_{K_2}^{K_b \left( \rho, \zeta \right)} \frac{K^2}{Z \left( K, \rho, \zeta \right)} d K - \int_{- i \rho_0}^{i \rho_0} \frac{H \left( K \right)}{Z_1 \left( K, \rho, \zeta \right)} K^2 d K \right\rbrace,
\end{equation}
with upper integration limits ${K_a \left( \rho, \zeta \right)}$ and ${K_b \left( \rho, \zeta \right)}$ defined as ultraelliptic functions via:

\begin{equation}
\int_{K_1}^{K_a \left( \rho, \zeta \right)} \frac{d K}{Z \left( K, \rho, \zeta \right)} + \int_{K_2}^{K_b \left( \rho, \zeta \right)} \frac{d K}{Z \left( K, \rho, \zeta \right)} = \int_{- i \rho_0}^{i \rho_0} \frac{H \left( K \right)}{Z_1 \left( K, \rho, \zeta \right)} d K,
\end{equation}
and:

\begin{equation}
\int_{K_1}^{K_a \left( \rho, \zeta \right)} \frac{K}{Z \left( K, \rho, \zeta \right)} d K + \int_{K_2}^{K_b \left( \rho, \zeta \right)} \frac{K}{Z \left( K, \rho, \zeta \right)} d K = \int_{- i \rho_0}^{i \rho_0} \frac{H \left( K \right)}{Z_1 \left( K, \rho, \zeta \right)} K d K,
\end{equation}
respectively, it is helpful to take ${K_b \left( \rho, \zeta \right)}$ to be on the other sheet of the Riemann surface, as compared to the previous setup. This yields the following equivalent form for the Ernst potential:

\begin{equation}
f \left( \rho, \zeta \right) = \exp \left\lbrace \int_{K_b \left( \rho, \zeta \right)}^{K_a \left( \rho, \zeta \right)} \frac{K^2}{Z \left( K, \rho, \zeta \right)} d K + \int_{K_1}^{K_2} \frac{K^2}{Z \left( K, \rho, \zeta \right)} d K - \int_{-i \rho_0}^{i \rho_0} \frac{H \left( K \right)}{Z_1 \left( K, \rho, \zeta \right)} K^2 d K \right\rbrace,
\end{equation}
with the corresponding equivalent forms for the integration limits:

\begin{equation}
\int_{K_b \left( \rho, \zeta \right)}^{K_a \left( \rho, \zeta \right)} \frac{d K}{Z \left( K, \rho, \zeta \right)} + \int_{K_1}^{K_2} \frac{d K}{Z \left( K, \rho, \zeta \right)} = \int_{- i \rho_0}^{i \rho_0} \frac{H \left( K \right)}{Z_1 \left( K, \rho, \zeta \right)} d K,
\end{equation}
and:

\begin{equation}
\int_{K_b \left( \rho, \zeta \right)}^{K_a \left( \rho, \zeta \right)} \frac{K}{Z \left( K, \rho, \zeta \right)} d K + \int_{K_1}^{K_2} \frac{K}{Z \left( K, \rho, \zeta \right)} d K = \int_{- i \rho_0}^{i \rho_0} \frac{H \left( K \right)}{Z_1 \left( K, \rho, \zeta \right)} K d K,
\end{equation}
respectively. From the original definitions of the ${\mu}$ parameter:

\begin{equation}
\mu = 2 \Omega^2 \left( \mu, \rho \right) \rho_{0}^{2} e^{-2 V_0 \left( \mu \right)},
\end{equation}
and the surface potential parameter ${V_0 \left( \mu \right)}$:

\begin{equation}
V_0 \left( \mu \right) = - \frac{1}{2} \sinh^{-1} \left\lbrace \mu + \frac{1 + \mu^2}{\wp \left[ I \left( \mu \right) ; \frac{4}{3} \mu^2 - 4, \frac{8}{3} \mu \left( 1 + \frac{\mu^2}{9} \right) \right] - \frac{2}{3} \mu} \right\rbrace,
\end{equation}
we obtain, in the limit as ${\mu \to \mu_0}$ (since ${\mu_0}$ is, by definition, the first zero of the elliptic Weierstrass function ${\wp \left[ I \left( \mu \right) ; \frac{4}{3} \mu^2 - 4, \frac{8}{3} \mu \left( 1 + \frac{\mu^2}{9} \right) \right]}$ appearing in the denominator), the following limiting expressions for the integrals above in terms of the spherical coordinate system ${\left( R, \theta \right)}$:

\begin{equation}
\int_{- i \rho_0}^{i \rho_0} \frac{H \left( K \right)}{Z_1 \left( K, \rho, \zeta \right)} d K - \int_{K_1}^{K_2} \frac{d K}{Z \left( K, \rho, \zeta \right)} = \frac{2 \Omega \left( \mu, \rho \right)}{R} - \frac{\pi i \cos \left( \theta \right)}{2 R^2},
\end{equation}

\begin{equation}
\int_{- i \rho_0}^{i \rho_0} \frac{H \left( K \right)}{Z_1 \left( K, \rho, \zeta \right)} K d K - \int_{K_1}^{K_2} \frac{K}{Z \left( K, \rho, \zeta \right)} d K = - \frac{\pi i}{2 R},
\end{equation}
and:

\begin{equation}
\int_{- i \rho_0}^{i \rho_0} \frac{H \left( K \right)}{Z_1 \left( K, \rho, \zeta \right)} K^2 d K - \int_{K_1}^{K_2} \frac{K^2}{Z \left( K, \rho, \zeta \right)} d K = 0,
\end{equation}
modulo considerations of periodicity, and under the hypothesis that the radial coordinate is strictly positive, i.e. ${R > 0}$. In the limit as ${\mu \to \mu_0}$, both ${K_1}$ and ${K_2}$ (i.e. the lower limits of integration) vanish:

\begin{equation}
\lim_{\mu \to \mu_0} \left[ K_1 \right] = \lim_{\mu \to \mu_0} \left[ - \bar{K_2} \right] = \rho_0 \lim_{\mu \to \mu_0} \left[ \sqrt{\frac{i - \mu}{\mu}} \right] = 0,
\end{equation}
under the hypothesis that ${\mathrm{Re} \left( K_1 \right) = - \mathrm{Re} \left( K_2 \right) < 0}$, and therefore the complex function ${Z \left( K, \rho, \zeta \right)}$ in the above integrals simplifies to:

\begin{equation}
Z \left( K, \rho, \zeta \right) = K^2 \sqrt{\left( K + i \left( \rho + i \zeta \right) \right) \left( K - i \left( \rho - i \zeta \right) \right)}.
\end{equation}
The unique solution to the above (now elementary) integral equations is therefore given by the complex Ernst potential for the extremal Kerr solution:

\begin{equation}
f \left( \rho, \zeta \right) = f \left( R, \theta \right) = \frac{2 \Omega \left( \mu, \rho_0 \right) R - 1 - i \cos \left( \theta \right)}{2 \Omega \left( \mu, \rho_0 \right) R + 1 - i \cos \left( \theta \right)},
\end{equation}
as required. This completes the proof.

\begin{figure}[ht]
\centering
\includegraphics[width=0.325\textwidth]{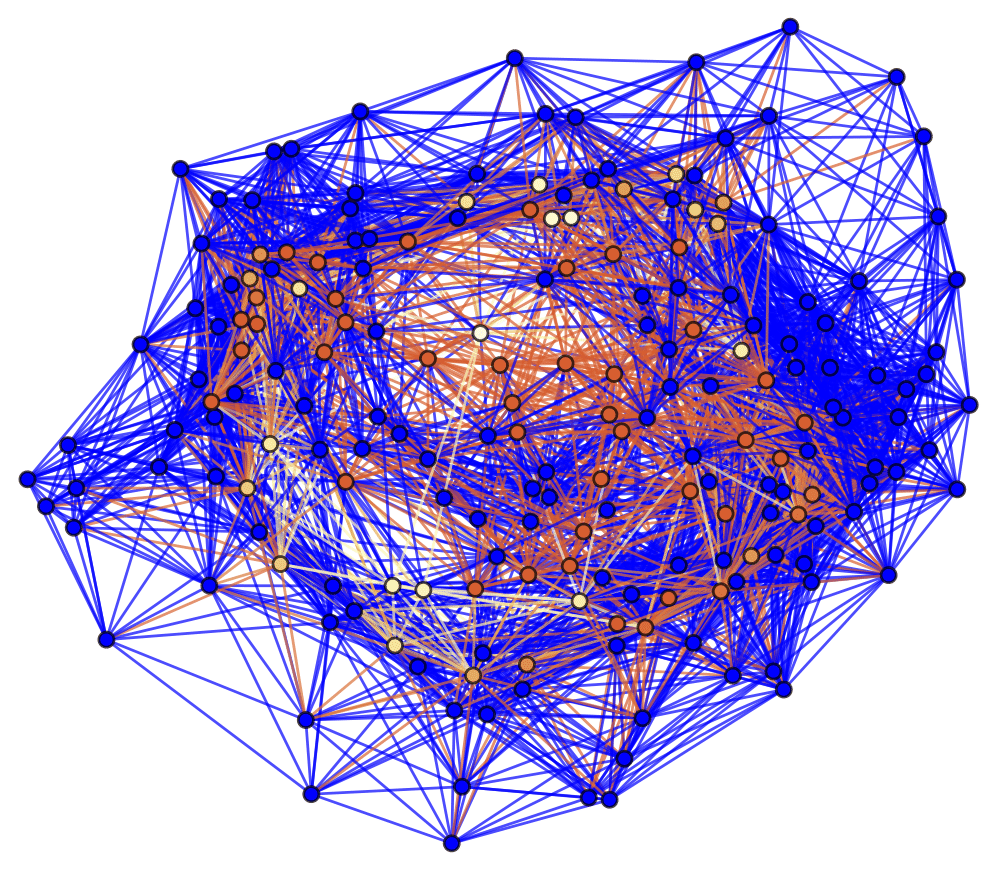}
\includegraphics[width=0.325\textwidth]{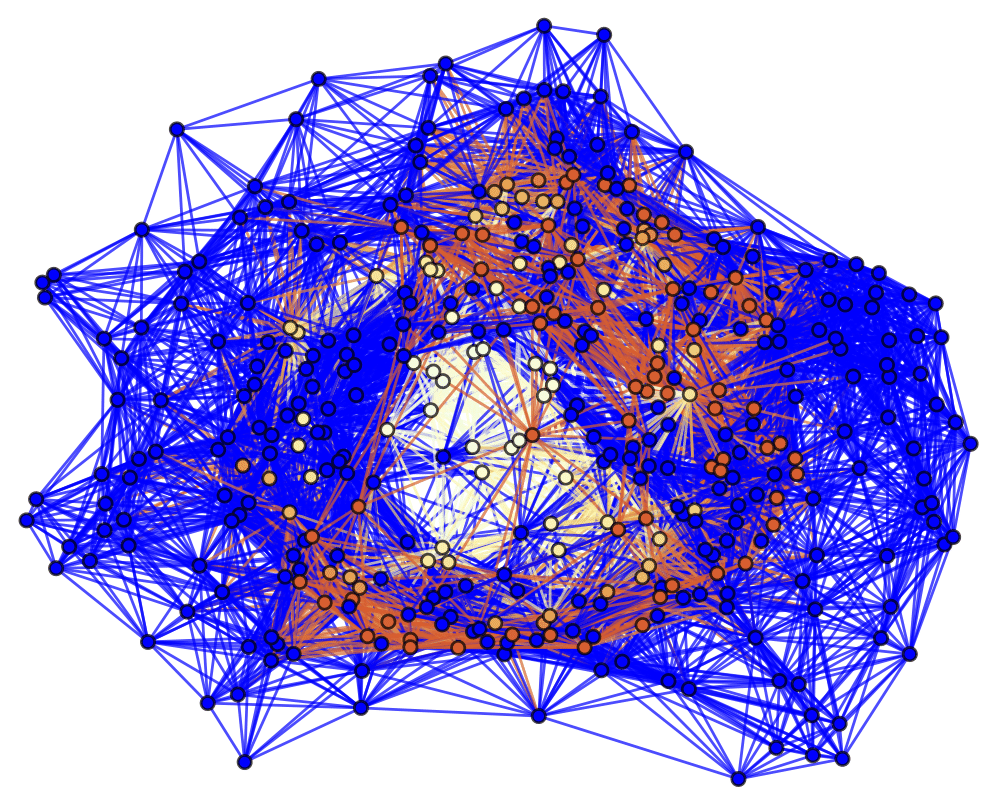}
\includegraphics[width=0.325\textwidth]{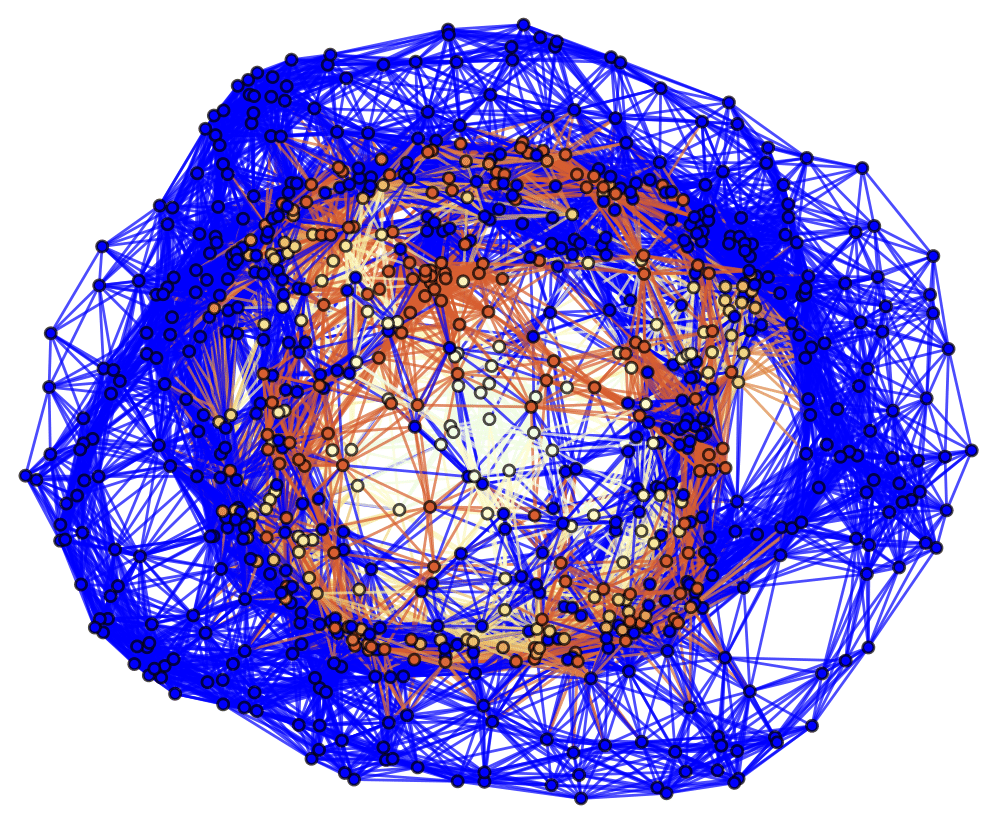}
\caption{Spatial hypergraphs corresponding to the initial hypersurface configuration of the massive scalar field ``bubble collapse'' to a maximally-rotating (extremal) Kerr black hole test, with a spinning (exponential) initial density distribution, at time ${t = 0 M}$, with resolutions of 200, 400 and 800 vertices, respectively. The hypergraphs have been adapted using the local curvature in the Boyer-Lindquist conformal factor ${\psi}$, and colored according to the value of the scalar field ${\Phi \left( t, R \right)}$.}
\label{fig:Figure18}
\end{figure}

\begin{figure}[ht]
\centering
\includegraphics[width=0.325\textwidth]{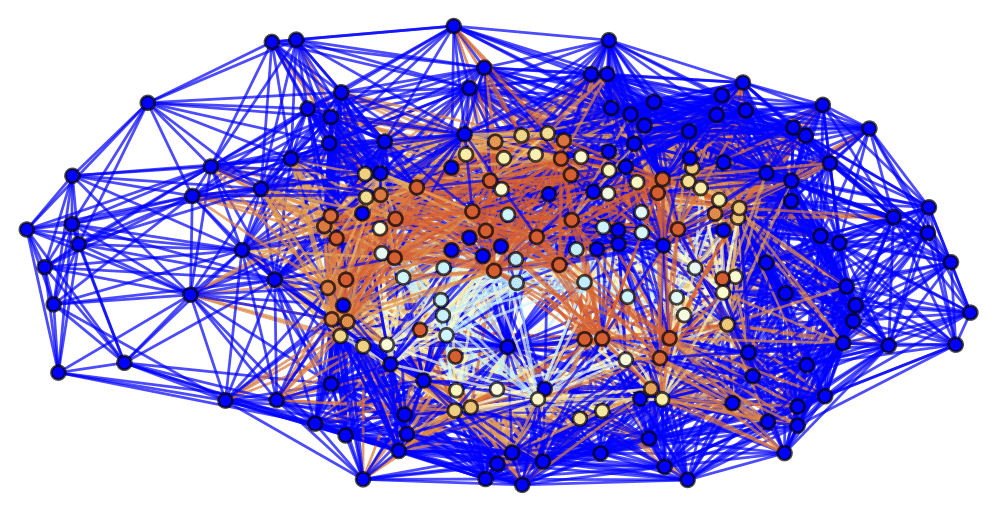}
\includegraphics[width=0.325\textwidth]{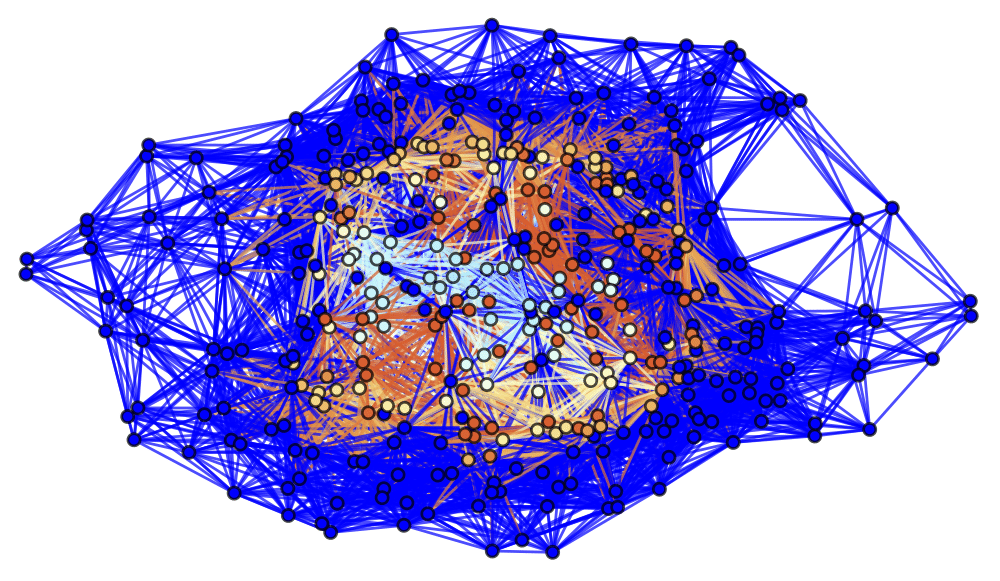}
\includegraphics[width=0.325\textwidth]{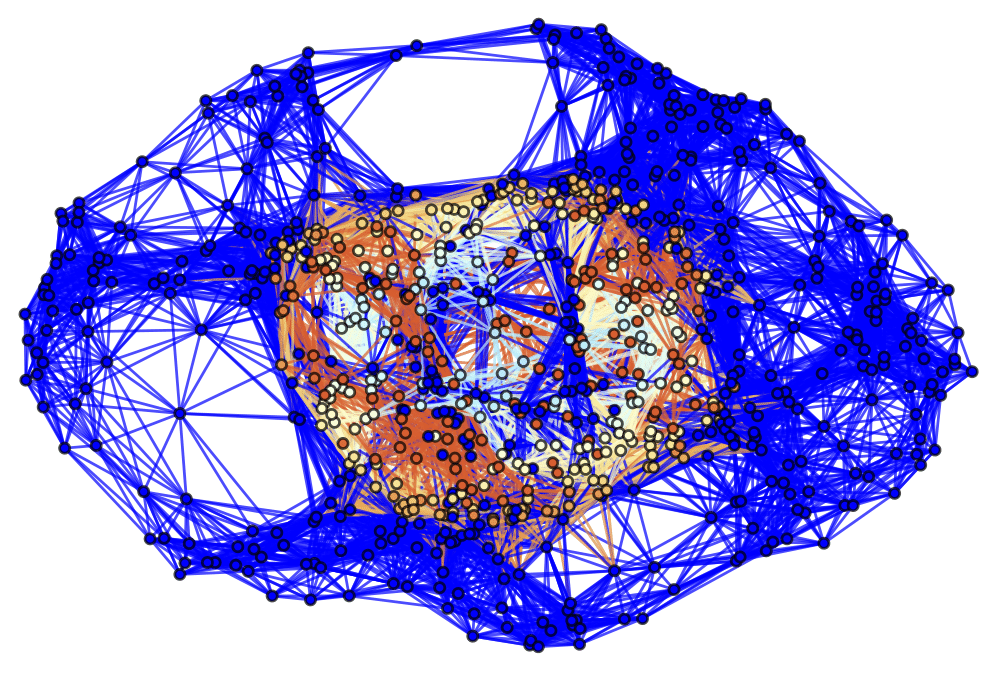}
\caption{Spatial hypergraphs corresponding to the first intermediate hypersurface configuration of the massive scalar field ``bubble collapse'' to a maximally-rotating (extremal) Kerr black hole test, with a spinning (exponential) initial density distribution, at time ${t = 1.5 M}$, with resolutions of 200, 400 and 800 vertices, respectively. The hypergraphs have been adapted using the local curvature in the Boyer-Lindquist conformal factor ${\psi}$, and colored according to the value of the scalar field ${\Phi \left( t, R \right)}$.}
\label{fig:Figure19}
\end{figure}

\begin{figure}[ht]
\centering
\includegraphics[width=0.325\textwidth]{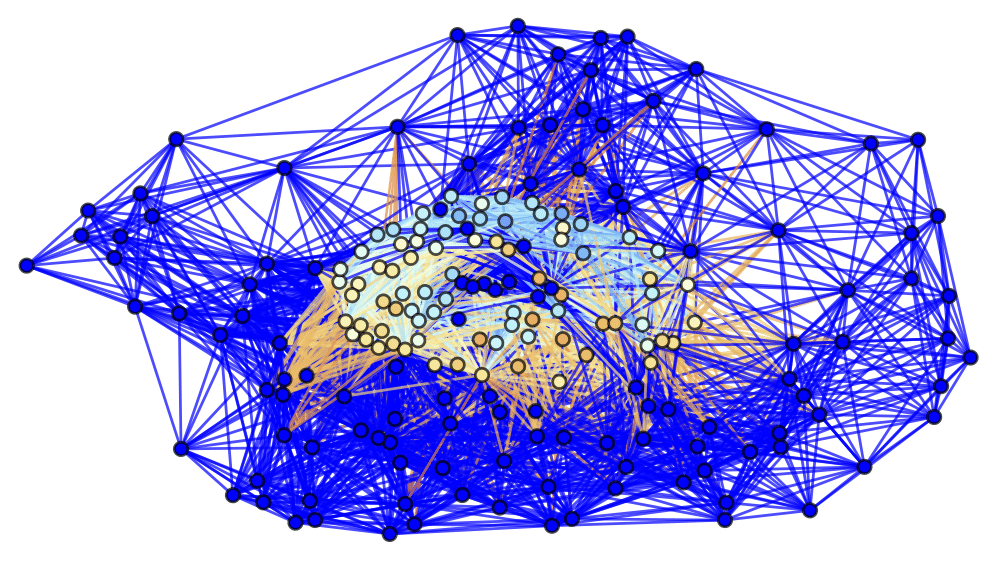}
\includegraphics[width=0.325\textwidth]{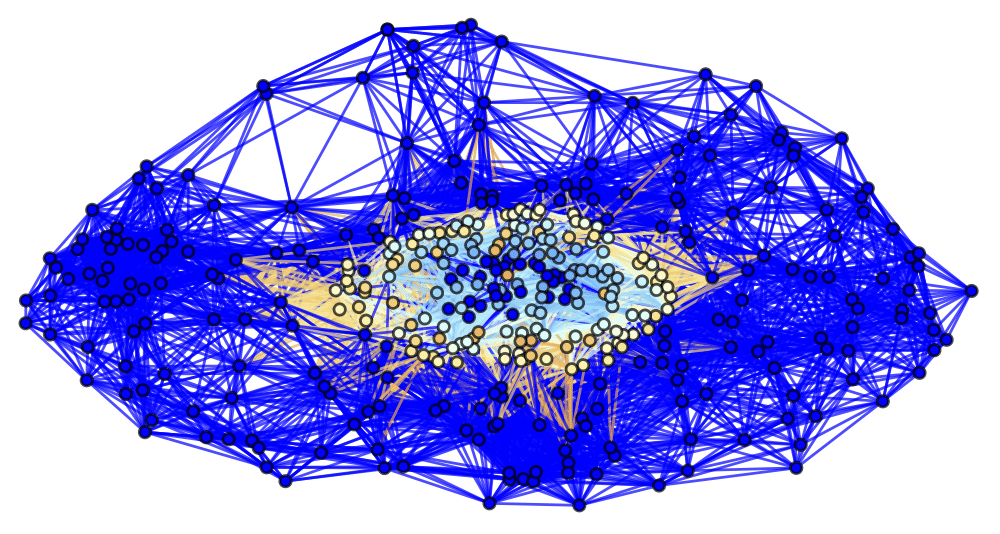}
\includegraphics[width=0.325\textwidth]{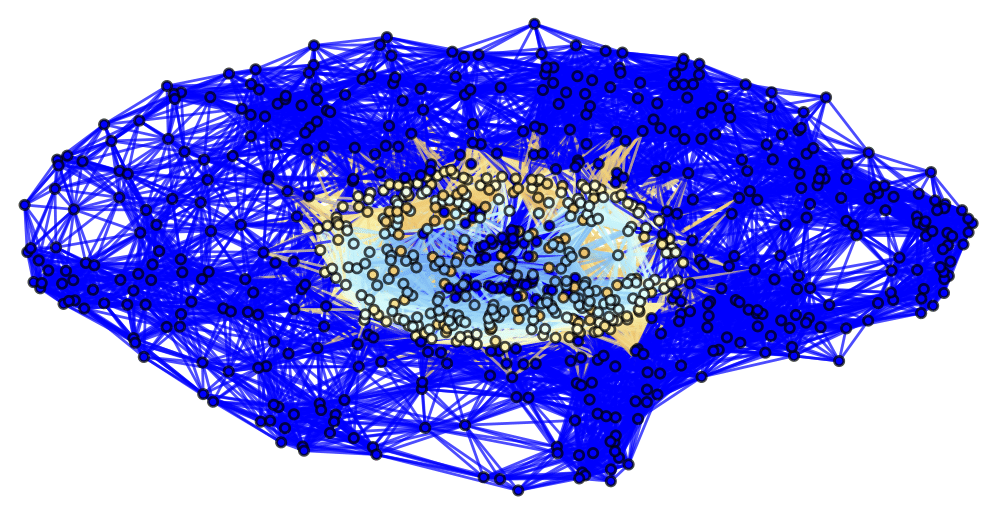}
\caption{Spatial hypergraphs corresponding to the second intermediate hypersurface configuration of the massive scalar field ``bubble collapse'' to a maximally-rotating (extremal) Kerr black hole test, with a spinning (exponential) initial density distribution, at time ${t = 3 M}$, with resolutions of 200, 400 and 800 vertices, respectively. The hypergraphs have been adapted using the local curvature in the Boyer-Lindquist conformal factor ${\psi}$, and colored according to the value of the scalar field ${\Phi \left( t, R \right)}$.}
\label{fig:Figure20}
\end{figure}

\begin{figure}[ht]
\centering
\includegraphics[width=0.325\textwidth]{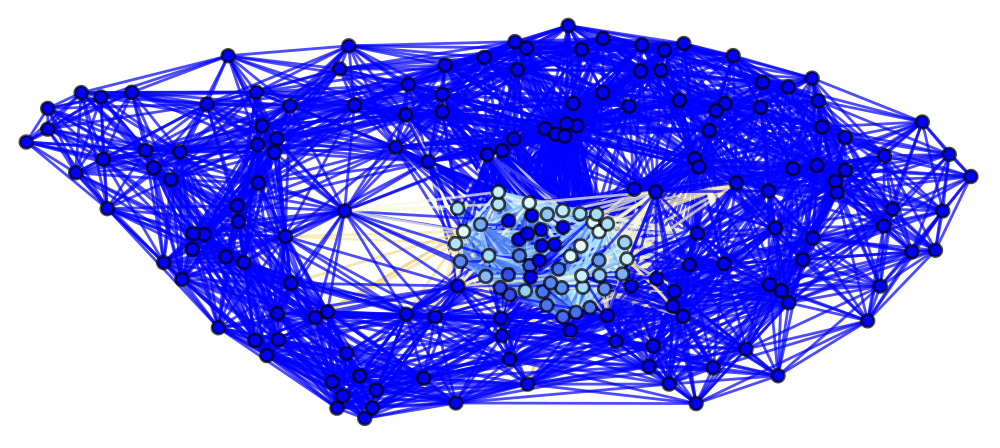}
\includegraphics[width=0.325\textwidth]{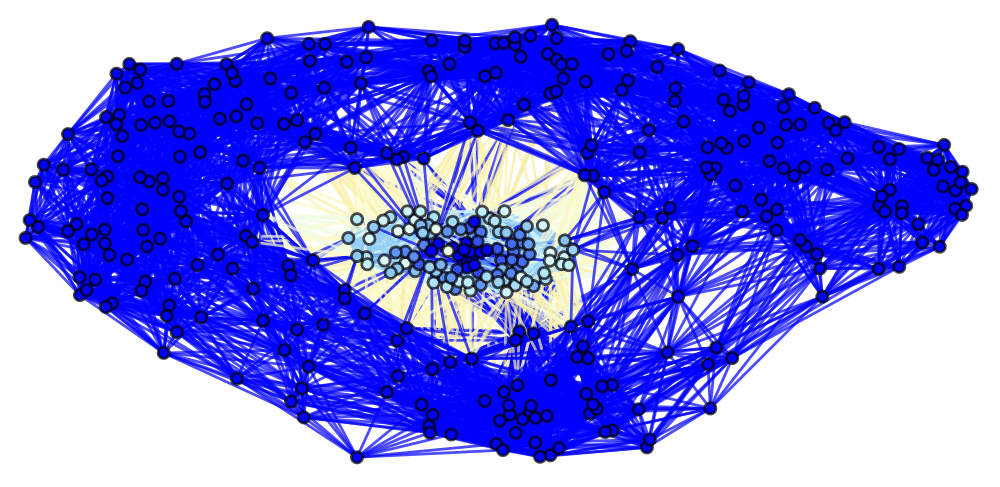}
\includegraphics[width=0.325\textwidth]{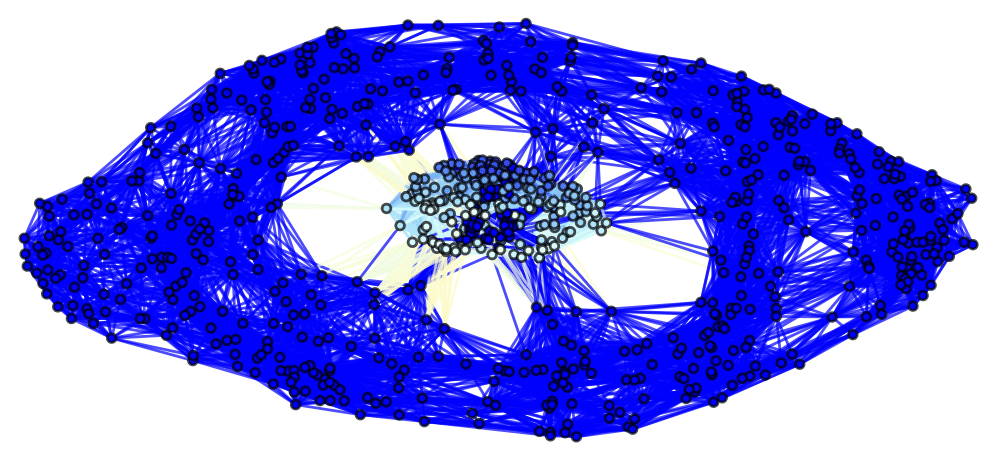}
\caption{Spatial hypergraphs corresponding to the final hypersurface configuration of the massive scalar field ``bubble collapse'' to a maximally-rotating (extremal) Kerr black hole test, with a spinning (exponential) initial density distribution, at time ${t = 4.5 M}$, with resolutions of 200, 400 and 800 vertices, respectively. The hypergraphs have been adapted using the local curvature in the Boyer-Lindquist conformal factor ${\psi}$, and colored according to the value of the scalar field ${\Phi \left( t, R \right)}$.}
\label{fig:Figure21}
\end{figure}

\begin{figure}[ht]
\centering
\includegraphics[width=0.325\textwidth]{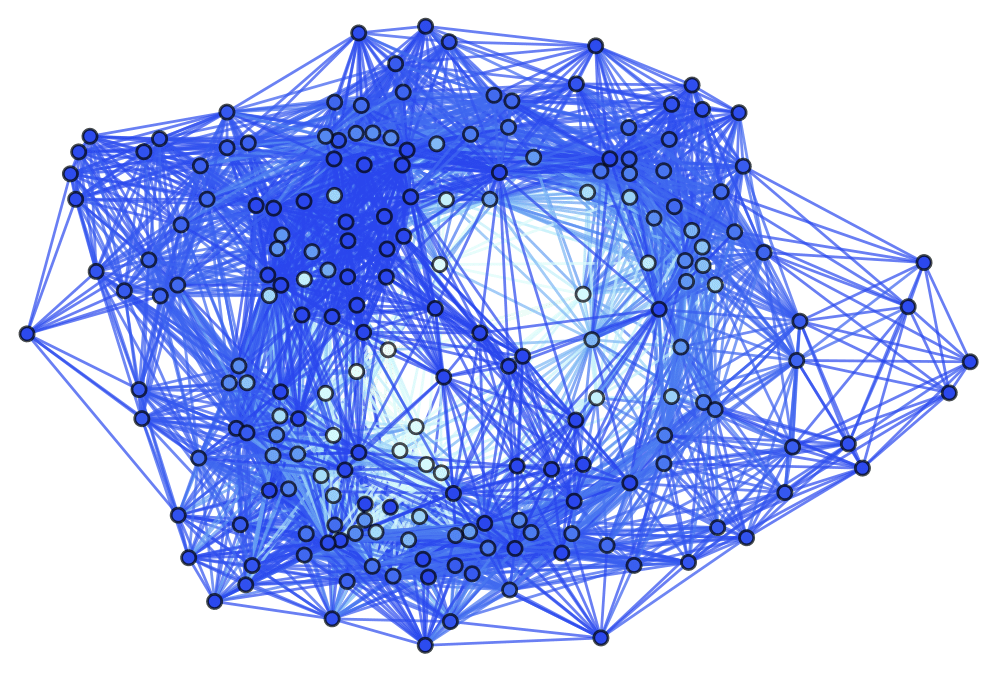}
\includegraphics[width=0.325\textwidth]{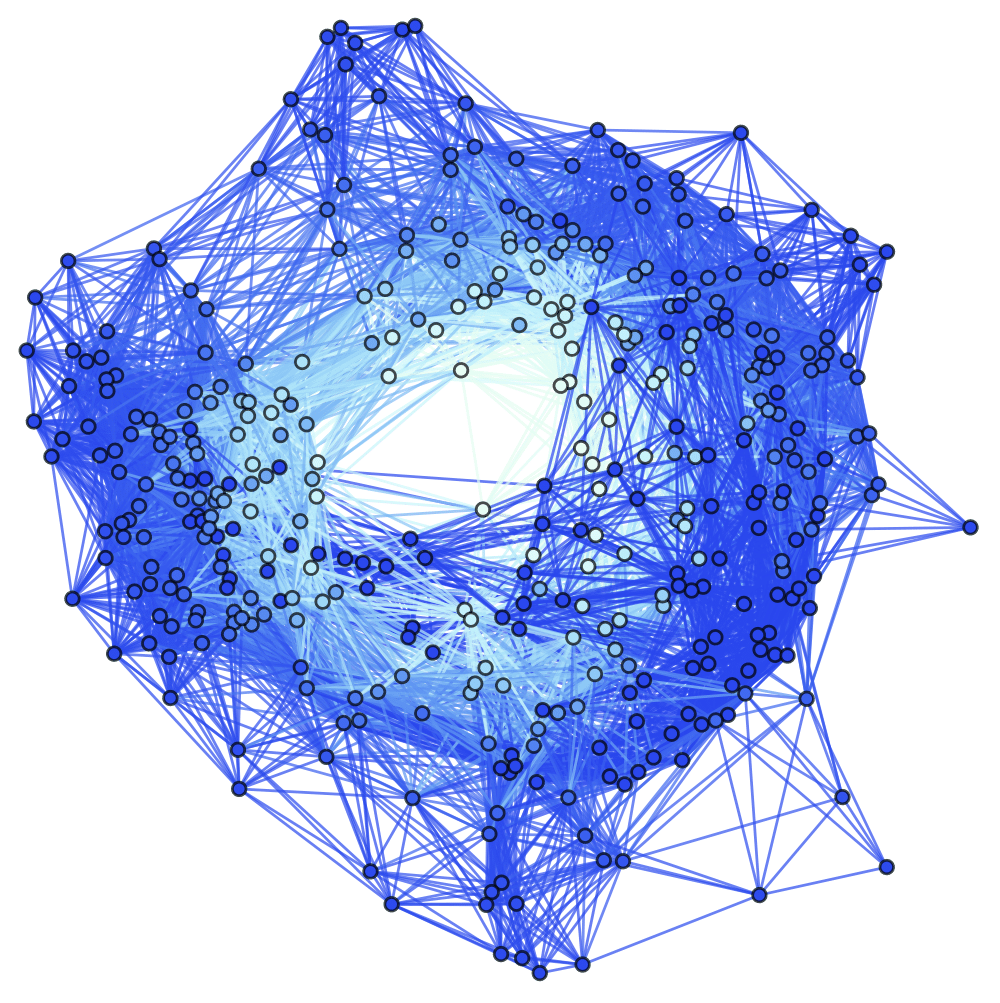}
\includegraphics[width=0.325\textwidth]{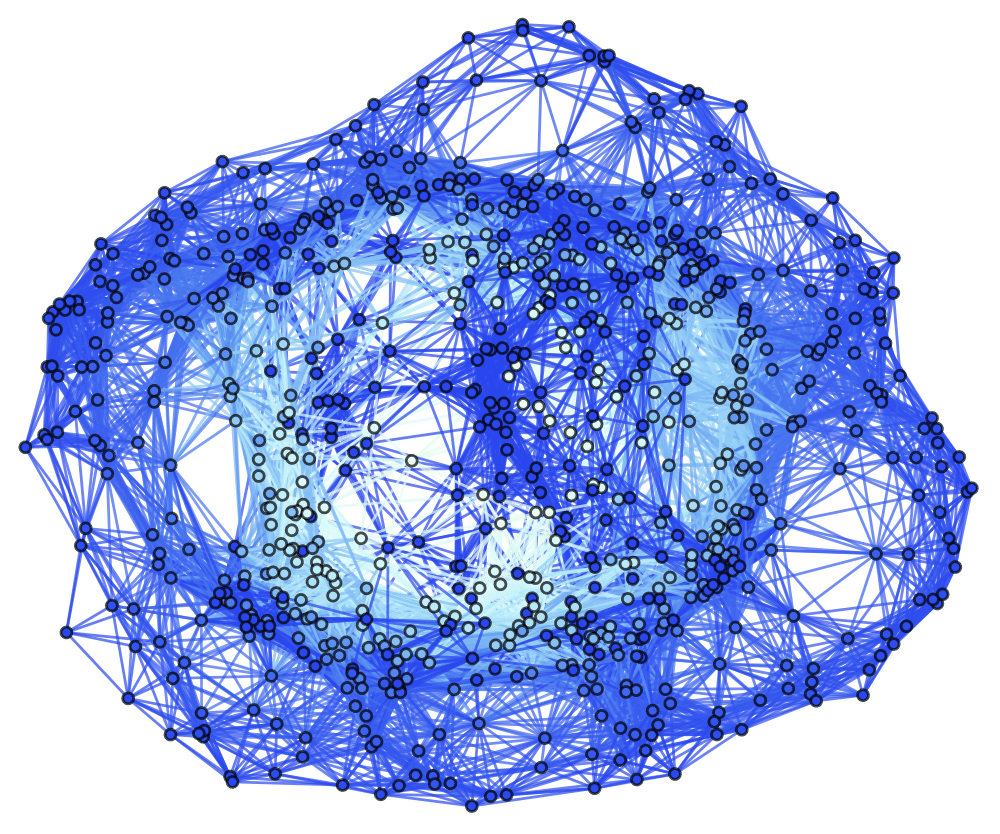}
\caption{Spatial hypergraphs corresponding to the initial hypersurface configuration of the massive scalar field ``bubble collapse'' to a maximally-rotating (extremal) Kerr black hole test, with a spinning (exponential) initial density distribution, at time ${t = 0 M}$, with resolutions of 200, 400 and 800 vertices, respectively. The hypegraphs have been adapted and colored using the local curvature in the Boyer-Lindquist conformal factor ${\psi}$.}
\label{fig:Figure22}
\end{figure}

\begin{figure}[ht]
\centering
\includegraphics[width=0.325\textwidth]{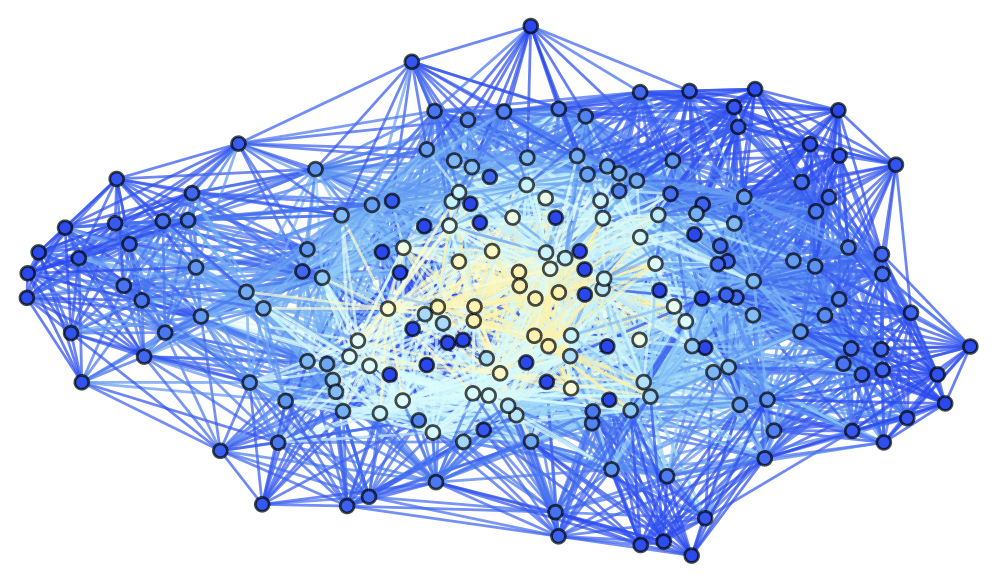}
\includegraphics[width=0.325\textwidth]{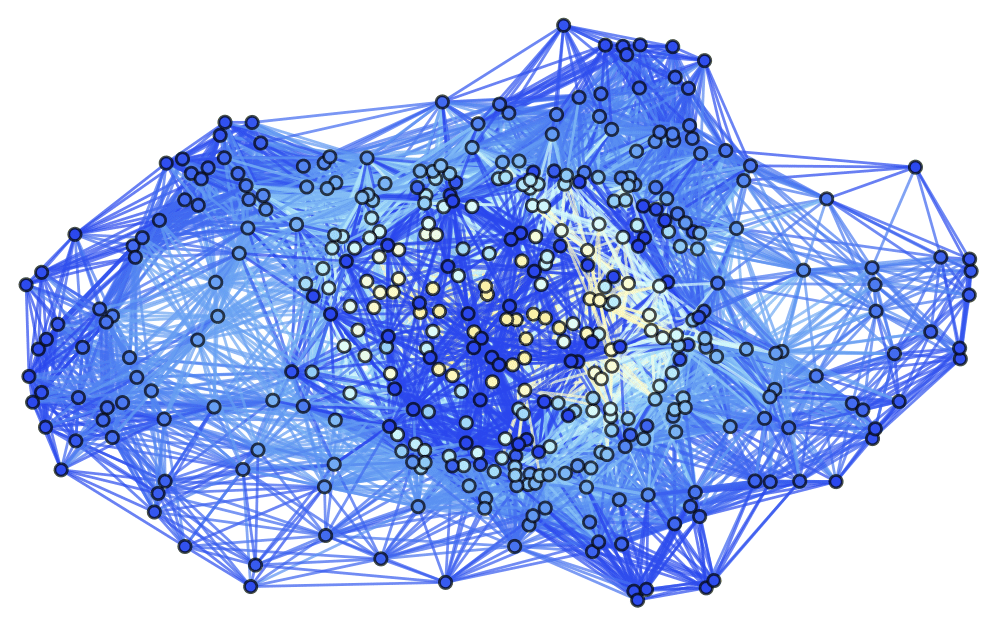}
\includegraphics[width=0.325\textwidth]{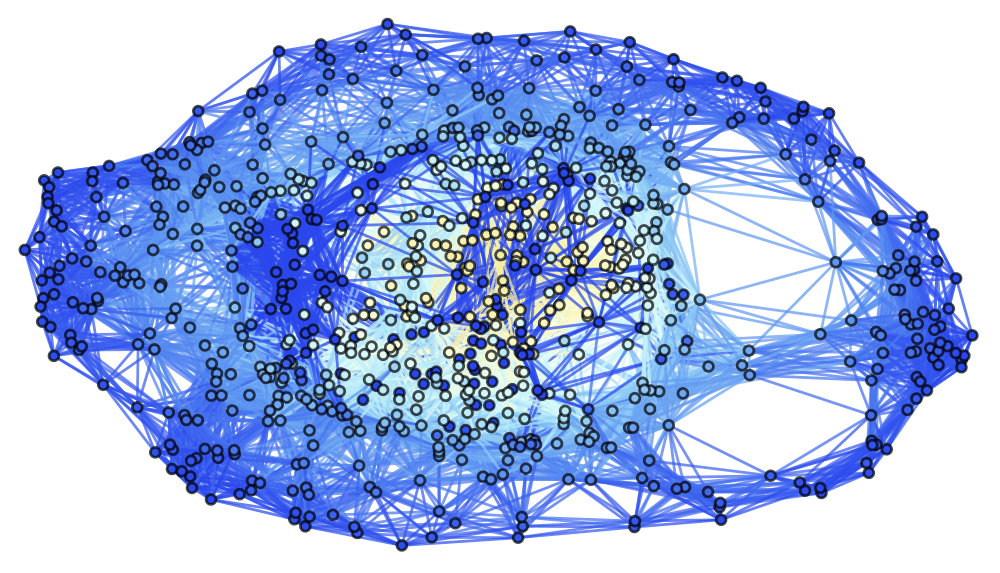}
\caption{Spatial hypergraphs corresponding to the first intermediate hypersurface configuration of the massive scalar field ``bubble collapse'' to a maximally-rotating (extremal) Kerr black hole test, with a spinning (exponential) initial density distribution, at time ${t = 1.5 M}$, with resolutions of 200, 400 and 800 vertices, respectively. The hypergraphs have been adapted and colored using the local curvature in the Boyer-Lindquist conformal factor ${\psi}$.}
\label{fig:Figure23}
\end{figure}

\begin{figure}[ht]
\centering
\includegraphics[width=0.325\textwidth]{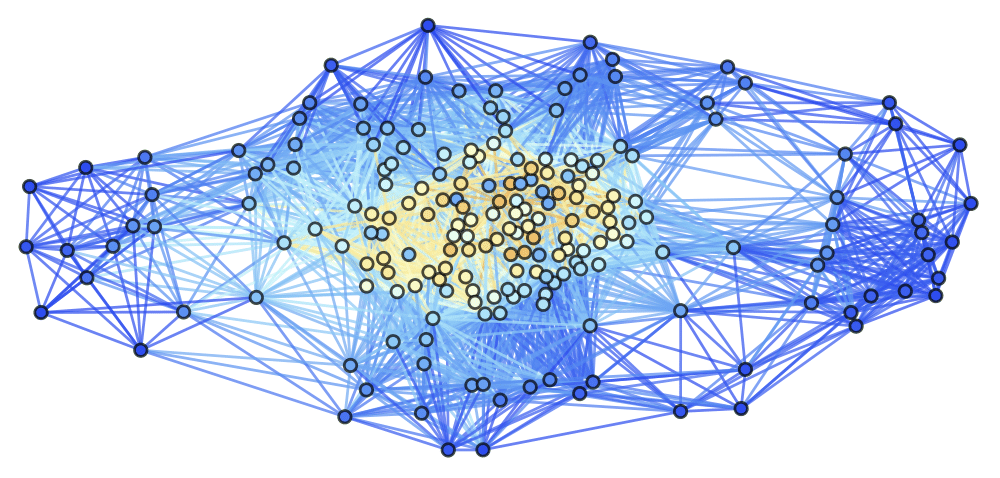}
\includegraphics[width=0.325\textwidth]{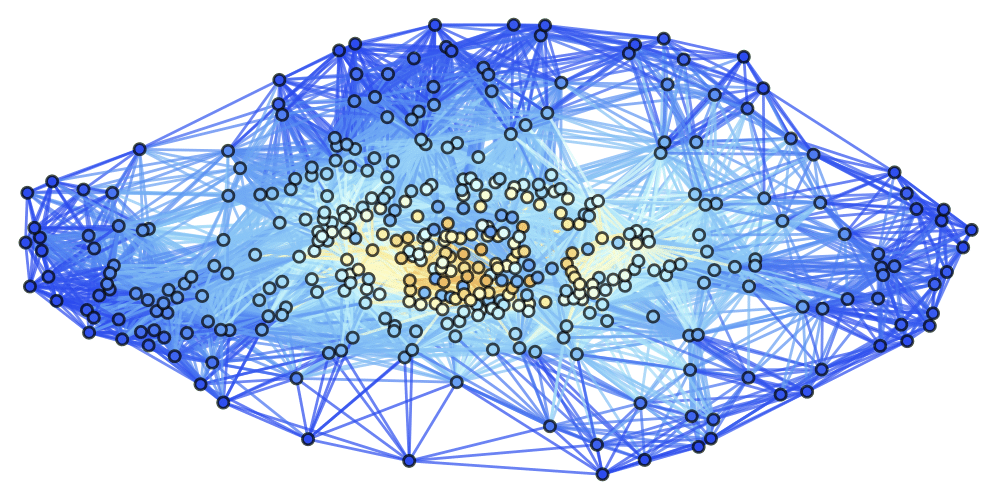}
\includegraphics[width=0.325\textwidth]{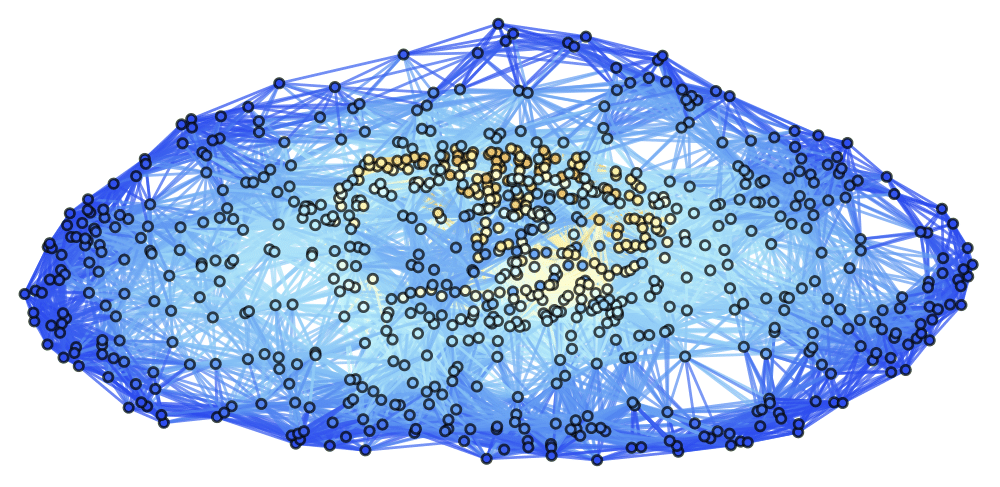}
\caption{Spatial hypergraphs corresponding to the second intermediate hypersurface configuration of the massive scalar field ``bubble collapse'' to a maximally-rotating (extremal) Kerr black hole test, with a spinning (exponential) initial density distribution, at time ${t = 3 M}$, with resolutions of 200, 400 and 800 vertices, respectively. The hypergraphs have been adapted and colored using the local curvature in the Boyer-Lindquist conformal factor ${\psi}$.}
\label{fig:Figure24}
\end{figure}

\begin{figure}[ht]
\centering
\includegraphics[width=0.325\textwidth]{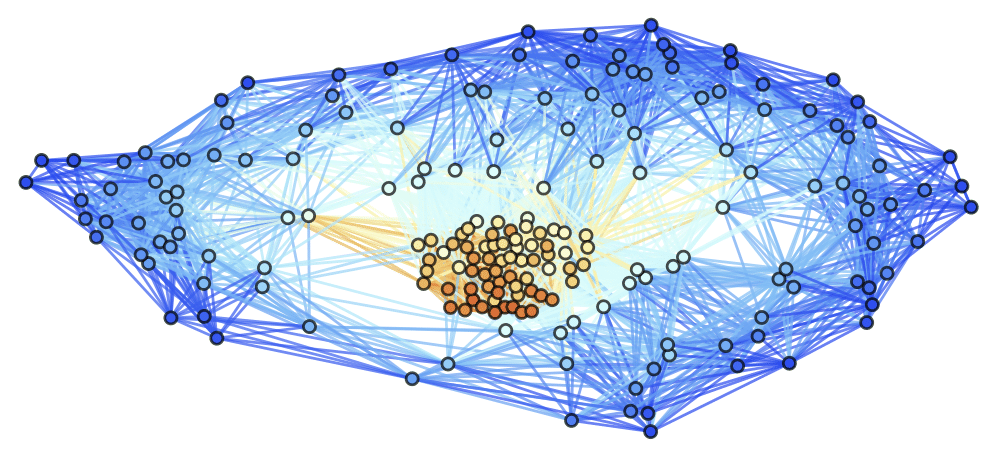}
\includegraphics[width=0.325\textwidth]{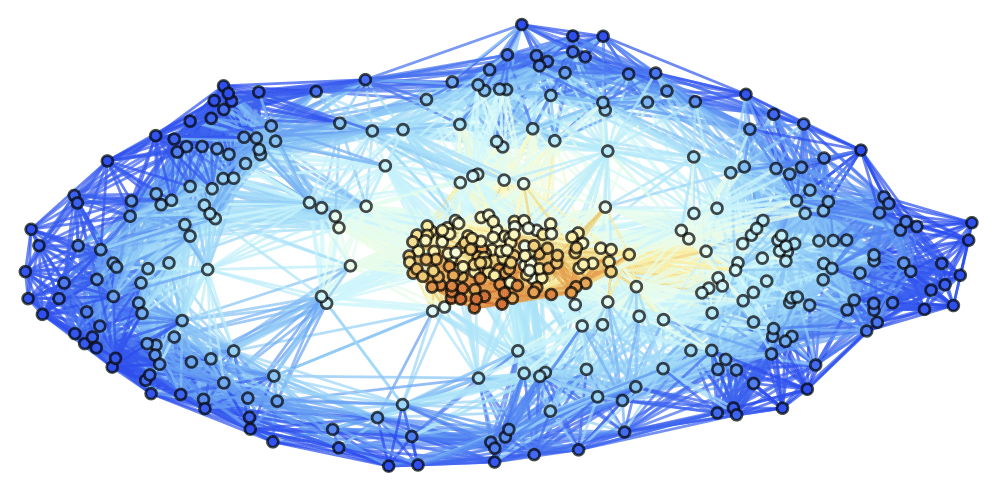}
\includegraphics[width=0.325\textwidth]{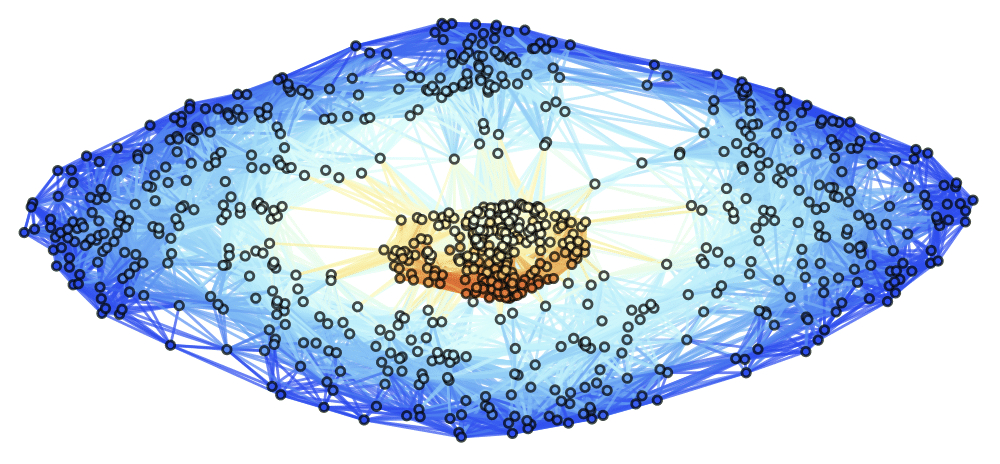}
\caption{Spatial hypergraphs corresponding to the final hypersurface configuration of the massive scalar field ``bubble collapse'' to a maximally-rotating (extremal) Kerr black hole test, with a spinning (exponential) initial density distribution, at time ${t = 4.5 M}$, with resolutions of 200, 400 and 800 vertices, respectively. The hypergraphs have been adapted and colored using the local curvature in the Boyer-Lindquist conformal factor ${\psi}$.}
\label{fig:Figure25}
\end{figure}

\begin{figure}[ht]
\centering
\includegraphics[width=0.325\textwidth]{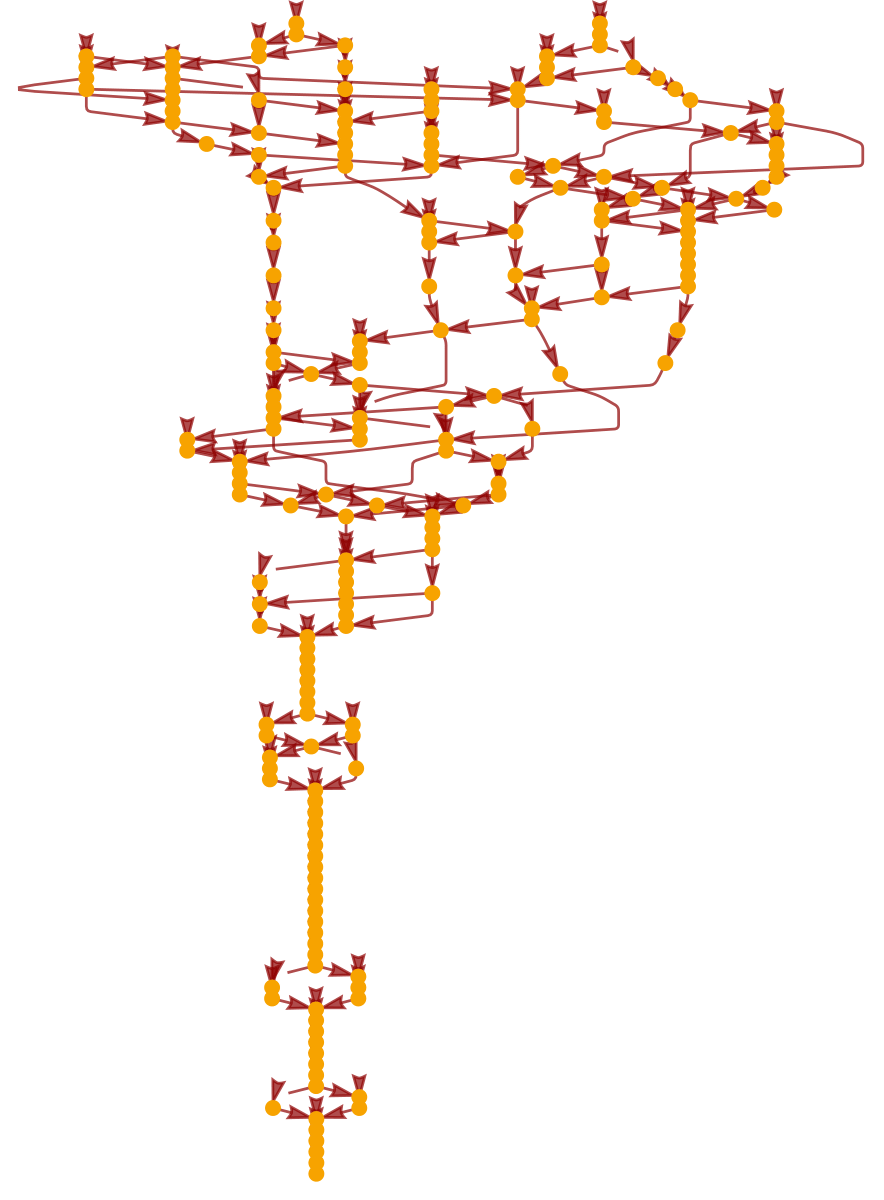}
\includegraphics[width=0.325\textwidth]{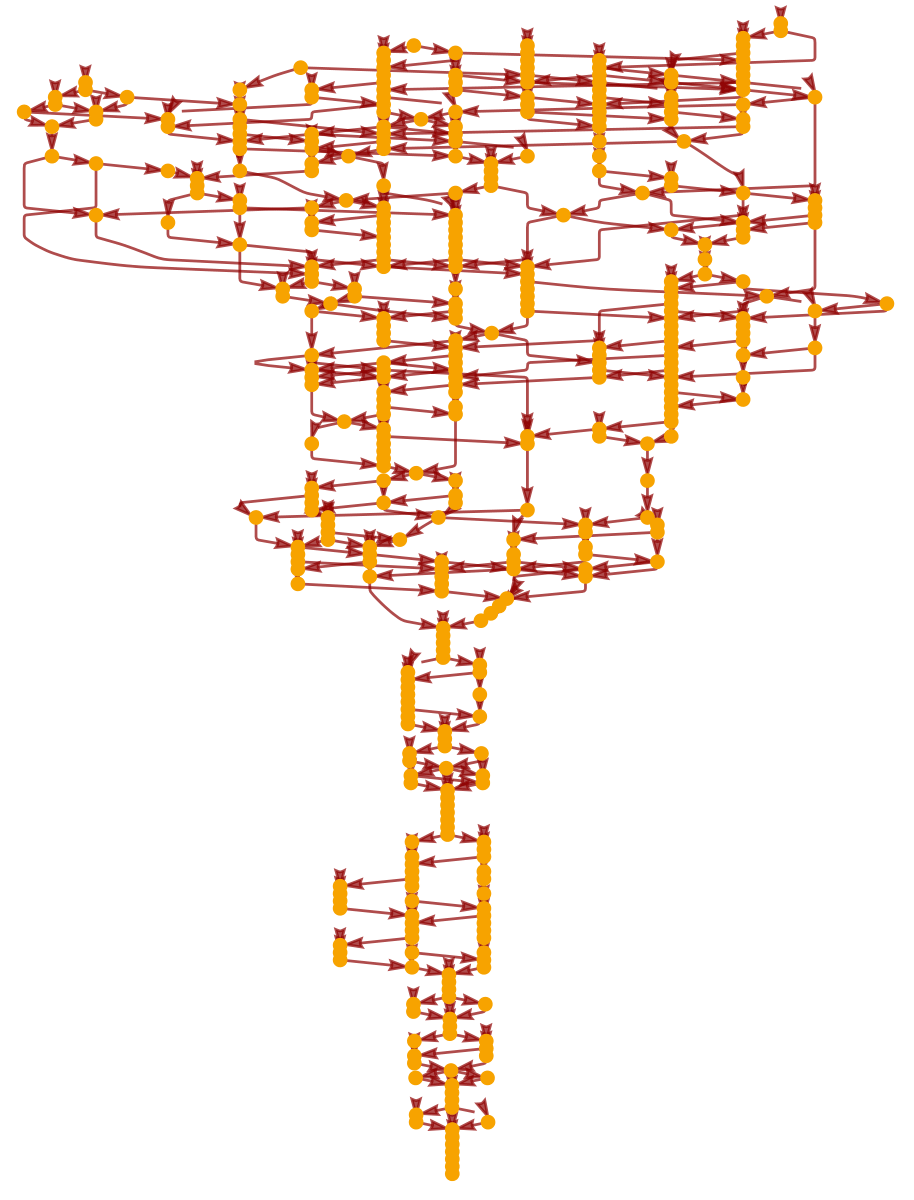}
\includegraphics[width=0.325\textwidth]{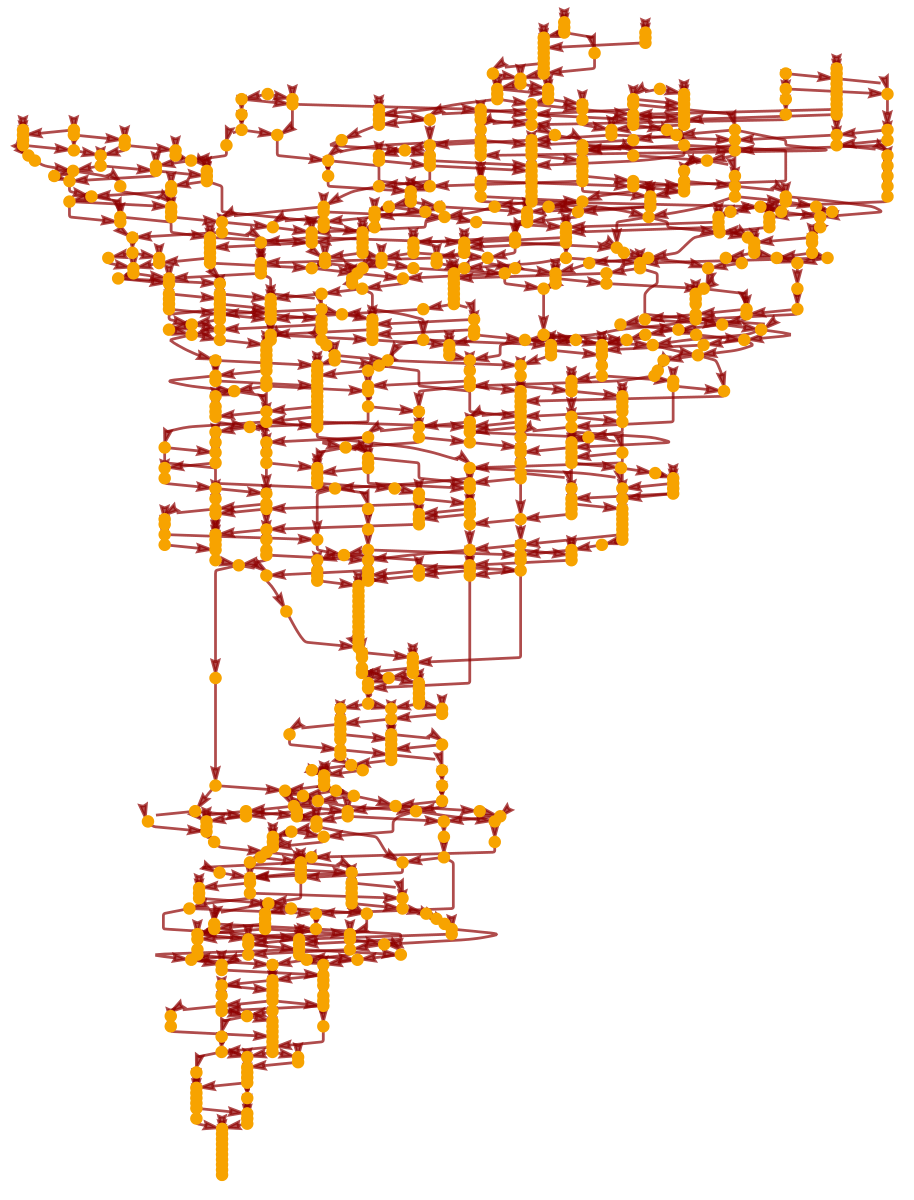}
\caption{Causal graphs corresponding to the discrete characteristic structure of the massive scalar field ``bubble collapse'' to a maximally-rotating (extremal) Kerr black hole test, with a spinning (exponential) initial density distribution, at time ${t = 4.5 M}$, with resolutions of 200, 400 and 800 hypergraph vertices, respectively.}
\label{fig:Figure26}
\end{figure}

\begin{figure}[ht]
\centering
\includegraphics[width=0.325\textwidth]{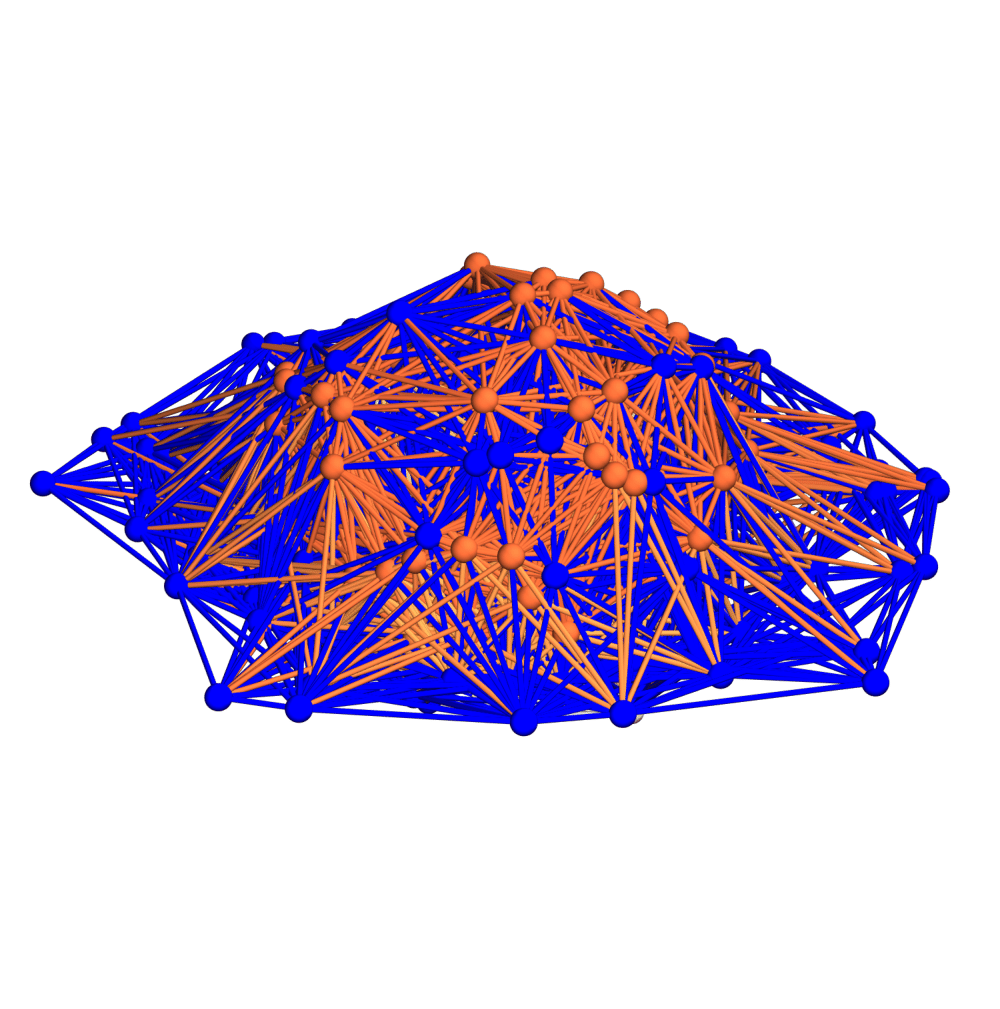}
\includegraphics[width=0.325\textwidth]{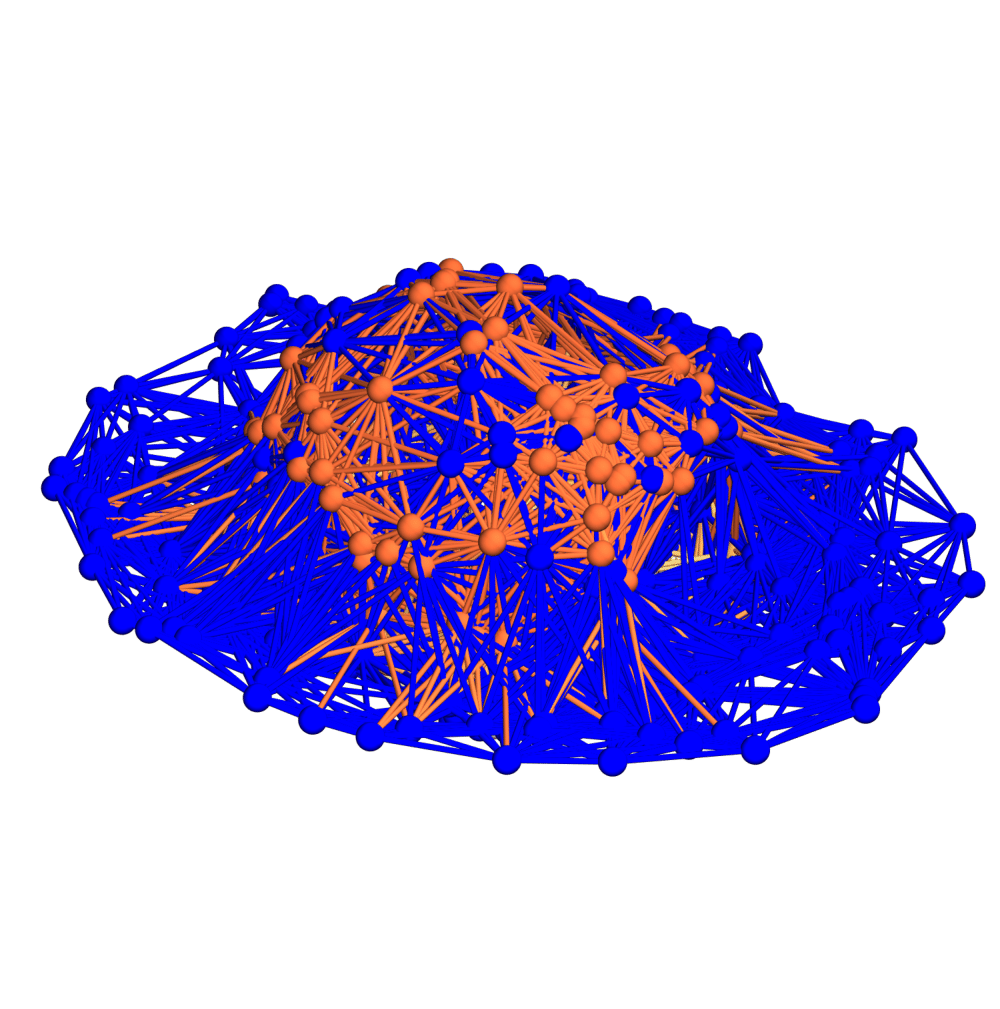}
\includegraphics[width=0.325\textwidth]{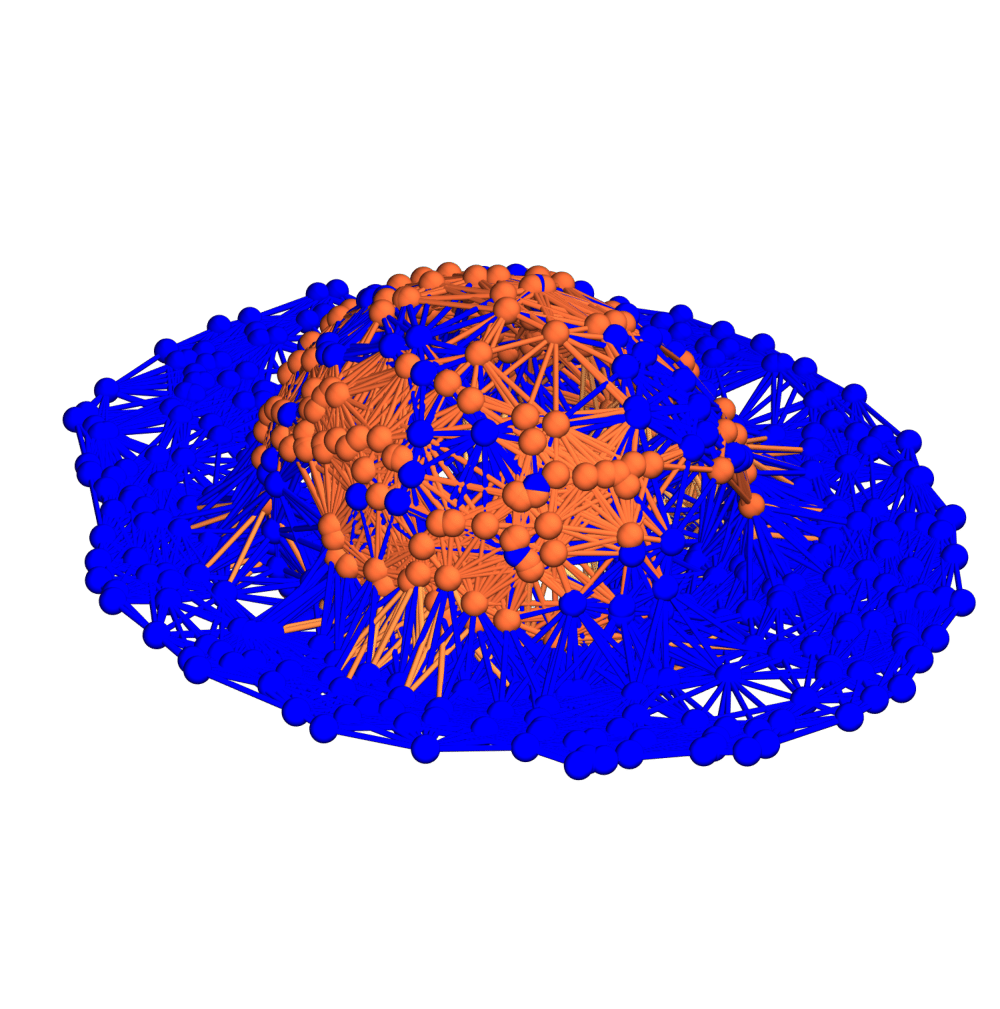}
\caption{Spatial hypergraphs corresponding to projections along the $z$-axis of the initial hypersurface configuration of the massive scalar field ``bubble collapse'' to a maximally-rotating (extremal) Kerr black hole test, with a spinning (exponential) initial density distribution, at time ${t = 0 M}$, with resolutions of 200, 400 and 800 vertices, respectively. The vertices have been assigned spatial coordinates according to the profile of the Boyer-Lindquist conformal factor ${\psi}$ through a spatial slice perpendicular to the $z$-axis, and the hypergraphs have been adapted using the local curvature in ${\psi}$, and colored according to the value of the scalar field ${\Phi \left( t, R \right)}$.}
\label{fig:Figure27}
\end{figure}

\begin{figure}[ht]
\centering
\includegraphics[width=0.325\textwidth]{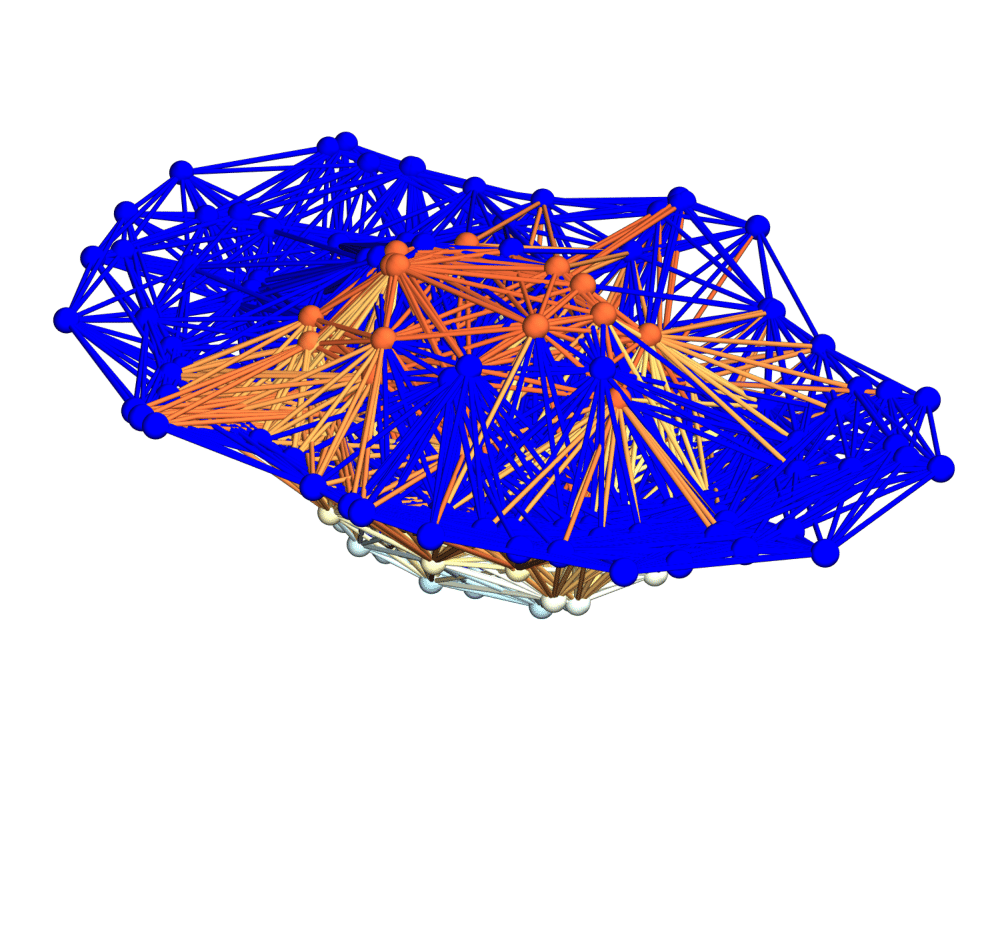}
\includegraphics[width=0.325\textwidth]{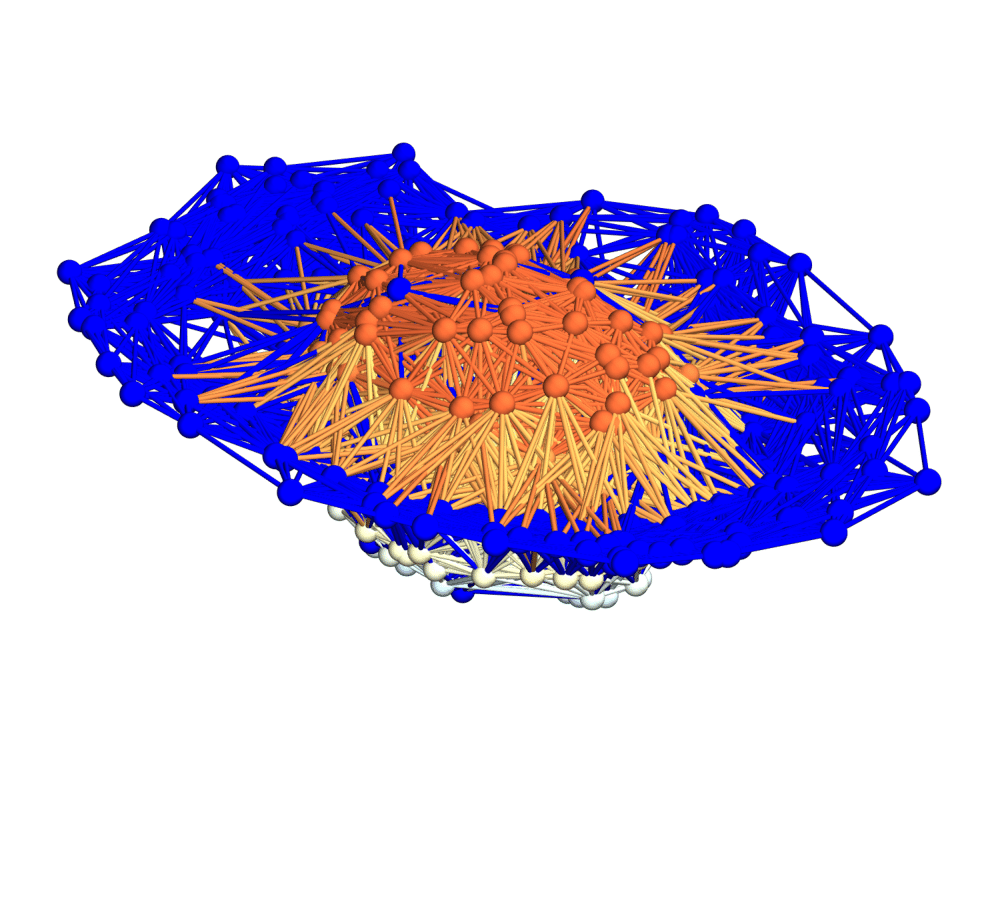}
\includegraphics[width=0.325\textwidth]{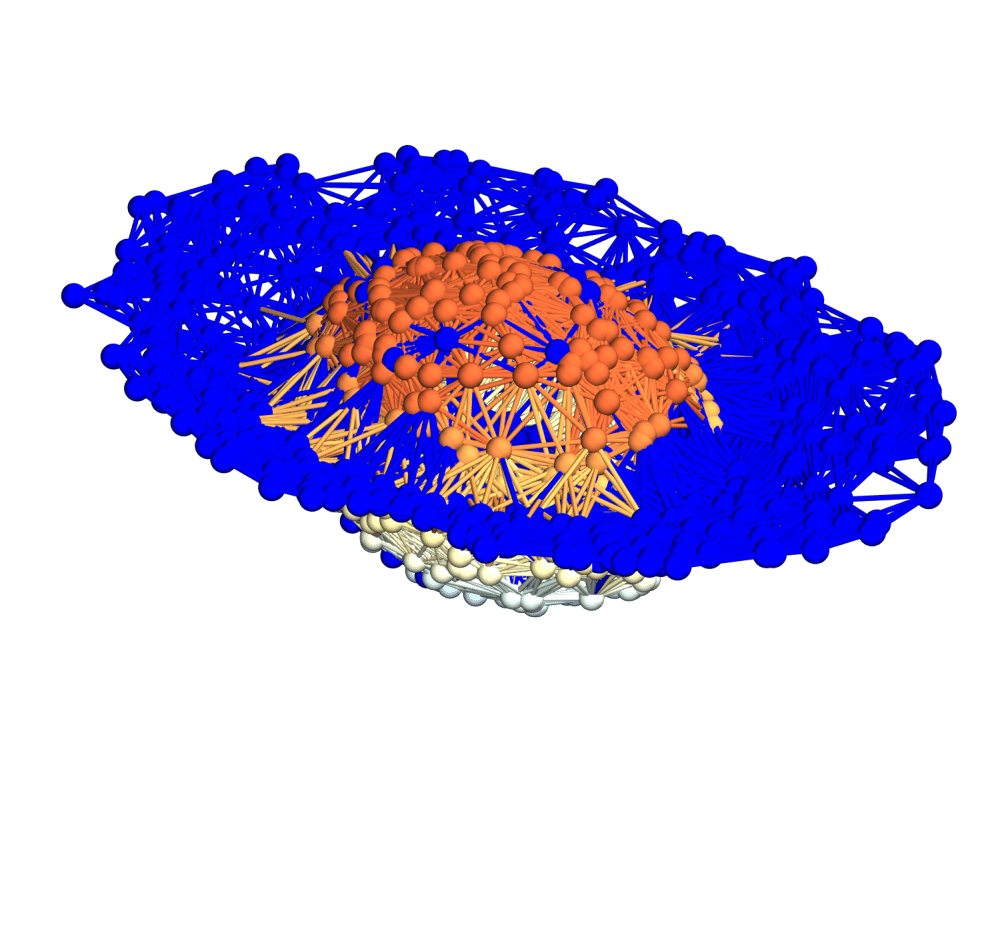}
\caption{Spatial hypergraphs corresponding to projections along the $z$-axis of the first intermediate hypersurface configuration of the massive scalar field ``bubble collapse'' to a maximally-rotating (extremal) Kerr black hole test, with a spinning (exponential) initial density distribution, at time ${t = 1.5 M}$, with resolutions of 200, 400 and 800 vertices, respectively. The vertices have been assigned spatial coordinates according to the profile of the Boyer-Lindquist conformal factor ${\psi}$ through a spatial slice perpendicular to the $z$-axis, and the hypergraphs have been adapted using the local curvature in ${\psi}$, and colored according to the value of the scalar field ${\Phi \left( t, R \right)}$.}
\label{fig:Figure28}
\end{figure}

\begin{figure}[ht]
\centering
\includegraphics[width=0.325\textwidth]{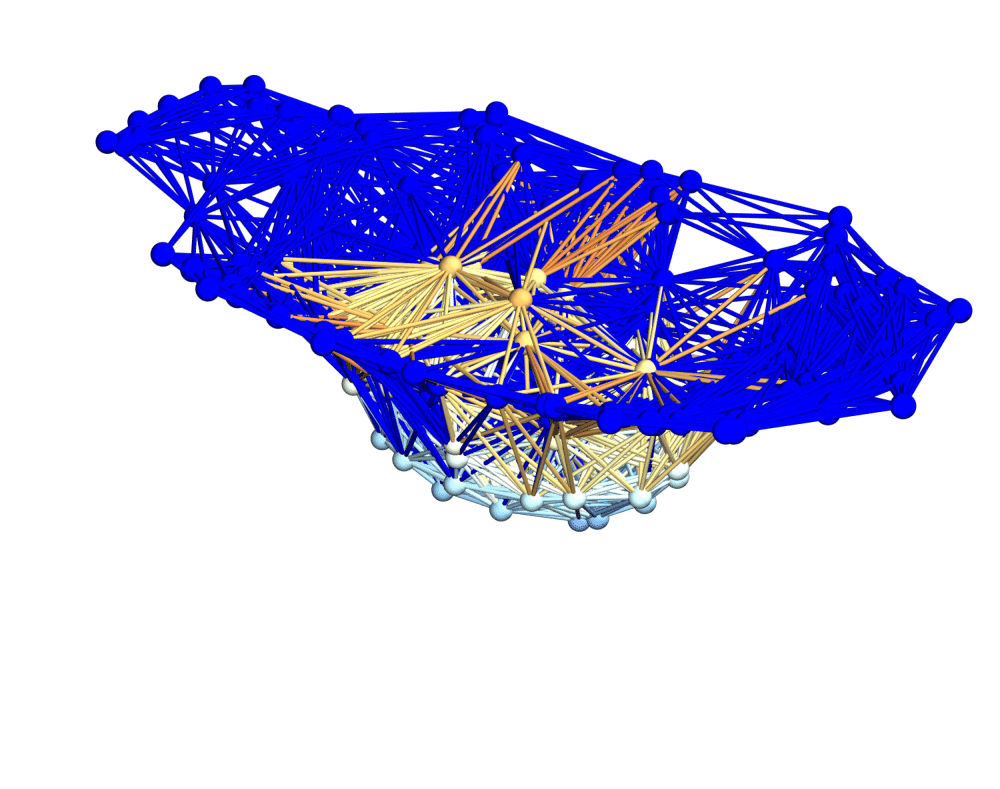}
\includegraphics[width=0.325\textwidth]{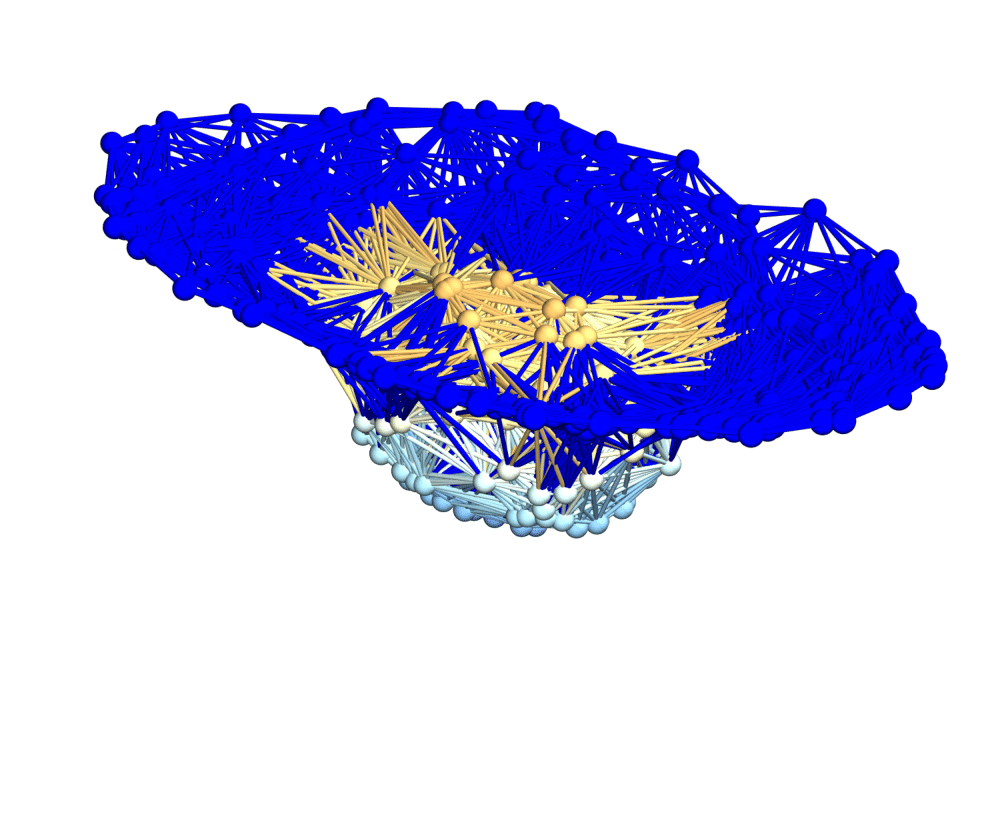}
\includegraphics[width=0.325\textwidth]{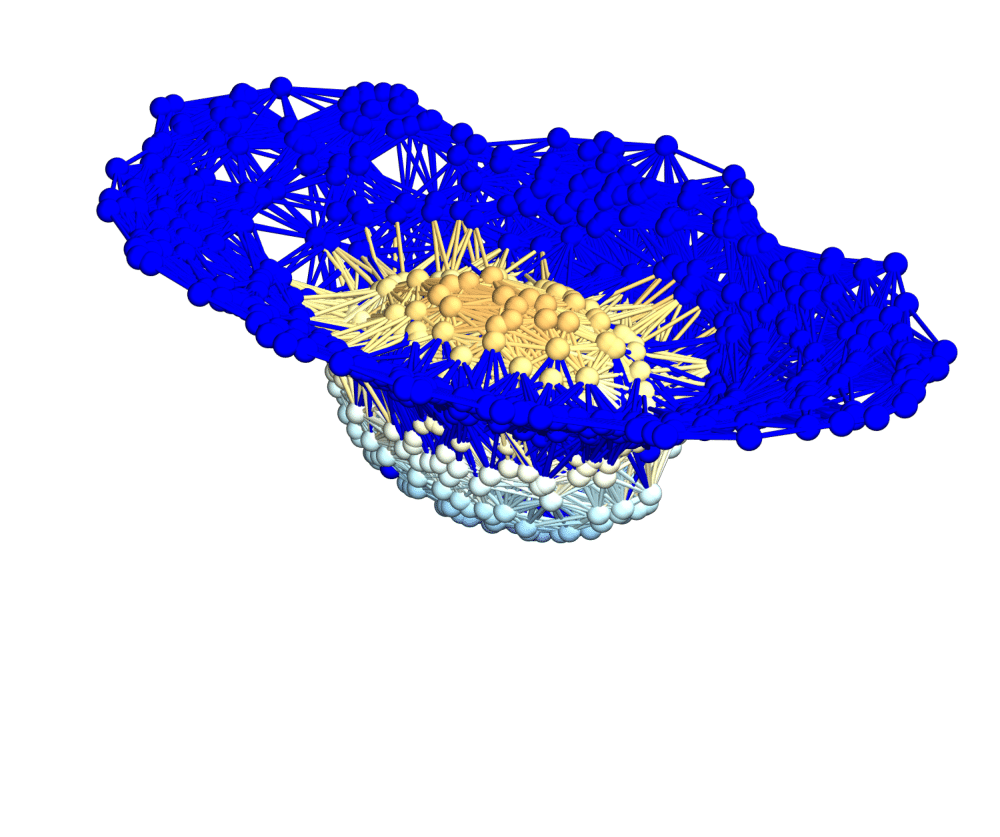}
\caption{Spatial hypergraphs corresponding to projections along the $z$-axis of the second intermediate hypersurface configuration of the massive scalar field ``bubble collapse'' to a maximally-rotating (extremal) Kerr black hole test, with a spinning (exponential) initial density distribution, at time ${t = 3 M}$, with resolutions of 200, 400 and 800 vertices, respectively. The vertices have been assigned spatial coordinates according to the profile of the Boyer-Lindquist conformal factor ${\psi}$ through a spatial slice perpendicular to the $z$-axis, and the hypergraphs have been adapted using the local curvature in ${\psi}$, and colored according to the value of the scalar field ${\Phi \left( t, R \right)}$.}
\label{fig:Figure29}
\end{figure}

\begin{figure}[ht]
\centering
\includegraphics[width=0.325\textwidth]{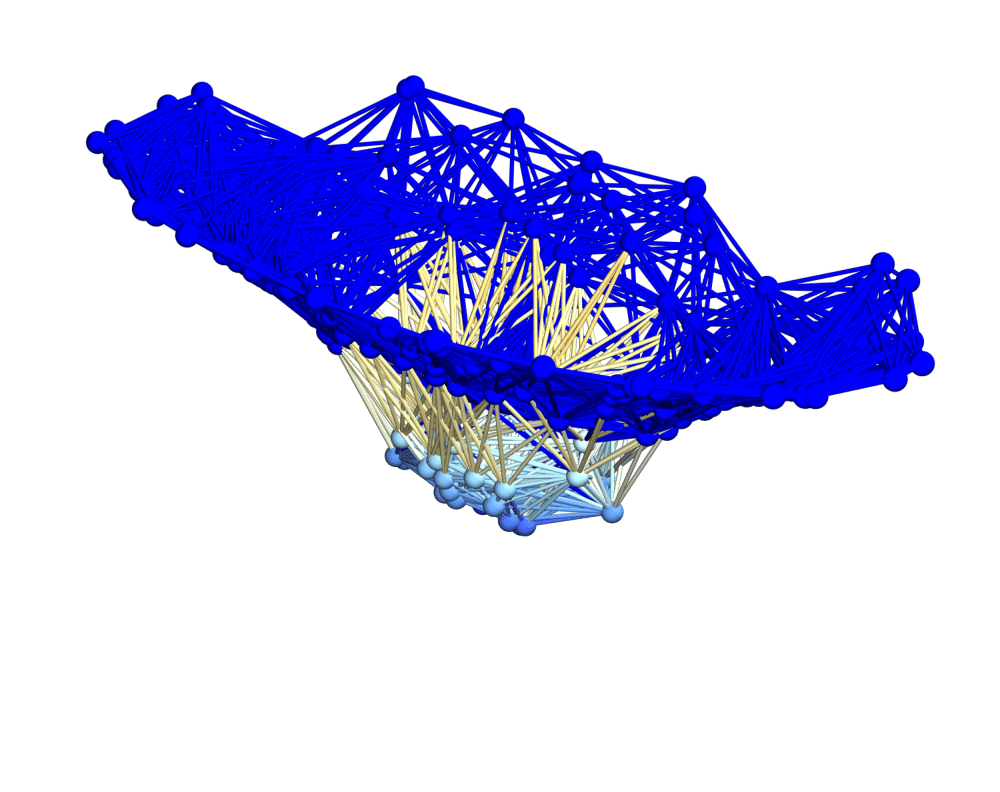}
\includegraphics[width=0.325\textwidth]{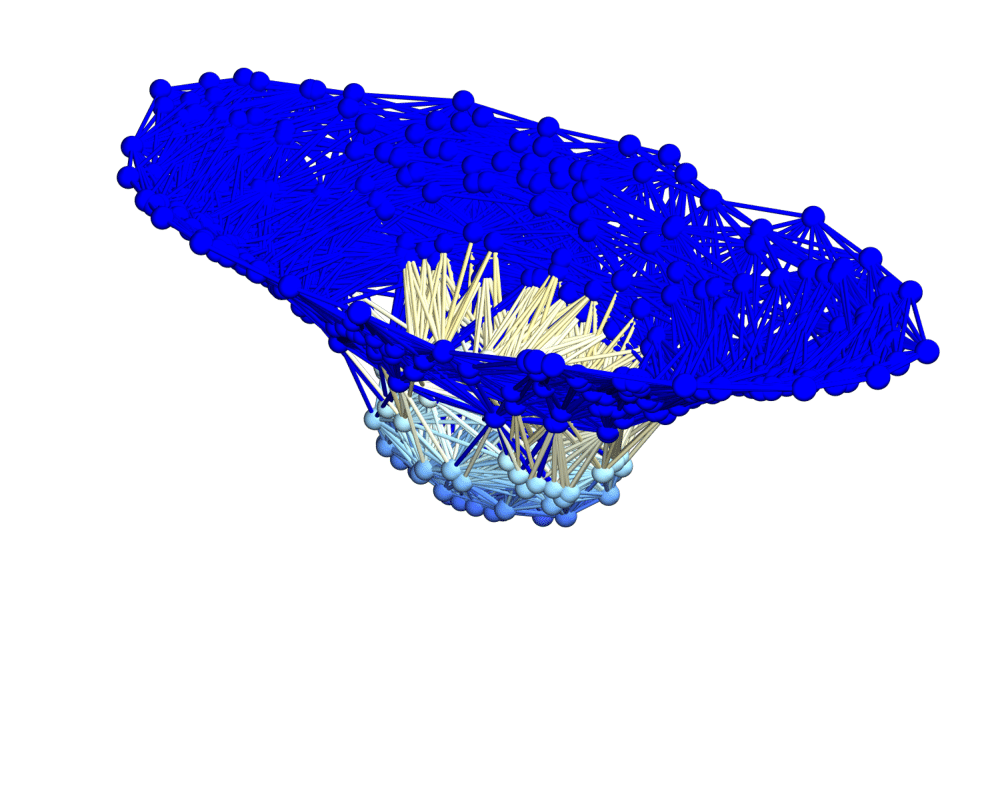}
\includegraphics[width=0.325\textwidth]{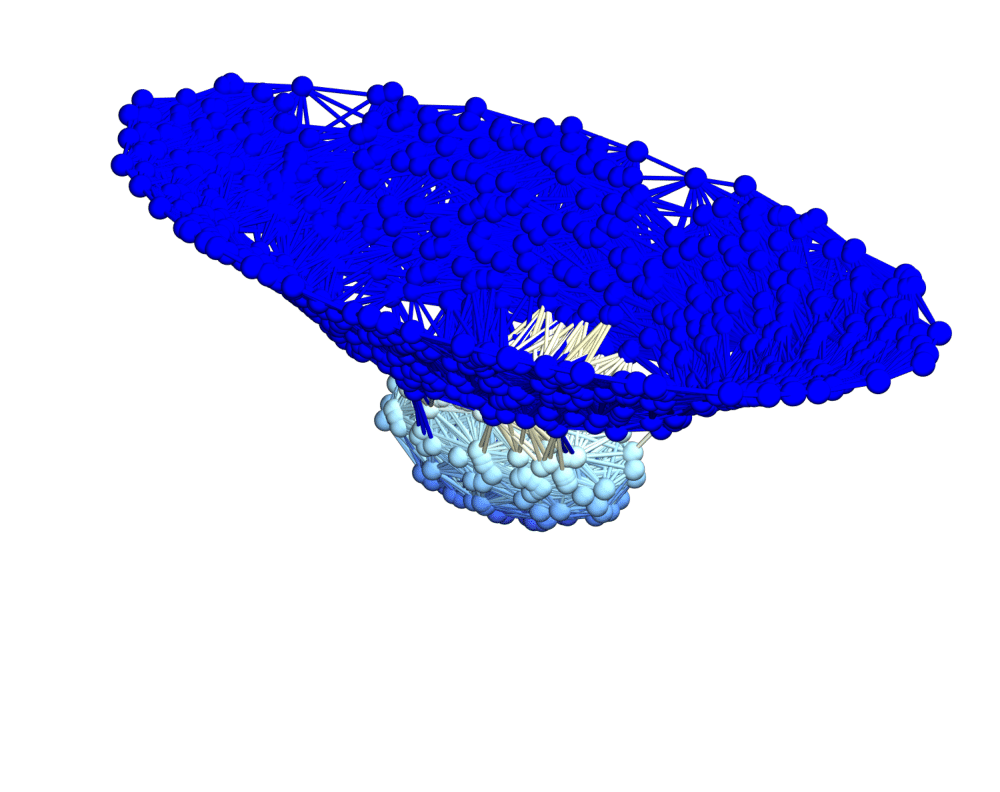}
\caption{Spatial hypergraphs corresponding to projections along the $z$-axis of the final hypersurface configuration of the massive scalar field ``bubble collapse'' to a maximally-rotating (extremal) Kerr black hole test, with a spinning (exponential) initial density distribution, at time ${t = 4.5 M}$, with resolutions of 200, 400 and 800 vertices, respectively. The vertices have been assigned spatial coordinates according to the profile of the Boyer-Lindquist conformal factor ${\psi}$ through a spatial slice perpendicular to the $z$-axis, and the hypergraphs have been adapted using the local curvature in ${\psi}$, and colored according to the value of the scalar field ${\Phi \left( t, R \right)}$.}
\label{fig:Figure30}
\end{figure}

\begin{figure}[ht]
\centering
\includegraphics[width=0.325\textwidth]{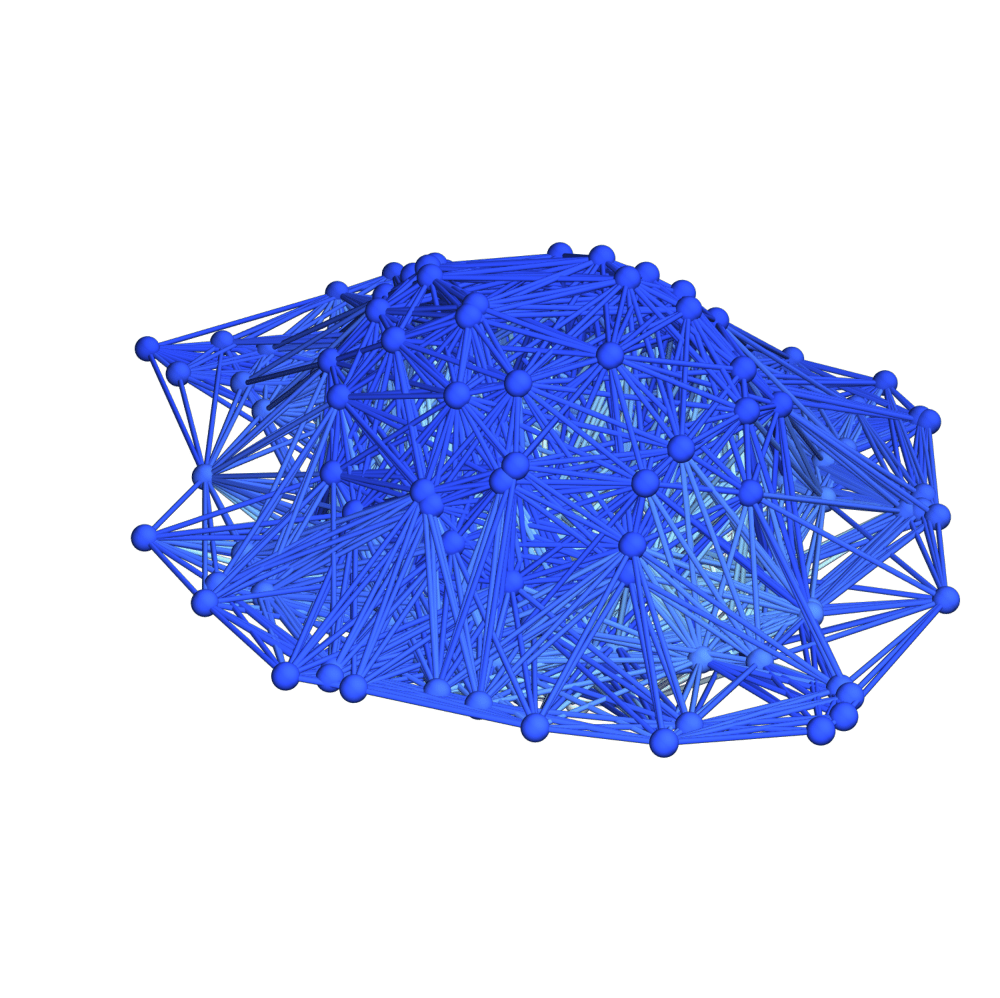}
\includegraphics[width=0.325\textwidth]{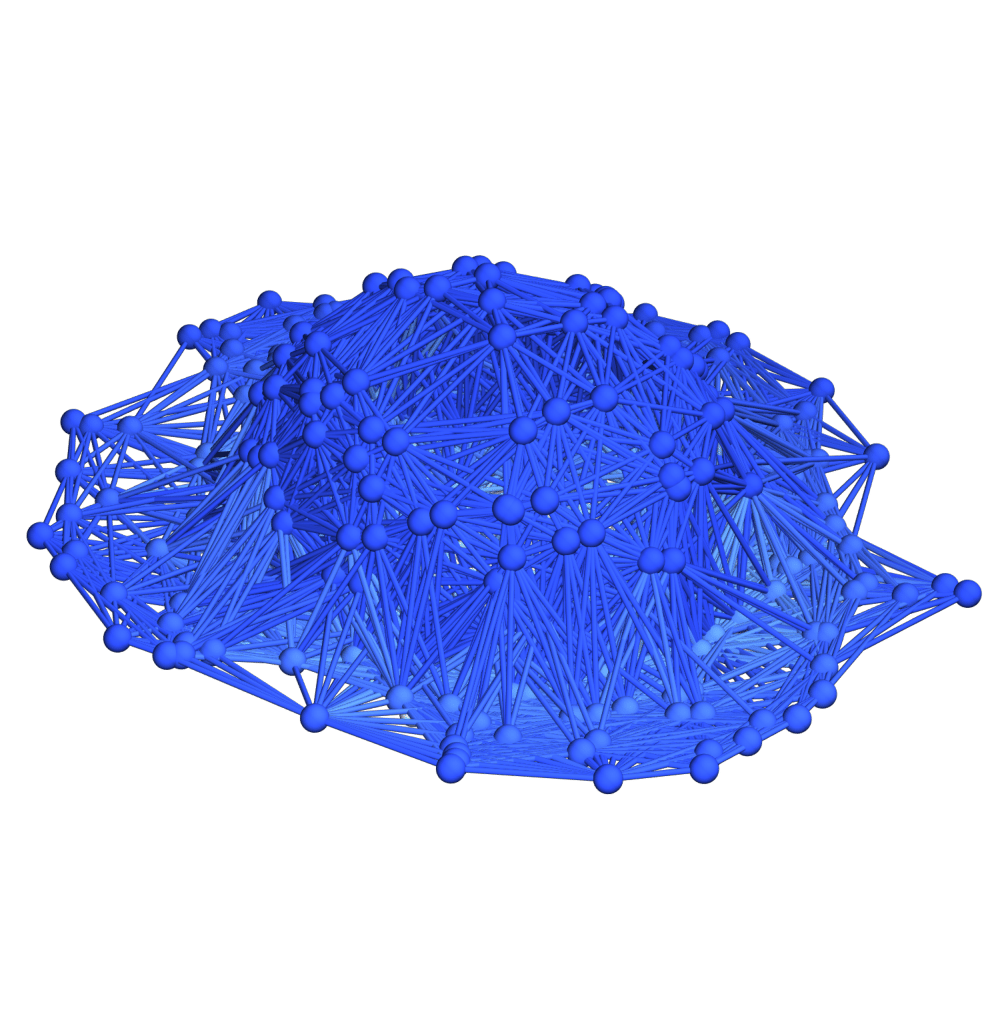}
\includegraphics[width=0.325\textwidth]{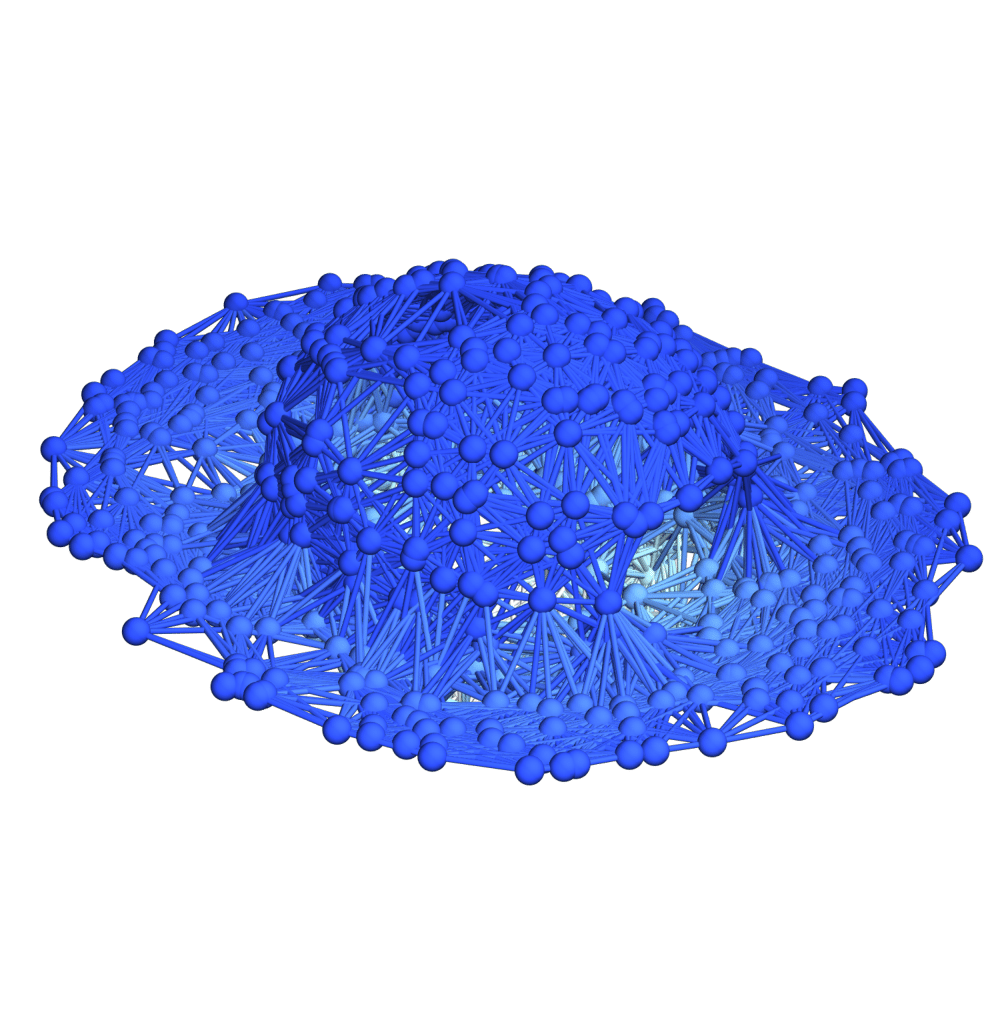}
\caption{Spatial hypergraphs corresponding to projections along the $z$-axis of the initial hypersurface configuration of the massive scalar field ``bubble collapse'' to a maximally-rotating (extremal) Kerr black hole test, with a spinning (exponential) initial density distribution, at time ${t = 0 M}$, with resolutions of 200, 400 and 800 vertices, respectively. The vertices have been assigned spatial coordinates according to the profile of the Boyer-Lindquist conformal factor ${\psi}$ through a spatial slice perpendicular to the $z$-axis, and the hypergraphs have been adapted and colored using the local curvature in ${\psi}$.}
\label{fig:Figure31}
\end{figure}

\begin{figure}[ht]
\centering
\includegraphics[width=0.325\textwidth]{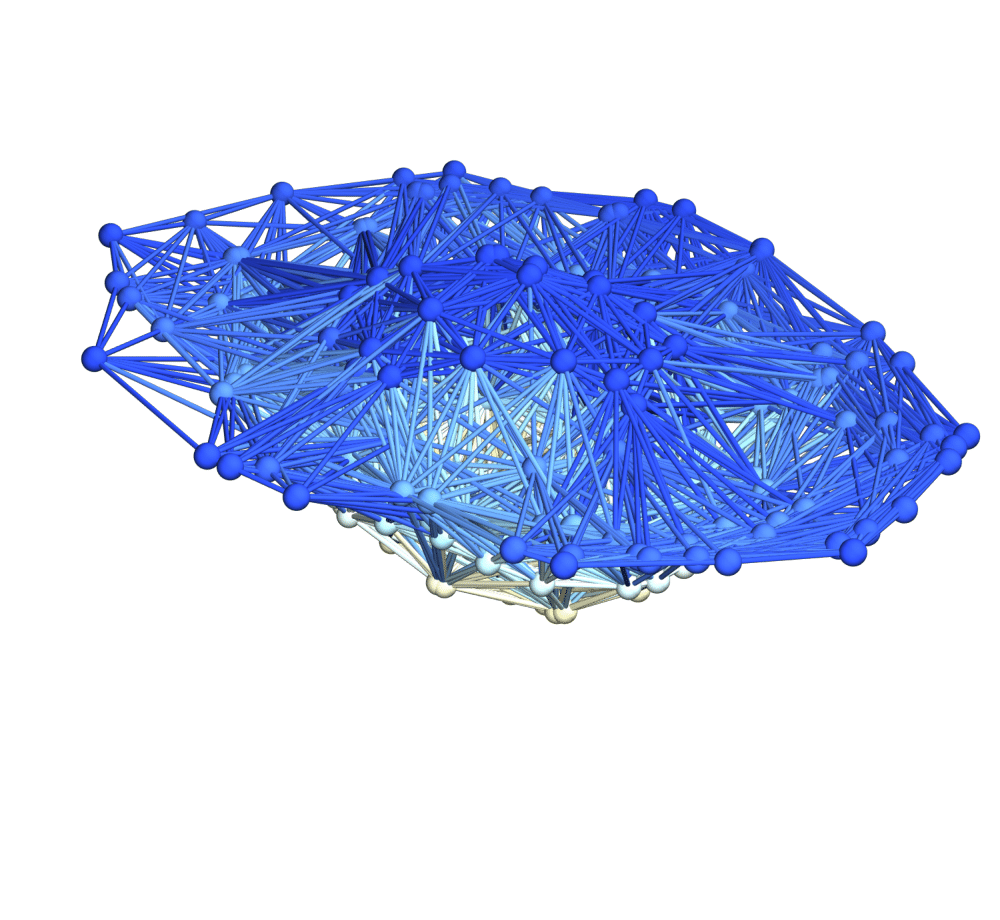}
\includegraphics[width=0.325\textwidth]{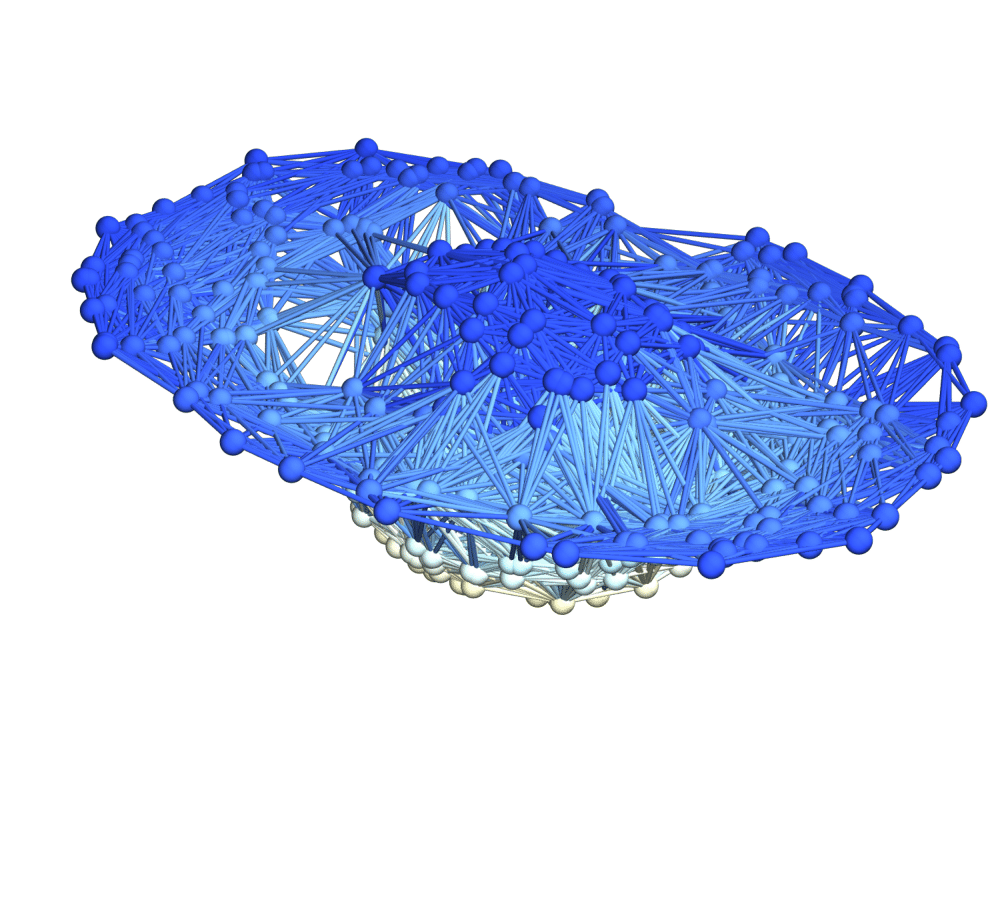}
\includegraphics[width=0.325\textwidth]{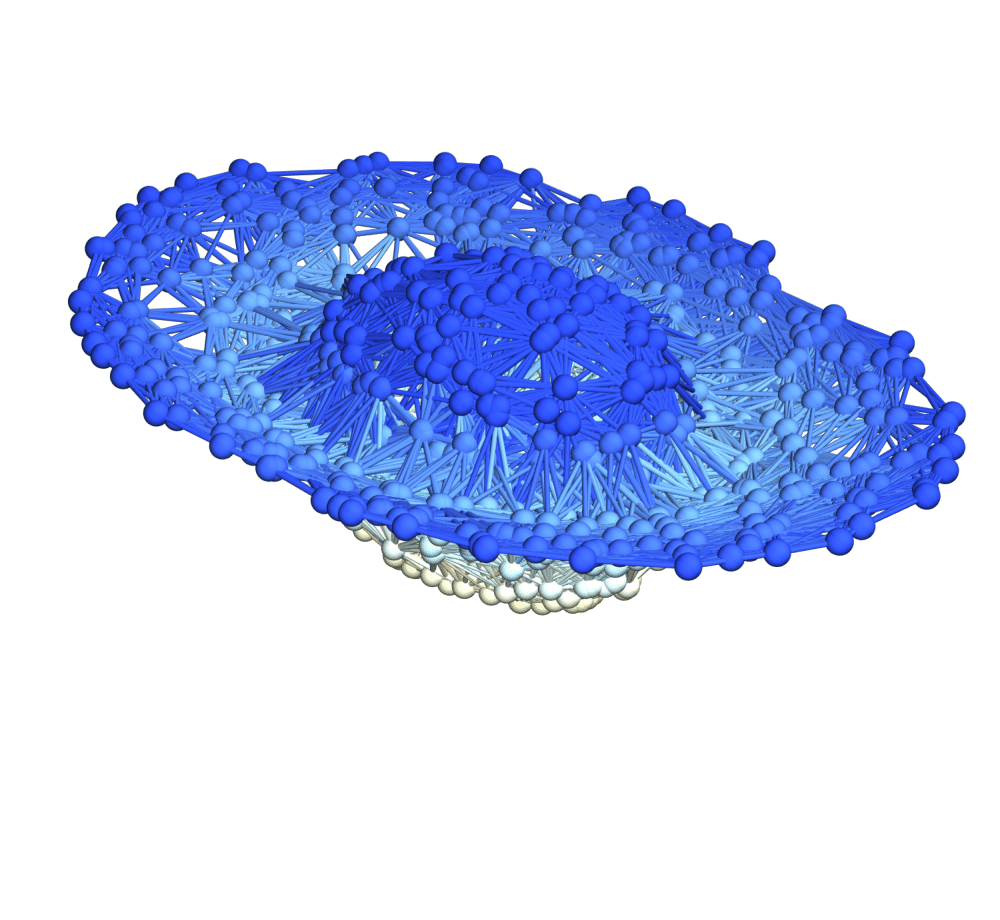}
\caption{Spatial hypergraphs corresponding to projections along the $z$-axis of the first intermediate hypersurface configuration of the massive scalar field ``bubble collapse'' to a maximally-rotating (extremal) Kerr black hole test, with a spinning (exponential) initial density distribution, at time ${t = 1.5 M}$, with resolutions of 200, 400 and 800 vertices, respectively. The vertices have been assigned spatial coordinates according to the profile of the Boyer-Lindquist conformal factor ${\psi}$ through a spatial slice perpendicular to the $z$-axis, and the hypergraphs have been adapted and colored using the local curvature in ${\psi}$.}
\label{fig:Figure32}
\end{figure}

\begin{figure}[ht]
\centering
\includegraphics[width=0.325\textwidth]{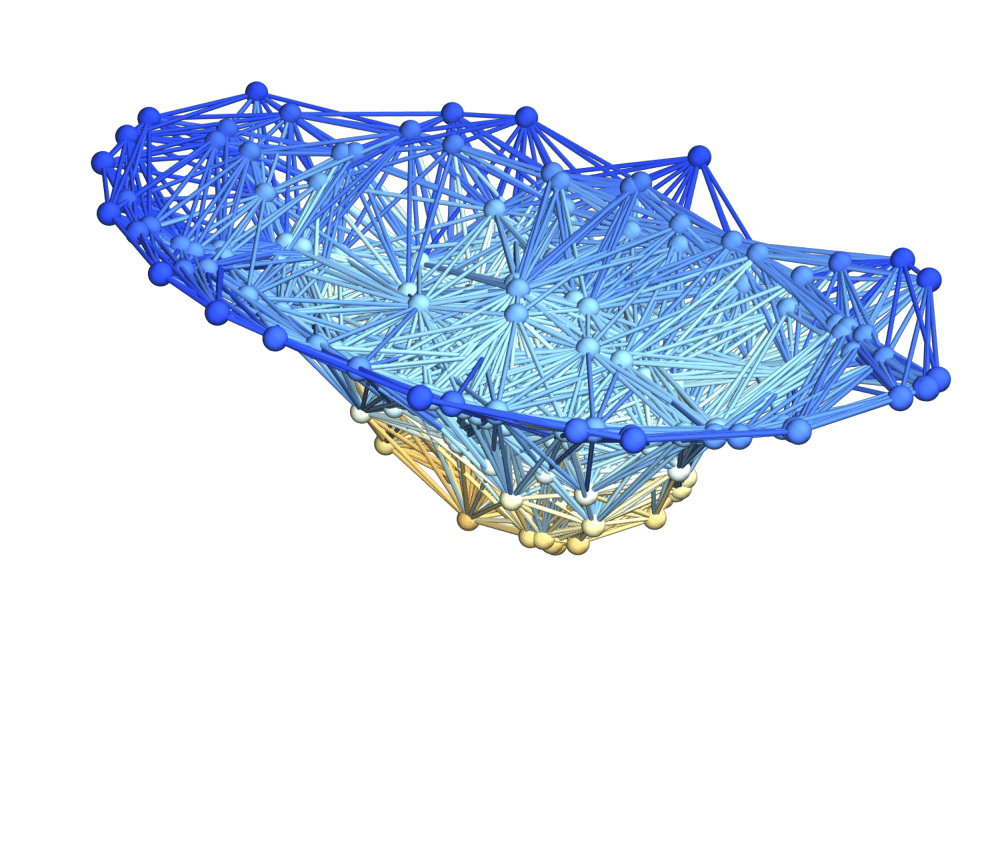}
\includegraphics[width=0.325\textwidth]{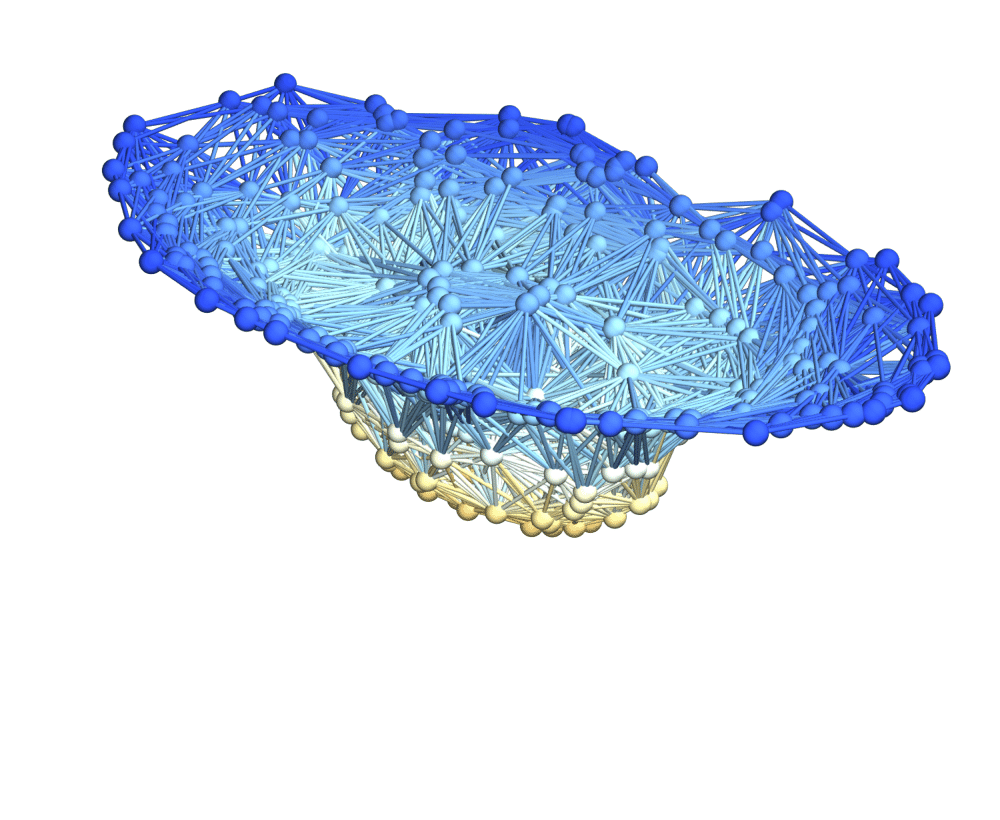}
\includegraphics[width=0.325\textwidth]{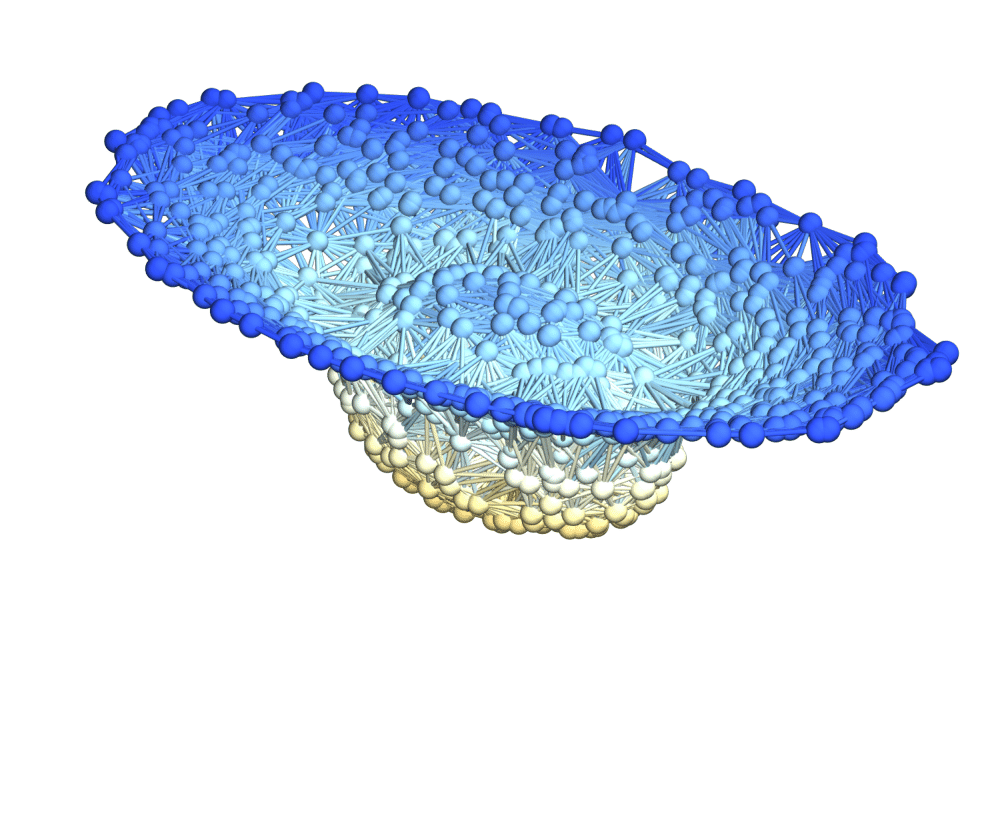}
\caption{Spatial hypergraphs corresponding to projections along the $z$-axis of the second intermediate hypersurface configuration of the massive scalar field ``bubble collapse'' to a maximally-rotating (extremal) Kerr black hole test, with a spinning (exponential) initial density distribution, at time ${t = 3 M}$, with resolutions of 200, 400 and 800 vertices, respectively. The vertices have been assigned spatial coordinates according to the profile of the Boyer-Lindquist conformal factor ${\psi}$ through a spatial slice perpendicular to the $z$-axis, and the hypergraphs have been adapted and colored using the local curvature in ${\psi}$.}
\label{fig:Figure33}
\end{figure}

\begin{figure}[ht]
\centering
\includegraphics[width=0.325\textwidth]{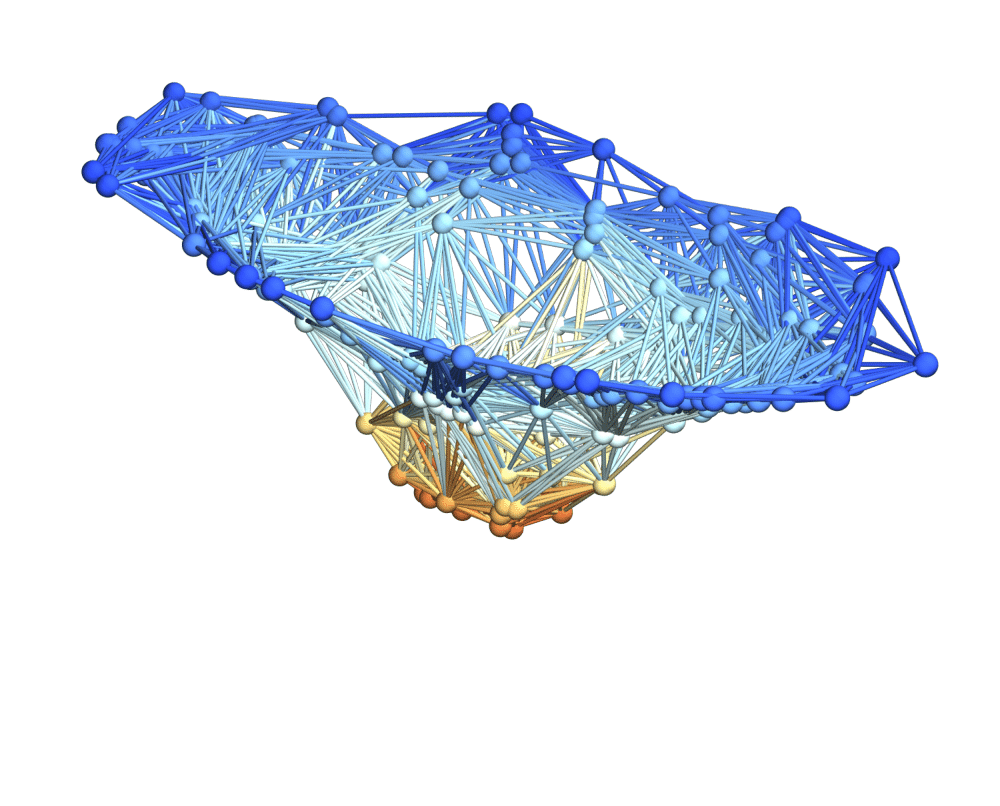}
\includegraphics[width=0.325\textwidth]{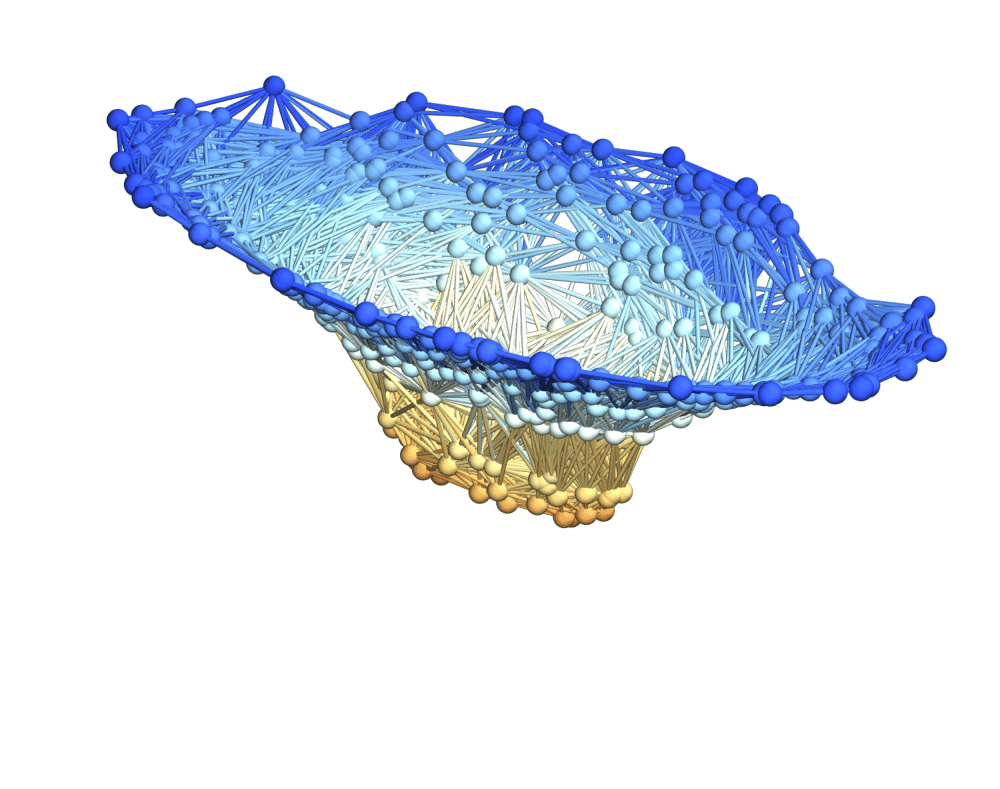}
\includegraphics[width=0.325\textwidth]{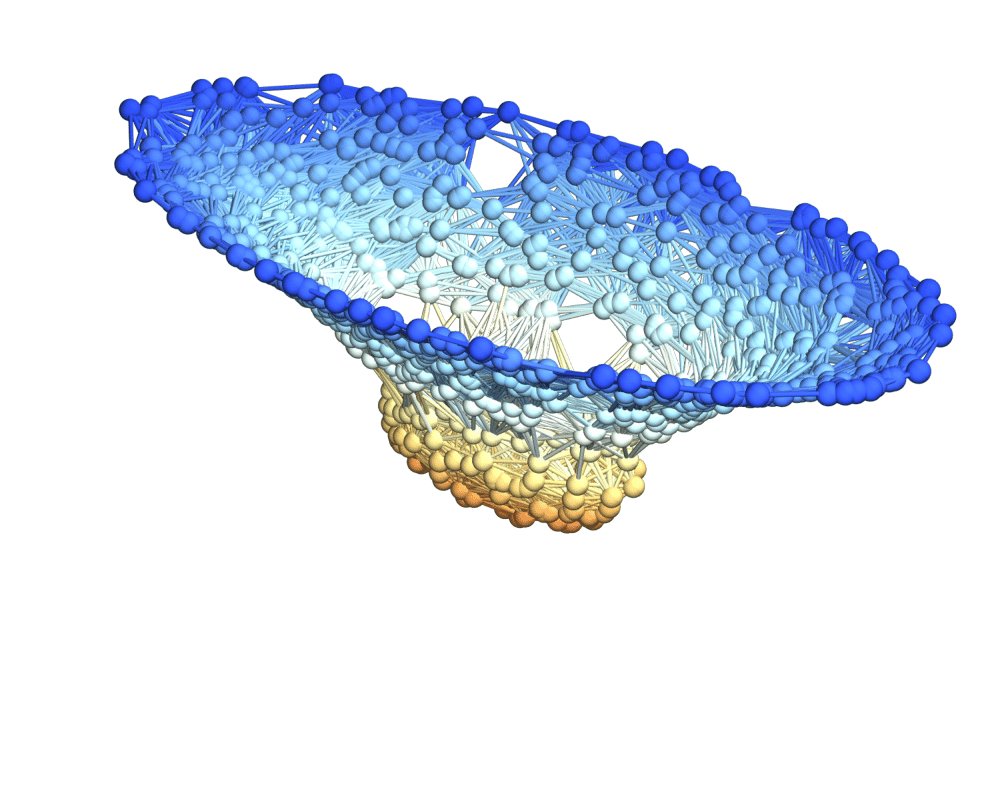}
\caption{Spatial hypergraphs corresponding to projections along the $z$-axis of the final hypersurface configuration of the massive scalar field ``bubble collapse'' to a maximally-rotating (extremal) Kerr black hole test, with a spinning (exponential) initial density distribution, at time ${t = 4.5 M}$, with resolutions of 200, 400 and 800 vertices, respectively. The vertices have been assigned spatial coordinates according to the profile of the Boyer-Lindquist conformal factor ${\psi}$ through a spatial slice perpendicular to the $z$-axis, and the hypergraphs have been adapted and colored using the local curvature in ${\psi}$.}
\label{fig:Figure34}
\end{figure}

We now proceed to simulate the evolution of an axially-symmetric massive scalar field ``bubble collapse'' problem numerically, corresponding to the collapse of a uniformly-rotating disk of dust to a maximally-rotating (extremal) Kerr black hole, using the same techniques as for the spherically-symmetric case analyzed previously. As before, we assume an exponential initial density profile for the massive scalar field ${\Phi \left( t, R\right)}$:

\begin{equation}
\rho \left( 0, R \right) = T_{t t} \left( 0, R \right) = \rho_0 \exp \left( - \left( \frac{R}{\lambda} \right)^3 \right),
\end{equation}
for initial density constant ${\rho_0}$ and radius of support ${\lambda}$, albeit replacing the spherically-symmetric background metric in Gaussian polar coordinates:

\begin{equation}
d s^2 = - d t^2 + e^{- 2 \Lambda \left( t, r \right)} d r^2 + R^2 \left( t, r \right) d \Omega^2,
\end{equation}
with the axially-symmetric background metric in (cylindrical) Weyl-Lewis-Papapetrou coordinates:

\begin{equation}
d s^2 = - e^{2 U \left( \rho, \zeta \right)} \left( d t + a \left( \rho, \zeta \right) d \varphi \right)^2 + e^{-2 U \left( \rho, \zeta \right)} \left[ e^{2 k \left( \rho, \zeta \right)} \left( d \rho^2 + d \zeta^2 \right) + W^2 \left( \rho, \zeta \right) d \varphi^2 \right],
\end{equation}
with ${\rho = R \sin \left( \theta \right)}$ and ${\zeta = R \cos \left( \theta \right)}$. As previously, we enforce Sommerfeld/radiative boundary conditions at the outermost boundary of the computational domain, set at radius ${60 M}$, and we evolve the solution until a final time of ${t = 4.5 M}$, with intermediate checks at times ${t = 1.5 M}$ and ${t = 3 M}$; the initial, first intermediate, second intermediate and final hypersurface configurations, with the hypergraphs adapted using the Boyer-Lindquist conformal factor ${\psi}$ and colored using the scalar field ${\Phi \left( t, R \right)}$ are shown in Figures \ref{fig:Figure18}, \ref{fig:Figure19}, \ref{fig:Figure20} and \ref{fig:Figure21}, respectively, with resolutions of 200, 400 and 800 vertices; similarly, Figures \ref{fig:Figure22}, \ref{fig:Figure23}, \ref{fig:Figure24} and \ref{fig:Figure25} show the initial, first intermediate, second intermediate and final hypersurface configurations, but with the hypergraphs \textit{both} adapted \textit{and} colored using the Boyer-Lindquist conformal factor ${\psi}$, respectively. Figure \ref{fig:Figure26} shows the discrete characteristic structure of the solutions after time ${t = 4.5 M}$ (using directed acyclic causal graphs to show discrete characteristic lines). Projections along the $z$-axis of the initial, first intermediate, second intermediate and final hypersurface configurations, with vertices assigned spatial coordinates according to the profile of the Boyer-Lindquist conformal factor ${\psi}$, are shown in Figures \ref{fig:Figure27}, \ref{fig:Figure28}, \ref{fig:Figure29} and \ref{fig:Figure30} (with hypergraphs colored using the scalar field ${\Phi \left( t, R \right)}$) and Figures \ref{fig:Figure31}, \ref{fig:Figure32}, \ref{fig:Figure33} and \ref{fig:Figure34} (with hypergraphs colored using the local curvature in ${\psi}$), respectively. The convergence rates for the Hamiltonian constraint after time ${t = 4.5 M}$, with respect to the ${L_1}$, ${L_2}$ and ${L_{\infty}}$-norms, illustrating approximately fourth-order convergence of the finite-difference scheme, are shown in Table \ref{tab:Table2}. We also confirm that the ADM mass of the spacetime remains approximately constant, that the linear momentum of the resulting extremal Kerr black hole converges to be approximately zero, and that the (ADM) angular momentum is approximately conserved (and therefore the angular momentum parameter $J$ of the limiting extremal Kerr black hole is approximately equal to the initial angular momentum of the rotating disk), as required.

\begin{table}[ht]
\centering
\begin{tabular}{|c|c|c|c|c|c|c|}
\hline
Vertices & ${\epsilon \left( L_1 \right)}$ & ${\epsilon \left( L_2 \right)}$ & ${\epsilon \left( L_{\infty} \right)}$ & ${\mathcal{O} \left( L_1 \right)}$ & ${\mathcal{O} \left( L_2 \right)}$ & ${\mathcal{O} \left( L_{\infty} \right)}$\\
\hline\hline
100 & ${9.43 \times 10^{-2}}$ & ${8.99 \times 10^{-2}}$ & ${8.14 \times 10^{-2}}$ & - & - & -\\
\hline
200 & ${5.22 \times 10^{-3}}$ & ${4.92 \times 10^{-3}}$ & ${3.65 \times 10^{-3}}$ & 4.17 & 4.19 & 4.48\\
\hline
400 & ${3.31 \times 10^{-4}}$ & ${3.75 \times 10^{-4}}$ & ${2.01 \times 10^{-4}}$ & 3.98 & 3.71 & 4.18\\
\hline
800 & ${2.82 \times 10^{-5}}$ & ${4.48 \times 10^{-5}}$ & ${1.98 \times 10^{-5}}$ & 3.55 & 3.06 & 3.35\\
\hline
1600 & ${2.44 \times 10^{-6}}$ & ${5.41 \times 10^{-6}}$ & ${1.87 \times 10^{-6}}$ & 3.53 & 3.05 & 3.40\\
\hline
\end{tabular}
\caption{Convergence rates for the massive scalar field ``bubble collapse'' problem to a maximally-rotating (extremal) Kerr black hole test, with respect to the ${L_1}$, ${L_2}$ and ${L_{\infty}}$-norms for the Hamiltonian constraint $H$ after time ${t = 4.5 M}$, showing approximately fourth-order convergence.}
\label{tab:Table2}
\end{table}

\clearpage

\section{Comparison with the Pure Wolfram Model Case}
\label{sec:Section4}

\begin{figure}[ht]
\centering
\includegraphics[width=0.295\textwidth]{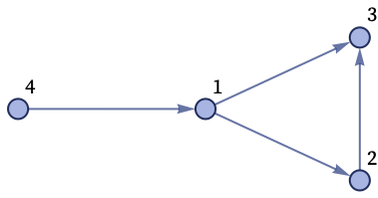}\hspace{0.2\textwidth}
\includegraphics[width=0.295\textwidth]{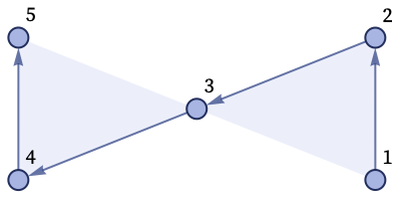}
\caption{Two elementary examples of directed spatial hypergraphs, corresponding to the collections of ordered relations ${\left\lbrace \left\lbrace 1, 2 \right\rbrace, \left\lbrace 1, 3 \right\rbrace, \left\lbrace 2, 3 \right\rbrace, \left\lbrace 4, 1 \right\rbrace \right\rbrace}$ and ${\left\lbrace \left\lbrace 1, 2, 3 \right\rbrace, \left\lbrace 3, 4, 5 \right\rbrace \right\rbrace}$, respectively.}
\label{fig:Figure58}
\end{figure}

\begin{figure}[ht]
\centering
\includegraphics[width=0.495\textwidth]{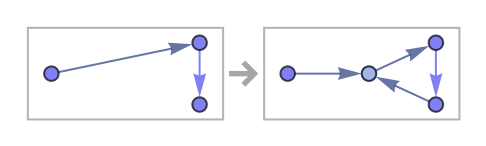}
\caption{An elementary example of a hypergraph rewriting rule, corresponding to the Wolfram model/set substitution system ${\left\lbrace \left\lbrace x, y \right\rbrace, \left\lbrace y, z \right\rbrace \right\rbrace \to \left\lbrace \left\lbrace w, y \right\rbrace, \left\lbrace y, z \right\rbrace, \left\lbrace z, w \right\rbrace, \left\lbrace x, w \right\rbrace \right\rbrace}$.}
\label{fig:Figure59}
\end{figure}

\begin{figure}[ht]
\centering
\includegraphics[width=0.695\textwidth]{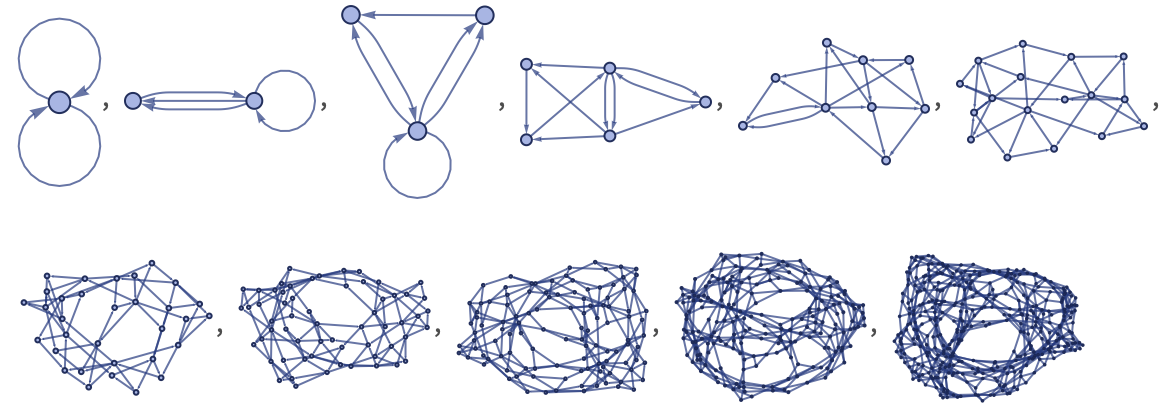}
\caption{An elementary example of a 10-step evolution history for the Wolfram model/hypergraph rewriting system corresponding to the set substitution rule ${\left\lbrace \left\lbrace x, y \right\rbrace, \left\lbrace y, z \right\rbrace \right\rbrace \to \left\lbrace \left\lbrace w, y \right\rbrace, \left\lbrace y, z \right\rbrace, \left\lbrace z, w \right\rbrace, \left\lbrace x, w \right\rbrace \right\rbrace}$, starting from the ``double self-loop'' initial condition ${\left\lbrace \left\lbrace 0, 0 \right\rbrace, \left\lbrace 0, 0 \right\rbrace \right\rbrace}$.}
\label{fig:Figure60}
\end{figure}

Our objective for this section is to demonstrate that the numerical results obtained via the \href{https://github.com/JonathanGorard/Gravitas}{\textsc{Gravitas}} framework within the preceding two sections may be reproduced via a pure hypergraph rewriting system that provably satisfies the Einstein field equations in the continuum limit (with an event selection function determined by the ADM gauge conditions presented within Section \ref{sec:Section1}), without recourse to discretizing an a priori continuous spacetime metric. Recall that the \textit{Wolfram model} hypergraph rewriting formalism\cite{wolfram}\cite{wolfram2}\cite{gorard}\cite{gorard2} may, at least in a large class of cases, be considered analogous to a full discretization of the Cauchy problem in general relativity\cite{gorard4}, with each spacelike hypersurface being given by a (finite, usually directed) hypergraph ${H = \left( V, E \right)}$, defined by:

\begin{equation}
E \subseteq \mathcal{P} \left( V \right) \setminus \left\lbrace \varnothing \right\rbrace,
\end{equation}
for power set function ${\mathcal{P}}$, i.e. each hyperedge ${e \in E}$ connects an arbitrary (non-empty) subset of vertices in $V$. Such \textit{spatial hypergraphs} may therefore be represented as finite collections of (ordered) relations between vertices, as illustrated in Figure \ref{fig:Figure58}. Once the Cauchy initial data have been specified through the construction of an  appropriate initial spatial hypergraph (usually by \textit{Poisson sprinkling}\cite{gorard3}), the Cauchy surface may then be evolved forwards in time by applying an \textit{abstract rewriting rule} $R$ defined over hypergraphs, of the general form:

\begin{equation}
R: \qquad H_1 = \left( V_1, E_1 \right) \to H_2 = \left( V_2, E_2 \right).
\end{equation}
Loosely speaking, such a rule specifies that a subhypergraph which matches the pattern defined by ${H_1 = \left( V_1, E_1 \right)}$ is to be replaced by a distinct subhypergraph matching the pattern defined by ${H_2 = \left( V_2, E_2 \right)}$, as illustrated in the example shown in Figure \ref{fig:Figure59}; the hypergraph rewriting semantics of Wolfram model evolution can be made fully rigorous using the formalism of double-pushout rewriting over selective adhesive categories\cite{gorard6}\cite{gorard7}\cite{gorard8} If we proceed to apply this rule to every possible/matching (non-overlapping) subhypergraph at each step, it defines an effective dynamics for the Wolfram model system, as demonstrated in Figure \ref{fig:Figure60} for the case of the rewriting rule ${\left\lbrace \left\lbrace x, y \right\rbrace, \left\lbrace y, z \right\rbrace \right\rbrace \to \left\lbrace \left\lbrace w, y \right\rbrace, \left\lbrace y, z \right\rbrace, \left\lbrace z, w \right\rbrace, \left\lbrace x, w \right\rbrace \right\rbrace}$, applied to the ``double self-loop'' initial condition ${\left\lbrace \left\lbrace 0, 0 \right\rbrace, \left\lbrace 0, 0 \right\rbrace \right\rbrace}$. Since there will, in general, exist multiple such choices of maximal non-overlapping sets of subhypergraphs to which to apply the rule at each step, there exists a certain degree of freedom over the evolution history of the system; this corresponds to the gauge freedom inherent to the solution of the general relativistic Cauchy problem\cite{gorard}\cite{gorard3}\cite{gorard4}.

\begin{figure}[ht]
\centering
\includegraphics[width=0.495\textwidth]{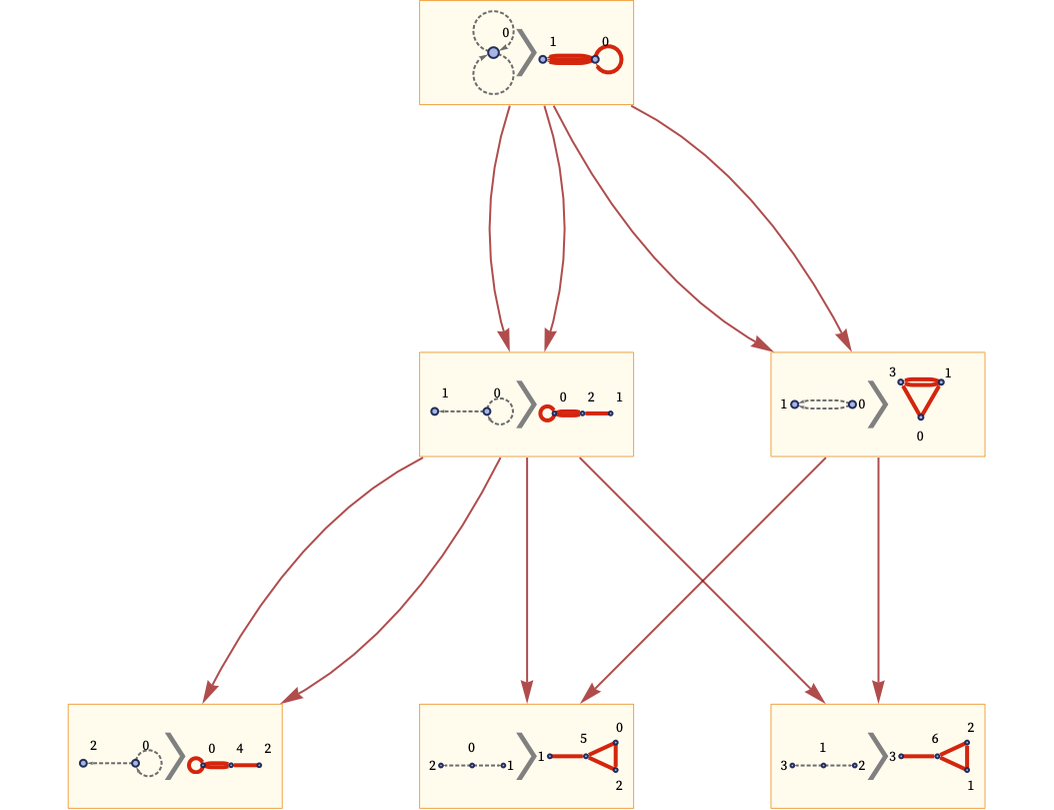}
\caption{An elementary example of a causal graph obtained after a 3-step evolution for the Wolfram model/hypergraph rewriting system corresponding to the set substitution rule ${\left\lbrace \left\lbrace x, y \right\rbrace, \left\lbrace x, z \right\rbrace \right\rbrace \to \left\lbrace \left\lbrace x, y \right\rbrace, \left\lbrace x, w \right\rbrace, \left\lbrace y, w \right\rbrace, \left\lbrace z, w \right\rbrace \right\rbrace}$, starting from the ``double self-loop'' initial condition ${\left\lbrace \left\lbrace 0, 0 \right\rbrace, \left\lbrace 0, 0 \right\rbrace \right\rbrace}$.}
\label{fig:Figure61}
\end{figure}

\begin{figure}[ht]
\centering
\includegraphics[width=0.495\textwidth]{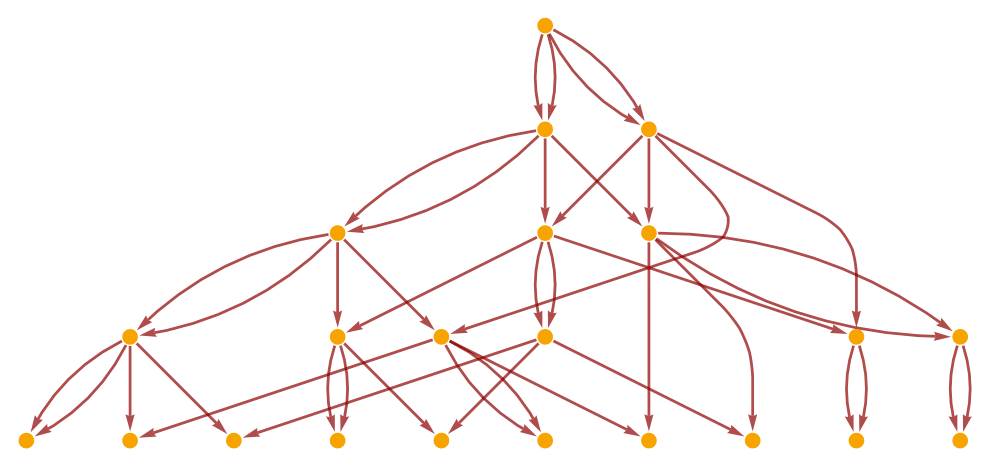}
\includegraphics[width=0.495\textwidth]{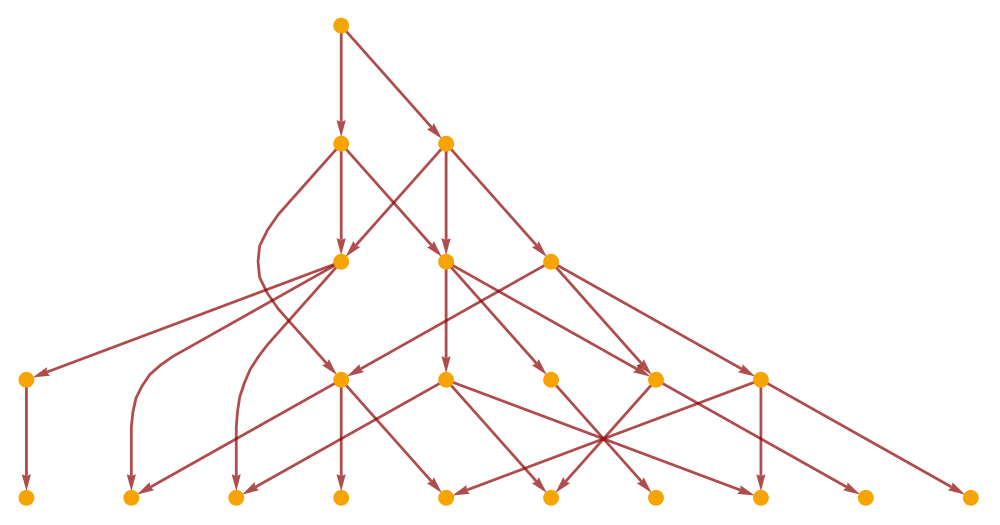}
\caption{The complete causal graph obtained after a 5-step evolution for the Wolfram model/hypergraph rewriting system corresponding to the set substitution rule ${\left\lbrace \left\lbrace x, y \right\rbrace, \left\lbrace x, z \right\rbrace \right\rbrace \to \left\lbrace \left\lbrace x, y \right\rbrace, \left\lbrace x, w \right\rbrace, \left\lbrace y, w \right\rbrace, \left\lbrace z, w \right\rbrace \right\rbrace}$ (left), shown together with its transitive reduction (right), starting from the ``double self-loop'' initial condition ${\left\lbrace \left\lbrace 0, 0 \right\rbrace, \left\lbrace 0, 0 \right\rbrace \right\rbrace}$.}
\label{fig:Figure62}
\end{figure}

If the sequence of spatial hypergraphs represents the foliation of the discrete spacetime into spacelike hypersurfaces, then the causal (i.e. conformally invariant) structure of that discrete spacetime may be represented by means of a directed acyclic graph, known as a \textit{causal graph}. Within such a causal graph, each vertex corresponds to an individual rewrite event, and each edge corresponds to a causal relationship between a pair of rewrite events, such that the directed edge ${A \to B}$ exists if and only if:

\begin{equation}
\mathrm{In} \left( B \right) \cap \mathrm{Out} \left( A \right) \neq \varnothing,
\end{equation}
i.e. if and only if the input for rewrite event $B$ makes use of hyperedges that were produced as part of the output of rewrite event $A$; again, the causal structure of Wolfram model systems may be made fully rigorous using the formalism of weak 2-categories and multiway evolution causal graphs\cite{gorard9}. An example illustrating how such a causal graph may be constructed algorithmically is shown in Figure \ref{fig:Figure61}, and it is shown in Figure \ref{fig:Figure62} how taking the \textit{transitive reduction} of such a causal graph yields the Hasse diagram for a causal set, i.e. it represents the conformal structure of a kind of ``skeletonized'' version of a Lorentzian manifold\cite{gorard3}. The condition of \textit{causal invariance} guarantees that the causal graphs produced by different choices of updating sequence in the hypergraph (and therefore different choices of relativistic gauge) will all ultimately be isomorphic, which in turn guarantees compatibility with general covariance in the continuum limit. The procedure by which Wolfram model/hypergraph rewriting rules may be constructed that provably satisfy the Einstein field equations in the continuum limit is somewhat complex and has been outlined in great detail elsewhere\cite{gorard}\cite{gorard3}\cite{gorard4}, and so in the interest of concision we do not seek to reproduce those details here. The salient features of that construction for our present purposes are as follows: the condition of causal invariance allows one to define a compatible notion of discrete intrinsic curvature (based on the Ollivier-Ricci curvature construction for arbitrary metric-measure spaces\cite{ollivier}\cite{eidi}) on both causal graphs and hypergraphs, and hence to construct a discrete analog of the Einstein-Hilbert action by taking the sum of the discrete Ricci scalar over the entire causal graph; if one further assumes that the hypergraph rewriting dynamics are \textit{weakly ergodic}, such that, in particular, the net flux of causal edges through any given hypersurface in the causal graph eventually converges to zero, then one is analytically justified in exchanging this discrete sum for an integral in the continuum limit; finally, if one assumes that the hypergraph rewriting rule is \textit{asymptotically dimension preserving}, such that the dimension of the limiting hypergraph converges to some fixed, finite value, then it follows that the discrete Einstein-Hilbert action must be extremized in this limit. Thus, any Wolfram model rule which is causal invariant, weakly ergodic and asymptotically dimension preserving will be consistent with the Einstein field equations in the continuum limit. Note that here, as elsewhere, isomorphisms between causal graphs and spatial hypergraphs are determined by means of a generalized form of the ``uniqueness trees'' algorithm\cite{gorard10}.

Although the majority of attention regarding the relativistic properties of the Wolfram model has thus far been focused on the case of vacuum spacetimes, in recent work\cite{gorard5} we showed how it is possible to equip an arbitrary casual graph (in particular, one generated via Wolfram model evolution) with the structure of a free, massless scalar field. In so doing, we broadly followed the approach of Johnston\cite{johnston3} in which one begins by applying the standard Fourier-analytic techniques described by Gel'fand and Shilov\cite{gelfand} and Egorov and Shubin\cite{egorov} in order to derive retarded Green's functions for arbitrary $d$-dimensional (continuous) spacetimes; for instance, in ${d = 1}$, ${d = 2}$, ${d = 3}$ and ${d = 4}$, these are given by:

\begin{equation}
\left( G_R \right)_{m}^{\left( 1 \right)} \left( x \right) = \theta \left( x \right) \frac{\sin \left( m x \right)}{m}, \qquad \left( G_R \right)_{m}^{\left( 2 \right)} \left( \mathbf{x} \right) = \theta \left( x^0 \right) \theta \left( \tau^2 \right) \frac{1}{2} J_0 \left( m \tau \right),
\end{equation}

\begin{equation}
\left( G_R \right)_{m}^{\left( 3 \right)} \left( \mathbf{x} \right) = \theta \left( x^0 \right) \theta \left( \tau^2 \right) \frac{1}{2 \pi} \frac{\cos \left( m \tau \right)}{\tau}, \qquad \left(G_R \right)_{m}^{\left( 4 \right)} \left( \mathbf{x} \right) = \theta \left( x^0 \right) \theta \left( \tau^2 \right) \left( \frac{1}{2 \pi} \delta \left( \tau^2 \right) - \frac{m}{4 \pi} \frac{J_1 \left( m \tau \right)}{\tau} \right),
\end{equation}
where ${\theta \left( \alpha \right)}$ is the standard Heaviside step function:

\begin{equation}
\theta \left( \alpha \right) = \begin{cases}
1, \qquad \text{ if } \alpha \geq 0,\\
0, \qquad \text{ if } \alpha < 0,
\end{cases}
\end{equation}
${\tau}$ denotes proper time distance:

\begin{equation}
\tau = \sqrt{- \left( x^0 \right)^2 + \left( x^1 \right)^2 + \left( x^2 \right)^2 + \cdots + \left( x^{d - 1} \right)^2},
\end{equation}
${J_{\alpha}}$ denote Bessel functions of the first kind (of order ${\alpha}$):

\begin{equation}
J_{\alpha} \left( x \right) = \sum_{m = 0}^{\infty} \frac{\left( -1 \right)^m}{m ! \Gamma \left( m + \alpha + 1 \right)} \left( \frac{x}{2} \right)^{2 m + \alpha},
\end{equation}
assuming validity of the series expansion around ${x = 0}$, and ${\delta}$ is the standard (1-dimensional) Dirac delta function, with the corresponding advanced Green's functions simply given by a slight modification of the boundary conditions:

\begin{equation}
\left( G_A \right)_{m}^{\left( d \right)} \left( \mathbf{x} \right) = \left( G_R \right)_{m}^{\left( d \right)} \left( - \mathbf{x} \right).
\end{equation}
By definition, such Green's functions correspond to solutions to the (massive) Klein-Gordon equation for a (real) scalar field:

\begin{equation}
\left( \Box + m^2 \right) G_{m}^{\left( d \right)} \left( \mathbf{x} \right) = \delta^d \left( \mathbf{x} \right),
\end{equation}
with ${\delta^ d}$ now being the $d$-dimensional generalization of the Dirac delta function. The massless retarded Green's functions ${\left( G_R \right)_{0}^{\left( d \right)} \left( \mathbf{x} \right)}$ were then obtained by evaluating the ${m \to 0}$ limit of the massive retarded Green's functions ${\left( G_R \right)_{m}^{\left( d \right)} \left( \mathbf{x} \right)}$, yielding for instance (using the ${d = 1}$, ${d = 2}$, ${d = 3}$ and ${d = 4}$ examples presented above):

\begin{equation}
\left( G_R \right)_{0}^{\left( d \right)} \left( x \right) = \lim_{m \to 0} \left[ \left( G_R \right)_{m}^{\left( 1 \right)} \left( x \right) \right] = \theta \left( x \right) x, \qquad \left( G_R \right)_{0}^{\left( 2 \right)} \left( \mathbf{x} \right) = \lim_{m \to 0} \left[ \left( G_R \right)_{m}^{\left( 2 \right)} \left( \mathbf{x} \right) \right] = \theta \left( x^0 \right) \theta \left( \tau^2 \right) \frac{1}{2},
\end{equation}

\begin{equation}
\left( G_R \right)_{0}^{\left( 3 \right)} \left( \mathbf{x} \right) = \lim_{m \to 0} \left[ \left( G_R \right)_{m}^{\left( 3 \right)} \left( \mathbf{x} \right) \right] = \theta \left( x^0 \right) \theta \left( \tau^2 \right) \frac{1}{2 \pi \tau},
\end{equation}
and:

\begin{equation}
\left( G_R \right)_{0}^{\left( 4 \right)} \left( \mathbf{x} \right) = \lim_{m \to 0} \left[ \left( G_R \right)_{m}^{\left( 4 \right)} \left( \mathbf{x} \right) \right] = \theta \left( x^0 \right) \theta \left( \tau^2 \right) \frac{1}{2 \pi} \delta \left( \tau^2 \right).
\end{equation}
When used in conjunction with Johnston's highly evocative ``hops and stops'' formalism for causal set Feynman propagators\cite{johnston3}, these massless retarded Green's functions in the continuum allow one to derive massless discrete \textit{propagator matrices} ${\left( K \right)_{0}^{\left( d \right)} \left( x, y \right)}$, such as:

\begin{equation}
\left( K_C \right)_{0}^{\left( 2 \right)} \left( x, y \right) = \frac{1}{2} C_0 \left( x , y \right),
\end{equation}
in the ${1 + 1}$-dimensional case, derived via \textit{summing over chains} (hence the $C$ subscript), or:

\begin{equation}
\left( K_P \right)_{0}^{\left( 4 \right)} \left( x,  y\right) = \frac{1}{2 \pi} \sqrt{\frac{1}{6}} L_0 \left( x, y \right),
\end{equation}
in the ${3 + 1}$-dimensional case, derived via \textit{summing over paths} (hence the $P$ subscript). In the above, ${C_0 \left( x, y \right)}$ designates the \textit{causal matrix}:

\begin{equation}
C_0 \left( x, y \right) = \begin{cases}
1, \qquad \text{ if } y \prec x,\\
0, \qquad \text{ otherwise},
\end{cases}
\end{equation}
i.e. the matrix designating which pairs of events are connected by directed edges in the \textit{transitive closure} of the causal graph (or, equivalently, which pairs of events are connected by directed paths in the full causal graph), whilst ${L_0 \left( x, y \right)}$ designates the \textit{link matrix}:

\begin{equation}
L_0 \left( x, y \right) = \begin{cases}
1, \qquad \text{ if } y \prec^{*} x,\\
0, \qquad \text{ otherwise},
\end{cases}
\end{equation}
i.e. the matrix designating which pairs of events are connected by directed edges in the \textit{transitive reduction} of the causal graph.

If one opts instead not to take the massless (i.e. ${m \to 0}$) limit, then one obtains a collection of slightly more complicated expressions for the massive discrete propagator matrices ${\left( K \right)_{M}^{\left( d \right)} \left( x, y \right)}$, for instance:

\begin{equation}
\left( K_C \right)_{M}^{\left( 2 \right)} \left( x, y \right) = \frac{1}{2} \sum_{k = 0}^{\infty} \left( -1 \right)^k \frac{M^{2 k}}{2^k} \left( C_0 \left( x, y \right) \right)^k,
\end{equation}
in the ${1 + 1}$-dimensional/chain-summing case, or:

\begin{equation}
\left( K_P \right)_{M}^{\left( 4 \right)} \left( x, y \right) = \frac{1}{2 \pi \sqrt{6}} \sum_{k = 0}^{\infty} \left( -1 \right)^k \left( \frac{M^2}{2 \pi \sqrt{6}} \right)^k \left( L_0 \left( x, y \right) \right)^k,
\end{equation}
in the ${3 + 1}$-dimensional/path-summing case. In the above, the matrix exponents ${\left( C_0 \left( x, y \right) \right)^k}$ and ${\left( L_0 \left( x, y \right) \right)^k}$ are defined in terms of a discrete convolution operation ${A * B}$ (which reduces simply to matrix multiplication in the case of causal and link matrices):

\begin{equation}
\left( A * B \right) \left( x, y \right) = \sum_{z} A \left( x, z \right) B \left( z, y \right).
\end{equation}
These general forms were proposed by Dowker et al.\cite{dowker2} as a generic ansatz for obtaining massive causal set Green's functions in arbitrary (integer) numbers of dimensions, based on the formal expansion of massive scalar field Green's functions ${G_{m}^{\left( d \right)} \left( \mathbf{x} \right)}$ in terms of massless ones ${G_{0}^{\left( d \right)} \left( \mathbf{x} \right)}$ in the continuum:

\begin{equation}
G_{m}^{\left( d \right)} \left( \mathbf{x} \right) = G_{0}^{\left( d \right)} \left( \mathbf{x} \right) - m^2 \left( G_{0}^{\left( d \right)} \left( \mathbf{x} \right) * G_{0}^{\left( d \right)} \left( \mathbf{x} \right) \right) + m^4 \left( G_{0}^{\left( d \right)} \left( \mathbf{x} \right) * G_{0}^{\left( d \right)} \left( \mathbf{x} \right) * G_{0}^{\left( d \right)} \left( \mathbf{x} \right) \right) - \cdots,
\end{equation}
with the continuum convolution operation ${A * B}$ here being given by:

\begin{equation}
\left( A * B \right) \left( x, y \right) = \int \sqrt{-g \left( x \right)} A \left( x,z \right) B \left( z, y \right) d^d z.
\end{equation}
Moreover, the mass parameter $M$ here is a ``normalized'' version of the scalar field mass $m$, defined by ${M^2 = \frac{m^2}{\rho_c}}$, where ${\rho_c}$ is the \textit{sprinkling density} of the causal graph/causal set, i.e. the ratio ${\rho_c = \frac{\left\langle \hat{n} \right\rangle}{v}}$ of the expected number of vertices in a causal subgraph ${\left\langle \hat{n} \right\rangle}$ to the spacetime volume of that subgraph $v$ in the appropriate continuum limit. As shown by Johnston\cite{johnston}\cite{johnston2}, in the limit ${\rho_c \to \infty}$ of infinite sprinkling density, these massive discrete propagator matrices converge (modulo multiplicative constants) to the appropriate massive scalar field Green's functions:

\begin{equation}
\lim_{\rho_c \to \infty} \left[ \left\langle \left( K_C \right)_{M}^{\left( 2 \right)} \left( x, y \right) \right\rangle \right] = G_{m}^{\left( 2 \right)} \left( x, y \right), \qquad \lim_{\rho_c \to \infty} \left[ \sqrt{\rho_c} \left\langle \left( K_P \right)_{M}^{\left( 4 \right)} \left( x, y \right) \right\rangle \right] = G_{m}^{\left( 4 \right)} \left( x, y \right),
\end{equation}
as required. Note that, for the sake of simplicity, we have presented the above analysis for the case of flat/Minkowski space, although (following Birrell and Davies\cite{birrell} and Fulling\cite{fulling}) by modifying the form of the Klein-Gordon equation such that the $d$-dimensional Green's function ${G_{m}^{\left( d \right)} \left( \mathbf{x}, \mathbf{y} \right)}$ now satisfies:

\begin{equation}
\left( \Box + m^2 + \zeta R \left( \mathbf{x} \right) \right) G_{m}^{\left( d \right)} \left( \mathbf{x}, \mathbf{y} \right) = \frac{\delta \left( \mathbf{x} - \mathbf{y} \right)}{\sqrt{g}}, \qquad \text{ where } \qquad g = s\sqrt{\mathrm{det} \left( g^{\mu \nu} \right)},
\end{equation}
where ${\zeta \in \mathbb{R}}$ is a real constant, ${\Box}$ is the curved spacetime analog of the d'Alembertian operator on scalar fields ${\phi \left( \mathbf{x} \right)}$:

\begin{equation}
\Box \phi \left( \mathbf{x} \right) = \frac{1}{\sqrt{g}} \partial_{\mu} \left[ g^{\mu \nu} \sqrt{g} \partial_{\nu} \phi \left( \mathbf{x} \right) \right],
\end{equation}
and ${R \left( \mathbf{x} \right)}$ is the (spacetime) Ricci scalar evaluated at point ${\mathbf{x}}$, we straightforwardly obtain the corresponding construction for massive scalar field Green's functions in arbitrary curved spacetimes (or at least those that permit Riemann normal neighborhoods). By replacing the continuum Ricci scalar ${R \left( \mathbf{x} \right)}$ with the discrete (Ollivier-Ricci) scalar curvature for Wolfram model hypergraphs/causal graphs, we obtain a corresponding construction for arbitrary Wolfram model systems. Full details may be found in \cite{gorard5}.

\begin{figure}[ht]
\centering
\includegraphics[width=0.325\textwidth]{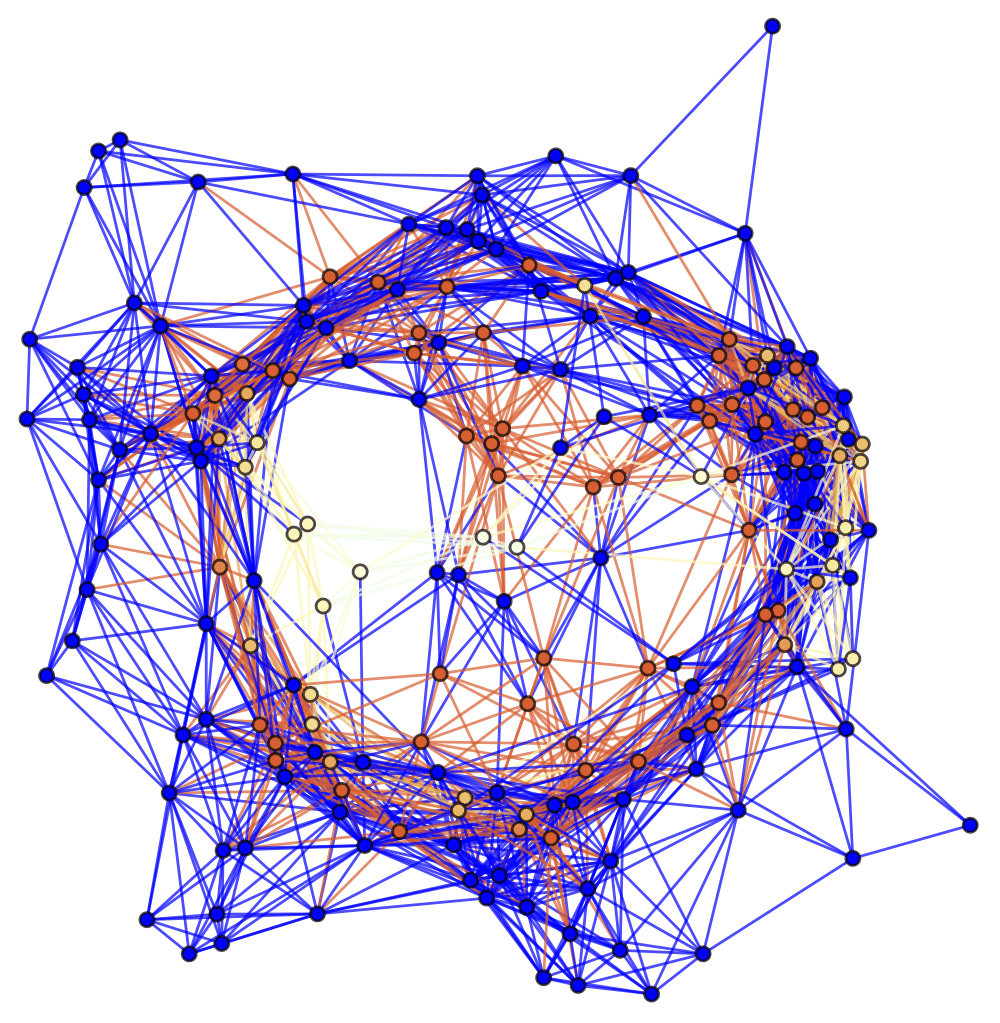}
\includegraphics[width=0.325\textwidth]{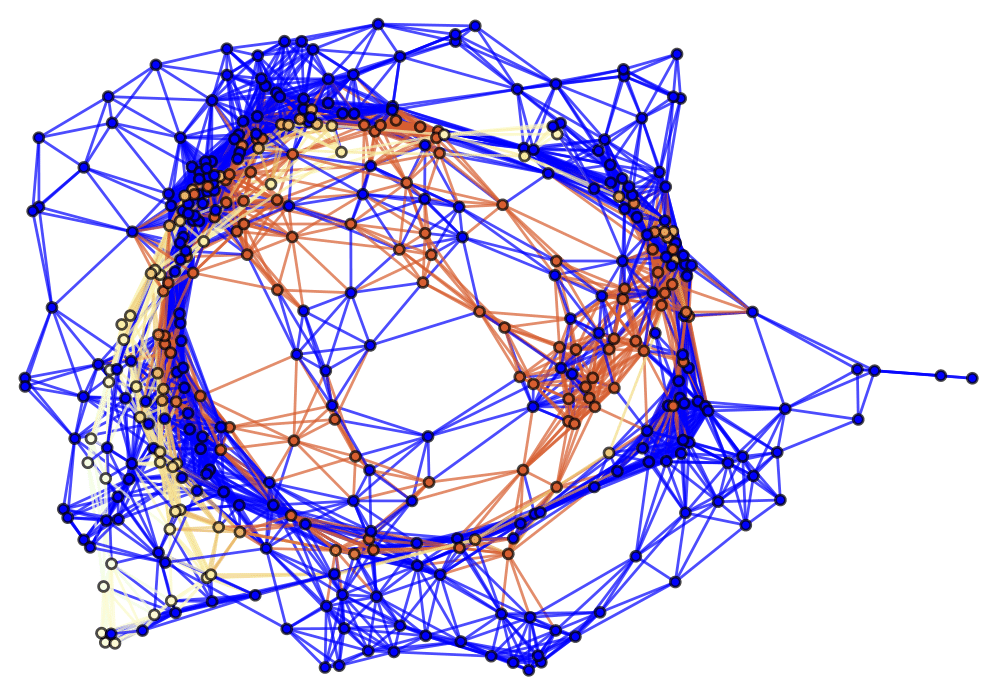}
\includegraphics[width=0.325\textwidth]{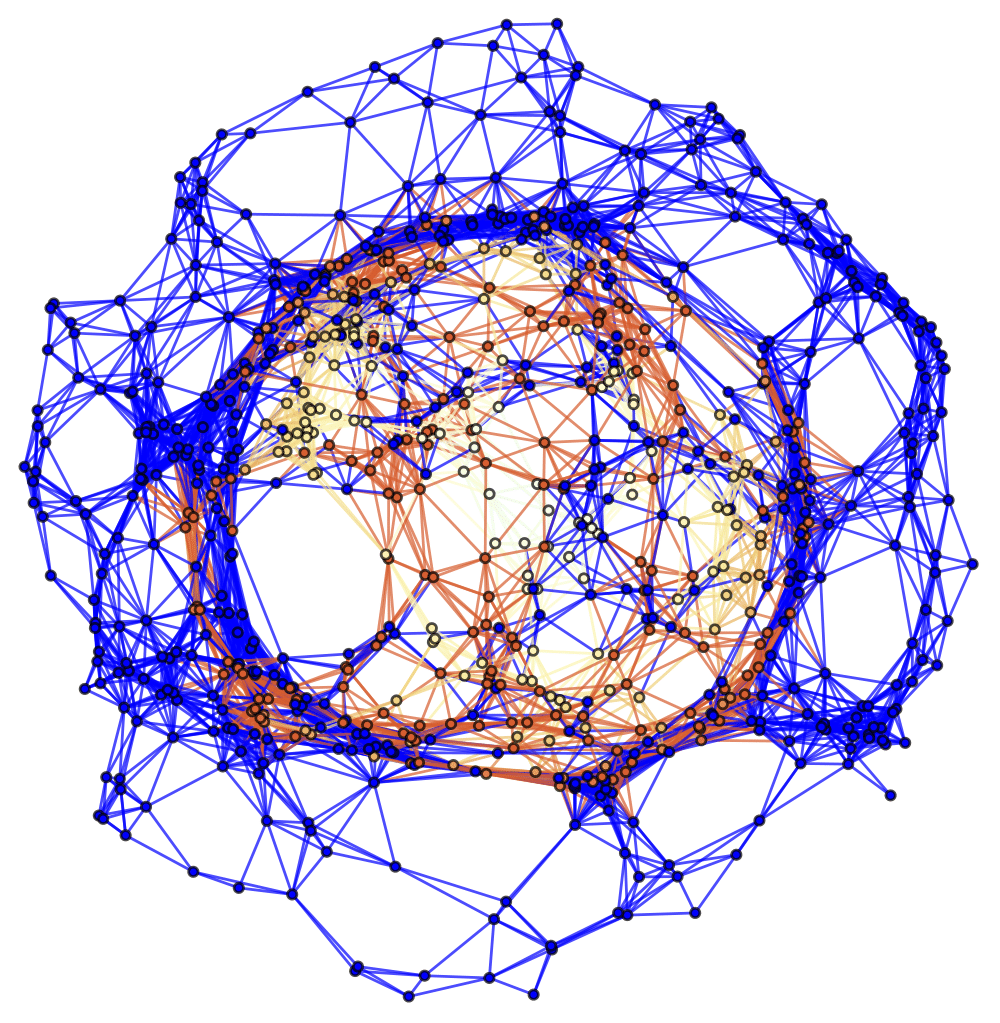}
\caption{Spatial hypergraphs corresponding to the initial hypersurface configuration of the massive scalar field ``bubble collapse'' to a non-rotating Schwarzschild black hole test, with an exponential initial density distribution, at time ${t = 0 M}$, produced via pure Wolfram model evolution (with set substitution rule ${\left\lbrace \left\lbrace x, y \right\rbrace, \left\lbrace y, z \right\rbrace, \left\lbrace z, w \right\rbrace, \left\lbrace w, v \right\rbrace \right\rbrace \to \left\lbrace \left\lbrace y, u \right\rbrace, \left\lbrace u, v \right\rbrace, \left\lbrace w, x \right\rbrace, \left\lbrace x, u \right\rbrace \right\rbrace}$),  with resolutions of 200, 400 and 800 vertices, respectively. The hypergraphs have been colored according to the value of the scalar field ${\Phi \left( t, r \right)}$.}
\label{fig:Figure40}
\end{figure}

\begin{figure}[ht]
\centering
\includegraphics[width=0.325\textwidth]{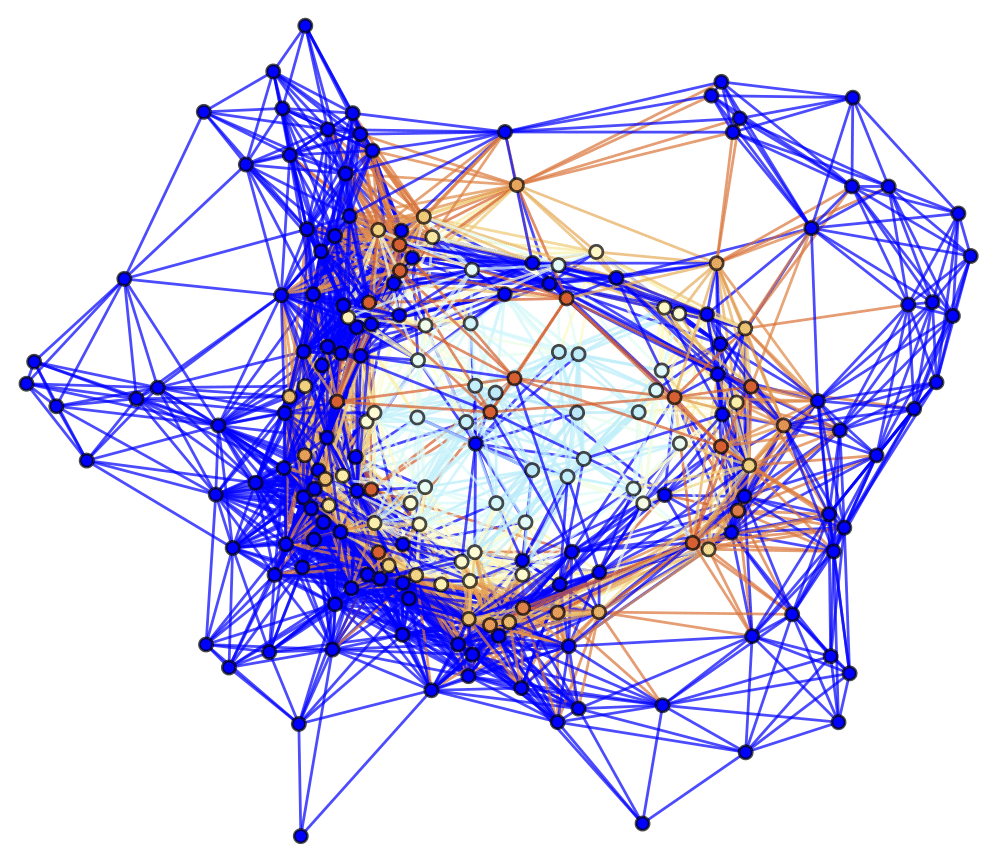}
\includegraphics[width=0.325\textwidth]{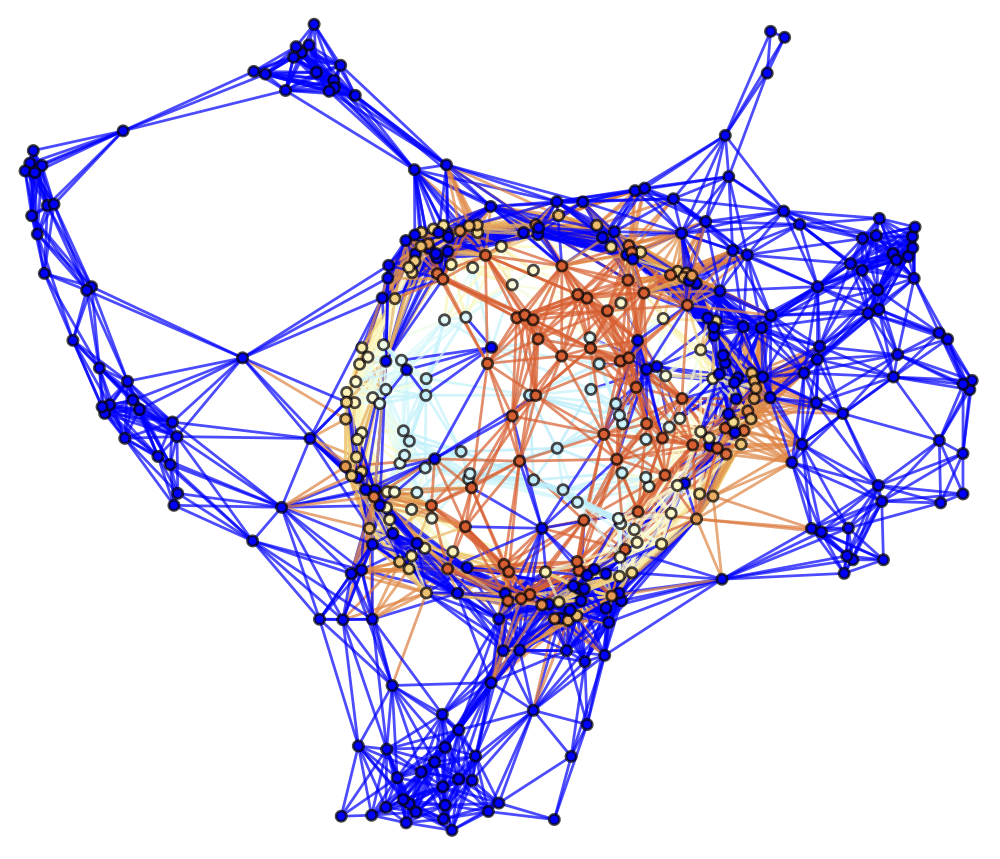}
\includegraphics[width=0.325\textwidth]{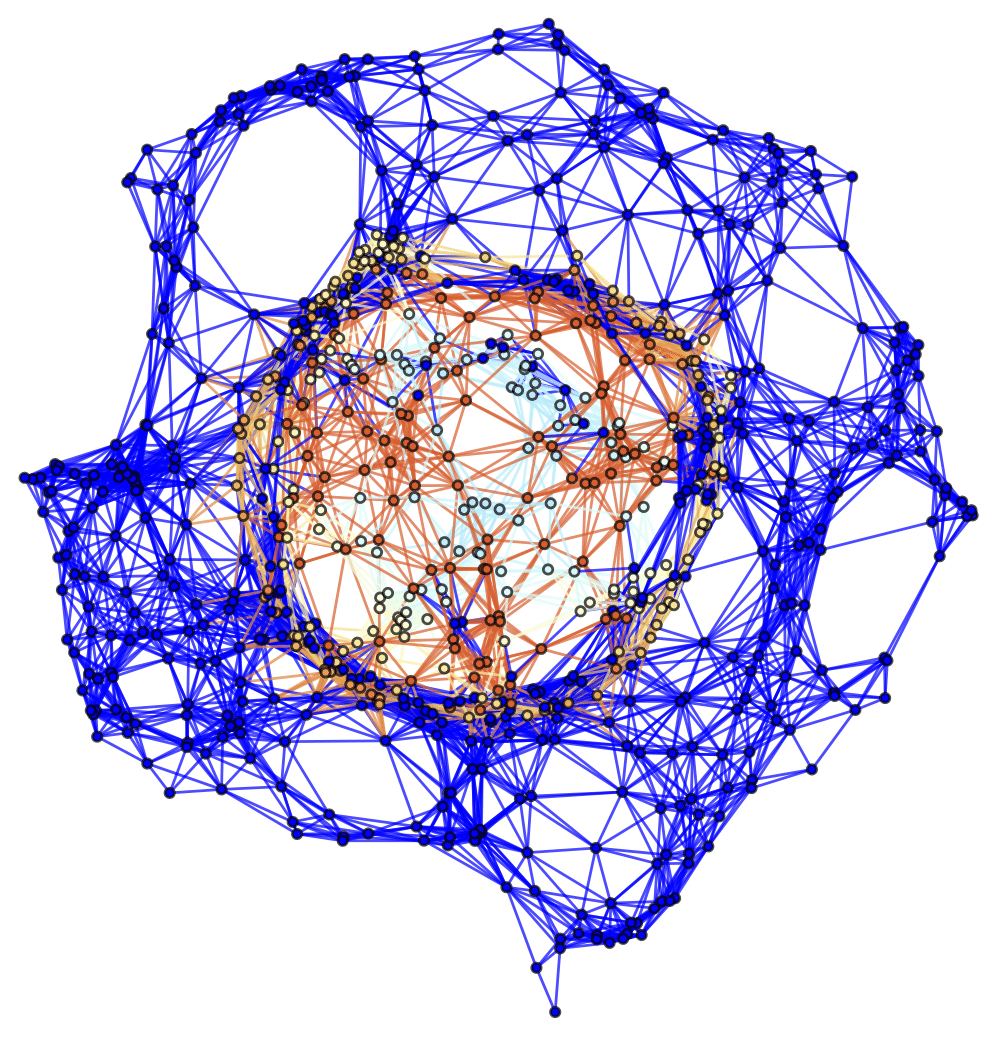}
\caption{Spatial hypergraphs corresponding to the first intermediate hypersurface configuration of the massive scalar field ``bubble collapse'' to a non-rotating Schwarzschild black hole test, with an exponential initial density distribution, at time ${t = 1.5 M}$, produced via pure Wolfram model evolution (with set substitution rule ${\left\lbrace \left\lbrace x, y \right\rbrace, \left\lbrace y, z \right\rbrace, \left\lbrace z, w \right\rbrace, \left\lbrace w, v \right\rbrace \right\rbrace \to \left\lbrace \left\lbrace y, u \right\rbrace, \left\lbrace u, v \right\rbrace, \left\lbrace w, x \right\rbrace, \left\lbrace x, u \right\rbrace \right\rbrace}$), with resolutions of 200, 400 and 800 vertices, respectively. The hypergraphs have been colored according to the value of the scalar field ${\Phi \left( t, r \right)}$.}
\label{fig:Figure41}
\end{figure}

\begin{figure}[ht]
\centering
\includegraphics[width=0.325\textwidth]{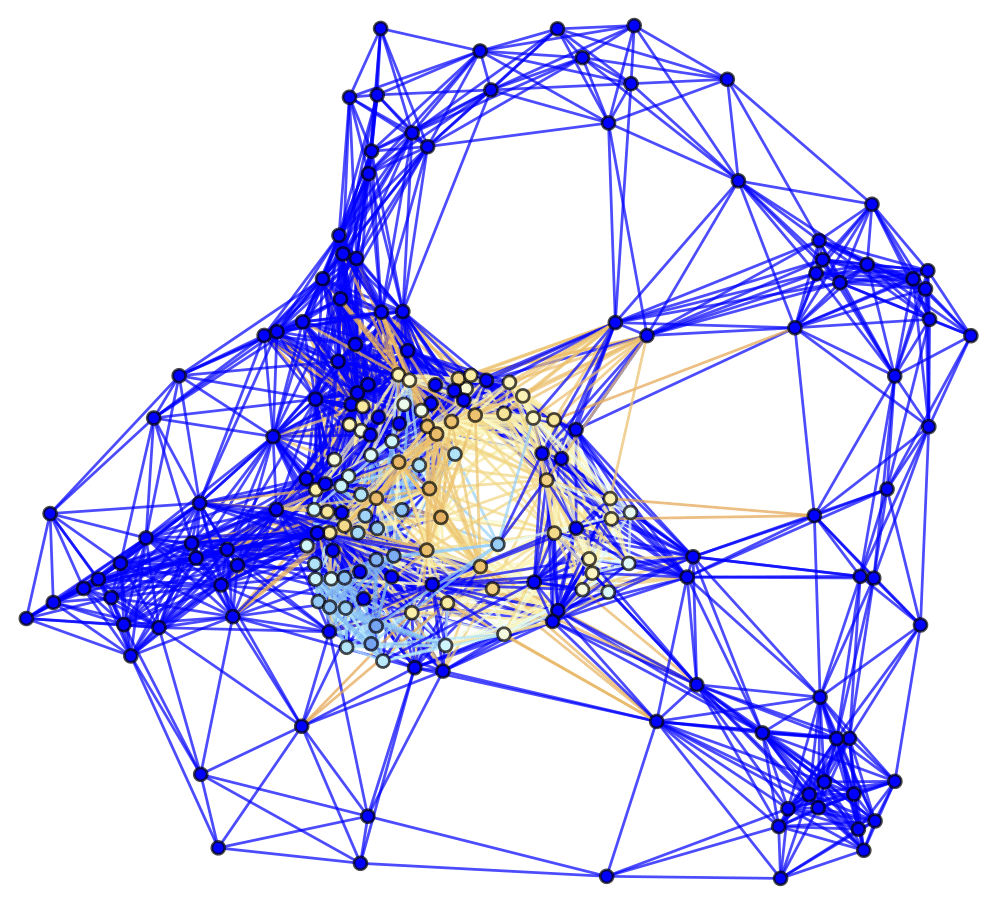}
\includegraphics[width=0.325\textwidth]{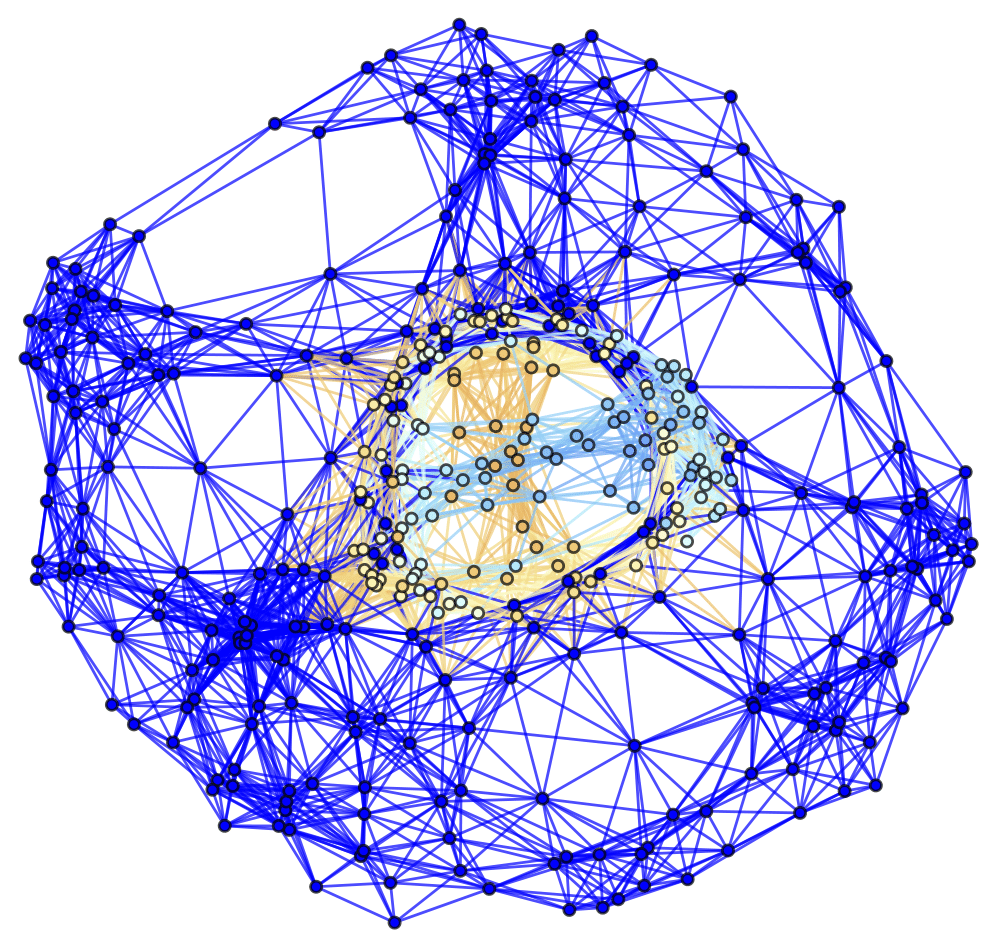}
\includegraphics[width=0.325\textwidth]{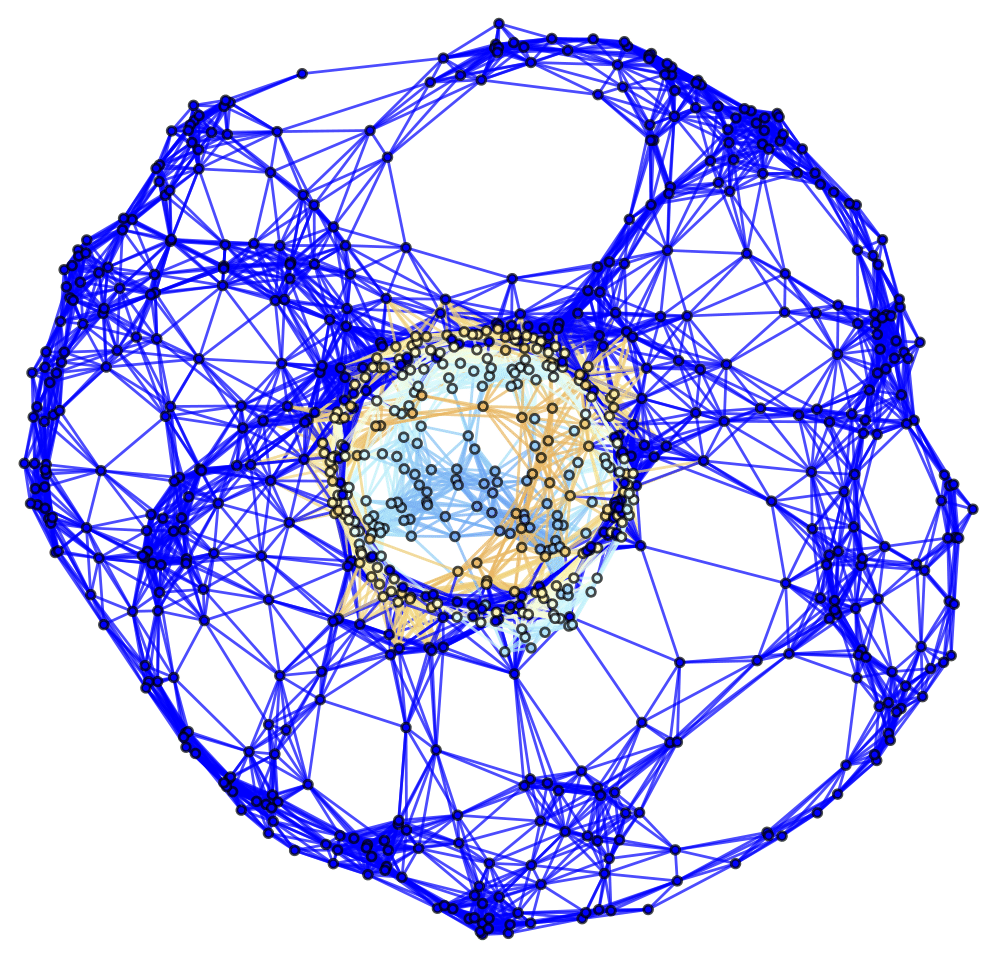}
\caption{Spatial hypergraphs corresponding to the second intermediate hypersurface configuration of the massive scalar field ``bubble collapse'' to a non-rotating Schwarzschild black hole test, with an exponential initial density distribution, at time ${t = 3 M}$, produced via pure Wolfram model evolution (with set substitution rule ${\left\lbrace \left\lbrace x, y \right\rbrace, \left\lbrace y, z \right\rbrace, \left\lbrace z, w \right\rbrace, \left\lbrace w, v \right\rbrace \right\rbrace \to \left\lbrace \left\lbrace y, u \right\rbrace, \left\lbrace u, v \right\rbrace, \left\lbrace w, x \right\rbrace, \left\lbrace x, u \right\rbrace \right\rbrace}$), with resolutions of 200, 400 and 800 vertices, respectively. The hypergraphs have been colored according to the value of the scalar field ${\Phi \left( t, r \right)}$.}
\label{fig:Figure42}
\end{figure}

\begin{figure}[ht]
\centering
\includegraphics[width=0.325\textwidth]{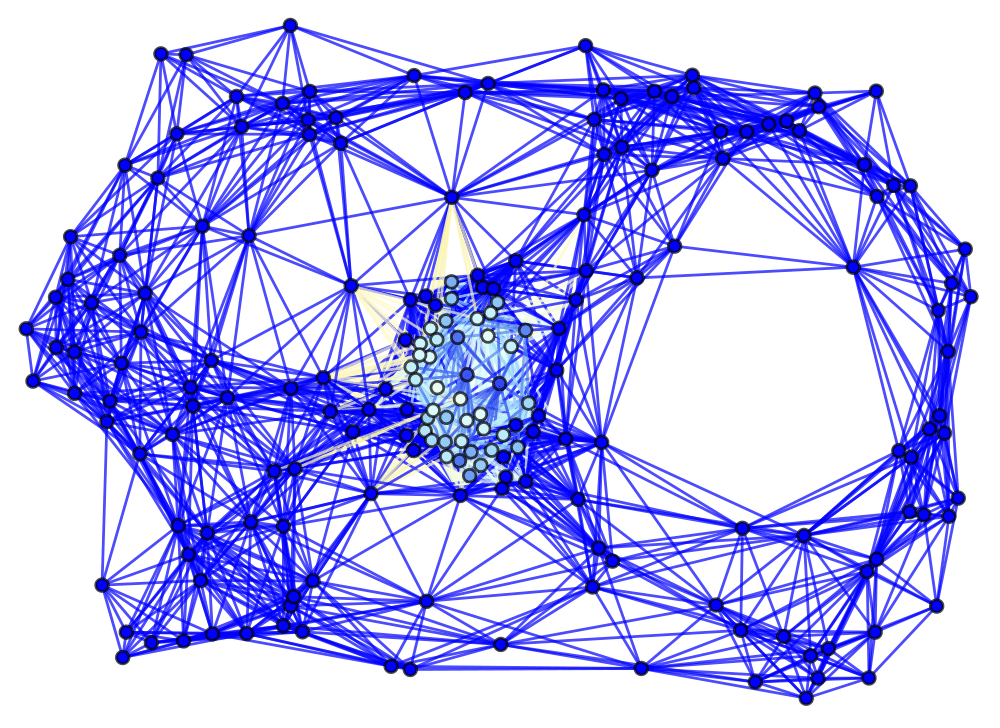}
\includegraphics[width=0.325\textwidth]{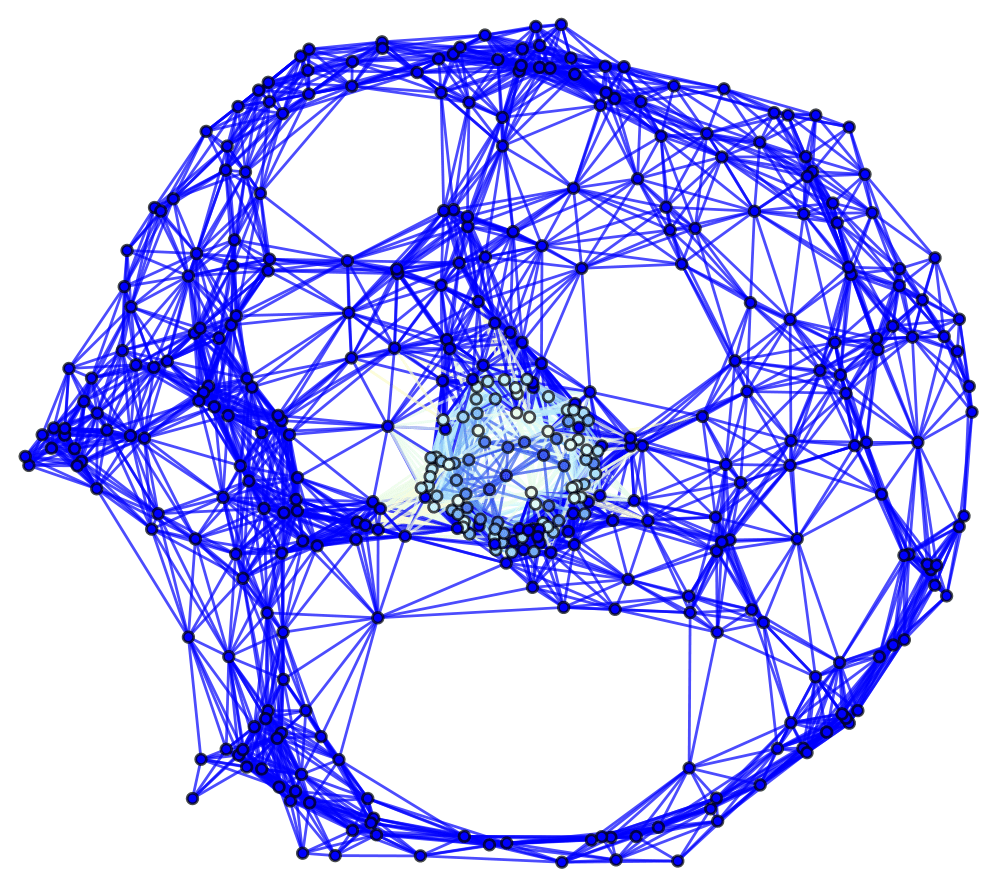}
\includegraphics[width=0.325\textwidth]{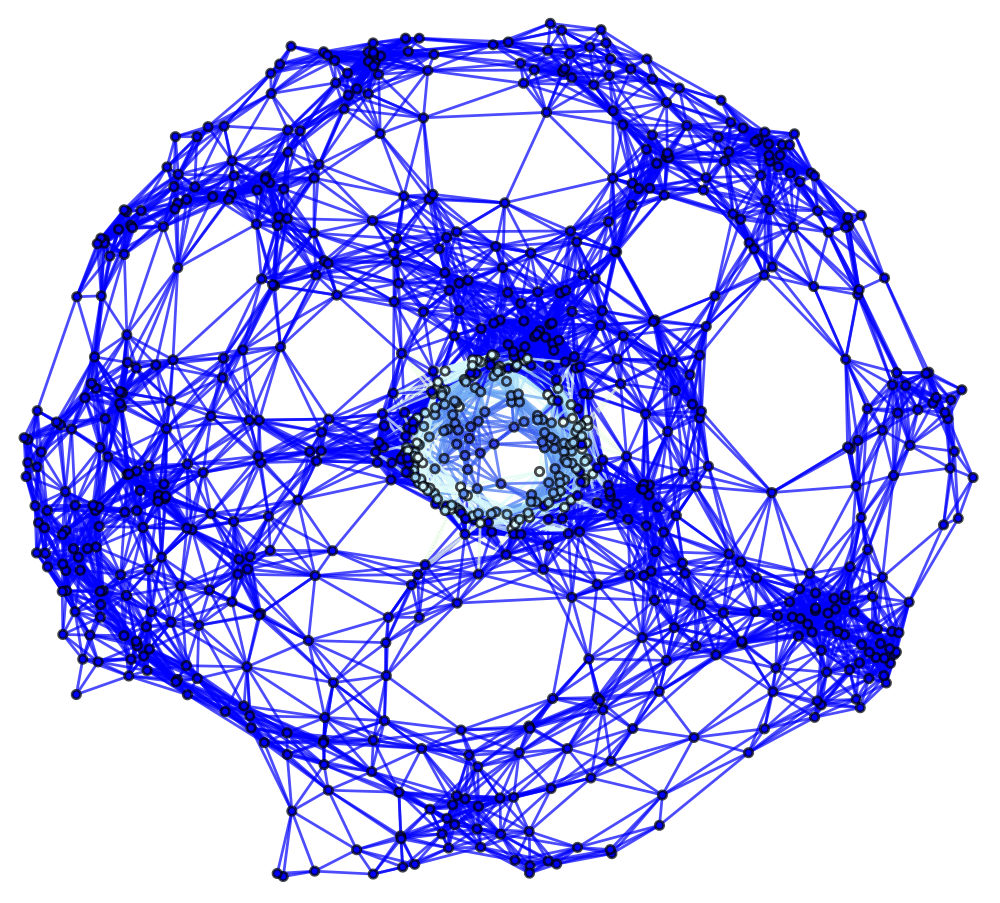}
\caption{Spatial hypergraphs corresponding to the final hypersurface configuration of the massive scalar field ``bubble collapse'' to a non-rotating Schwarzschild black hole test, with an exponential initial density distribution, at time ${t = 4.5 M}$, produced via pure Wolfram model evolution (with set substitution rule ${\left\lbrace \left\lbrace x, y \right\rbrace, \left\lbrace y, z \right\rbrace, \left\lbrace z, w \right\rbrace, \left\lbrace w, v \right\rbrace \right\rbrace \to \left\lbrace \left\lbrace y, u \right\rbrace, \left\lbrace u, v \right\rbrace, \left\lbrace w, x \right\rbrace, \left\lbrace x, u \right\rbrace \right\rbrace}$), with resolutions of 200, 400 and 800 vertices, respectively. The hypergraphs have been colored according to the value of the scalar field ${\Phi \left( t, r \right)}$.}
\label{fig:Figure43}
\end{figure}

\begin{figure}[ht]
\centering
\includegraphics[width=0.325\textwidth]{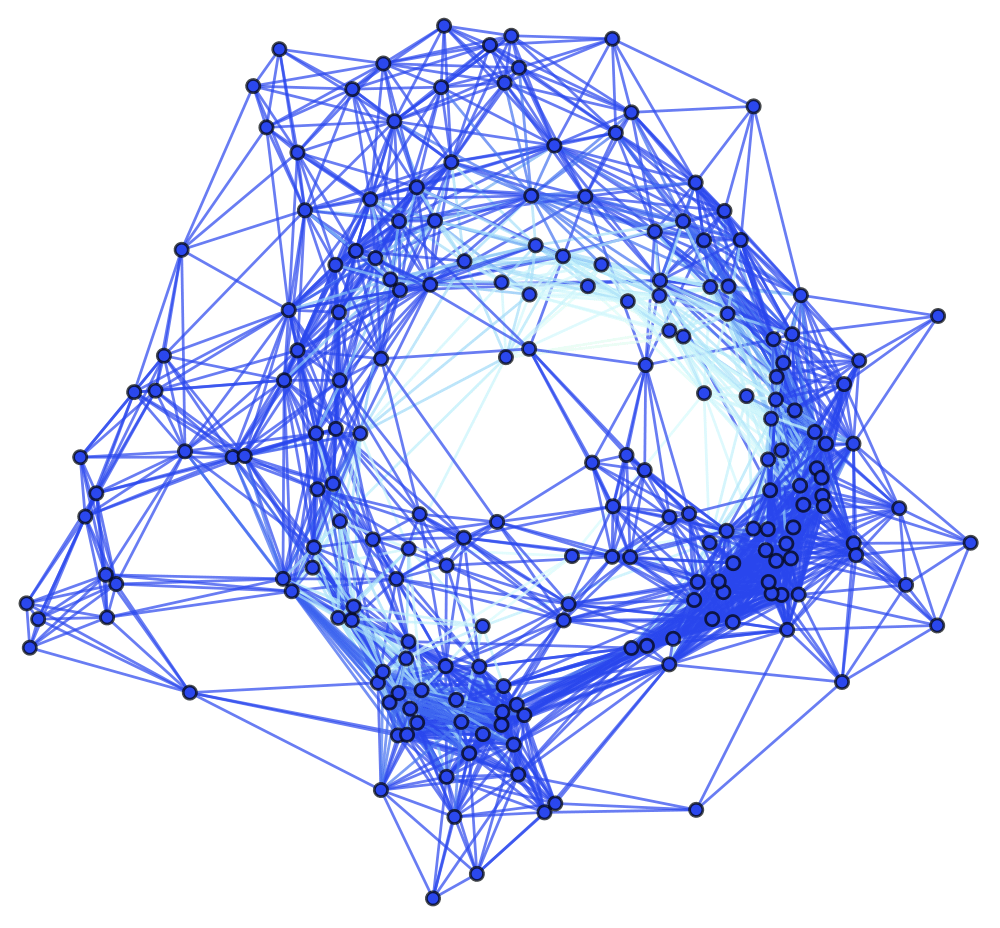}
\includegraphics[width=0.325\textwidth]{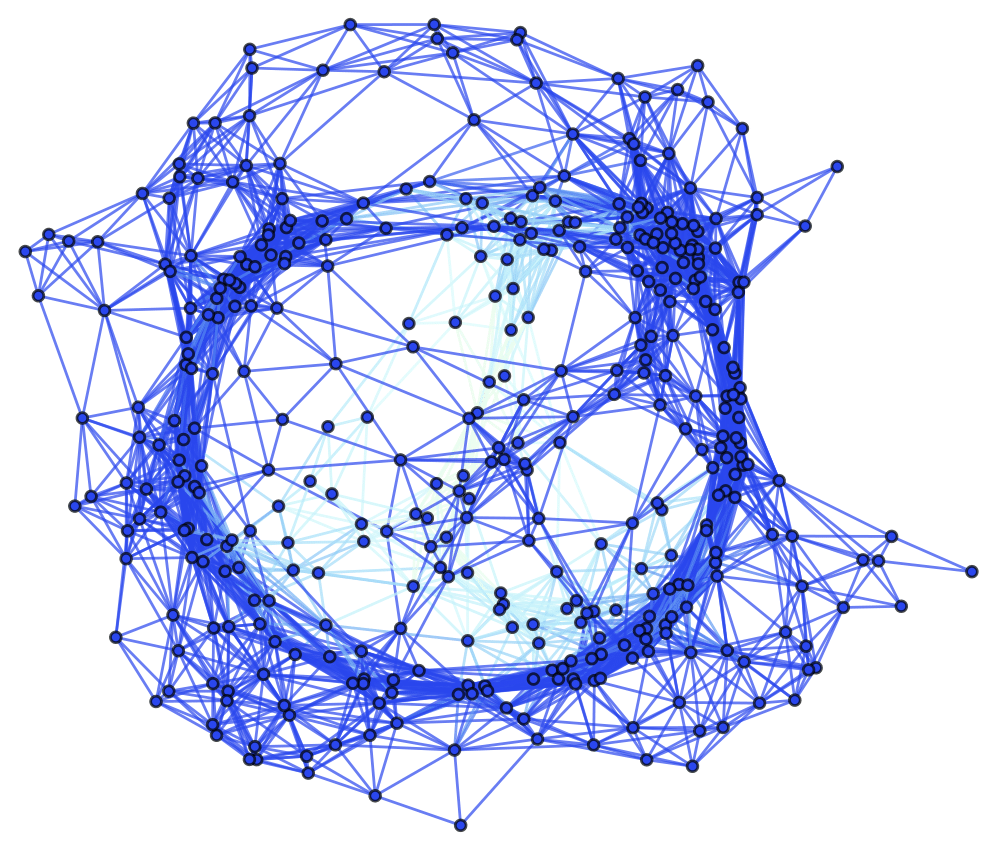}
\includegraphics[width=0.325\textwidth]{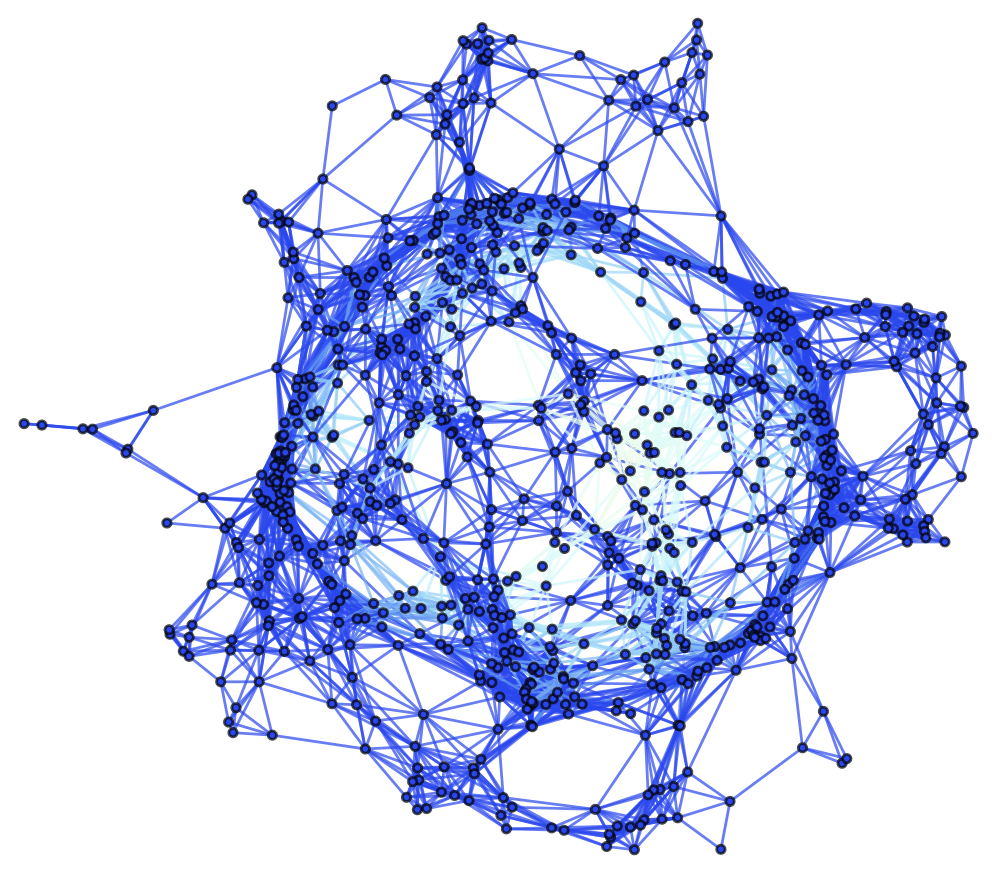}
\caption{Spatial hypergraphs corresponding to the initial hypersurface configuration of the massive scalar field ``bubble collapse'' to a non-rotating Schwarzschild black hole test, with an exponential initial density distribution, at time ${t = 0 M}$, produced via pure Wolfram model evolution (with set substitution rule ${\left\lbrace \left\lbrace x, y \right\rbrace, \left\lbrace y, z \right\rbrace, \left\lbrace z, w \right\rbrace, \left\lbrace w, v \right\rbrace \right\rbrace \to \left\lbrace \left\lbrace y, u \right\rbrace, \left\lbrace u, v \right\rbrace, \left\lbrace w, x \right\rbrace, \left\lbrace x, u \right\rbrace \right\rbrace}$), with resolutions of 200, 400 and 800 vertices, respectively. The hypergraphs have been colored according to the local curvature in the Schwarzschild conformal factor ${\psi}$.}
\label{fig:Figure44}
\end{figure}

\begin{figure}[ht]
\centering
\includegraphics[width=0.325\textwidth]{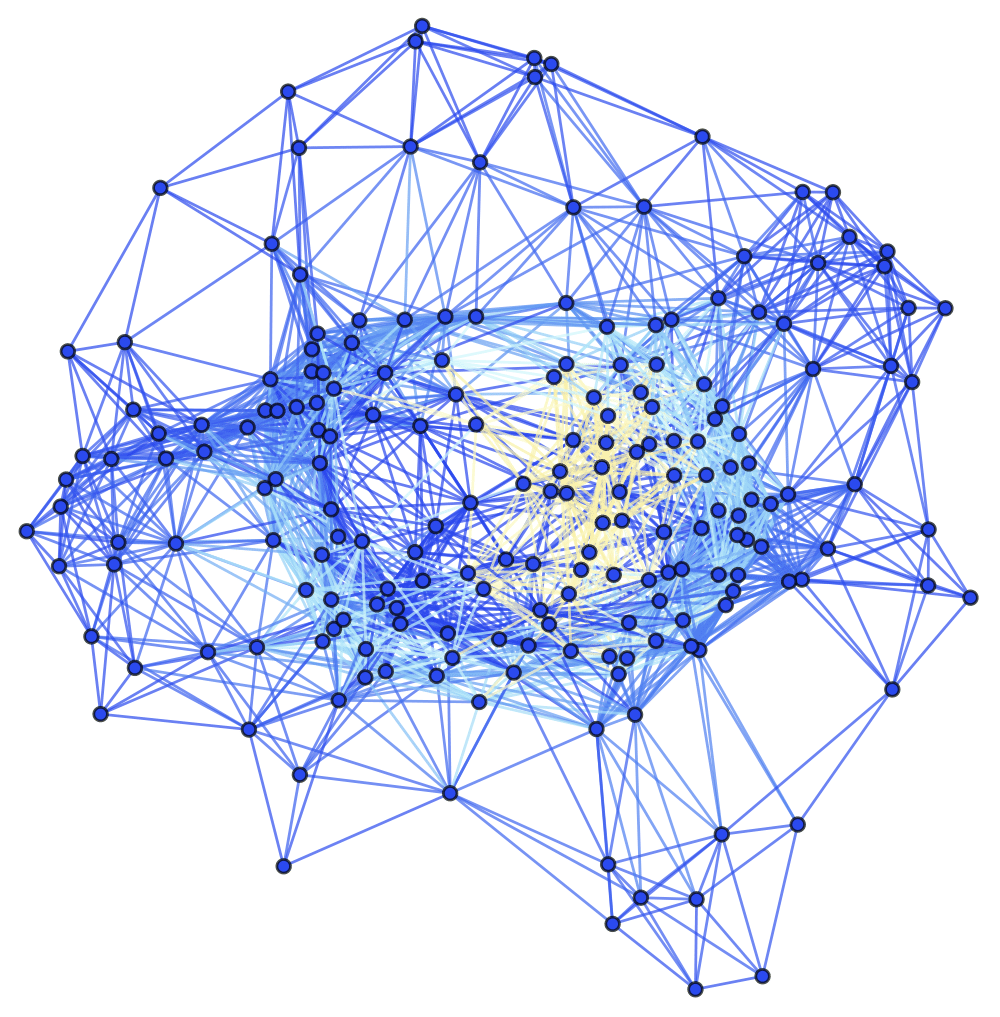}
\includegraphics[width=0.325\textwidth]{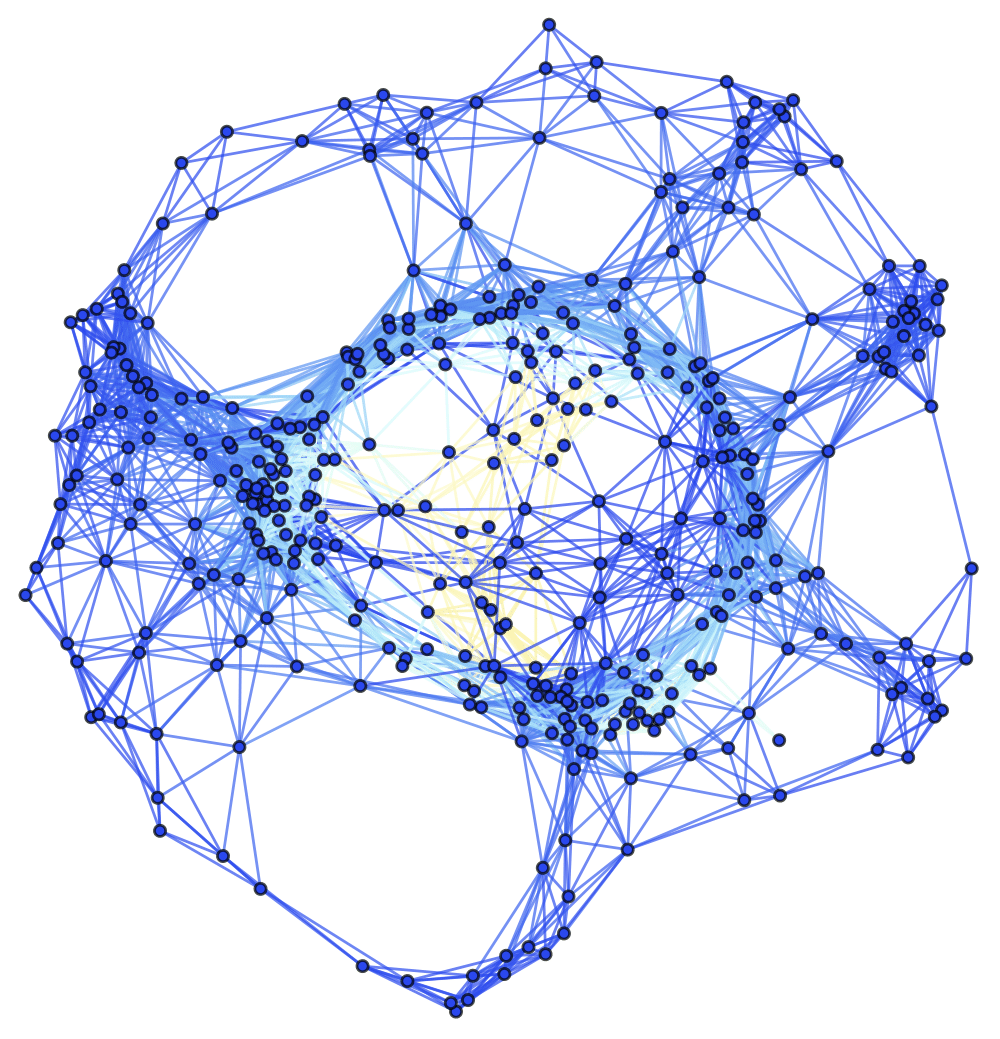}
\includegraphics[width=0.325\textwidth]{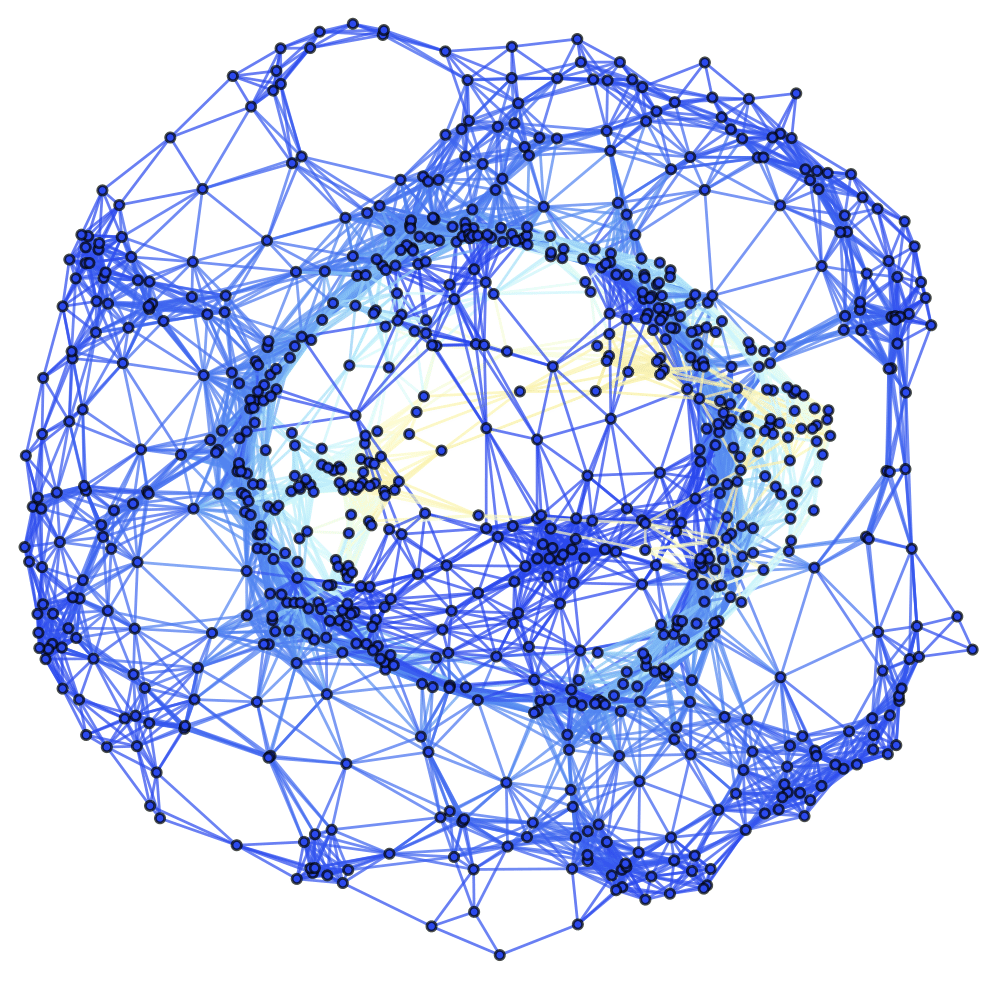}
\caption{Spatial hypergraphs corresponding to the first intermediate hypersurface configuration of the massive scalar field ``bubble collapse'' to a non-rotating Schwarzschild black hole test, with an exponential initial density distribution, at time ${t = 1.5 M}$, produced via pure Wolfram model evolution (with set substitution rule ${\left\lbrace \left\lbrace x, y \right\rbrace, \left\lbrace y, z \right\rbrace, \left\lbrace z, w \right\rbrace, \left\lbrace w, v \right\rbrace \right\rbrace \to \left\lbrace \left\lbrace y, u \right\rbrace, \left\lbrace u, v \right\rbrace, \left\lbrace w, x \right\rbrace, \left\lbrace x, u \right\rbrace \right\rbrace}$), with resolutions of 200, 400 and 800 vertices, respectively. The hypergraphs have been colored according to the local curvature in the Schwarzschild conformal factor ${\psi}$.}
\label{fig:Figure45}
\end{figure}

\begin{figure}[ht]
\centering
\includegraphics[width=0.325\textwidth]{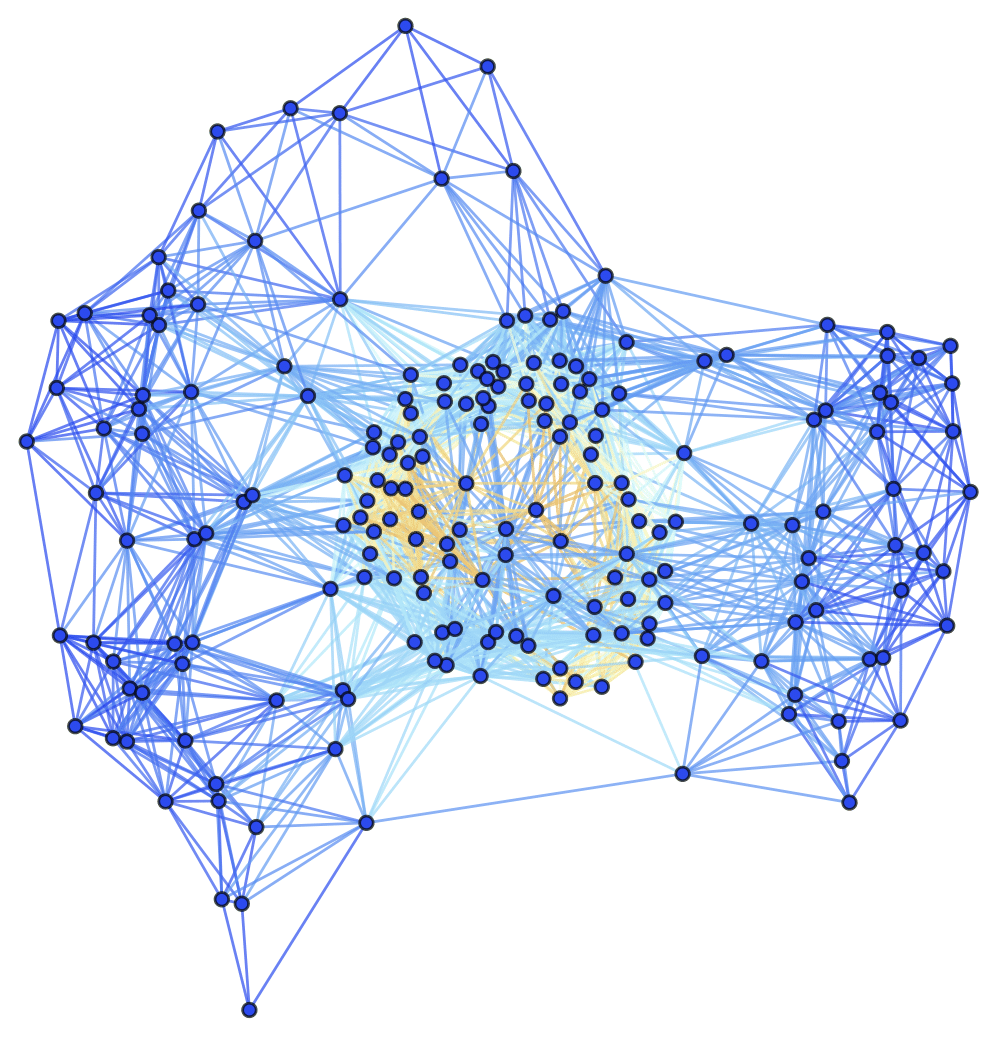}
\includegraphics[width=0.325\textwidth]{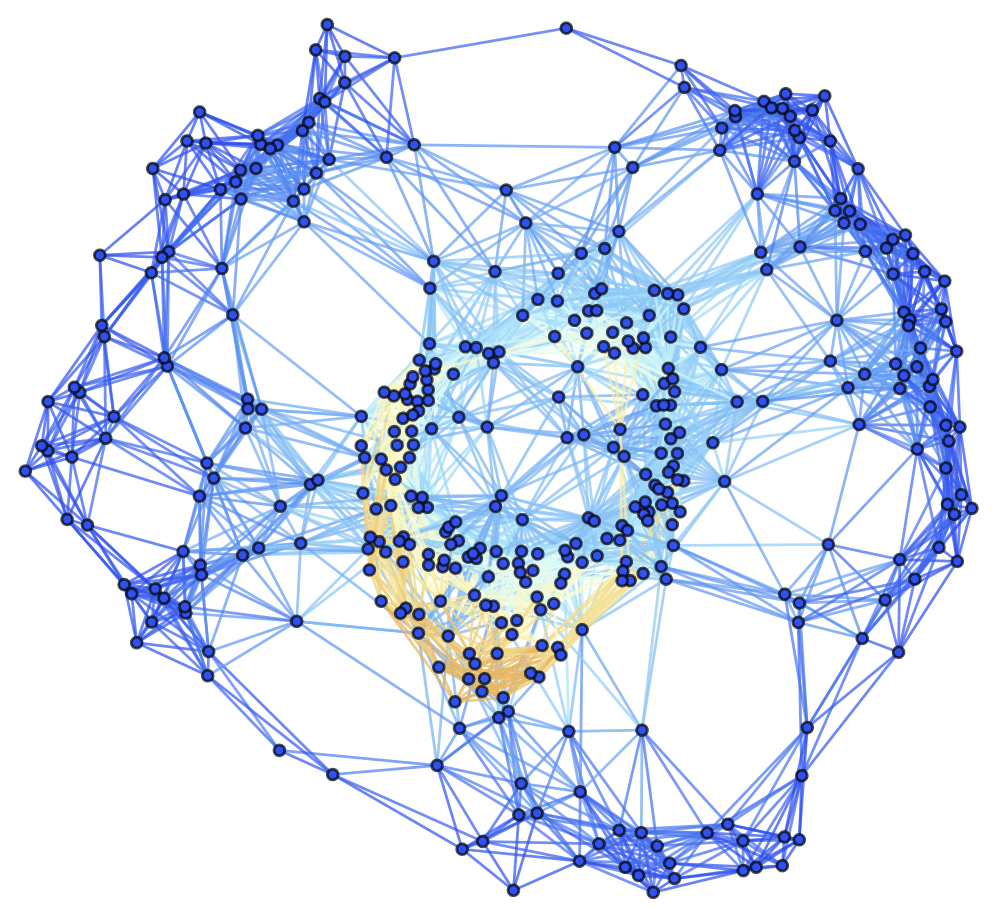}
\includegraphics[width=0.325\textwidth]{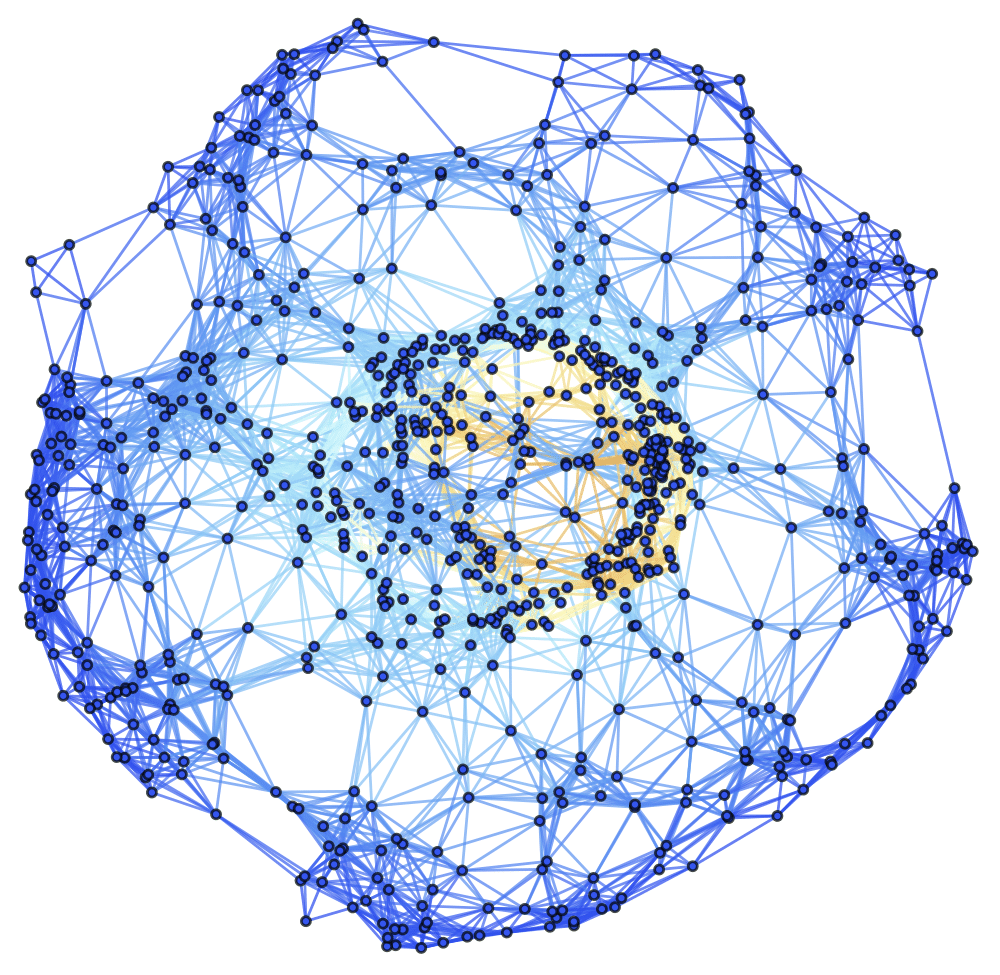}
\caption{Spatial hypergraphs corresponding to the second intermediate hypersurface configuration of the massive scalar field ``bubble collapse'' to a non-rotating Schwarzschild black hole test, with an exponential initial density distribution, at time ${t = 3 M}$, produced via pure Wolfram model evolution (with set substitution rule ${\left\lbrace \left\lbrace x, y \right\rbrace, \left\lbrace y, z \right\rbrace, \left\lbrace z, w \right\rbrace, \left\lbrace w, v \right\rbrace \right\rbrace \to \left\lbrace \left\lbrace y, u \right\rbrace, \left\lbrace u, v \right\rbrace, \left\lbrace w, x \right\rbrace, \left\lbrace x, u \right\rbrace \right\rbrace}$), with resolutions of 200, 400 and 800 vertices, respectively. The hypergraphs have been colored according to the local curvature in the Schwarzschild conformal factor ${\psi}$.}
\label{fig:Figure46}
\end{figure}

\begin{figure}[ht]
\centering
\includegraphics[width=0.325\textwidth]{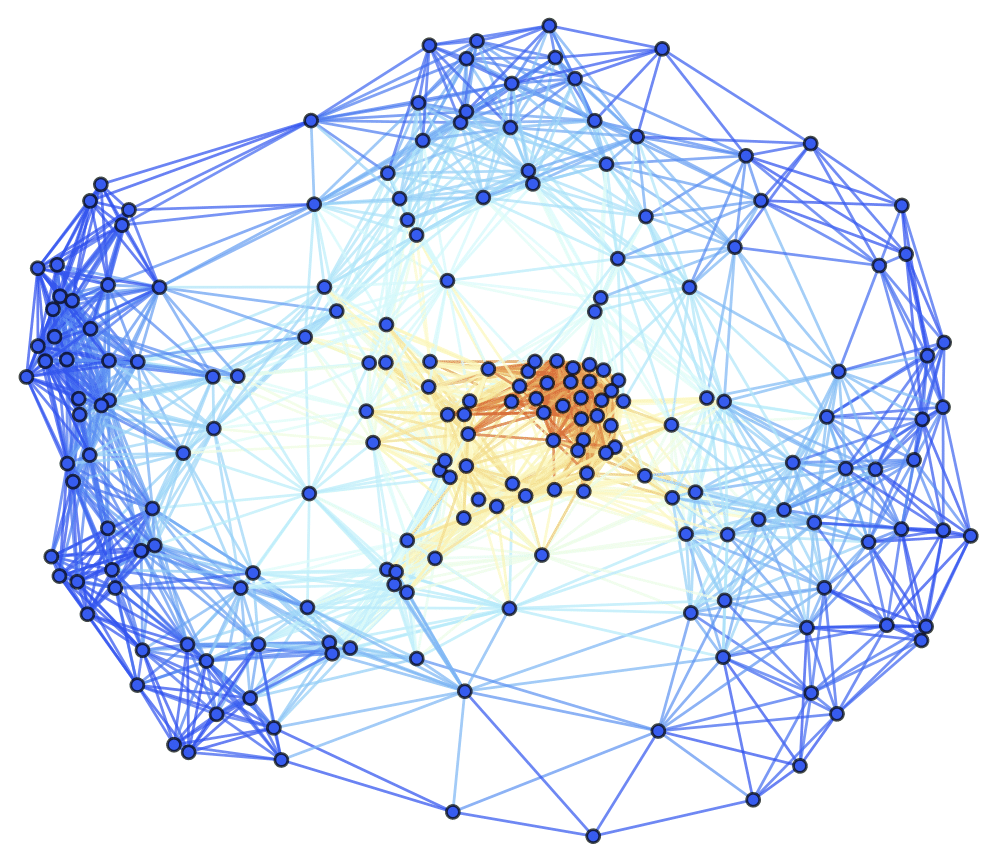}
\includegraphics[width=0.325\textwidth]{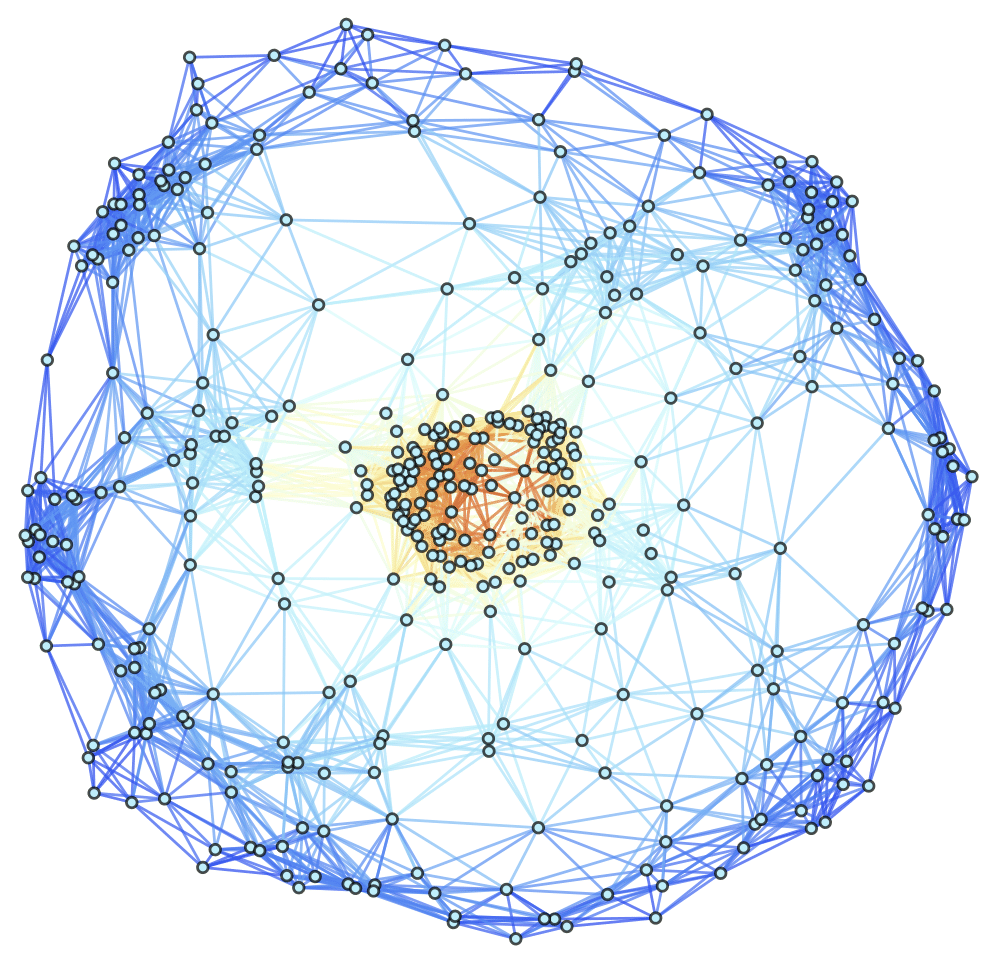}
\includegraphics[width=0.325\textwidth]{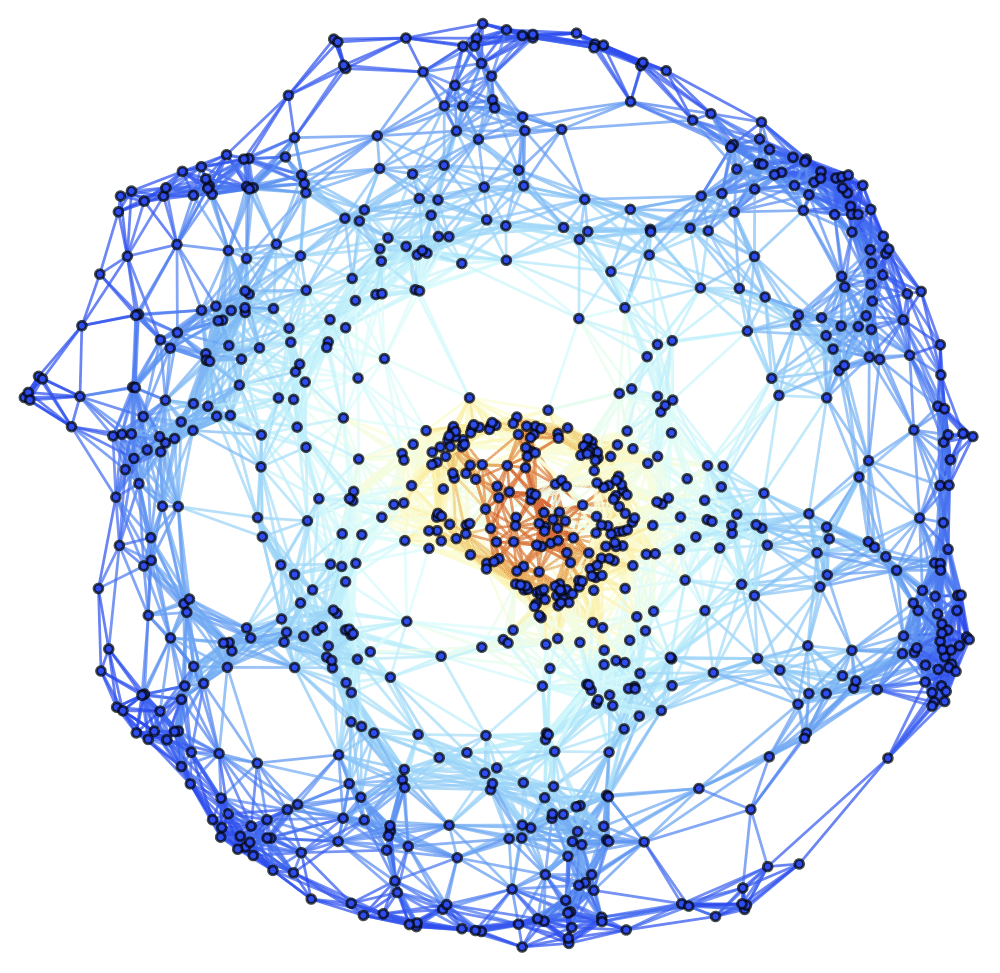}
\caption{Spatial hypergraphs corresponding to the final hypersurface configuration of the massive scalar field ``bubble collapse'' to a non-rotating Schwarzschild black hole test, with an exponential initial density distribution, at time ${t = 4.5 M}$, produced via pure Wolfram model evolution (with set substitution rule ${\left\lbrace \left\lbrace x, y \right\rbrace, \left\lbrace y, z \right\rbrace, \left\lbrace z, w \right\rbrace, \left\lbrace w, v \right\rbrace \right\rbrace \to \left\lbrace \left\lbrace y, u \right\rbrace, \left\lbrace u, v \right\rbrace, \left\lbrace w, x \right\rbrace, \left\lbrace x, u \right\rbrace \right\rbrace}$), with resolutions of 200, 400 and 800 vertices, respectively. The hypergraphs have been colored according to the local curvature in the Schwarzschild conformal factor ${\psi}$.}
\label{fig:Figure47}
\end{figure}

\begin{figure}[ht]
\centering
\includegraphics[width=0.325\textwidth]{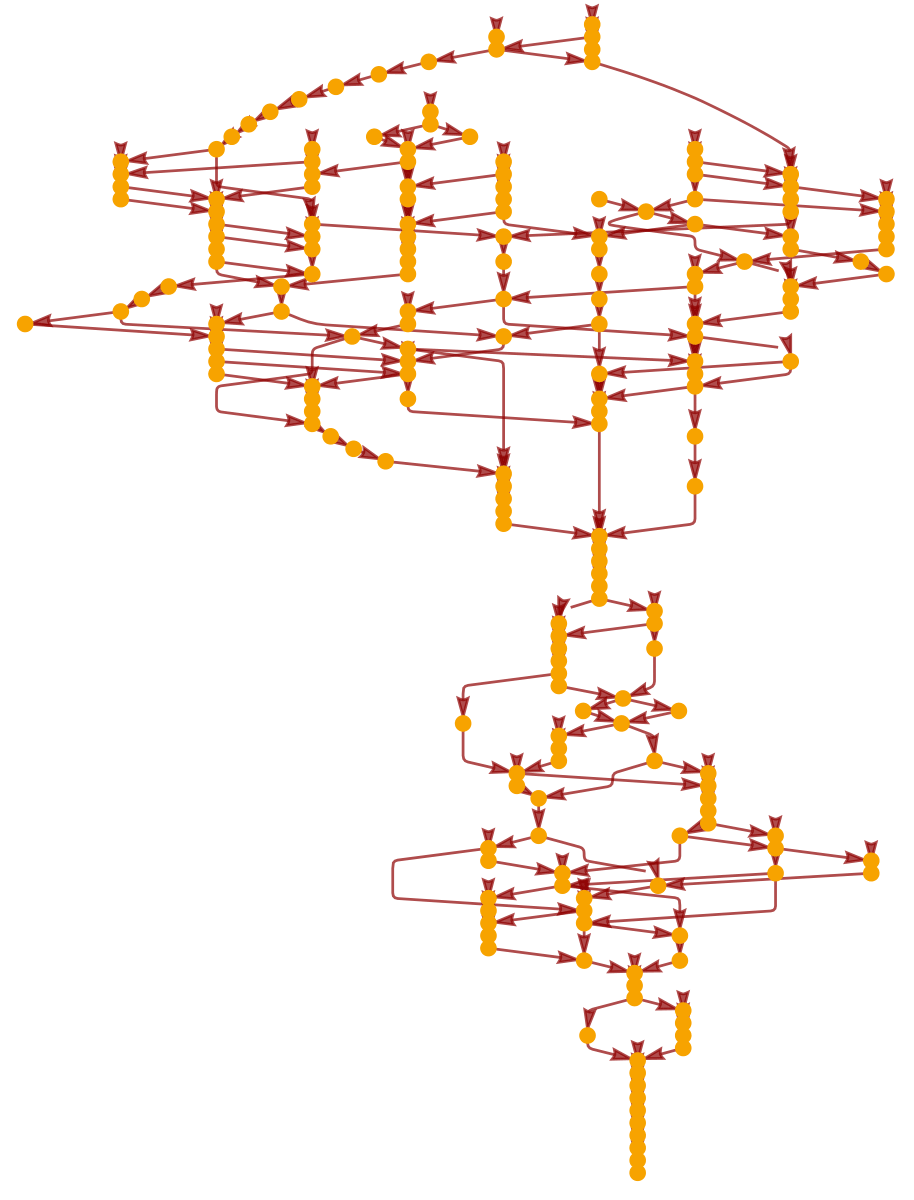}
\includegraphics[width=0.325\textwidth]{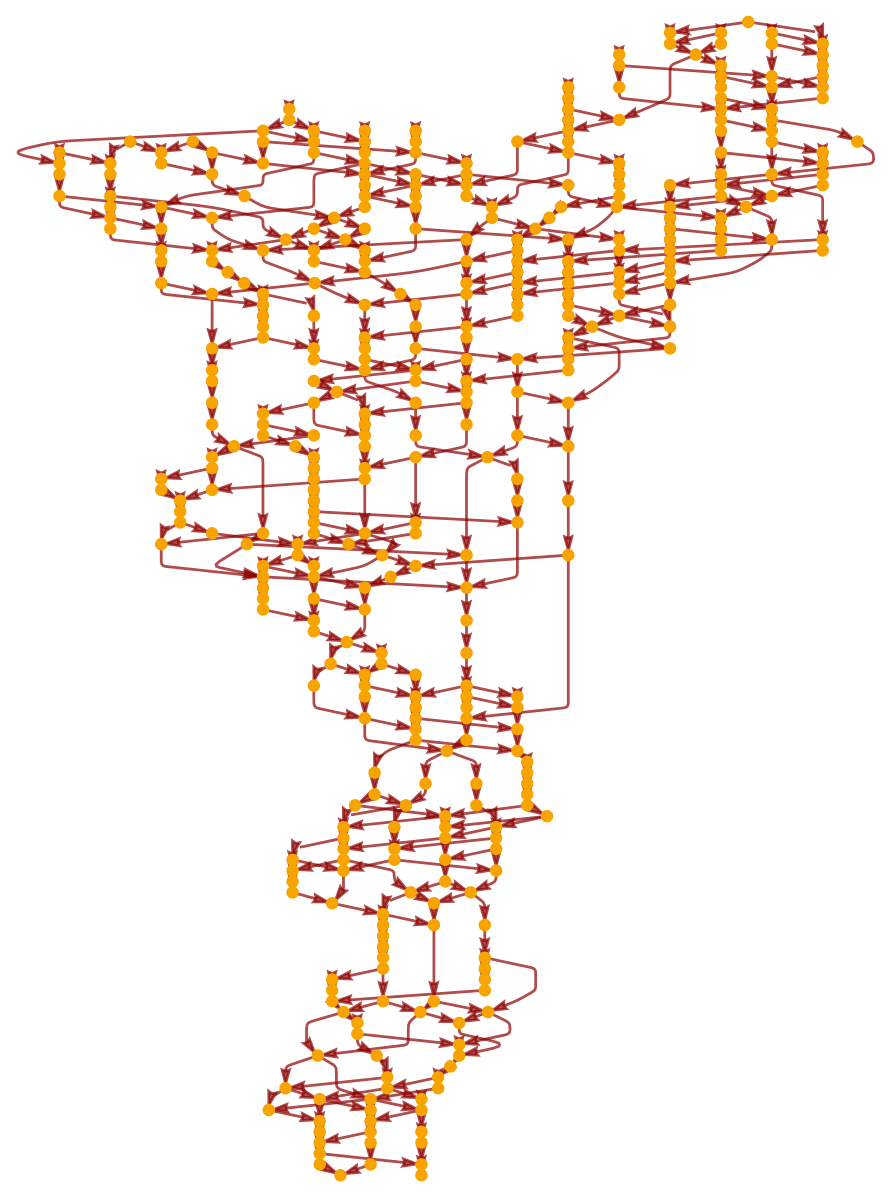}
\includegraphics[width=0.325\textwidth]{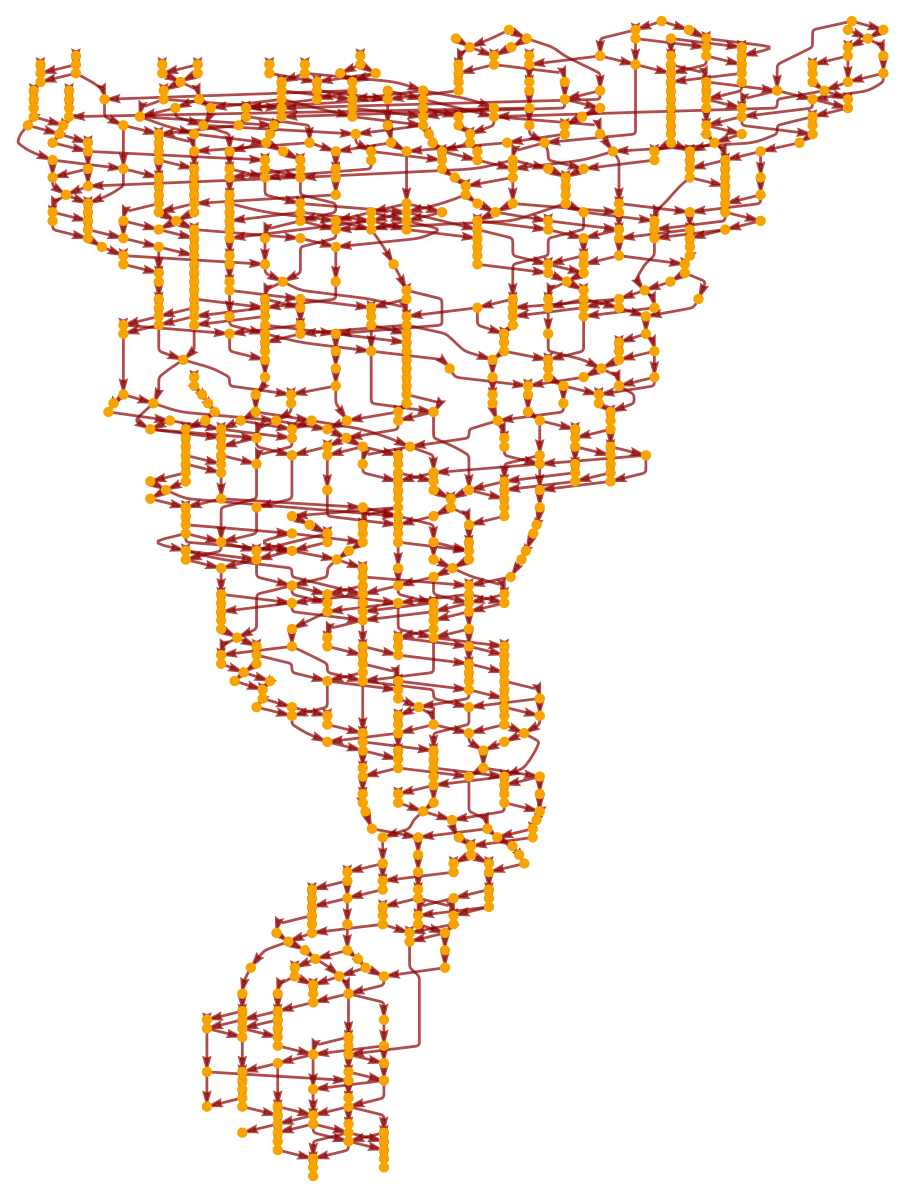}
\caption{Causal graphs corresponding to the discrete characteristic structure of the massive scalar field ``bubble collapse'' to a non-rotating Schwarzschild black hole test, with an exponential initial density distribution, at time ${t = 4.5 M}$, produced via pure Wolfram model evolution (with set substitution rule ${\left\lbrace \left\lbrace x, y \right\rbrace, \left\lbrace y, z \right\rbrace, \left\lbrace z, w \right\rbrace, \left\lbrace w, v \right\rbrace \right\rbrace \to \left\lbrace \left\lbrace y, u \right\rbrace, \left\lbrace u, v \right\rbrace, \left\lbrace w, x \right\rbrace, \left\lbrace x, u \right\rbrace \right\rbrace}$), with resolutions of 200, 400 and 800 hypergraph vertices, respectively.}
\label{fig:Figure48}
\end{figure}

\begin{figure}[ht]
\centering
\includegraphics[width=0.325\textwidth]{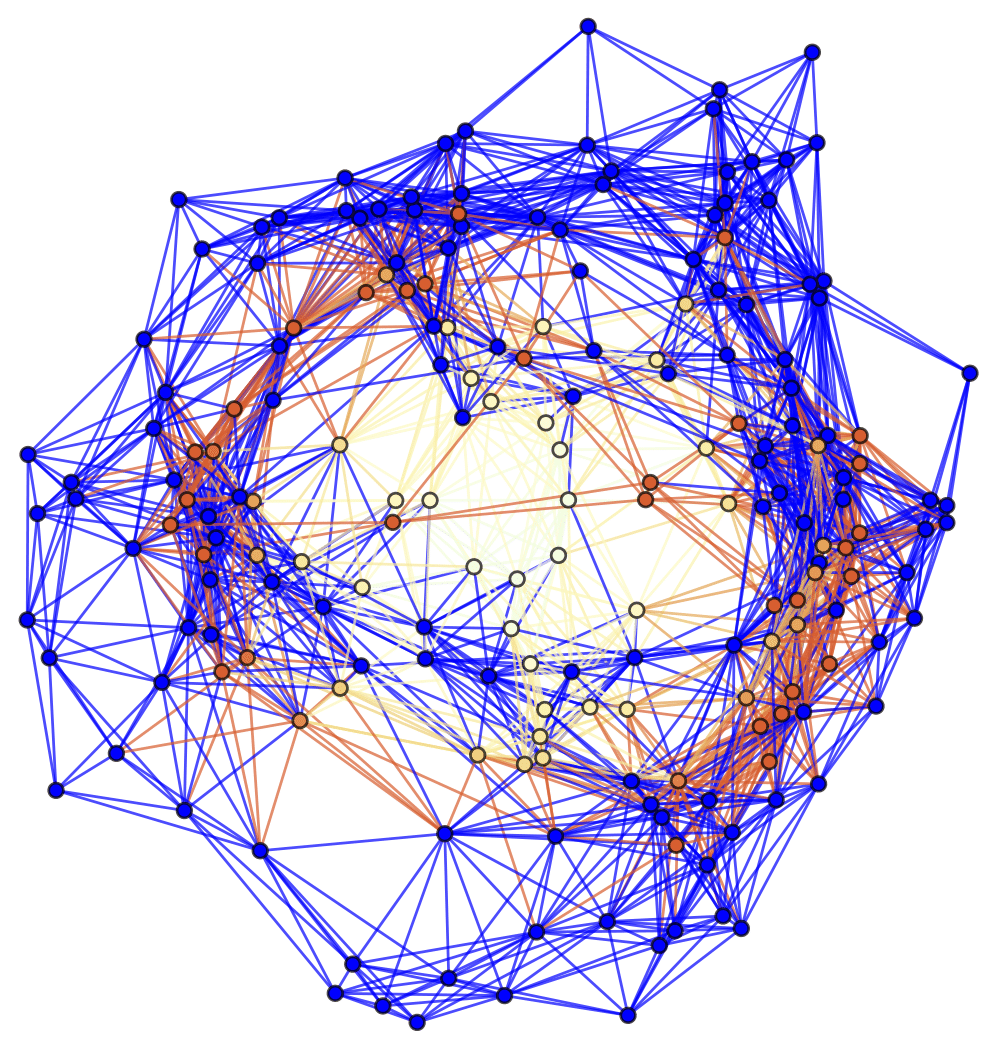}
\includegraphics[width=0.325\textwidth]{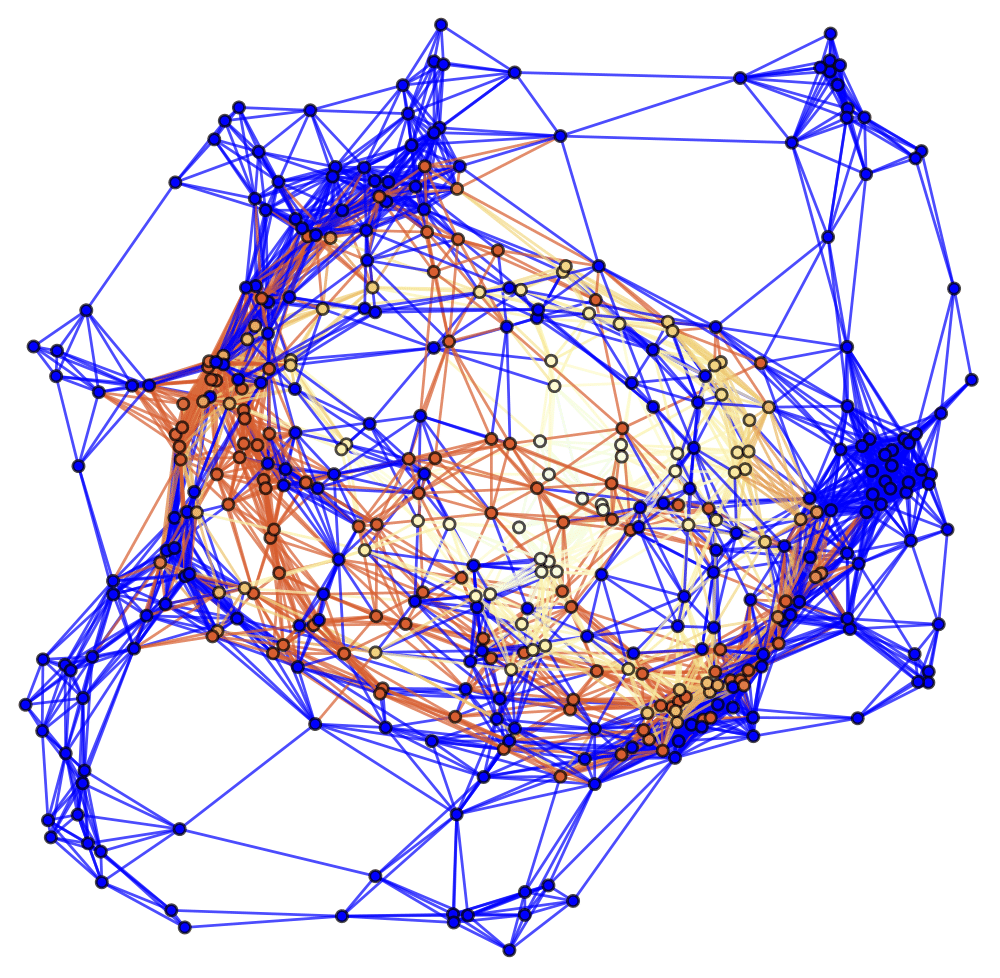}
\includegraphics[width=0.325\textwidth]{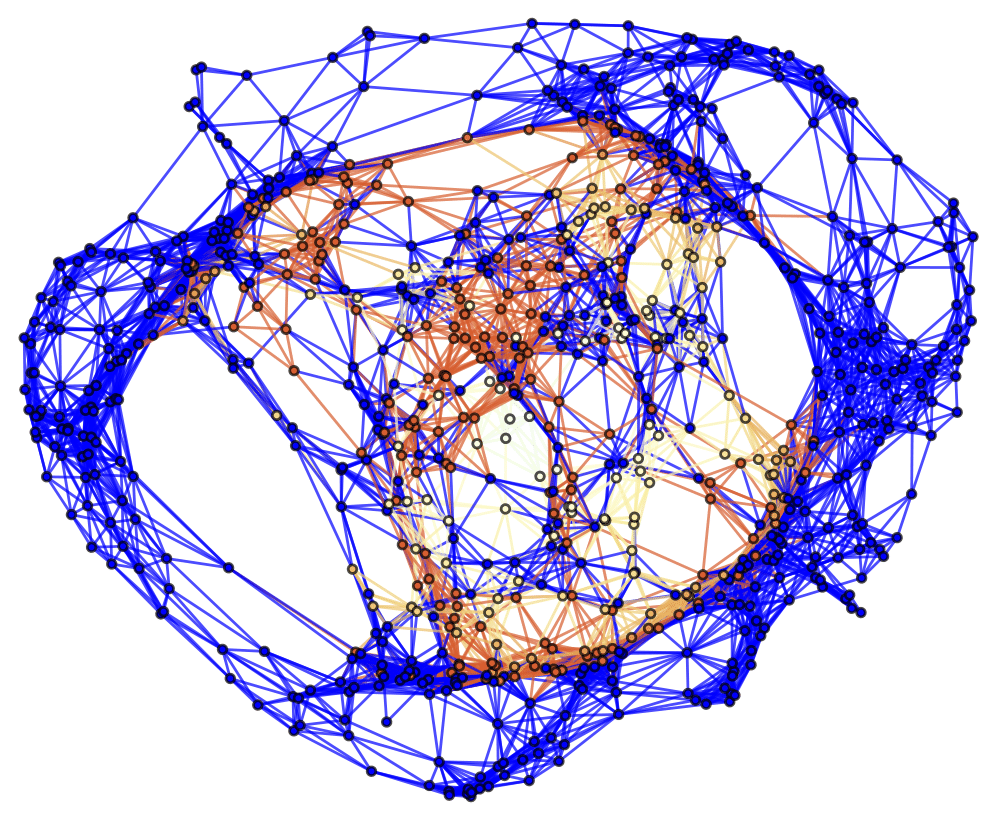}
\caption{Spatial hypergraphs corresponding to the initial hypersurface configuration of the massive scalar field ``bubble collapse'' to a maximally-rotating (extremal) Kerr black hole test, with a spinning (exponential) initial density distribution, at time ${t = 0 M}$, produced via pure Wolfram model evolution (with set substitution rule ${\left\lbrace \left\lbrace x, y \right\rbrace, \left\lbrace y, z \right\rbrace, \left\lbrace z, w \right\rbrace, \left\lbrace w, v \right\rbrace \right\rbrace \to \left\lbrace \left\lbrace y, u \right\rbrace, \left\lbrace u, v \right\rbrace, \left\lbrace w, x \right\rbrace, \left\lbrace x, u \right\rbrace \right\rbrace}$), with resolutions of 200, 400 and 800 vertices, respectively. The hypergraphs have been colored according to the value of the scalar field ${\Phi \left( t, R \right)}$.}
\label{fig:Figure49}
\end{figure}

\begin{figure}[ht]
\centering
\includegraphics[width=0.325\textwidth]{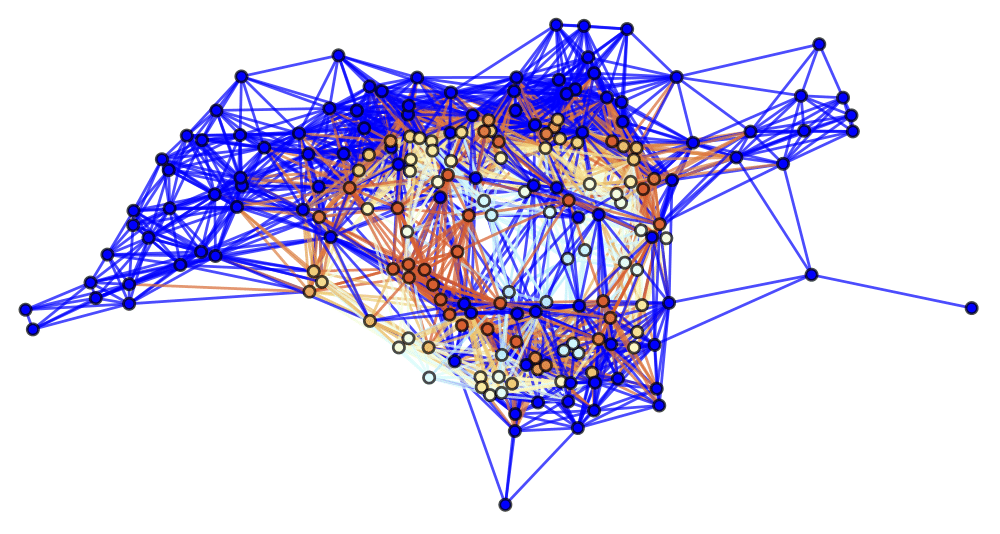}
\includegraphics[width=0.325\textwidth]{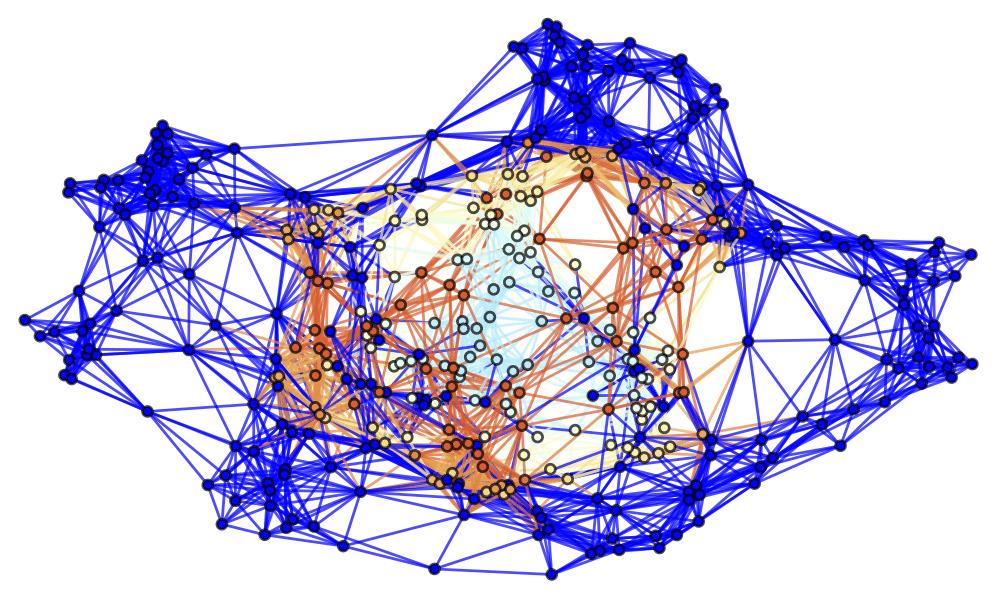}
\includegraphics[width=0.325\textwidth]{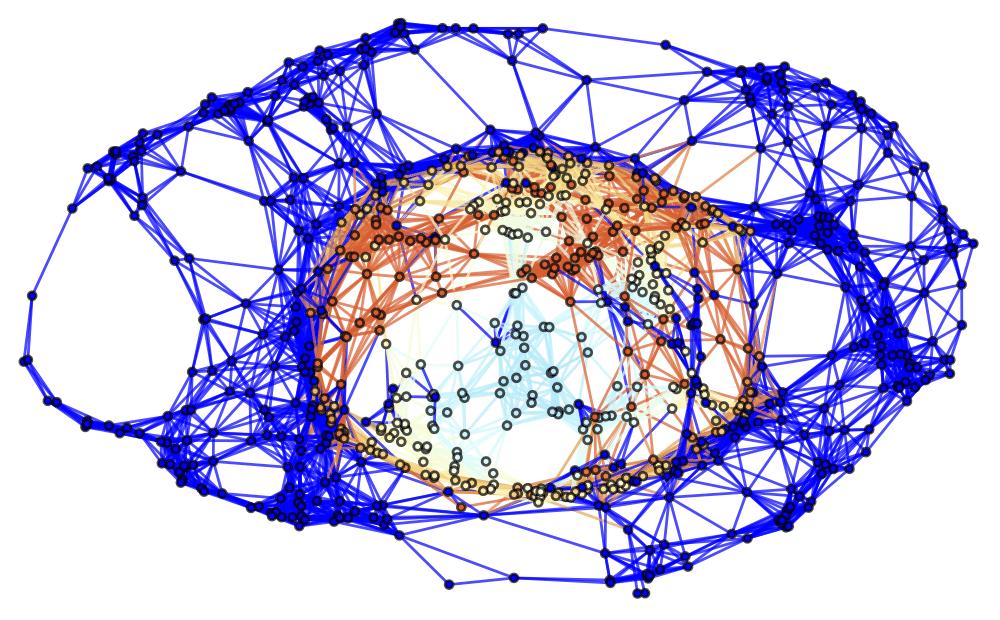}
\caption{Spatial hypergraphs corresponding to the first intermediate hypersurface configuration of the massive scalar field ``bubble collapse'' to a maximally-rotating (extremal) Kerr black hole test, with a spinning (exponential) initial density distribution, at time ${t = 1.5 M}$, produced via pure Wolfram model evolution (with set substitution rule ${\left\lbrace \left\lbrace x, y \right\rbrace, \left\lbrace y, z \right\rbrace, \left\lbrace z, w \right\rbrace, \left\lbrace w, v \right\rbrace \right\rbrace \to \left\lbrace \left\lbrace y, u \right\rbrace, \left\lbrace u, v \right\rbrace, \left\lbrace w, x \right\rbrace, \left\lbrace x, u \right\rbrace \right\rbrace}$), with resolutions of 200, 400 and 800 vertices, respectively. The hypergraphs have been colored according to the value of the scalar field ${\Phi \left( t, R \right)}$.}
\label{fig:Figure50}
\end{figure}

\begin{figure}[ht]
\centering
\includegraphics[width=0.325\textwidth]{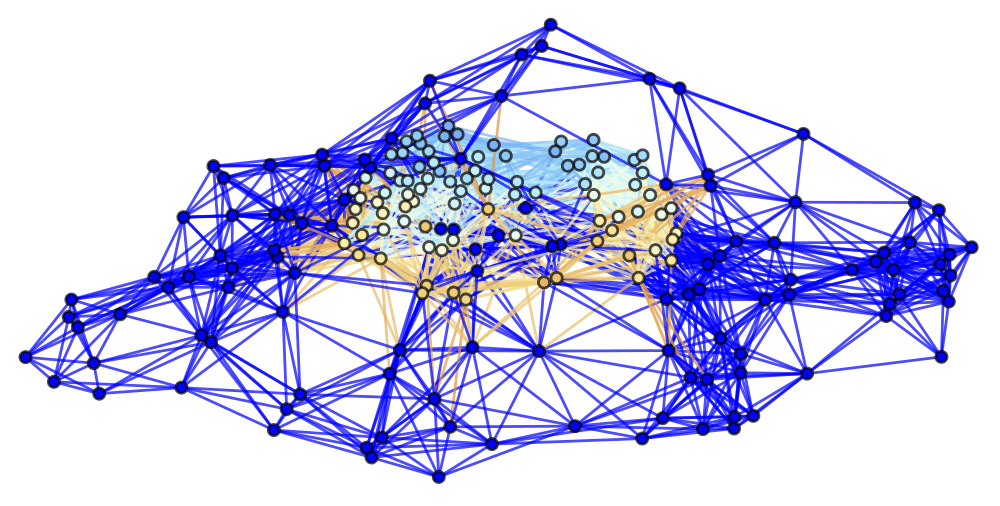}
\includegraphics[width=0.325\textwidth]{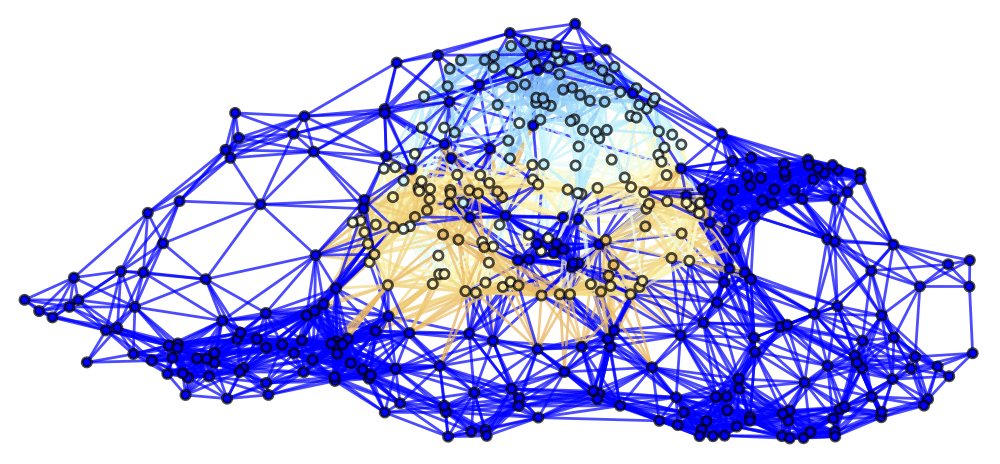}
\includegraphics[width=0.325\textwidth]{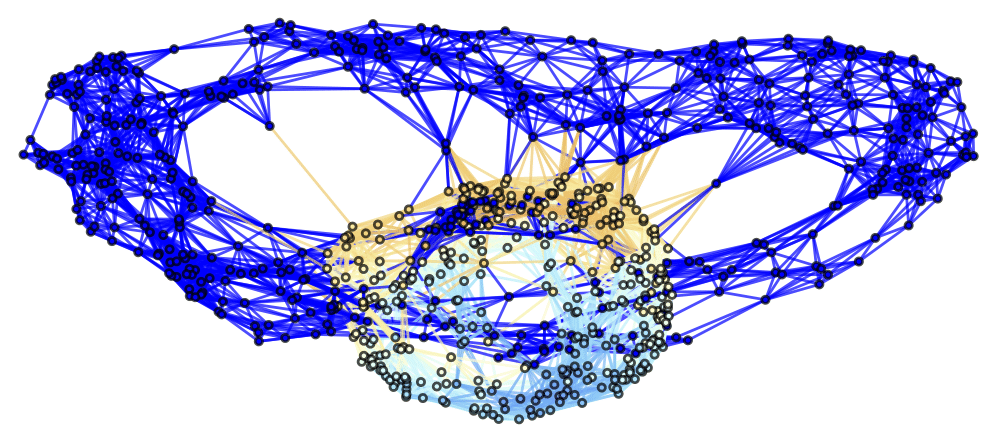}
\caption{Spatial hypergraphs corresponding to the second intermediate hypersurface configuration of the massive scalar field ``bubble collapse'' to a maximally-rotating (extremal) Kerr black hole test, with a spinning (exponential) initial density distribution, at time ${t = 3 M}$, produced via pure Wolfram model evolution (with set substitution rule ${\left\lbrace \left\lbrace x, y \right\rbrace, \left\lbrace y, z \right\rbrace, \left\lbrace z, w \right\rbrace, \left\lbrace w, v \right\rbrace \right\rbrace \to \left\lbrace \left\lbrace y, u \right\rbrace, \left\lbrace u, v \right\rbrace, \left\lbrace w, x \right\rbrace, \left\lbrace x, u \right\rbrace \right\rbrace}$), with resolutions of 200, 400 and 800 vertices, respectively. The hypergraphs have been colored according to the value of the scalar field ${\Phi \left( t, R \right)}$.}
\label{fig:Figure51}
\end{figure}

\begin{figure}[ht]
\centering
\includegraphics[width=0.325\textwidth]{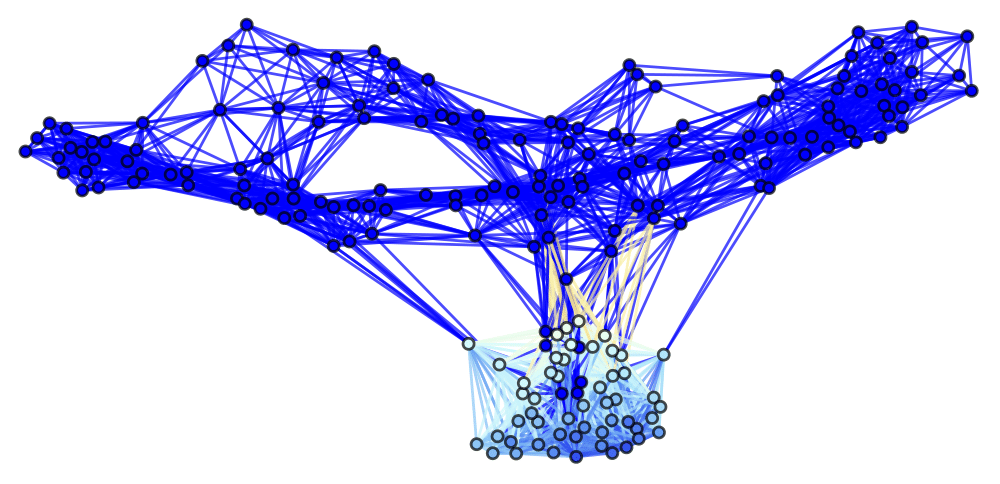}
\includegraphics[width=0.325\textwidth]{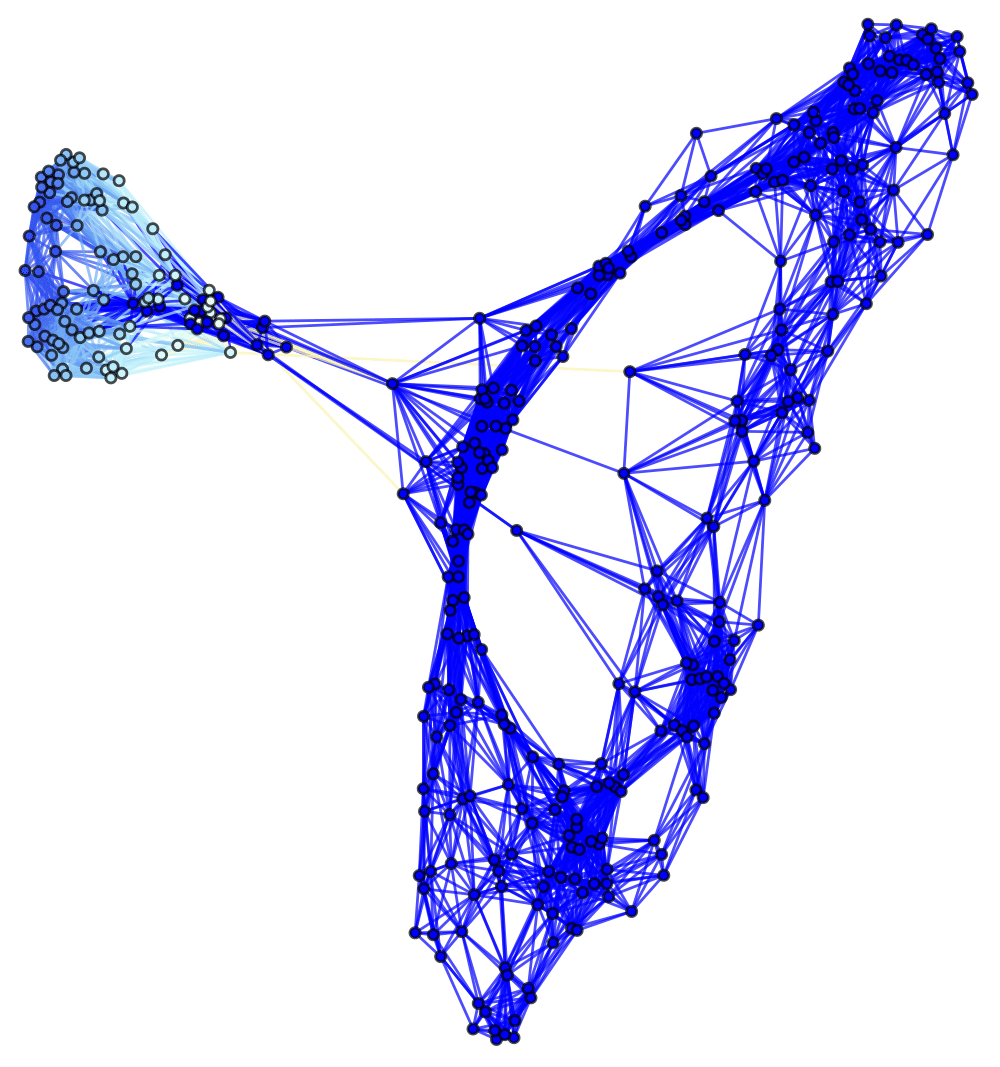}
\includegraphics[width=0.325\textwidth]{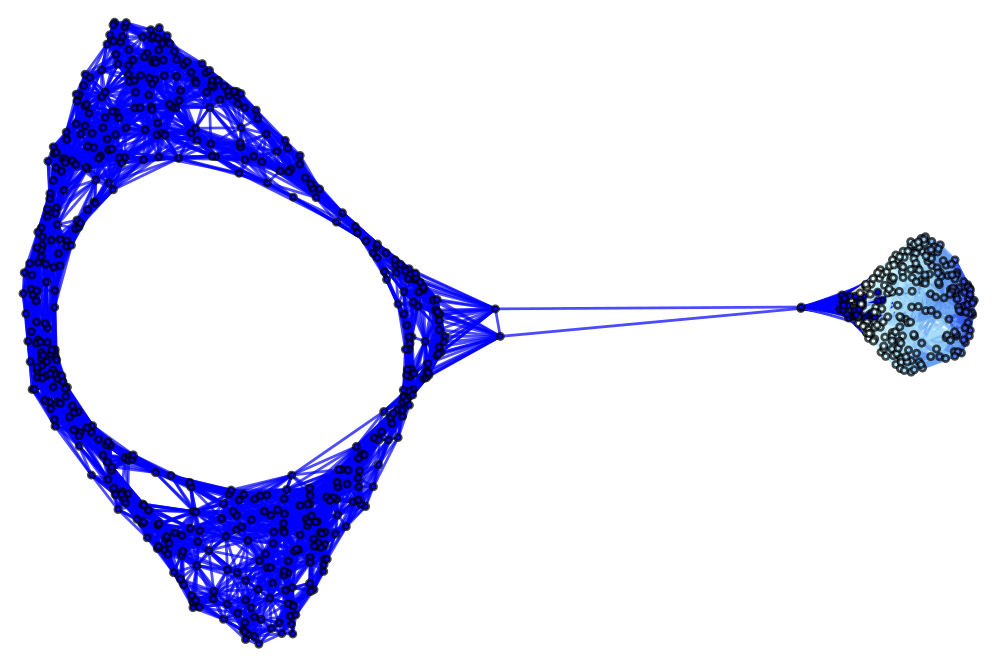}
\caption{Spatial hypergraphs corresponding to the final hypersurface configuration of the massive scalar field ``bubble collapse'' to a maximally-rotating (extremal) Kerr black hole test, with a spinning (exponential) initial density distribution, at time ${t = 4.5 M}$, produced via pure Wolfram model evolution (with set substitution rule ${\left\lbrace \left\lbrace x, y \right\rbrace, \left\lbrace y, z \right\rbrace, \left\lbrace z, w \right\rbrace, \left\lbrace w, v \right\rbrace \right\rbrace \to \left\lbrace \left\lbrace y, u \right\rbrace, \left\lbrace u ,v \right\rbrace, \left\lbrace w, x \right\rbrace, \left\lbrace x, u \right\rbrace \right\rbrace}$), with resolutions of 200, 400 and 800 vertices, respectively. The hypergraphs have been colored according to the value of the scalar field ${\Phi \left( t, R \right)}$.}
\label{fig:Figure52}
\end{figure}

\begin{figure}[ht]
\centering
\includegraphics[width=0.325\textwidth]{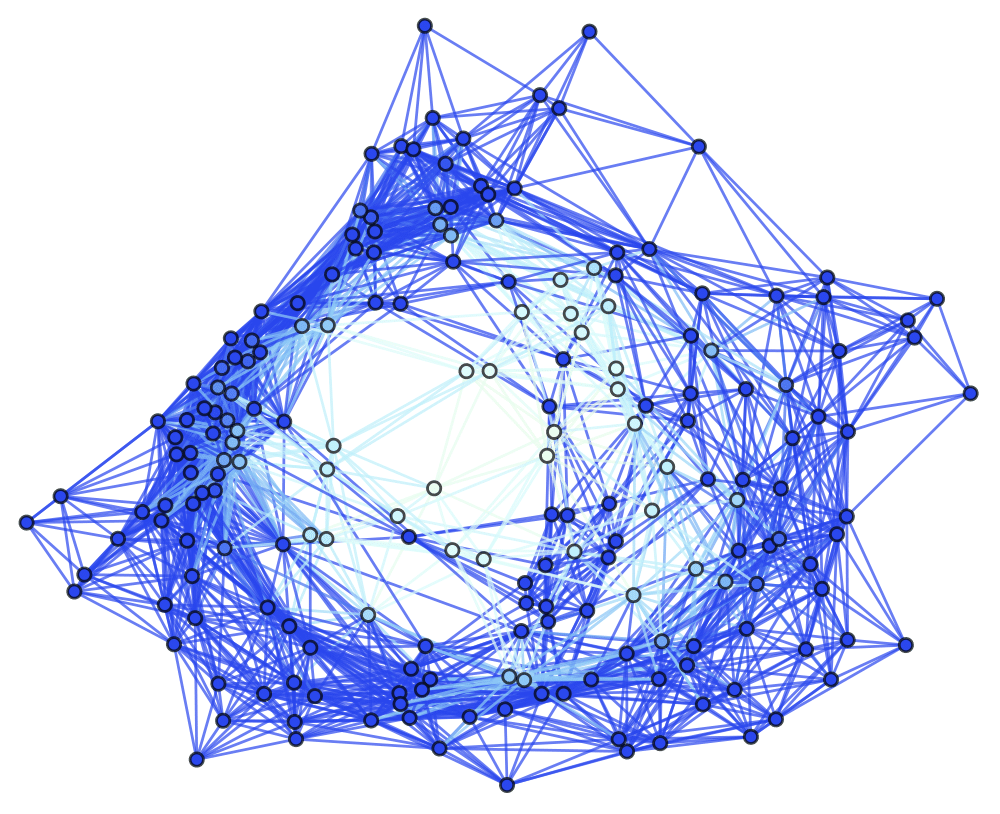}
\includegraphics[width=0.325\textwidth]{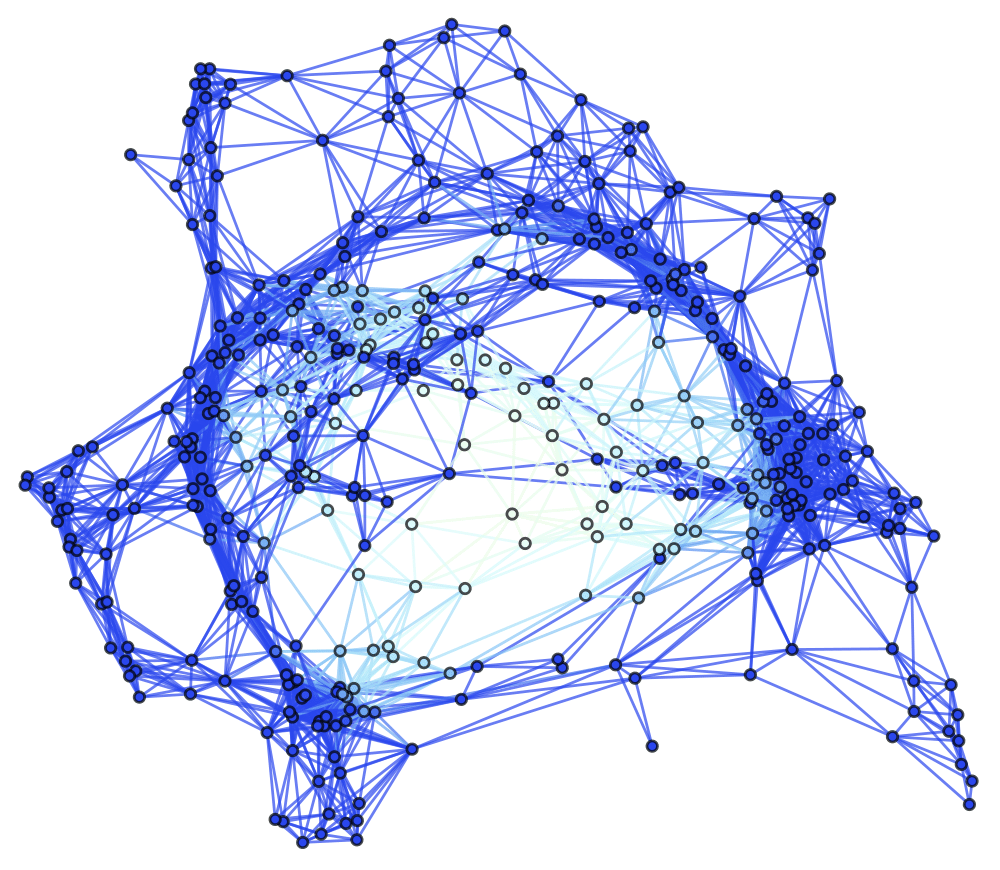}
\includegraphics[width=0.325\textwidth]{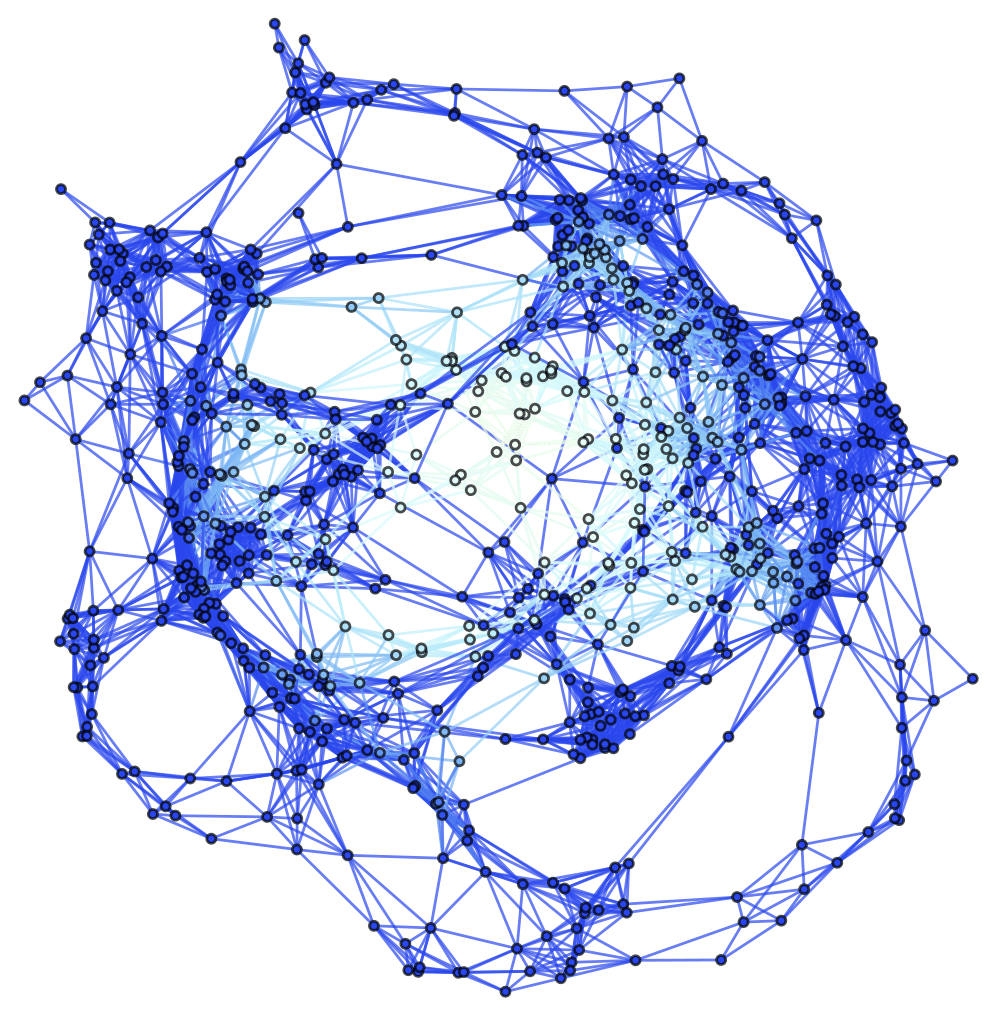}
\caption{Spatial hypergraphs corresponding to the initial hypersurface configuration of the massive scalar field ``bubble collapse'' to a maximally-rotating (extremal) Kerr black hole test, with a spinning (exponential) initial density distribution, at time ${t = 0 M}$, produced via pure Wolfram model evolution (with set substitution rule ${\left\lbrace \left\lbrace x, y \right\rbrace, \left\lbrace y, z \right\rbrace, \left\lbrace z, w \right\rbrace, \left\lbrace w, v \right\rbrace \right\rbrace \to \left\lbrace \left\lbrace y, u \right\rbrace, \left\lbrace u, v \right\rbrace, \left\lbrace w, x \right\rbrace, \left\lbrace x, u \right\rbrace \right\rbrace}$), with resolutions of 200, 400 and 800 vertices, respectively. The hypergraphs have been colored according to the local curvature in the Boyer-Lindquist conformal factor ${\psi}$.}
\label{fig:Figure53}
\end{figure}

\begin{figure}[ht]
\centering
\includegraphics[width=0.325\textwidth]{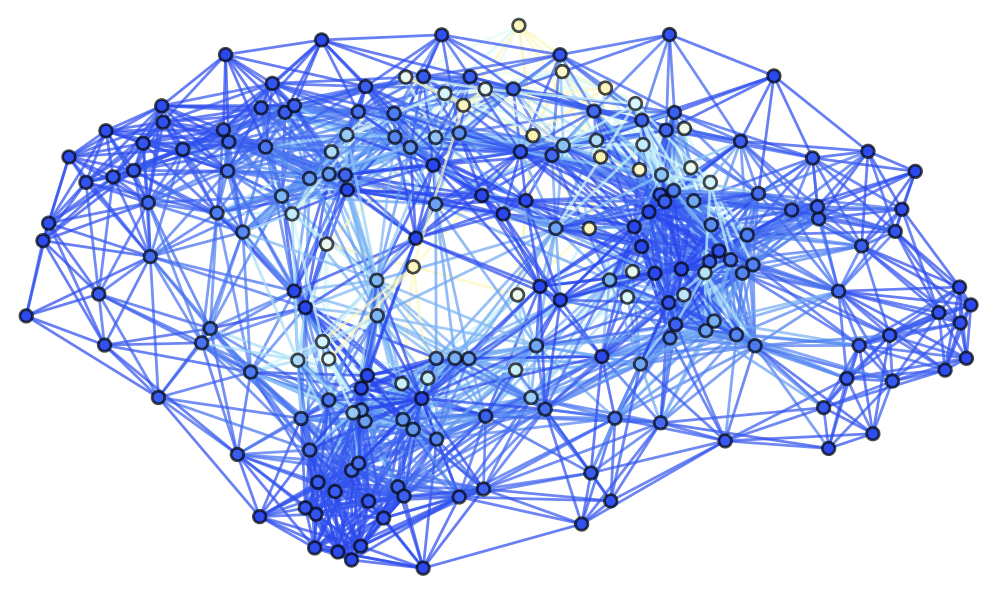}
\includegraphics[width=0.325\textwidth]{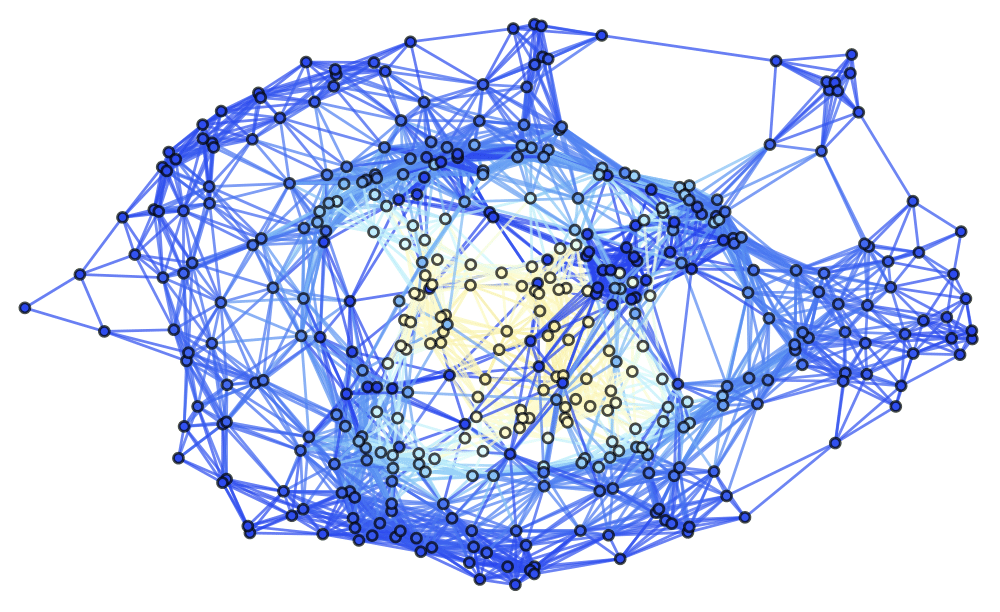}
\includegraphics[width=0.325\textwidth]{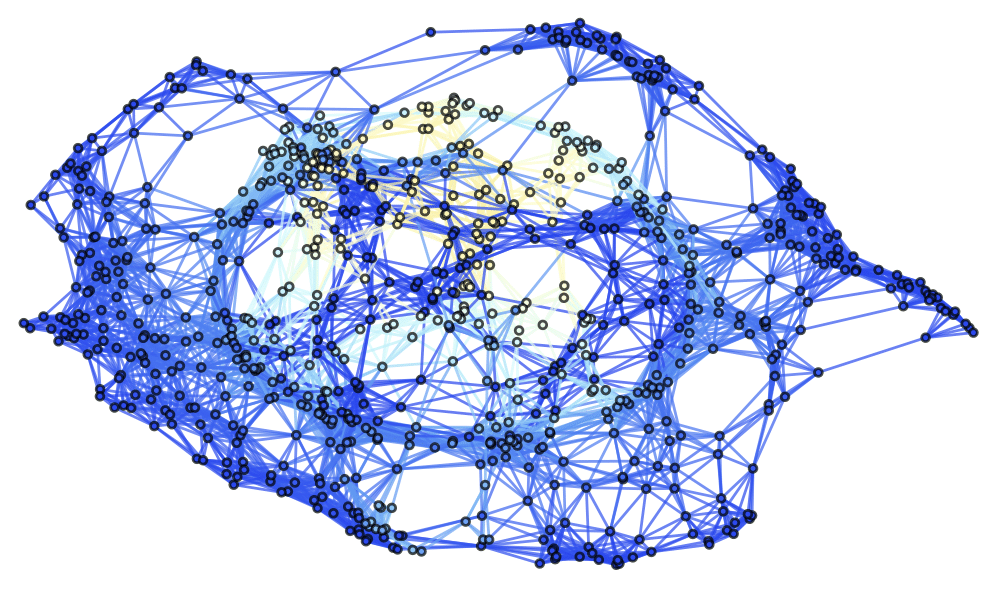}
\caption{Spatial hypergraphs corresponding to the first intermediate hypersurface configuration of the massive scalar field ``bubble collapse'' to a maximally-rotating (extremal) Kerr black hole test, with a spinning (exponential) initial density distribution, at time ${t = 1.5 M}$, produced via pure Wolfram model evolution (with set substitution rule ${\left\lbrace \left\lbrace x, y \right\rbrace, \left\lbrace y, z \right\rbrace, \left\lbrace z, w \right\rbrace, \left\lbrace w, v \right\rbrace \right\rbrace \to \left\lbrace \left\lbrace y, u \right\rbrace, \left\lbrace u, v \right\rbrace, \left\lbrace w, x \right\rbrace, \left\lbrace x, u \right\rbrace \right\rbrace}$), with resolutions of 200, 400 and 800 vertices, respectively. The hypergraphs have been colored according to the local curvature in the Boyer-Lindquist conformal factor ${\psi}$.}
\label{fig:Figure54}
\end{figure}

\begin{figure}[ht]
\centering
\includegraphics[width=0.325\textwidth]{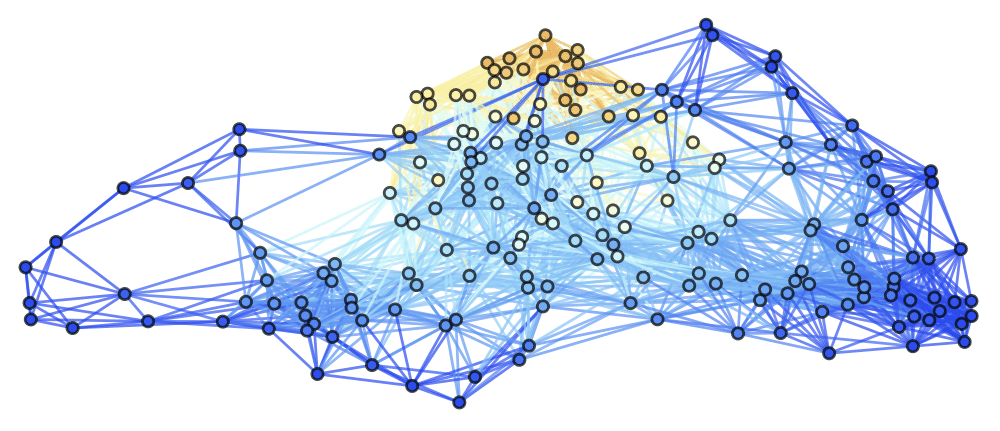}
\includegraphics[width=0.325\textwidth]{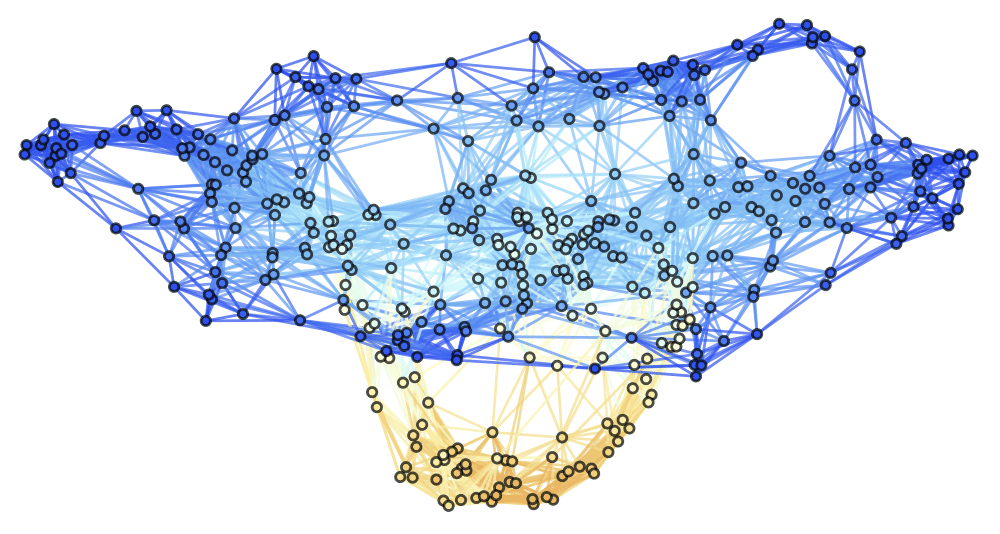}
\includegraphics[width=0.325\textwidth]{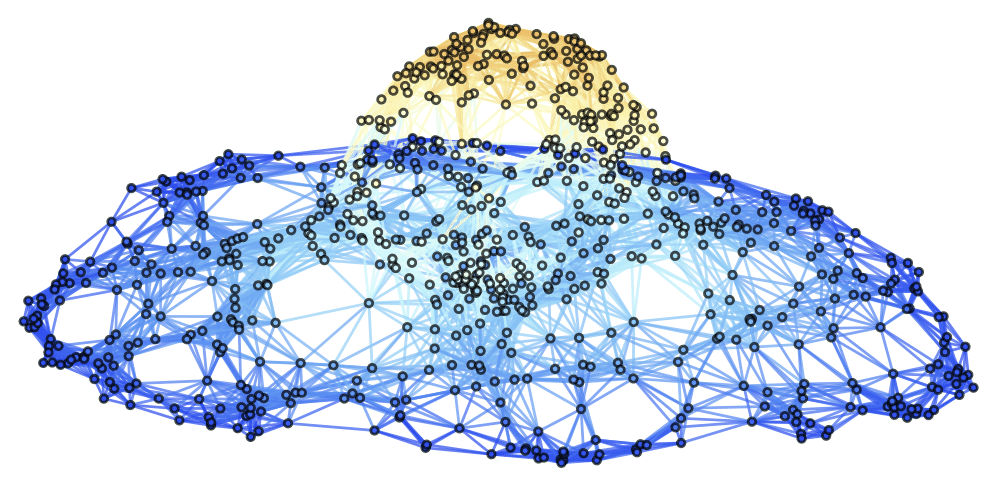}
\caption{Spatial hypergraphs corresponding to the second intermediate hypersurface configuration of the massive scalar field ``bubble collapse'' to a maximally-rotating (extremal) Kerr black hole test, with a spinning (exponential) initial density distribution, at time ${t = 3 M}$, produced via pure Wolfram model evolution (with set substitution rule ${\left\lbrace \left\lbrace x, y \right\rbrace, \left\lbrace y, z \right\rbrace, \left\lbrace z, w \right\rbrace, \left\lbrace w, v \right\rbrace \right\rbrace \to \left\lbrace \left\lbrace y, u \right\rbrace, \left\lbrace u, v \right\rbrace, \left\lbrace w, x \right\rbrace, \left\lbrace x, u \right\rbrace \right\rbrace}$), with resolutions of 200, 400 and 800 vertices, respectively. The hypergraphs have been colored according to the local curvature in the Boyer-Lindquist conformal factor ${\psi}$.}
\label{fig:Figure55}
\end{figure}

\begin{figure}[ht]
\centering
\includegraphics[width=0.325\textwidth]{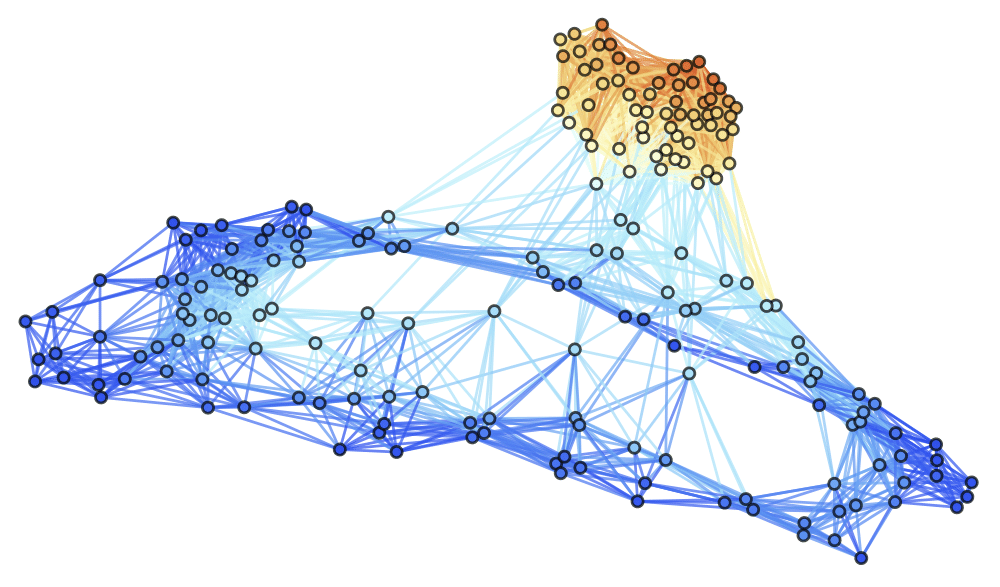}
\includegraphics[width=0.325\textwidth]{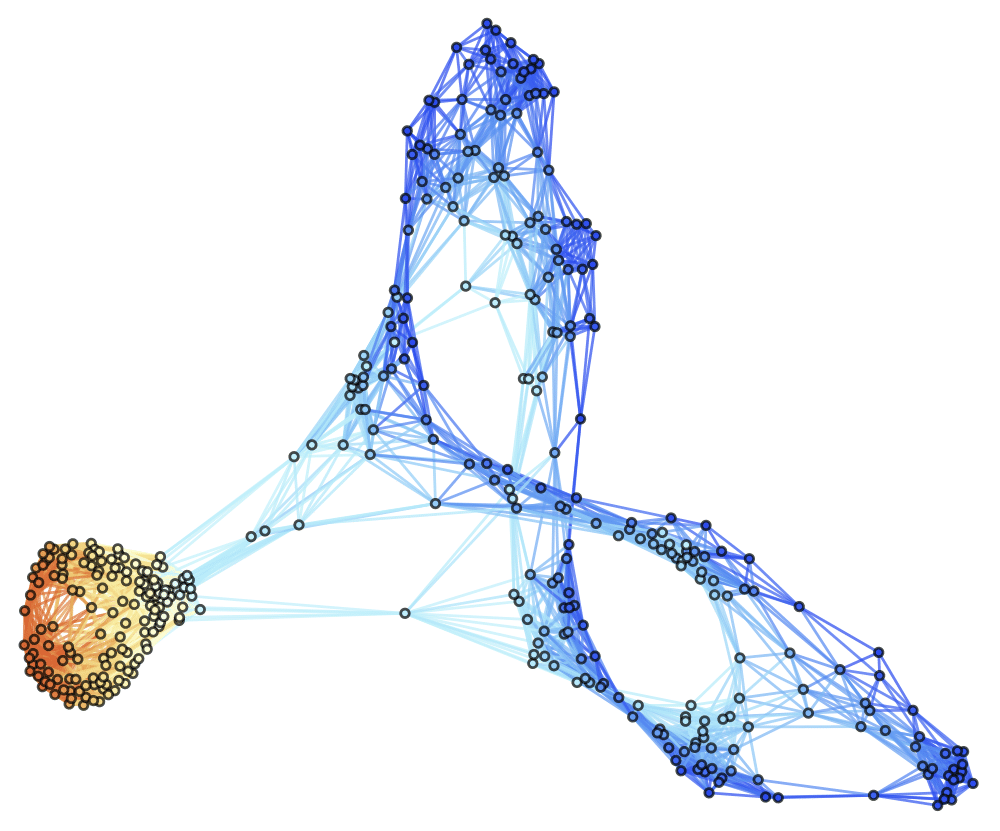}
\includegraphics[width=0.325\textwidth]{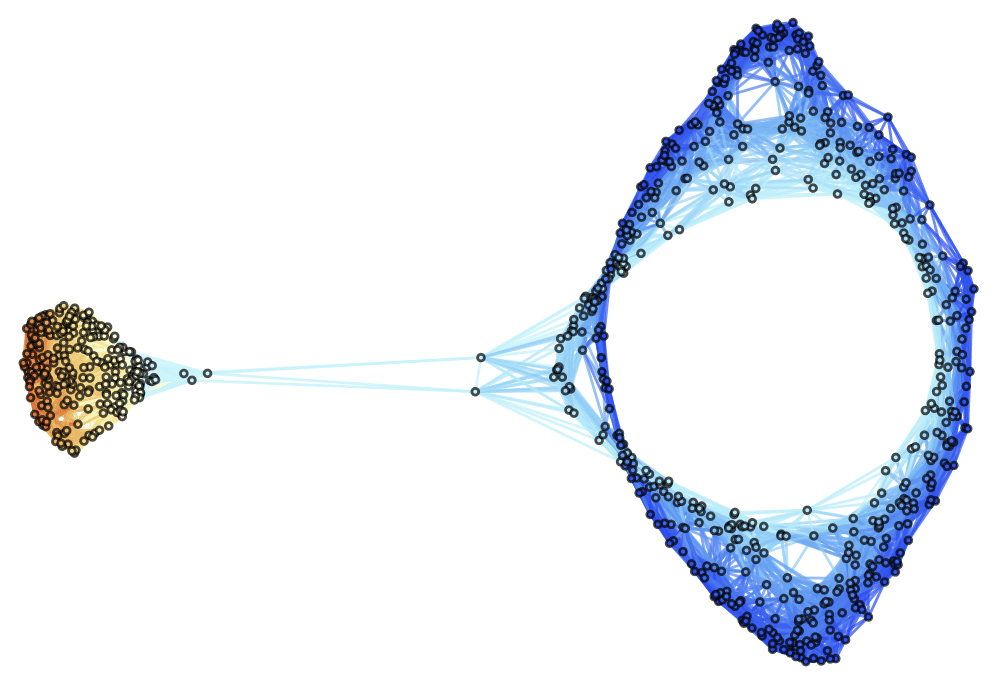}
\caption{Spatial hypergraphs corresponding to the final hypersurface configuration of the massive scalar field ``bubble collapse'' to a maximally-rotating (extremal) Kerr black hole test, with a spinning (exponential) initial density distribution, at time ${t = 4.5 M}$, produced via pure Wolfram model evolution (with set substitution rule ${\left\lbrace \left\lbrace x, y \right\rbrace, \left\lbrace y, z \right\rbrace, \left\lbrace z,w \right\rbrace, \left\lbrace w, v \right\rbrace \right\rbrace \to \left\lbrace \left\lbrace y, u \right\rbrace, \left\lbrace u, v \right\rbrace, \left\lbrace w, x \right\rbrace, \left\lbrace x, u \right\rbrace \right\rbrace}$), with resolutions of 200, 400 and 800 vertices, respectively. The hypergraphs have been colored according to the local curvature in the Boyer-Lindquist conformal factor ${\psi}$.}
\label{fig:Figure56}
\end{figure}

\begin{figure}[ht]
\centering
\includegraphics[width=0.325\textwidth]{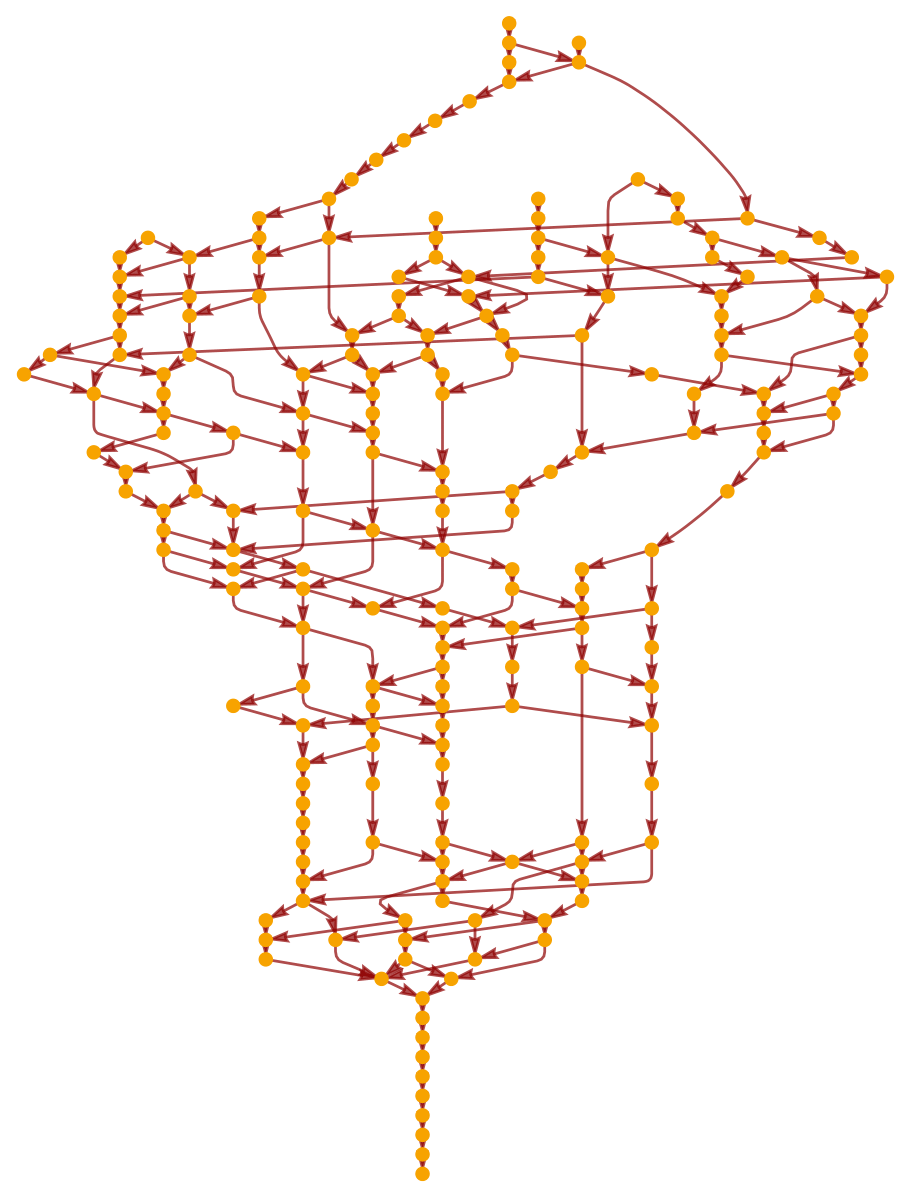}
\includegraphics[width=0.325\textwidth]{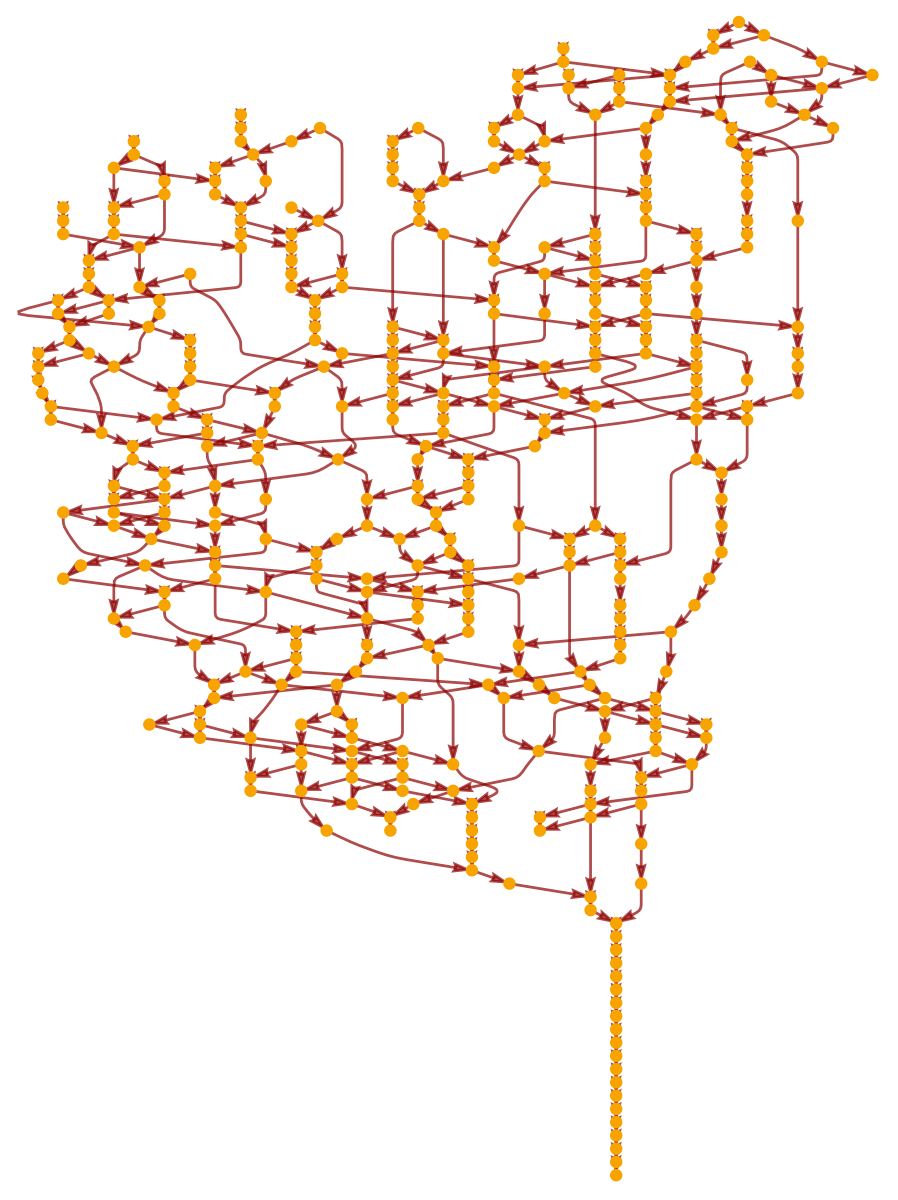}
\includegraphics[width=0.325\textwidth]{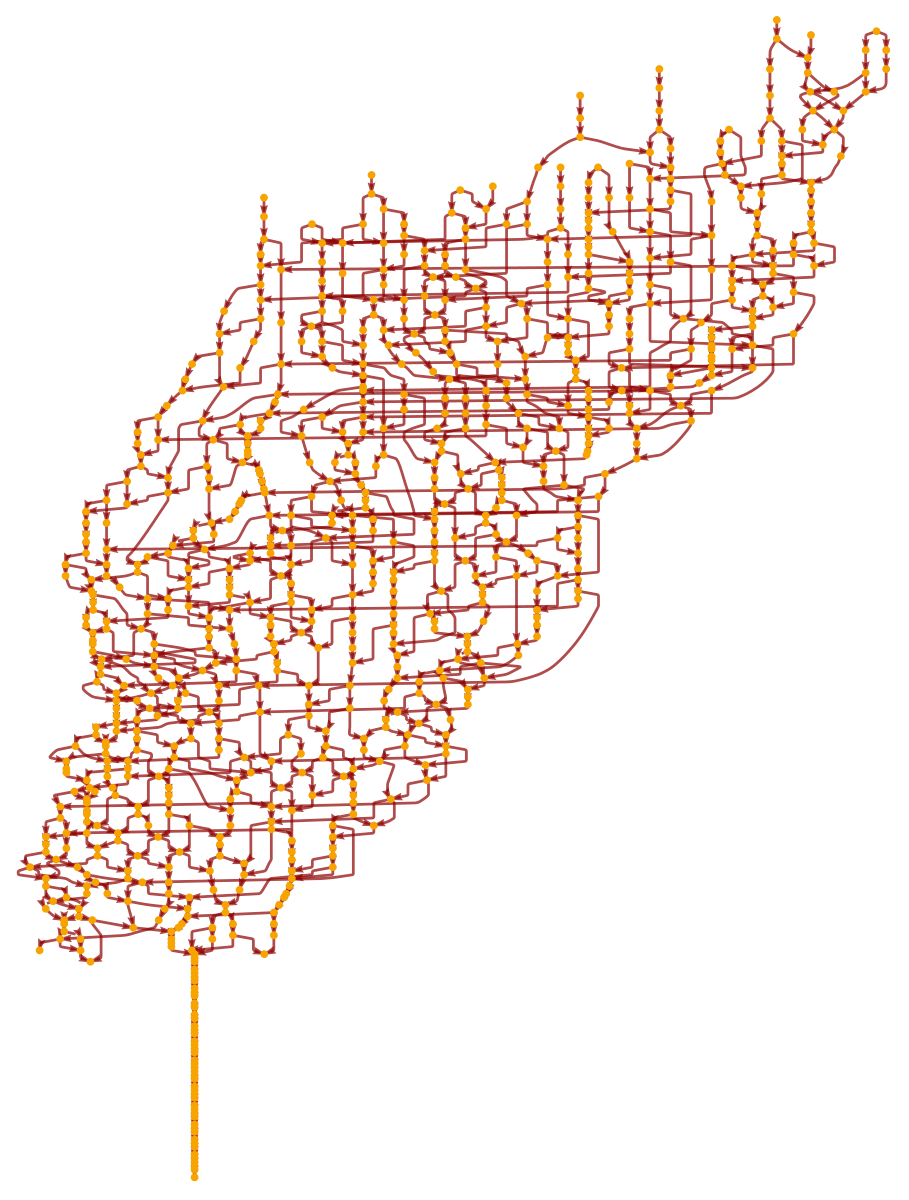}
\caption{Causal graphs corresponding to the discrete characteristic structure of the massive scalar field ``bubble collapse'' to a maximally-rotating (extremal) Kerr black hole test, with a spinning (exponential) initial density distribution, at time ${t = 4.5 M}$, produced via pure Wolfram model evolution (with set substitution rule ${\left\lbrace \left\lbrace x, y \right\rbrace, \left\lbrace y, z \right\rbrace, \left\lbrace z, w \right\rbrace, \left\lbrace w, v \right\rbrace \right\rbrace \to \left\lbrace \left\lbrace y, u \right\rbrace, \left\lbrace u, v \right\rbrace, \left\lbrace w, x \right\rbrace, \left\lbrace x, u \right\rbrace \right\rbrace}$), with resolutions of 200, 400 and 800 hypergraph vertices, respectively.}
\label{fig:Figure57}
\end{figure}

Using this overall approach, we construct an exponential initial density profile for a massive scalar field ${\Phi \left( t, r \right)}$ defined over Wolfram model hypergraphs:

\begin{equation}
\rho \left( 0, r \right) = T_{t t} \left( 0, r \right) = \rho_0 \exp \left( - \left( \frac{r}{\lambda} \right)^3 \right),
\end{equation}
with initial density constant ${\rho_0}$ and radius of support ${\lambda}$. The Cauchy initial data is set up by taking a spherically-symmetric background metric in Gaussian polar coordinates:

\begin{equation}
d s^2 = - d t^2 + e^{-2 \Lambda \left( t, r \right)} d r^2 + R^2 \left( t, r \right) d \Omega^2,
\end{equation}
and constructing a corresponding spatial hypergraph via Poisson sprinkling (with fixed sprinkling density ${\rho_c}$). We evolve the resulting hypergraph until a final time of ${t = 4.5 M}$, with intermediate checks at ${t = 1.5 M}$ and ${t = 3 M}$; the initial, first intermediate, second intermediate and final hypersurface configurations, with the hypergraphs colored using the scalar field ${\Phi \left( t,r \right)}$, are shown in Figures \ref{fig:Figure40}, \ref{fig:Figure41}, \ref{fig:Figure42} and \ref{fig:Figure43}, respectively, with resolutions of 200, 400 and 800 vertices; similarly, Figures \ref{fig:Figure44}, \ref{fig:Figure45}, \ref{fig:Figure46} and \ref{fig:Figure47} show the initial, first intermediate, second intermediate and final hypersurface configurations, but with the hypergraphs colored using the Schwarzschild conformal factor ${\psi}$, respectively. Figure \ref{fig:Figure48} shows the corresponding causal graphs for the resulting Wolfram model evolutions, with the convergence rates for the Hamiltonian constraint after time ${t = 4.5 M}$, with respect to the ${L_1}$, ${L_2}$ and ${L_{\infty}}$-norms, illustrating approximately second-order convergence, shown in Table \ref{tab:Table3}. Furthermore, we perform the analogous construction (consisting of an exponential initial density profile for the scalar field ${\Phi \left( t, R \right)}$, embedded within a Poisson-sprinkled initial hypergraph with fixed sprinkling density ${\rho_c}$) for an axially-symmetric background metric in (cylindrical) Weyl-Lewis-Papapetrou coordinates:

\begin{equation}
d s^2 = - e^{2 U \left( \rho, zeta \right)} \left( d t  + a \left( \rho, \zeta \right) d \varphi \right)^2 + e^{-2 U \left( \rho, \zeta \right)} \left[ e^{2 k \left( \rho, \zeta \right)} \left( d \rho^2 + d \zeta^2 \right) + W^2 \left(\rho, \zeta \right) d \varphi^2 \right],
\end{equation}
with ${\rho = R \sin \left( \theta \right)}$ and ${\zeta = R \cos \left( \theta \right)}$. Once again, the initial, first intermediate, second intermediate and final hypersurface configurations (corresponding to times ${t = 0M}$, ${t = 1.5 M}$, ${t = 3M}$ and ${t = 4.5 M}$, respectively), with hypergraphs colored using the scalar field ${\Phi \left( t, R \right)}$, are shown in Figures \ref{fig:Figure49}, \ref{fig:Figure50}, \ref{fig:Figure51} and \ref{fig:Figure52}, respectively, with resolutions of 200, 400 and 800 vertices; similarly, Figures \ref{fig:Figure53}, \ref{fig:Figure54}, \ref{fig:Figure55} and \ref{fig:Figure56} show the initial, first intermediate, second intermediate and final hypersurface configurations, but with the hypergraphs colored using the Boyer-Lindquist conformal factor ${\psi}$, respectively. Figure \ref{fig:Figure57} shows the corresponding causal graphs for the resulting Wolfram model evolutions, with the convergence rates for the Hamiltonian constraint after time ${t = 4.5 M}$, with respect to the ${L_1}$, ${L_2}$ and ${L_{\infty}}$-norms, illustrating approximately second-order convergence, shown in Table \ref{tab:Table4}. In both cases we confirm approximate conservation of the ADM mass, linear momentum and angular momentum of the overall spacetime for the duration of the evolution. Full details regarding the construction and evolution of general relativistic Cauchy data via Wolfram model systems, as well as the extraction of all associated curvature and stress-energy properties of the resulting discrete spacetimes, can be found in \cite{gorard4}.

\begin{table}[ht]
\centering
\begin{tabular}{|c|c|c|c|c|c|c|}
\hline
Vertices & ${\epsilon \left( L_1 \right)}$ & ${\epsilon \left( L_2 \right)}$ & ${\epsilon \left( L_{\infty} \right)}$ & ${\mathcal{O} \left( L_1 \right)}$ & ${\mathcal{O} \left( L_2 \right)}$ & ${\mathcal{O} \left( L_{\infty} \right)}$\\
\hline\hline
100 & ${1.50 \times 10^{-1}}$ & ${1.79 \times 10^{-1}}$ & ${1.76 \times 10^{-2}}$ & - & - & -\\
\hline
200 & ${4.53 \times 10^{-2}}$ & ${6.10 \times 10^{-2}}$ & ${5.53 \times 10^{-3}}$ & 1.73 & 1.55 & 1.67\\
\hline
400 & ${1.55 \times 10^{-2}}$ & ${1.90 \times 10^{-2}}$ & ${1.88 \times 10^{-3}}$ & 1.54  & 1.68 & 1.56\\
\hline
800 & ${3.57 \times 10^{-3}}$ & ${5.22 \times 10^{-3}}$ & ${6.36 \times 10^{-4}}$ & 2.12 & 1.87  & 1.56\\
\hline
1600 & ${1.07 \times 10^{-3}}$ & ${1.47 \times 10^{-3}}$ & ${1.74 \times 10^{-4}}$ & 1.74 & 1.83 & 1.86\\
\hline
\end{tabular}
\caption{Convergence rates for the massive scalar field ``bubble collapse'' to a non-rotating Schwarzschild black hole test, with respect to the ${L_1}$, ${L_2}$ and ${L_{\infty}}$-norms for the Hamiltonian constraint $H$ after time ${t = 4.5 M}$, produced via pure Wolfram model evolution (with set substitution rule ${\left\lbrace \left\lbrace x, y \right\rbrace, \left\lbrace y, z \right\rbrace, \left\lbrace z, w \right\rbrace, \left\lbrace w, v \right\rbrace \right\rbrace \to \left\lbrace \left\lbrace y, u \right\rbrace, \left\lbrace u, v \right\rbrace, \left\lbrace w, x \right\rbrace, \left\lbrace x, u \right\rbrace \right\rbrace}$), showing approximately second-order convergence.}
\label{tab:Table3}
\end{table}

\begin{table}[ht]
\centering
\begin{tabular}{|c|c|c|c|c|c|c|c|}
\hline
Vertices & ${\epsilon \left( L_1 \right)}$ & ${\epsilon \left( L_2 \right)}$ & ${\epsilon \left( L_{\infty} \right)}$ & ${\mathcal{O} \left( L_1 \right)}$ & ${\mathcal{O} \left( L_2 \right)}$ & ${\mathcal{O} \left( L_{\infty} \right)}$\\
\hline\hline
100 & ${2.56 \times 10^{-1}}$ & ${2.34 \times 10^{-2}}$ & ${4.35 \times 10^{-1}}$ & - & - & -\\
\hline
200 & ${8.75 \times 10^{-2}}$ & ${4.82 \times 10^{-3}}$ & ${1.47 \times 10^{-1}}$ & 1.55 & 2.28 & 1.57\\
\hline
400 & ${2.36 \times 10^{-2}}$ & ${1.63 \times 10^{-3}}$ & ${3.60 \times 10^{-2}}$ & 1.89 & 1.57 & 2.03\\
\hline
800 & ${5.46 \times 10^{-3}}$ & ${2.49 \times 10^{-4}}$ & ${7.56 \times 10^{-3}}$ & 2.11 & 2.49 & 2.25\\
\hline
1600 & ${1.18 \times 10^{-3}}$ & ${8.22 \times 10^{-5}}$ & ${2.02 \times 10^{-3}}$ & 2.21 & 1.81 & 1.90\\
\hline
\end{tabular}
\caption{Convergence rates for the massive scalar field ``bubble collapse'' to a maximally-rotating (extremal) Kerr black hole test, with respect to the ${L_1}$, ${L_2}$ and ${L_{\infty}}$-norms for the Hamiltonian constraint $H$ after time ${t = 4.5 M}$, produced via pure Wolfram model evolution (with set substitution rule ${\left\lbrace \left\lbrace x, y \right\rbrace, \left\lbrace y, z \right\rbrace, \left\lbrace z, w \right\rbrace, \left\lbrace w, v \right\rbrace \right\rbrace \to \left\lbrace \left\lbrace y, u \right\rbrace, \left\lbrace u, v \right\rbrace, \left\lbrace w, x \right\rbrace, \left\lbrace x, u \right\rbrace \right\rbrace}$), showing approximately second-order convergence.}
\label{tab:Table4}
\end{table}

\clearpage

\section{Concluding Remarks}
\label{sec:Section5}

This article has succeeded in demonstrating, through a combination of rigorous mathematical analysis and explicit numerical simulation using the \href{https://github.com/JonathanGorard/Gravitas}{\textsc{Gravitas}} framework, that if one starts with an analytical solution to the massive scalar field ``bubble collapse'' problem in either spherical or axial symmetry (yielding convergence a non-rotating Schwarzschild black hole or a maximally-rotating/extremal Kerr black hole, respectively), and proceeds to apply microscopic perturbations away from these exact symmetries, at least of the kind that are consistent with discretization of the underlying spacetime, then the solution structure in both cases remains stable under appropriate assumptions. We have, in addition, shown that both solutions may be recovered as limiting cases of a pure Wolfram model/hypergraph rewriting evolution, without any a priori continuity assumptions on the underlying spacetime. To the best of our knowledge, this represents the first systematic analysis of the evolution of a Wolfram model system representing a true \textit{non-vacuum solution} to the Einstein field equations in the continuum limit, as well as the first robust analysis of gravitational collapse and singularity-formation behavior within a reasonably generic discrete spacetime setting. However, the techniques employed here are clearly exceedingly limited in scope, depending as they do upon an unphysically high degree of a priori symmetry, as well as an extremely idealized matter model (i.e. a massive scalar field obeying the discrete Klein-Gordon equation) for the stress-energy tensor. Recent work investigating global homotopic aspects of Wolfram model systems\cite{arsiwalla}\cite{arsiwalla2} provides one plausible direction by which it might conceivably be possible to formulate a more global topological theorem regarding geodesic incompleteness in generic Wolfram model systems, in a similar style to Penrose's original 1964 singularity theorem\cite{penrose}. Moreover, there is an ongoing effort to integrate more sophisticated stress-energy models (e.g. relativistic dust, relativistic radiation, perfect fluids, electromagnetic fields, etc.) into the \href{https://github.com/JonathanGorard/Gravitas}{\textsc{Gravitas}} framework, and thereby to make such stress-energy distributions compatible with Wolfram model evolution in broadly the same manner as we have done here with massive scalar fields; it would be highly interesting to investigate the analogs of relativistic energy conditions and equations of state within arbitrary Wolfram model causal graphs, and to probe the effects that such properties and constraints might have on the dynamics and singularity structure of the resulting discrete spacetimes. Such an extension might also afford one the opportunity to extend certain, presently highly idealized, quantum mechanical computations regarding (e.g.) black hole entropies\cite{maldacena}\cite{shah} to the case of more astrophysically realistic black holes, such as those produced through the gravitational collapse of stellar matter. One might potentially even be able to use such techniques to study various open cosmological questions, such as the formation of supermassive primordial black holes within certain inflationary models\cite{zeldovitch}, in the discrete spacetime setting. Finally, we note that it would be particularly exciting to study certain relativistic critical phenomena, such as the famous Choptuik scalar field collapse solution\cite{choptuik}, by means of such methods, in order to assess whether and to what extent to which the scaling relations and universality properties observed within standard simulations of continuous spacetimes\cite{gundlach2} also extend to discrete ones.

\section*{Acknowledgments}

The author would like to thank the DAMTP GR journal club at the University of Cambridge (and particularly Justin Ripley) for providing the opportunity to present an early version of these results and receive useful feedback, Salvatore Vultaggio for clarifying discussions regarding the physical significance of scalar fields in general relativity and scalar-tensor gravity theories, Jos\'e M. Mart\'in-Garc\'ia for providing enlightening pointers to recent literature on relativistic critical phenomena in gravitational collapse, and Stephen Wolfram for general encouragement and helpful discussions throughout.

\end{document}